\documentclass{aa}

\usepackage{graphicx}
\usepackage{txfonts}
\usepackage{amssymb}

\usepackage{lscape}
\bibpunct{(}{)}{;}{a}{}{,}

\begin{document}

\title{Chemical evolution in the early phases of massive star formation {\sc II}: Deuteration}

\author{T. Gerner\inst{1}, Y. L. Shirley\inst{2}, H. Beuther\inst{1}, D. Semenov\inst{1},
H. Linz\inst{1}, T. Albertsson\inst{1}, \and Th. Henning\inst{1}}
\institute{Max-Planck-Institut f\"ur Astronomie, K\"onigstuhl 17, D-69117 Heidelberg, Germany \and Steward Observatory, University of Arizona, Tucson, AZ 85721, USA\\ \email{gerner@mpia.de}}

\authorrunning{Gerner et al.}
\titlerunning{Early phases of massive star formation {\sc II}: Deuteration}

\abstract {The chemical evolution in high-mass star-forming regions is
  still poorly constrained. Studying the evolution of deuterated
  molecules allows to differentiate between subsequent stages of
  high-mass star formation regions due to the strong temperature
  dependence of deuterium isotopic fractionation. We observed a sample
  of 59 sources including 19 infrared dark clouds, 20 high-mass
  protostellar objects, 11 hot molecular cores and 9 ultra-compact
  H{\sc ii} regions in the (3-2) transitions of the four deuterated
  molecules, DCN, DNC, DCO$^+$ and N$_2$D$^+$ as well as their
  non-deuterated counterpart.\\ The overall detection fraction of DCN,
  DNC and DCO$^+$ is high and exceeds 50\% for most of the
  stages. N$_2$D$^+$ was only detected in a few infrared dark clouds
  and high-mass protostellar objects. It can be related to problems in
  the bandpass at the frequency of the transition and to low
  abundances in the more evolved, warmer stages. We find median D/H
  ratios of $~0.02$ for DCN, $~0.005$ for DNC, $~0.0025$ for DCO$^+$
  and $~0.02$ for N$_2$D$^+$. While the D/H ratios of DNC, DCO$^+$ and
  N$_2$D$^+$ decrease with time, DCN/HCN peaks at the hot molecular
  core stage. We only found weak correlations of the D/H ratios for
  N$_2$D$^+$ with the luminosity of the central source and the FWHM of
  the line, and no correlation with the H$_2$ column density.  In
  combination with a previously observed set of 14 other molecules
  (Paper {\sc I}) we fitted the calculated column densities with an
  elaborate 1D physico-chemical model with time-dependent D-chemistry
  including ortho- and para-H$_2$ states. Good overall fits to the
  observed data have been obtained the model. It is one of the first
  times that observations and modeling have been combined to derive
  chemically based best-fit models for the evolution of high-mass star
  formation including deuteration.}


\maketitle

\section{Introduction}\label{sec:introduction}
The chemical evolution in high-mass star formation is still poorly
understood and a field of intense investigations. The question of
which molecules are to be used to distinguish between different
evolutionary stages is of great interest. Those so called chemical
clocks could be used to derive lifetimes of the different stages and
help to infer the absolute ages of those objects. In addition,
studying deuterium chemistry is also very useful to constrain physical
parameters as for example the ionization fraction, temperature and
density \citep[e.g.,][]{crapsi2005,chen2011}.  In particular,
deuterated molecules are very prominent candidates for probing this
evolutionary sequence, since their chemistry depends highly on the
temperature and the thermal history of an object
\citep{caselli2012,albertsson2013}. The deuteration fraction (the
ratio between the column density of a deuterated molecule and its
non-deuterated counterpart) is therefore an important parameter in
order to study these evolutionary effects.

Theoretical and observational deuteration studies of low-mass
star-forming regions revealed a large increase by several orders of
magnitude in the deuteration fraction of starless cores with respect
to the cosmic atomic D/H ratio of $\sim 10^{-5}$
\citep{linsky2003,oliveira2003} and discussed possible trends with the
evolutionary state of the star-forming region
\citep[e.g.,][]{caselli2002c,crapsi2005,bourke2012,friesen2013}. Correlations
of the deuteration fraction were seen, e.g., with the dust temperature
and the level of CO depletion \citep{emprechtinger2009} or with the
density \citep{daniel2007}. Whether the deuterium chemistry during
high-mass star formation behaves similar to that of low-mass cores is
poorly constrained by observations so far. The current studies mostly
target single deuterated species instead of a larger set of molecules,
or are focused on a limited number of sources. \citet{miettinen2011}
found deuteration fractions in a sample of seven massive clumps
associated with IRDCs that are lower than the values found in low-mass
starless cores. Early studies of very young IRDCs by
\citet{pillai2007} and more evolved HMPOs by \citet{fontani2006}
indicated a trend of higher deuteration fractions for the younger,
cooler sources. A recent attempt to systematically study a larger
sample of different evolutionary stages by \citet{fontani2011}
revealed that the N$_2$D$^+$/N$_2$H$^+$ column density ratio is an
indicator for the evolutionary stage in high-mass star
formation. \citet{chen2011} observed several dense cores covering
different evolutionary stages in three massive star-forming clouds and
studied the deuteration fraction of N$_2$H$^+$ and the role of CO
depletion in this context. They found a clear trend of decreasing
deuteration fraction with increasing gas temperature tracing different
evolutionary stages. They also found an increasing trend of the
deuteration fraction with the CO depletion factor, which is similarly
seen in low-mass protostellar cores.

In order to study the deuteration in high-mass star forming regions in
an evolutionary sense, we divide the high-mass star formation sequence
into different stages \citep[see
  also][]{gerner2014}. \citet{beuther2006b,zinnecker2007,tan2014}
divided the evolutionary sequence into different phases based on their
physical conditions. We describe the evolutionary picture from an
observationally point of view and distinguish between 4
observationally motivated stages based on the underlying physical
sequence. First, our picture starts with an initially starless
infrared dark cloud phase (IRDC). At this point these objects are
close to isothermal and consist of cold and dense gas and dust. In
this approach we do not consider a long living pre-IRDC phase, which
is proposed in theoretical works
\citep[e.g.,][]{narayanan2008,heitsch2008} and also supported by
observations \citep[e.g.,][]{barnes2011}. This phase should be less
dense and in our model we define the year zero of our evolutionary
sequence when the densities start to be higher than $10^4$~cm$^{-3}$.
%According to our modeling, these densities are enough to transform almost all atomic gas within several thousand years into molecular gas and the cloud reaches the molecular gas IRDC phase.
While starless IRDCs only emit in the (sub-)millimeter regime, places
of beginning star-formation start to show up as point sources at
$\mu$m-wavelengths. Eventually, the overdensities within the IRDC
begin to collapse and form one or several accreting protostars with
$>8\,M_{\odot}$ in the next phase, that is a high-mass protostellar
object (HMPO). The internal sources of HMPOs emit at mid-infrared
wavelengths and their radiation starts to heat up the environment,
leading to non-isothermal temperature profiles. The higher
temperatures boost the molecular complexity leading to the hot
molecular core phase (HMC). This phase is from a physical point a
sub-group of the HMPO phase, but clearly distinguishes from a chemical
point of view, driven by the higher temperatures that liberate
molecules from molecular-rich ices and increase the molecular
complexity of the source. Finally, the UV-radiation of the central
star(s) ionizes the ambient gas and create an ultra-compact H{\sc ii}
region (UCH{\sc ii}). That is the last stage considered in our
evolutionary picture. These objects presumably have stopped accreting
and complex molecules seen in the HMC phase are not longer
detectable. It is possible that overlaps occur between these stages,
leading to HMCs associated with UCH{\sc ii} regions and even still
accreting protostars within UCH{\sc ii} regions. High-mass
star-forming sites are complex objects. In order to circumvent the
problem of the coexistence of different stages in one object, we want
to statistically characterize the evolution along the different
stages.

In this work we continue and extend an investigation of the chemical
evolution in 59 high-mass star forming regions in different
evolutionary stages \citep{gerner2014} towards deuterated
molecules. In the previous work we measured the beam averaged column
densities of 14 different molecular species and derived a chemical
evolutionary picture across the evolutionary sequence in high-mass
star formation starting with IRDCs via HMPOs to HMCs and finally
UCH{\sc ii} regions. We found that overall the chemical complexity,
column densities and abundances increase with evolutionary stage. We
fitted the data with a 1D physico-chemical modeling approach and found
good agreement with the observations. Here we want to measure the
deuteration fractions of the four deuterated molecules DCN, DNC,
DCO$^+$ and N$_2$D$^+$ and test their correlations with evolutionary
stage and with physical parameters such as the luminosity of the
objects. Furthermore, we model the derived column densities with a
state-of-the-art 1D deuterium chemical model along the evolutionary
sequence from IRDCs via HMPOs to HMCs and UCH{\sc ii} regions.

The structure of the paper is the following. In
Section~\ref{sec:sourcesample} we introduce the source sample,
followed by a description of the observations in
Section~\ref{sec:observations} and an introduction to deuterium
fractionation in Section~\ref{sec:d_chemistry}. In
Section~\ref{sec:results} we present the results of the analyzed
observational data. In Section~\ref{sec:discussion} we introduce the
model used to fit the data and discuss the modeling results as well as
their implications. We conclude with a summary in
Section~\ref{sec:conclusion}.

\section{Source sample} \label{sec:sourcesample}
\begin{figure}
\includegraphics[width=0.5\textwidth]{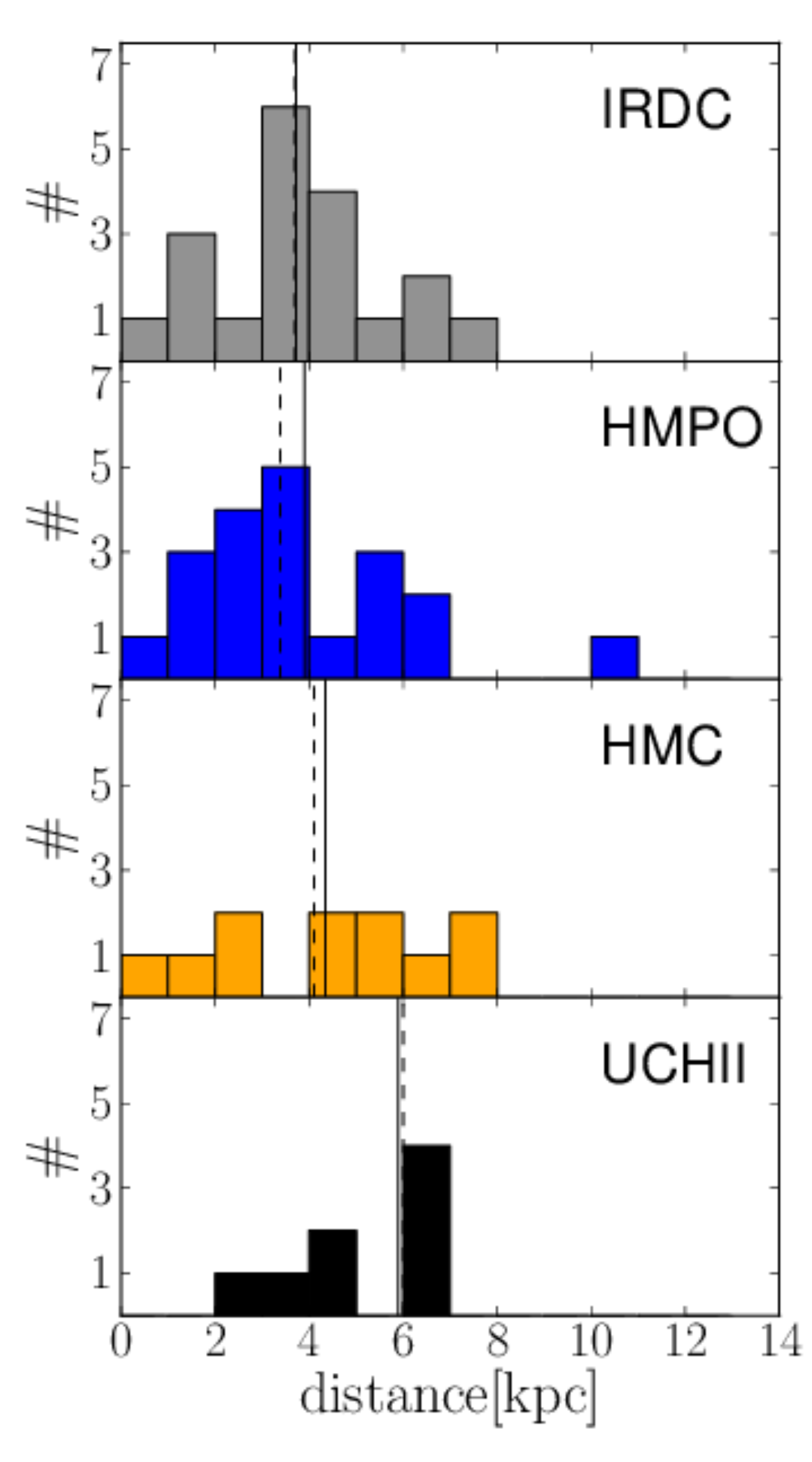}
\caption{From top to bottom: distance distribution for each of the
  subsamples, IRDCs, HMPOs, HMCs and UCH{\sc ii} regions. The bins
  have widths of $1$~kpc and show the number of sources per distance
  range. The vertical dashed line shows the median distance and the
  vertical solid line the mean distance of the subsample.
\label{fig:distances}}
\end{figure}

The sources are taken from \citet{gerner2014} and initially selected
from different source lists. The total sample contains 59 high-mass
star-forming regions, consisting of 19 IRDCs and 20 HMPOs as well as
11 HMCs and 9 UCH{\sc ii}s. The sources were selected from well known
source catalogs of the literature without specific selection criteria
such as spherical symmetry. The lists of the IRDCs were first
presented in \citet{carey2000} and \citet{sridharan2005} and are part
of the {\it Herschel} guaranteed time key project EPOS \citep[The
  Early Phase of Star Formation, ][]{ragan2012}. This sample consists
of 6 IRDCs showing no internal point sources below 70$\mu$m and 13
IRDCs that have internal point sources at 24$\mu$m and 70$\mu$m. The
HMPOs are taken from the well-studied sample by \citet{sridharan2002}
and \citet{beuther2002a,beuther2002b}. HMC sources are selected from
the line-rich sample of \citet{hatchell1998b}, including a few
additional well-known HMCs: W3IRS5, W3H$_2$O and Orion-KL. For the
UCH{\sc ii} regions, we selected line-poor high-mass star-forming
regions from \citet{hatchell1998b}, and additional sources from
\citet{wc1989b}. Recent studies towards single sources of the ones
included in this work overall confirm the evolutionary classification
given in the referenced papers. The full source list is given in Table
\ref{tbl:sourcelist}. The table gives also the distance of each
  source. In the cases that the distance ambiguity of near and far
  kinematic solution could not be resolved we used the near kinematic
  solution. In Figure~\ref{fig:distances} we show the distribution of
  the distances for all four subsamples including their median and
  mean value. The median and mean distance for the IRDCs, HMPOs and
  HMCs are very similar with values around $~4$kpc. Thus, the larger
  range of distances of the total sample should not affect the
  comparison between the median or mean properties of the different
  subsamples, e.g., their detection statistics or their derived median
  column densities (e.g., due to uncertain beam filling
  fractions). Only the UCH{\sc ii} subsample shows significantly
  larger distances. However, since this subsample is not of our main
  interest in this work, this does not dramatically influence the
  results of this study.

\section{Observations}\label{sec:observations}
The 59 sources were observed with the Arizona Radio Observatory
Submillimeter Telescope (SMT) in 2013 between February 12-15, March
10-13 and on March 31 and April 1 with $\sim 100$~h total observing
time. For the observations we used the ALMA type 1.3~mm dual
polarization sideband separating heterodyne receiver and the
filterbanks as backends with a resolution of 250~kHz which corresponds
to $\sim0.3$\,km\,s$^{-1}$ resolution in velocity. The beam size of
the SMT at 1.3~mm is $\sim30\arcsec$. One single integration took
5\,min in position-switching mode, with 2.5 minutes on-source
time. The emission in the deuterated lines were integrated 2-3 times
and the emission in the non-deuterated lines 1-2 times, depending on
the observing conditions. The data was calibrated using data of
Jupiter from the same observation runs assuming a sideband rejection
of 13db. The mean system temperatures of the spectra $T_{\rm sys}$ is
$~380$~K. The data reduction was conducted with the standard
GILDAS\footnote{http://www.  iram.fr/IRAMFR/GILDAS} software package
CLASS. All spectra from each source were baseline subtracted,
calibrated to $T_{\rm {mb}}$ scale with typical beam efficiencies of
$~0.6$ and averaged. In few cases we used the $1$~MHz filterbanks
spectra due to very broad lines (e.g., in Orion-KL) or when the line
was not detected in the 250~kHz filterbanks (with 128 channels) but in
the $1$~MHz filterbanks (with 512 channels).  The line integrals were
measured by summing the line emission channel by channel for detected
lines. The detection criteria is S/N $> 3$.

The line parameters of the observed molecules are given in
Table~\ref{tbl:moleculeslist}.

\begin{table}
\tiny
\caption{List of analyzed molecules with transitions, frequencies,
  energies of the upper level, and Einstein coefficients $A_{\rm
    ul}$.}
\label{tbl:moleculeslist}
\centering
\begin{tabular}{lrccc}
\hline\hline
Molecule & Transition & Frequency & E$_u$/k & $A_{\rm ul}$ \\
 &  & $[{\rm GHz}]$ & $[\rm K]$ & $[10^{-3} \times \rm s^{-1}]$  %rho_{crit}=A_ul/n_coll
\\
\hline
DCO$^+$       &   3-2     &   216.1126    &     20.74      &  0.772      \\
DCN           &   3-2     &   217.2385    &     20.85      &  0.457      \\
DNC           &   3-2     &   228.9105    &     21.97      &  0.557      \\
N$_2$D$^+$    &   3-2     &   231.3218    &     22.20      &  0.712      \\
H$^{13}$CO$^+$&   3-2     &   260.2553    &     24.98      &  1.337     \\
HCN           &   3-2     &   265.8864    &     25.52      &  0.836      \\
HCO$^+$       &   3-2     &   267.5576    &     25.68      &  1.476     \\
HNC           &   3-2     &   271.9811    &     26.11      &  0.934      \\
N$_2$H$^+$    &   3-2     &   279.5117    &     26.82      &  1.259     \\
\hline
\end{tabular}
\tablefoot{Values are taken from LAMDA database \citep{schoeier2005}.}
\end{table}
%mention optically thick lines ?

\subsection{RMS} \label{sec:rms}
\begin{figure}
\includegraphics[width=0.5\textwidth]{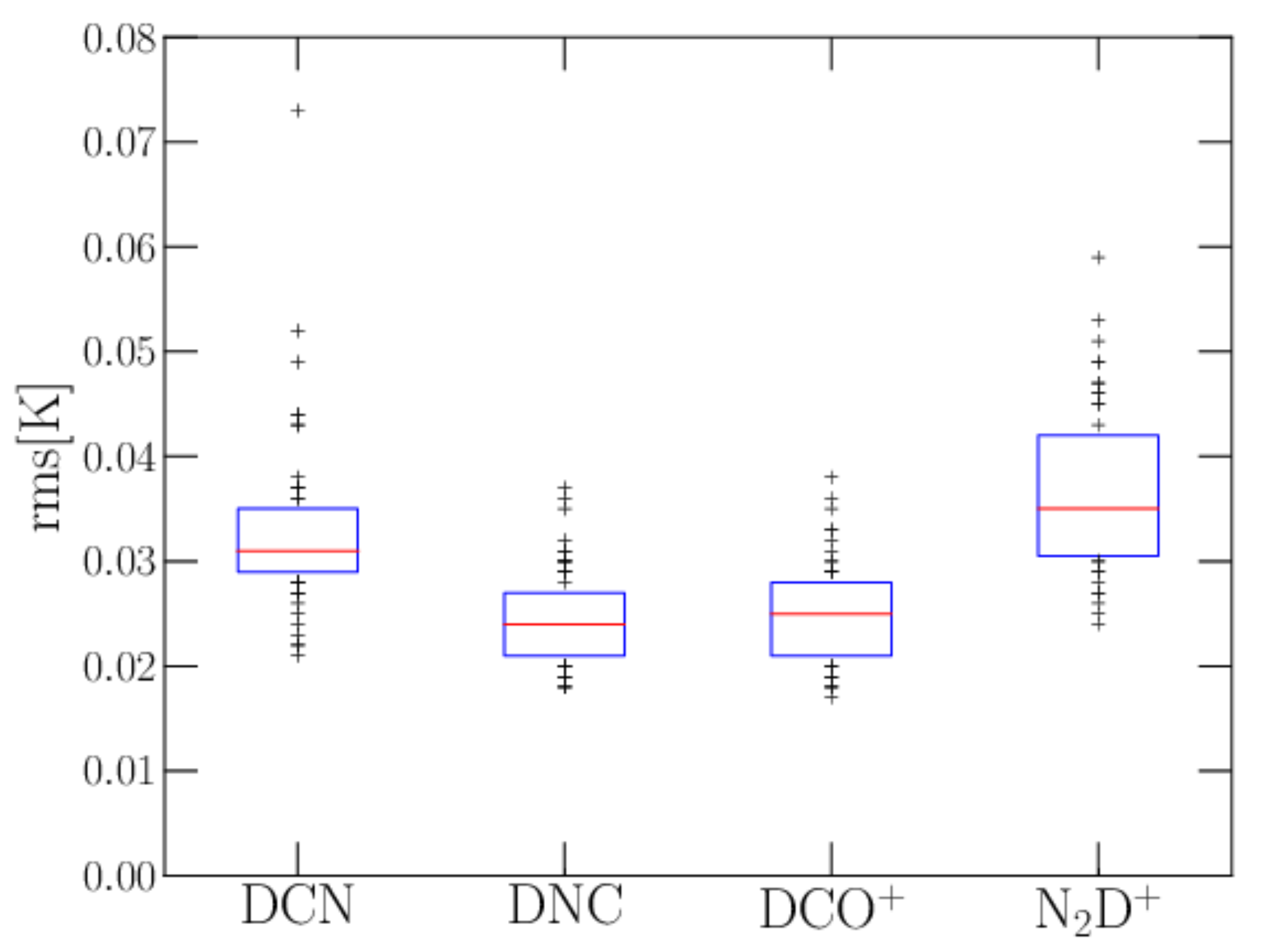}
\caption{$1\sigma$ rms values of spectra of the deuterated
  molecules. The red solid line shows the median, the blue box the
  25\%-75\% range and the crosses mark the outliers.
\label{fig:rmsplot}}
\end{figure}

The median $1\sigma$ rms values for the deuterated molecules are
$0.031$~K for DCN (3-2), $0.024$~K for DNC (3-2), $0.025$~K for
DCO$^+$ (3-2) and $0.035$~K for N$_2$D$^+$ (3-2). DCN (3-2) was
observed in the later shifts and the higher rms value is due to worse
observing conditions during that period. The higher rms value for
N$_2$D$^+$(3-2) is due to problems at this specific transition
frequency and described in the following Section. An overview of all
rms values is given in Figure~\ref{fig:rmsplot}. The strong outlier in
DCN (3-2) is from Orion-KL which is still a detection.

\subsection{Problems with N$_2$D$^+$ spectra} \label{sec:problems}
Reducing and analyzing the spectra of the N$_2$D$^+$ (3-2) transition
at 231.32~GHz was problematic due to blending by an overlaying
pressure broadened ozone line in the atmosphere. This ozone line
depends on the elevation of the source and time of the observation and
affects the different sources with varying strength. Due to this
contamination, the mean $1\sigma$ rms value of the N$_2$D$^+$ line is
40\% higher than for the other deuterated spectral lines and thus the
threshold for a detection is higher. 
%Another factor leading to higher
%rms-value might be the higher frequency of the N$_2$D$^+$ (3-2),
%although this effect is likely of minor importance. 
In the 5 sources IRDC18454.1, IRDC18454.3, Orion-KL, HMC010.47 and
HMC031.41 the ozone line makes it impossible to discriminate between a
detection or a non-detection.

%\begin{figure}
%\epsscale{.80}
%\plotone{figures/n2dp_example.eps}
%\caption{Spectrum of N$_2$D$^+$ blended with ozone line of .
%\label{fig:deutfraction}}
%\end{figure}

\section{Deuterium fractionation}\label{sec:d_chemistry}
Under cold conditions, the deuteration fraction of molecules can be
enhanced through reactions with, e.g., deuterated H$_3^+$,
H$_2$D$^+$. While the chemical reaction that deuterates H$_3^+$
proceeds without a thermal barrier, the backward reaction is
endothermic and has a thermal energy barrier
\citep{watson1974,caselli2012,albertsson2013}. This energy barrier
leads to an enhanced formation fraction at low temperatures, but
becomes negligible for higher temperatures at which the backward
reaction becomes as efficient as the forward reaction. The chemical
reactions and their energy barriers that introduce deuterium into some
of the key molecules are the following:
\begin{eqnarray}\label{equ:reaction1}
{\rm H}_3^+ + {\rm HD} & \rightleftarrows & {\rm H}_2{\rm D}^+ \ \, \ + {\rm H}_2 + 230{\rm K} \\\label{equ:reaction2}
{\rm CH}_3^+ + {\rm HD}   & \rightleftarrows &  {\rm CH}_2{\rm D}^+ + {\rm H}_2 + 370{\rm K} \\\label{equ:reaction3}
{\rm C}_2{\rm H}_2^+ + {\rm HD} & \rightleftarrows & {\rm C}_2{\rm HD}^+ + {\rm H}_2 + 550{\rm K}
\end{eqnarray}

The temperature range for an effective D-enhancement for the pathway
via H$_3^+$ isotopologues is $\sim 10-30$~K, whereas it is $\sim
10-80$~K for pathways via light hydrocarbons
\citep{millar1989,albertsson2013}. This enhances the deuteration
fraction of the key molecules H$_3^+$ and CH$_3^+$ in different
temperature regimes. Another effect that increases the
H$_2$D$^+$/H$_3^+$ ratio is the depletion of neutral gas-phase species
(e.g., CO, N$_2$) \citep[see][]{dalgarno1984,roberts2000}. This
enhancement is then imprinted on the molecules formed through these
reaction partners.

In the literature different models exist describing the formation
routes of deuterated molecules and their relative importance. The
dominant formation pathways of DCO$^+$ and N$_2$D$^+$ are via a
low-temperature route through H$_3^+$ isotopologues, whereas DCN can
be formed via a fractionation route involving light hydrocarbons. The
back-reaction of the route via H$_3^+$ isotopologues sets in at
temperatures around $30$~K and their deuteration fractions decrease
with rising temperature. According to \citet{roueff2007}, both DCN and
DNC can be formed efficiently at low temperatures via deuteration of
HNC or HCN. However, at higher temperatures DNC gets destroyed by
reactions with atomic oxygen, unlike DCN. The chemical models of
\citet{turner2001} indicate that only DCN is also formed via reactions
with light hydrocarbons involved, but not DNC. If the
\citet{turner2001} scheme is correct, then molecules such as C$_2$D,
HDCO or C$_3$HD should show a similar behavior like DCN with
temperature, since they are also formed via CH$_2$D$^+$ and
C$_2$HD$^+$.

\citet{parise2009} found a low DCO$^+$/HCO$^+$ column density ratio
but significant deuteration fractions for HCN and H$_2$CO under
temperature conditions of $\sim 70$~K towards the Orion Bar PDR. Model
calculations by \citet{roueff2007} predicted that in general the
DNC/HNC, DCO$^+$/HCO$^+$ and N$_2$D$^+$/N$_2$H$^+$ column density
ratios decrease with temperature, but are almost constant with
density. The DCN/HCN column density ratio shows a more complex
behavior with temperature, reaching the largest ratio for $\sim 30$~K,
and shows a stronger increase with density. They found that the reason
for this behavior is twofold. First, due to the enhanced abundance of
radicals (e.g., CHD and CD$_2$) that form DCN. Second, the main
destruction pathways of DCN are reactions with the ions HCO$^+$ and
H$_3$O$^+$, that leads to DCNH$^+$ which subsequently returns to DCN
via dissociative recombination. The deuteration fractions also
strongly depend on the assumed elemental abundances.

\section{Results}\label{sec:results}

\subsection{Detection fractions for DCN, DNC, DCO$^+$ and N$_2$D$^+$}
\begin{figure*}
\includegraphics[width=1.0\textwidth]{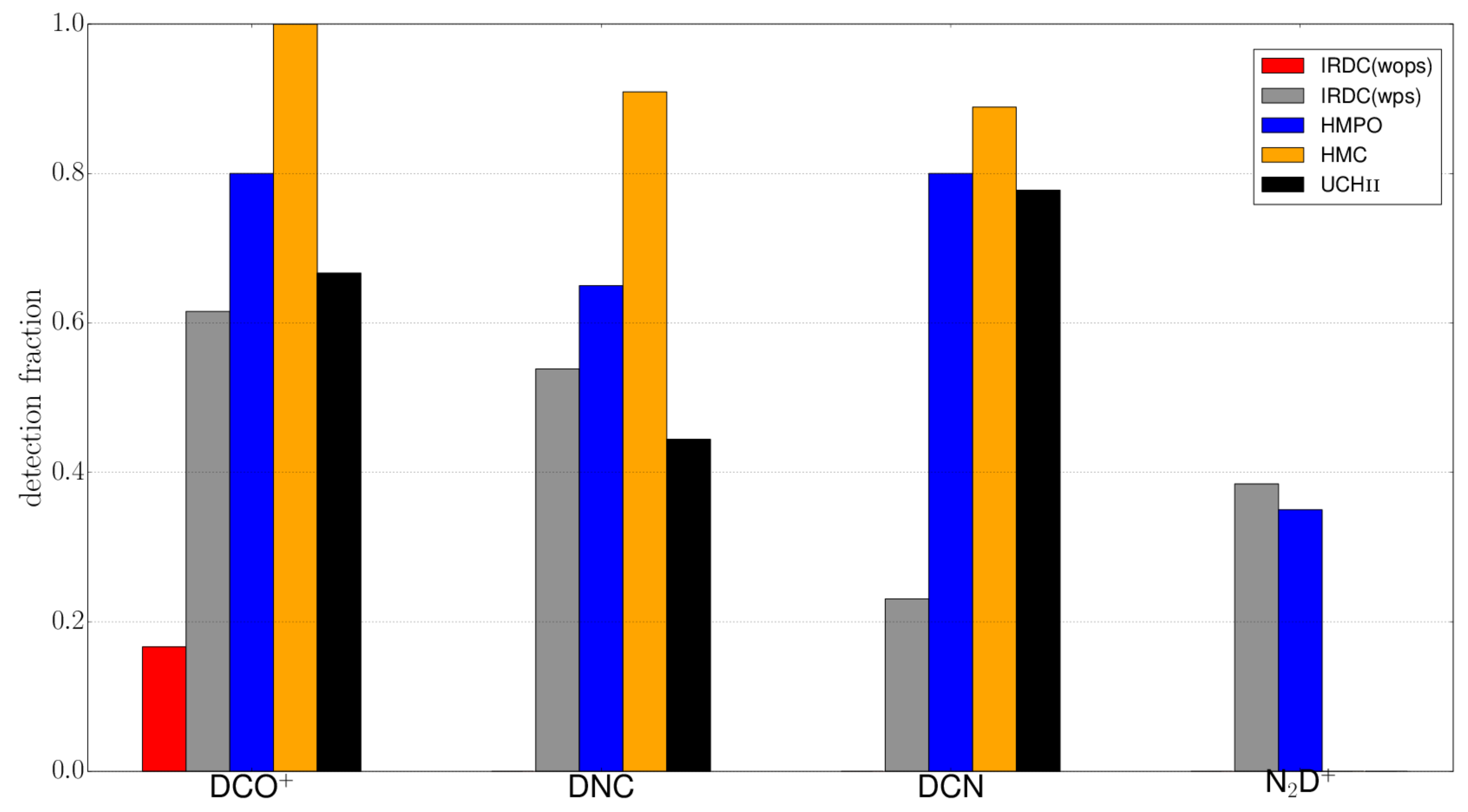}
\caption{Detection fraction of the 4 observed deuterated species in
  the different evolutionary stages. Here the infrared dark cloud
    (IRDC) stage is divided into the two subsamples without and with
    an embedded point source at 24 or 70\,$\mu$m, indicated in the
    caption as IRDC(wops) and IRDC(wps), respectively. The detection
    fractions are shown from left to right with infrared dark clouds
    without embedded point sources in red (IRDC(wops)), infrared dark
    clouds with embedded point sources (IRDC(wps)) in grey, high-mass
    protostellar objects (HMPO) in blue, hot molecular cores (HMC) in
    yellow and ultracompact H{\sc ii} regions (UCH{\sc ii}) in black.
\label{fig:deutfraction}}
\end{figure*}

The detection fractions of DCN, DNC, DCO$^+$ and N$_2$D$^+$ in the
single stages are shown in Figure~\ref{fig:deutfraction}. In this
figure, the IRDC stage is split into the two categories of sources
mentioned in Sect.~\ref{sec:sourcesample}, without and with an
embedded point source. From that figure we identify three trends:

1) N$_2$D$^+$(3-2) is only detected towards IRDCs and HMPOs. Since the
N$_2$H$^+$ abundance is almost constant over the full evolutionary
sequence (Paper {\sc I}), most likely a strong temperature dependence
of the N$_2$D$^+$ production leading to its disappearance at $T
\gtrsim 30$~K is the answer to its apparent absence in the HMC and
UCH{\sc ii} stages. Another complication is problems with the
particular bandpass of its (3-2) transition mentioned in
Sect.~\ref{sec:problems} which lead to a higher mean $1\sigma$ rms
value. That does not mean that N$_2$D$^+$ is not present, since some
of our sources with non-detections in N$_2$D$^+$(3-2) are detected by
\citet{fontani2011} in N$_2$D$^+$(2-1). This could indicate the
importance of excitation conditions for the detection of the (3--2)
transitions. Nonetheless, the detection rates of N$_2$D$^+$(3-2) are
clearly lower than the detection rates of the other observed molecular
transitions of this work. 2) The only deuterated molecule detected in
the presumably coldest IRDCs without an embedded point source at
24$\mu$m (Spitzer/MIPSGAL) or 70$\mu$m (Herschel/PACS) is DCO$^+$
which has also the highest detection fraction in the more evolved
IRDCs.  3) The detection fractions of DCN, DNC and DCO$^+$ in warm
IRDCs up to UCH{\sc ii} regions are comparable and peak at the HMC
stage.  In general the detection fractions towards the observed
high-mass star-forming regions are high, $\gtrsim 60$\%.  It is
important to mention here that the strength of a transition is the
product of column density and temperature and thus detections at the
lower temperatures present in the earlier stages are more difficult. A
lower detection fraction does not necessarily mean that the column
densities in the earlier stages are lower compared to the later
stages. The differences in column densities along the evolutionary
sequence is discussed in Sect~\ref{sec:ratios_observed}.

\subsection{Excitation temperatures and final column densities}\label{sec:excitation_temp}
In this section we discuss the excitation temperatures used to derive
the column densities and how we combined the data from Paper {\sc I}
and this work to obtain the final column densities. The derivation of
the column densities is shown in the appendix in
Section~\ref{sec:column_density}.

\begin{figure*}
\includegraphics[width=0.5\textwidth]{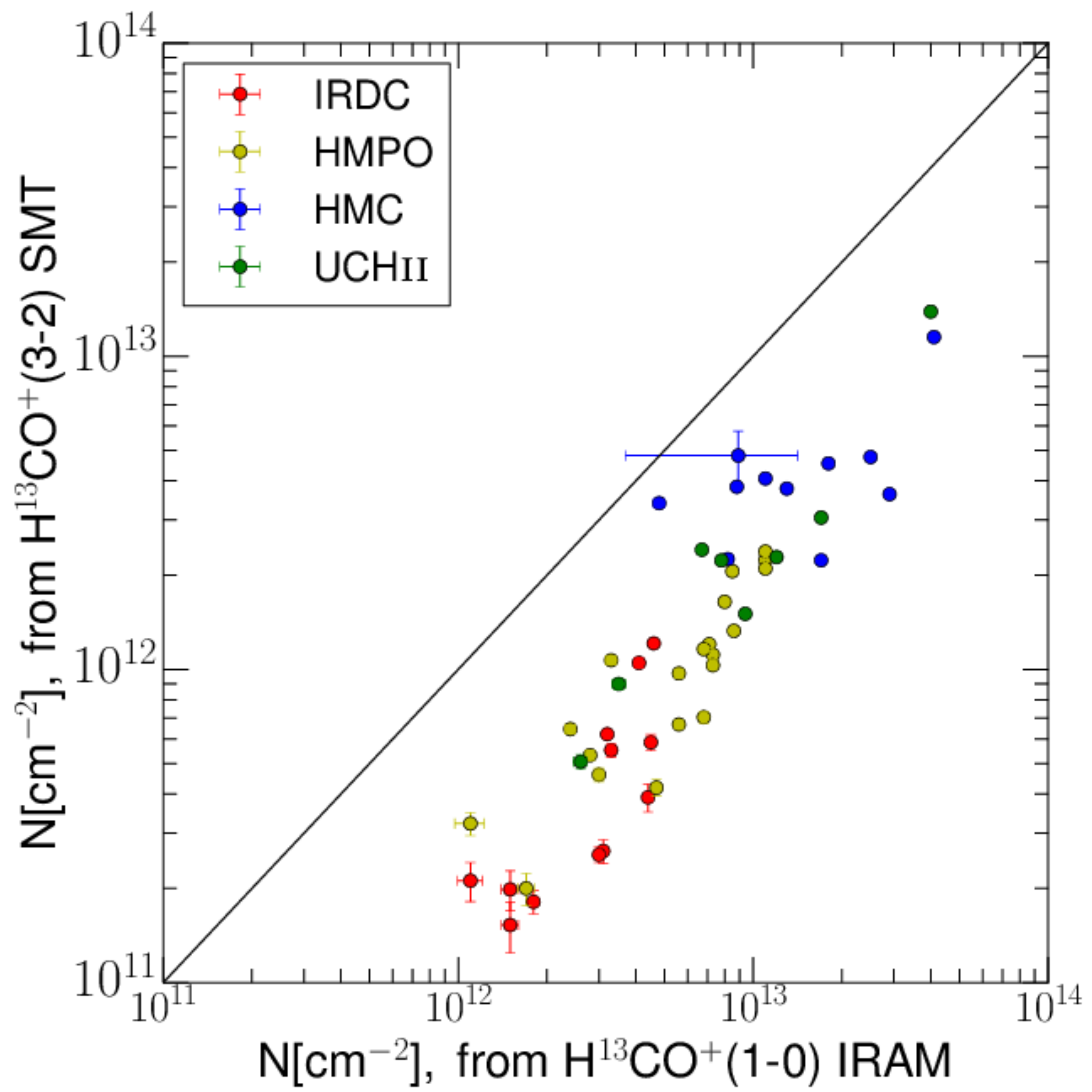}
\includegraphics[width=0.5\textwidth]{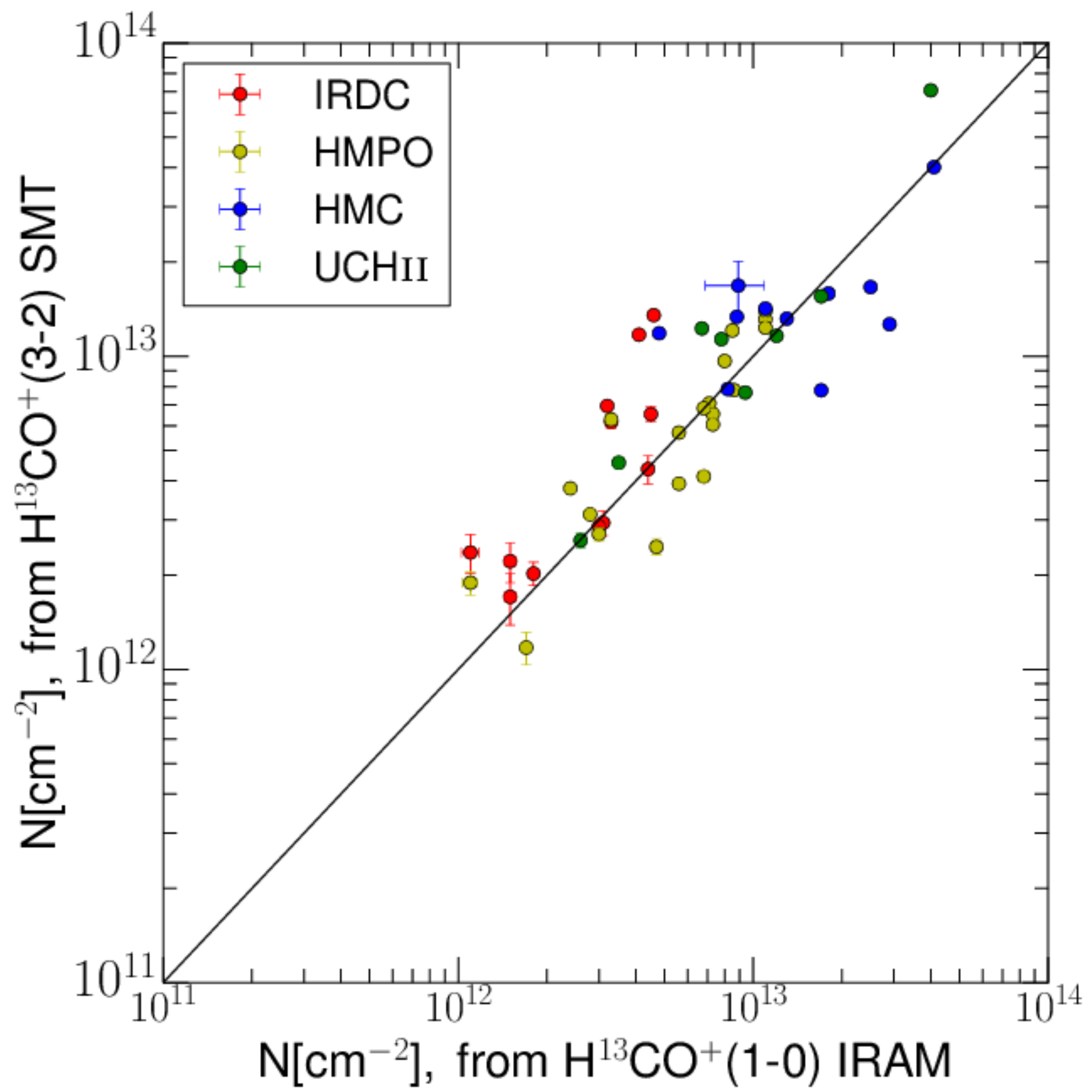}
\caption{Left: Comparison of column densities of H$^{13}$CO$^+$(1-0)
  and H$^{13}$CO$^+$(3-2) assuming the same excitation temperatures
  for both transitions. Right: Comparison of column densities of
  H$^{13}$CO$^+$(1-0) and H$^{13}$CO$^+$(3-2) using the excitation
  temperatures discussed in section \ref{sec:excitation_temp}. The
  errors show the uncertainties in the measured integrated flux. In
  some cases that uncertainty is too small to be seen on this plot.}
\label{fig:h13cop_vergleich} 
\end{figure*}

The column densities for the non-deuterated species were partly taken
from \citet{gerner2014} and partly calculated from the new
observations. A detailed overview on the exact combination of the
previous work and this one is given below. In the previous study we
observed the (1-0)-transitions of several molecules with the IRAM 30m
telescope. The size of the beam for those observed (1-0)-transitions
is very close to the size of the beam of the (3-2)-transitions
observed for this work with the SMT. In the previous work we used
likely optically thin (1-0)-transitions of H$^{13}$CO$^+$ and
HN$^{13}$C and, due to their hyperfine structure, optical depth
corrected N$_2$H$^+$ and HCN to derive their column densities. The
spectra of the first study were partly affected by source emission at
the present off-positions and strong optical depth effects. In these
cases the optical depth and the integrated intensity could not be
reliably determined. Thus we refrained from using those lines in the
analysis. In this work we have additional information from the (3-2)
transitions to complement the missing column densities. In order to
obtain consistent results from both observation runs we compared the
column densities of H$^{13}$CO$^+$(1-0) with the column densities of
H$^{13}$CO$^+$(3-2) derived with $T_{\rm ex}=20.9$~K (IRDC), $T_{\rm
  ex}=29.5$~K (HMPO), $T_{\rm ex}=40.2$~K (HMC) and $T_{\rm
  ex}=36.0$~K (UCH {\sc ii}). Those are the mean temperatures of the
best-fit models from \citet{gerner2014}. The comparison is shown in
Figure~\ref{fig:h13cop_vergleich}.

The derived column densities for the IRAM data are clearly all
higher. Excluding problems in the quality of the data and the
calibration leads to the assumption that, while the (1-0) transitions
are in LTE, the upper levels are sub-thermally populated and thus the
(3-2) transitions are not in LTE. This implies that the average
densities of the gas should be below $\sim 10^6-10^7$cm$^{-3}$. Since
the two different lines trace gas with different excitation
conditions, the spatial extend probed by both transitions might be
different. However, we do not know the beam filling factors for the
different molecules and their different transitions. In addition, we
calculated the column densities from the H$^{13}$CO$^+$(1-0) and (3-2)
transitions with the non-LTE radiative transfer code
RADEX\footnote{\url{https://www.sron.rug.nl/~vdtak/radex/index.shtml}}
\citep{vandertak2007}. The derived column densities for H$^{13}$CO$^+$
were within a factor of $~2$ in agreement with the column densities
derived under the assumption of LTE from the H$^{13}$CO$^+$(1-0)
transition. This is well within the uncertainties and validates the
assumptions made for the (1-0) transitions. In order to compensate for
the differences in the excitation conditions between (1-0) and (3-2)
transitions, we calculated the excitation temperature of the (3-2)
transition for each source which would be necessary to derive the
total column densities derived from the (1-0) transition. From those
excitation temperatures we computed the median value for each
stage. That resulted in $T_{\rm ex}=5.2$ (IRDC), $T_{\rm ex}=6.2$
(HMPO), $T_{\rm ex}=7.2$ (HMC) and $T_{\rm ex}=6.4$ (UCH {\sc ii}) for
the (3-2) transitions. The comparison of the calibrated
H$^{13}$CO$^+$(3-2) column densities derived with the lower excitation
temperatures and the (1-0) data is shown in
Figure~\ref{fig:h13cop_vergleich}. This correction should also reduce
the error in the beam filling fraction and make the different
transitions comparable.

For the four different molecules we derived the column densities in
the following way.

\subsubsection{HCO$^+$ and HNC}
For HCO$^+$ and HNC we used H$^{13}$CO$^+$(1-0) and HN$^{13}$C
transitions, respectively, assuming the relative isotopic ratio
depending on the galactocentric distance described in
\citet{wilson1994}. In \citet{gerner2014} we used the representative
value for the Sun and the local ISM of $^{12}$C/$^{13}$C=89
\citep{lodders2003}. This lowers the derived column densities for this
work.

\subsubsection{N$_2$H$^+$ and HCN}
For N$_2$H$^+$ we took the column densities from the (1-0) transitions
from \citet{gerner2014}, for which we used, in case of optically thick
lines, the optical depth and the excitation temperature from the
hyperfine fits made with
CLASS\footnote{\url{http://www.iram.fr/IRAMFR/GILDAS/doc/pdf/class.pdf}}. The
hyperfine structure fitting routine (METHOD HFS) assumes that all
components of the hyperfine structure have the same excitation
temperature and the same width and that the components are separated
in frequency by the laboratory values. From the comparison of the
ratios of the line intensities of the hyperfine components to the
theoretically expected ratios, the fit routine estimates the optical
depth of the line. In the optically thick case that allows us to
determine the excitation temperature. In the optically thin case we
assumed the excitation temperatures given in
Section~\ref{sec:excitation_temp}.

For HCN we used the optical depth corrected column densities from the
(1-0) transitions when available. In all other cases we computed the
mean difference between column densities derived from the (1-0)
transitions and the (3-2) transitions and applied this factor of
$\sim1.7$ to the derived column densities from the (3-2) transitions.

All column densities are beam averaged quantities. The resulting
median abundances including all detections and upper limits for each
subsequent evolutionary phase are given in
Tables~\ref{tbl:obsmedianabun}. In the derivation of the column
densities we assumed the excitation temperatures of the deuterated and
non-deuterated molecule of the same transition to be equal. This
assumption is valid for high densities and high excitation
temperatures, but might lead to an underestimation of the deuteration
column density ratios $D_{\rm {frac}}$ by a factor of $\sim2-4$
(Shirley 2015, in prep.).
%\citep[see][]{shirley2013}.
%Referenz shirley!

%Merged Tables
\begin{table*}
%\tiny
\caption{Observed median column densities ($N$) and the standard
  deviation ($SD$) for IRDCs, HMPOs, HMCs and UCH{\sc ii} regions as
  $a(x)=a \times 10^x$. The median includes detections and upper
  limits of non-detections. Species detected in less than $50\%$ of
  the sources are indicated by an upper limit.}
\label{tbl:obsmedianabun}
\centering
\begin{tabular}{lcccccccc}
\hline \hline

         & \multicolumn{2}{c}{IRDC} & \multicolumn{2}{c}{HMPO} & \multicolumn{2}{c}{HMC} & \multicolumn{2}{c}{UCH{\sc ii}} \\ 
         & $N$ & $SD$ & $N$ & $SD$ & $N$ & $SD$ & $N$ & $SD$ \\
         & cm$^{-2}$ & cm$^{-2}$ &  cm$^{-2}$ &  cm$^{-2}$ &  cm$^{-2}$ &  cm$^{-2}$ &  cm$^{-2}$ &  cm$^{-2}$ \\
\hline
HCO$^+$       & 7.9(13) & 6.8(13) & 3.7(14) &  1.7(14) & 7.6(14) &  4.3(14)   & 3.8(14) &  5.9(14) \\
HCN           & 7.2(13) & 5.7(13) & 1.3(14) &  1.7(14) & 3.5(14) &  3.4(15)   & 3.4(14) &  3.7(14) \\
HNC           & 9.7(13) & 8.4(13) & 2.9(14) &  1.6(14) & 5.5(14) &  3.5(14)   & 2.0(14) &  7.7(14) \\
N$_2$H$^+$    & 2.2(13) & 1.5(13) & 4.6(13) &  3.2(13) & 5.5(13) &  4.6(13)   & 3.7(13) &  6.4(13) \\
DCO$^+$       & $\leq$3.1(11) & 1.1(12) & 6.7(11) &  6.5(11) & 2.0(12) &  7.6(12)   & 4.6(11) &  3.4(11) \\
DCN           & $\leq$6.1(11) & 7.3(11) & 2.0(12) &  2.2(12) & 8.0(12) &  3.6(13)   & 1.9(12) &  6.0(12) \\
DNC           & $\leq$6.1(11) & 9.2(11) & 1.0(12) &  1.2(12) & 8.7(11) &  7.1(11)   & $\leq$3.0(11) &  1.9(12) \\
N$_2$D$^+$    & $\leq$6.8(11) & 4.3(11) & $\leq$3.8(11) &  4.0(11) & $\leq$2.2(11) & 6.6(10) & $\leq$3.3(11) &  4.6(10) \\
\hline
\end{tabular}
%\tablefoottext{}{{}}
\end{table*}

\section{Discussion}\label{sec:discussion}

\subsection{Deuteration fractions}\label{sec:ratios_observed}
\begin{figure*}
\includegraphics[width=0.5\textwidth]{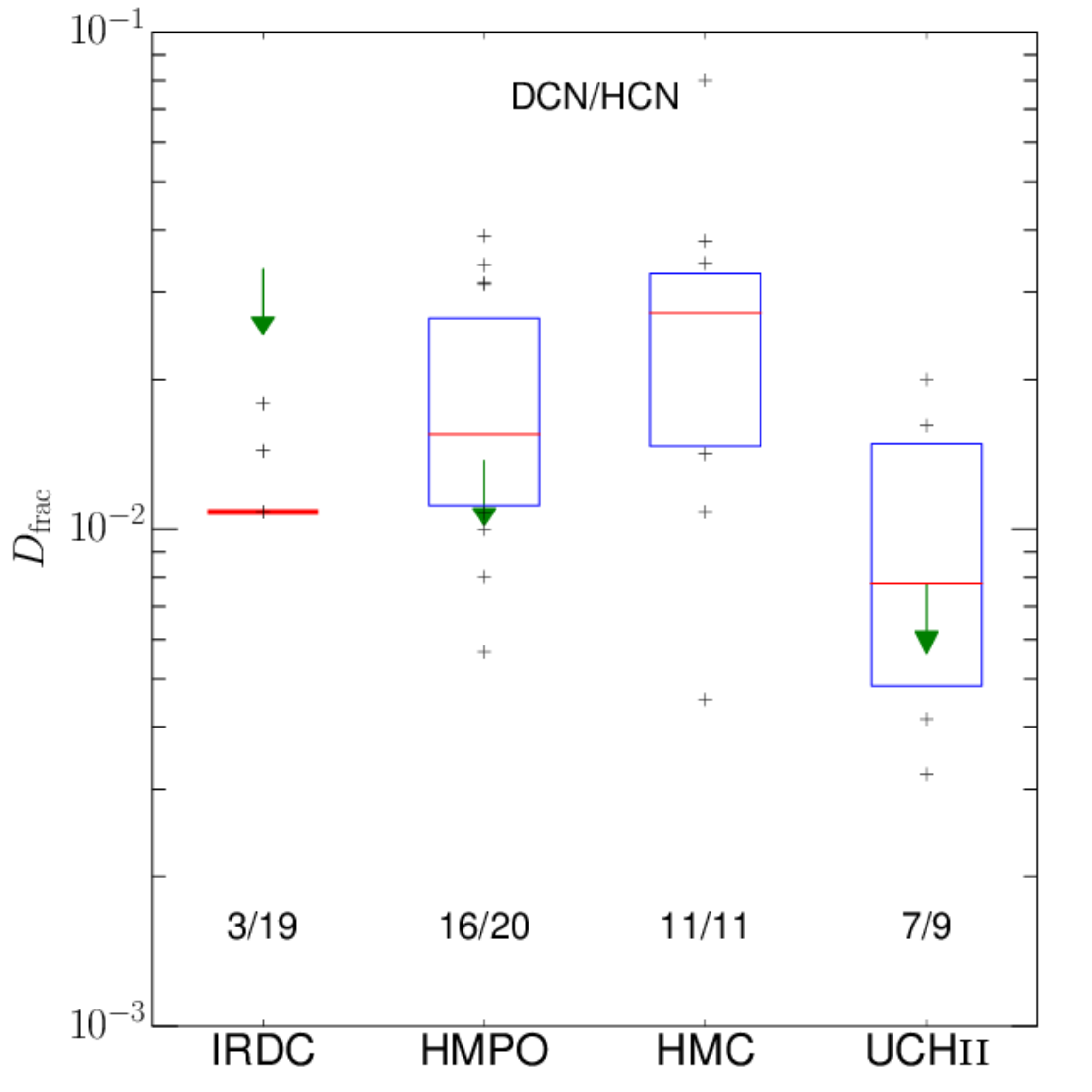}
\includegraphics[width=0.5\textwidth]{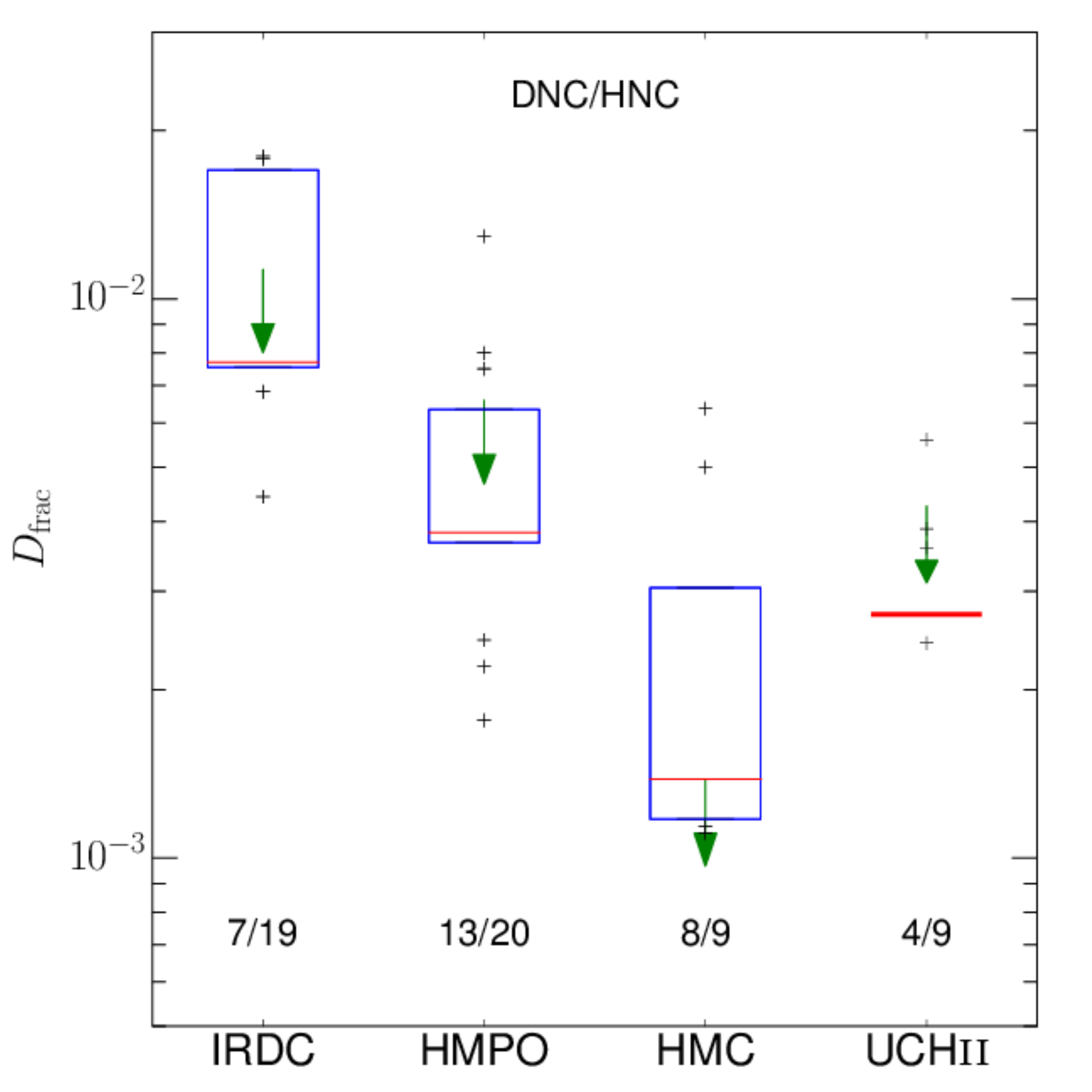}\\
\includegraphics[width=0.5\textwidth]{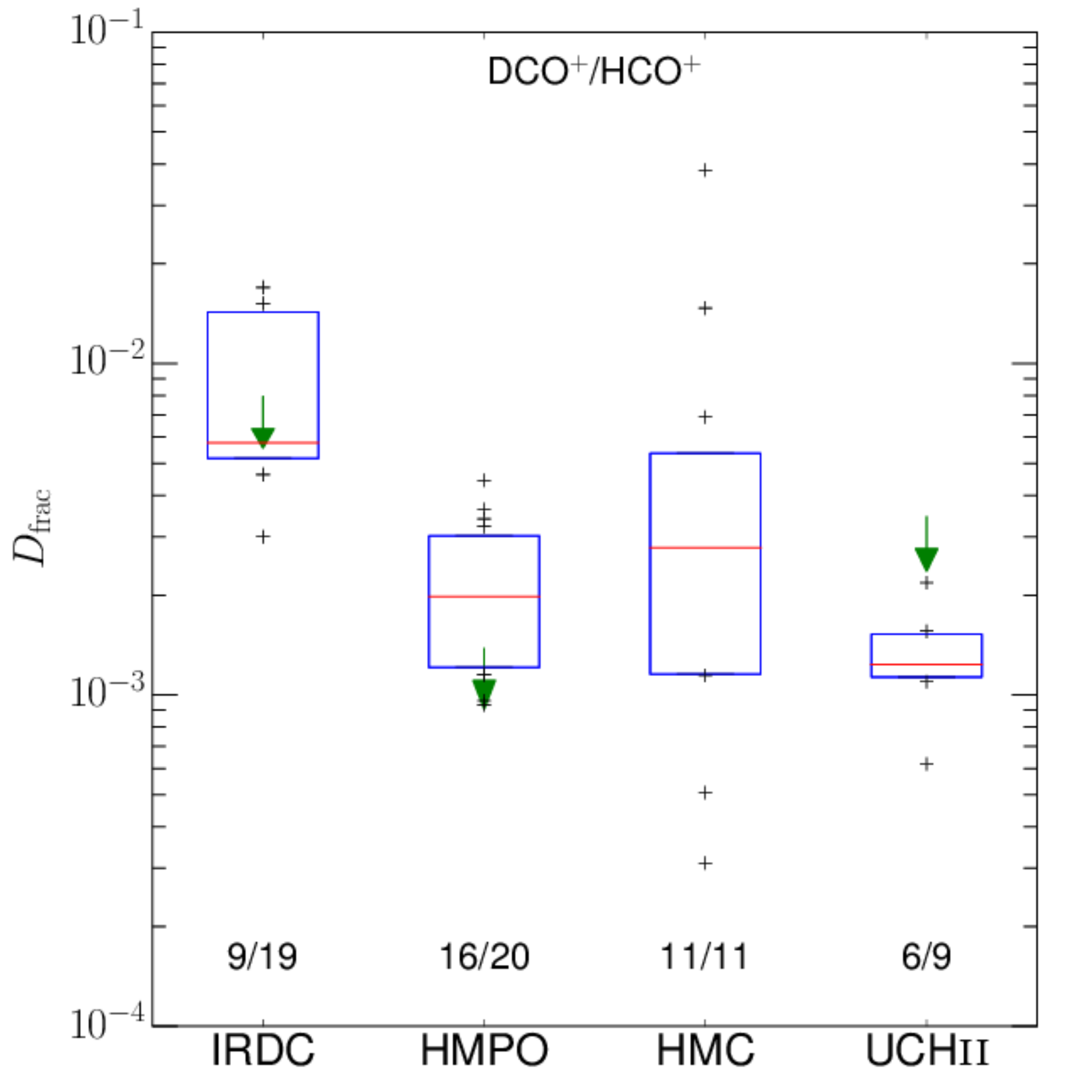}
\includegraphics[width=0.5\textwidth]{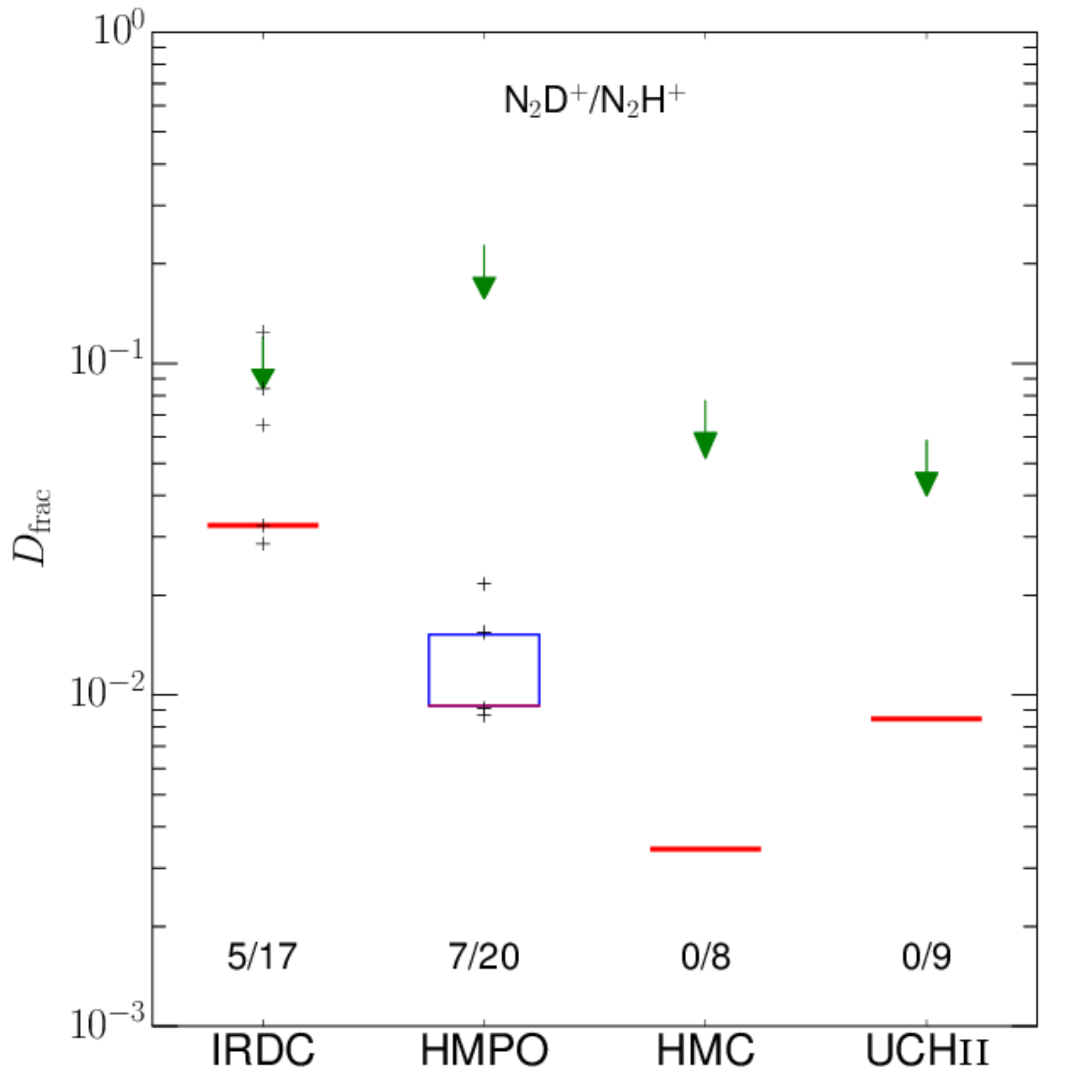}
\caption{Spread in deuteration fractions $D_{\rm {frac}}$ among
    the 4 evolutionary stages for HCN, HNC, HCO$^+$ and N$_2$H$^+$. In
    cases of 6 or more detections within one stage, the blue box
    shows the 25\%-75\% range and the crosses mark the remaining 50\%
    of the data points. In cases of less than 6 data points, only the
    individual data points are shown. The red solid line shows the
    median of all detections and upper limits. The downward arrow shows the
    maximum value of all upper limits within that stage. The fractions
    at the bottom of each stage give the detection rate.}
\label{fig:coldensspread}
\end{figure*}

The spread of the deuteration fractions $D_{\rm {frac}}$ as well
  as the median value for each stage is shown in
  Figure~\ref{fig:coldensspread}. These values include detections and
  upper limits. In the case of N$_2$D$^+$ the median $3\sigma$-limits
  of non-detections for the HMC-stage and UCH{\sc ii}-stage are given,
  since the detection fraction is zero. These values can be considered
  as a sensitivity limit.

\begin{figure*}
\includegraphics[width=0.5\textwidth]{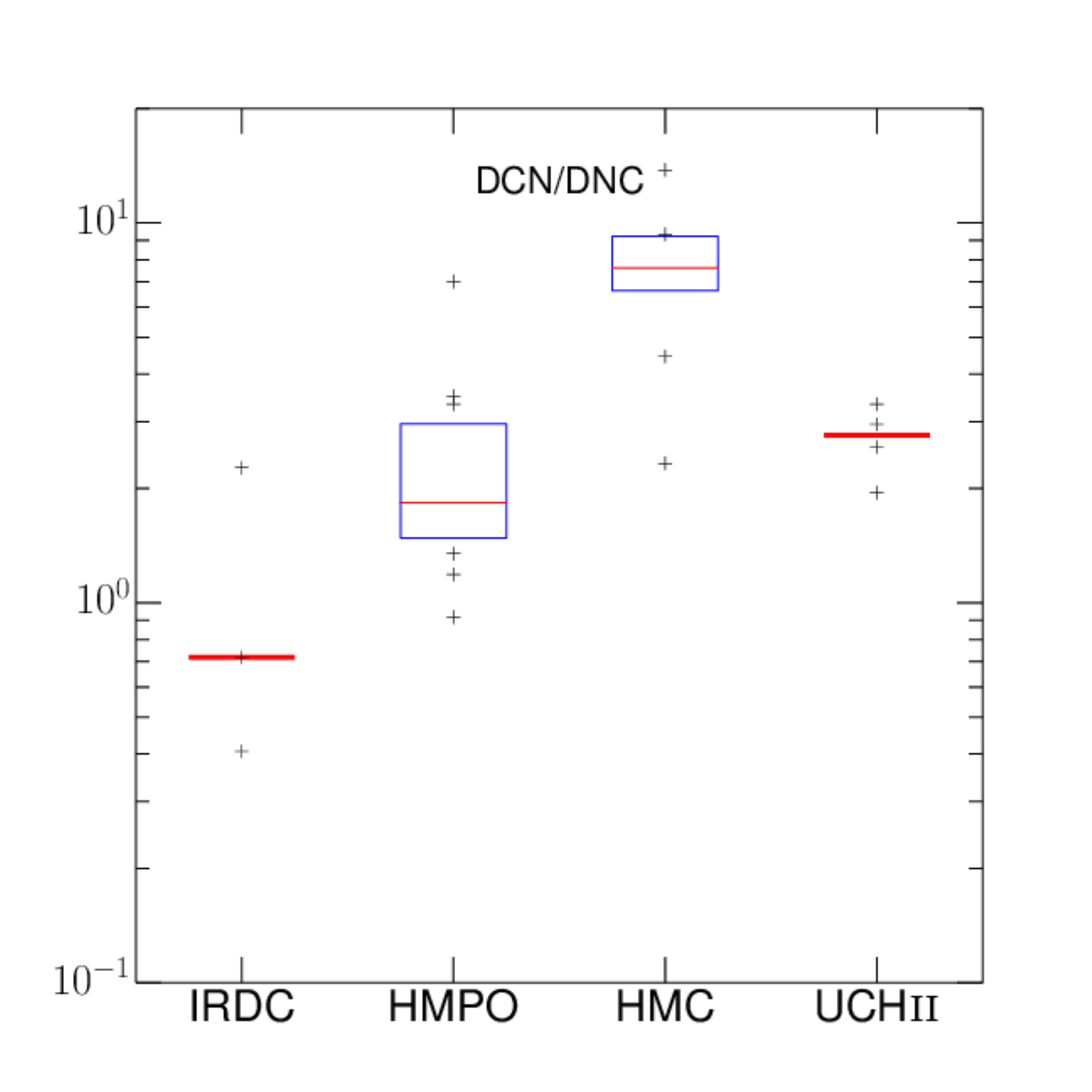}
\includegraphics[width=0.5\textwidth]{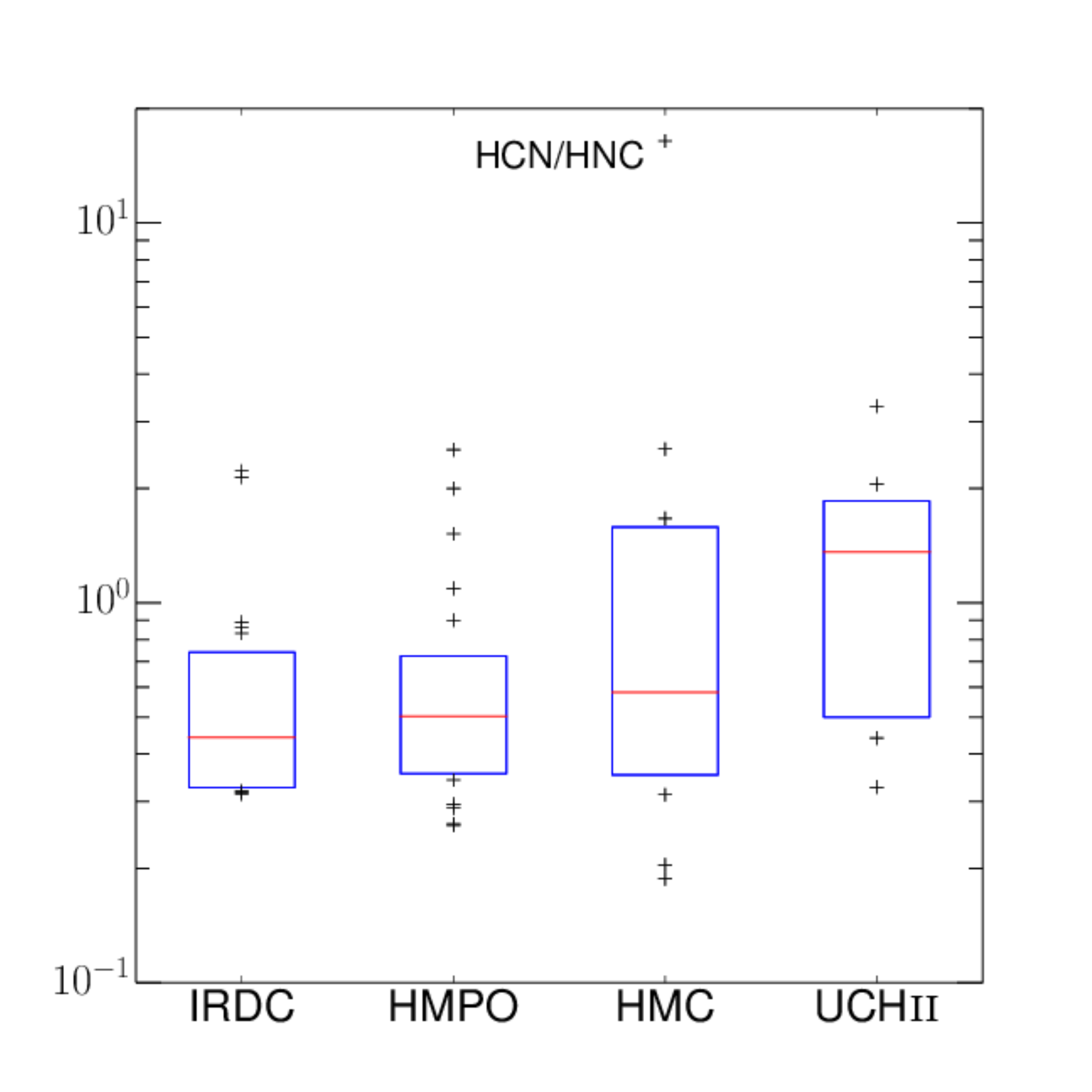}
\caption{Spread in fractions among the 4 evolutionary stages for
    DCN/DNC and HCN/HNC for detections in both molecules only. In case
    of 6 or more detections within one stage, the blue box shows the
    25\%-75\% range and the crosses mark the remaining 50\% of the
    data points. In cases of less than 6 data points, only the
    individual data points are shown. The red solid line shows the
    median of the ratio of detections in both molecules.
\label{fig:morefraction}}
\end{figure*}

The ratios of the column densities of DNC, DCO$^+$ and N$_2$D$^+$ with
their non-deuterated counterparts HNC, HCO$^+$ and N$_2$H$^+$ all show
decreasing trends with evolutionary stage, despite the large spread of
ratios within individual stages. Only HCN shows an increase in the
ratio with the maximum reached in the HMC-stage.  In some cases the
consecutive stages appear not very different from each other within
the error budget (e.g., HMPO-HMC of DCO$^+$/HCO$^+$, HMC-UCH{\sc ii}
of DNC/HNC). The likely reason for this is that the evolution along
the sequence is continuous, and there might be overlaps between
consecutive stages like, for example, some of the HMCs are associated
with UCH{\sc ii} regions. One can distinguish between two different
scenarios that can lead to an overlap between consecutive stages. In
the cases like HMC029.96 (G29.96), high spatial resolution
observations show a separation of the actual UCH{\sc ii} region and
the neighboring HMC of $2\arcsec$
\citep{cesaroni1994,beuther2007d}. The HMC and the UCH{\sc ii} region
are two (or more) distinct entities. The second scenario are cases
like HMC009.62 (G9.62), where also a separate UCH{\sc ii} region
exists, but in addition a small hypercompact H{\sc ii} region is
present at the location of the HMC itself \citep[][]{testi2000}. This
is a general caveat in single-dish observations of HMCs that might be
contaminated by UCH{\sc ii} regions in the beam. Here, the sources are
classified based on their chemistry and that is likely dominated by
the HMC. However, that does not totally cancel this effect and
differences in the measured ratios between different stages are in
some cases weaker. Nevertheless, the global trends are clearly
visible.

Furthermore, Figure~\ref{fig:morefraction} shows the observed column
density ratios of DCN/DNC and HCN/HNC for sources detected in both
molecules, respectively. While the ratio of the non-deuterated
molecules is almost constant, the ratio of the deuterated molecules
shows clearly a peak at the HMC stage. %That qualitative behavior is
reflected by the results of the one-way ANOVA tests that yield
probability values $p<10^{-4}$ for DCN/DNC and $p=0.17$ in the case of
HCN/HNC . The KS2 test of the consecutive stages confirms these
findings.  This leads to the conclusion that DCN can be formed more
efficiently or is destroyed less efficiently than DNC in the more
evolved and thus warmer sources. The observed behavior can be
understood according to the dominant formation pathways scenario
stated by \citet{turner2001}. He claimed that all four deuterated
molecules can be efficiently formed at low temperatures via
H$_2$D$^+$, but only DCN can be formed at higher temperature via
CH$_2$D$^+$. That would lead to the observed trends in the deuteration
fractions, since DCN can still be formed in the more evolved stages,
in contrary to the other observed deuterated molecules.  However,
newer chemical models, e.g., from \citet{roueff2007} or the updated
model we use, do not support this scenario of a difference in dominant
formation pathways between DCN and DNC. According to these new models,
both isomers are formed from light hydrocarbons and thus exhibit a
similar evolution in the abundance ratios with
temperature. \citet{hiraoka2006} were studying the association
reactions of CN with D in laboratory experiments and found at a
temperature of $10$~K an intensity ratio of DNC/DCN of $\sim3$. At
temperatures $>20$~K, the formation of DNC and DCN became
negligible. This is consistent with the derived value for the IRDC
stage in this work on the order of unity, but contradicts the higher
DCN/DNC ratios found in the HMC stage, suggesting another possible
formation route. The HCN/HNC median ratios is between $~0.4-1.4$ in
the different stages and slightly increases with stage. This is
consistent within the observational uncertainties with theoretical
expectations of the ratio of $~1$ in cold clouds and $>1$ in warmer
clouds \citep{sarrasin2010}.

\subsection{Relation between deuteration and other parameters}\label{sec:deuteration_relations}
\begin{figure*}
\includegraphics[width=0.5\textwidth]{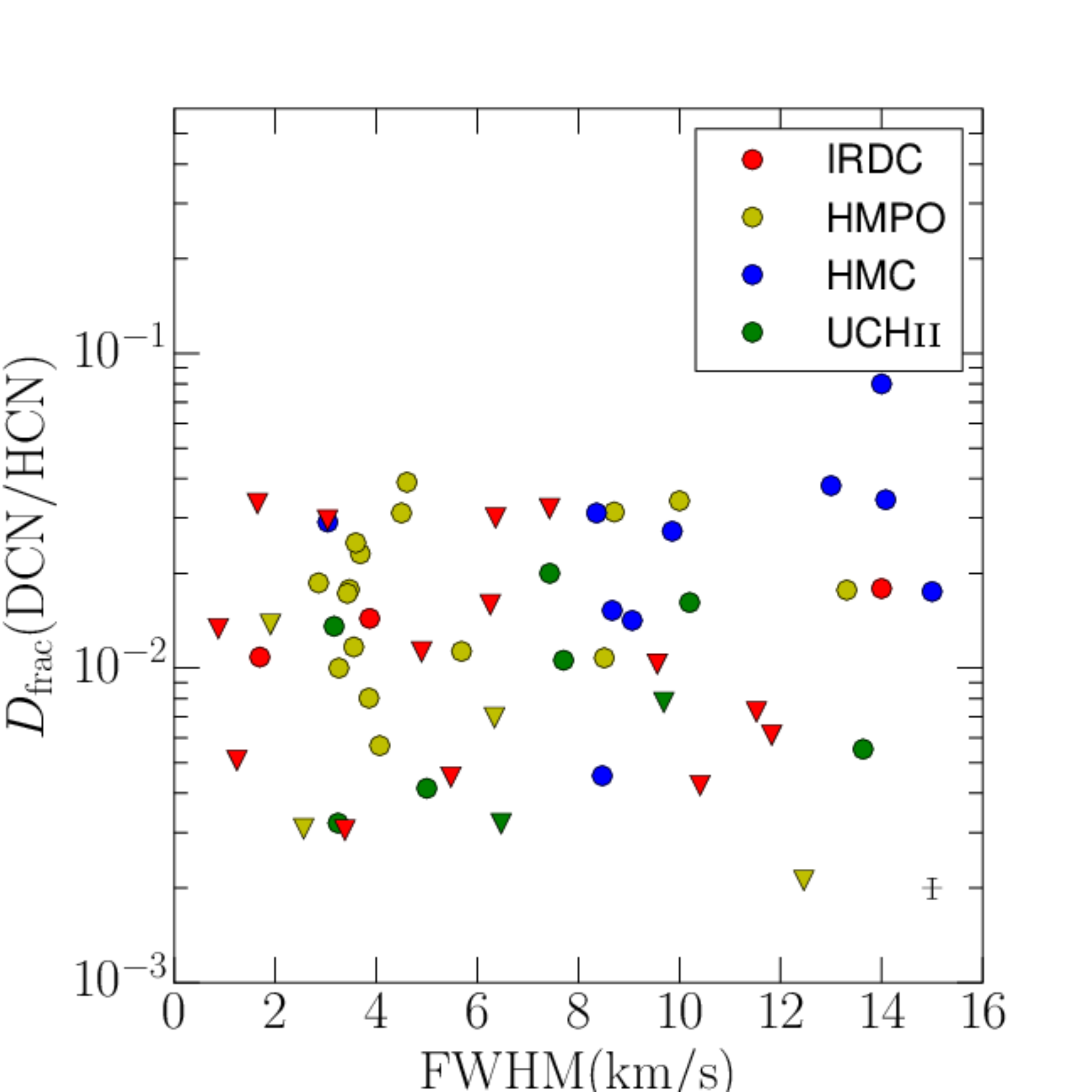}
\includegraphics[width=0.5\textwidth]{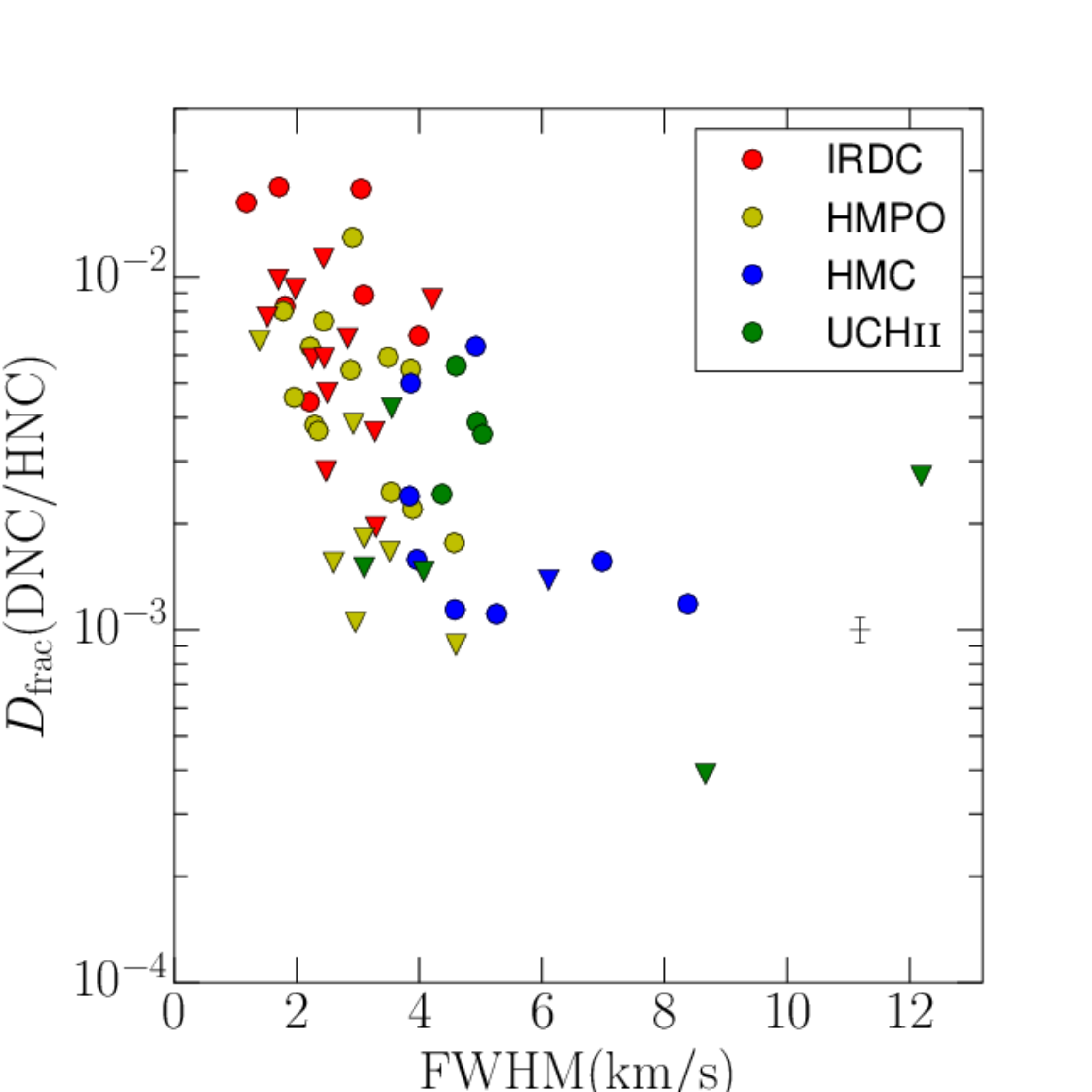}\\
\includegraphics[width=0.5\textwidth]{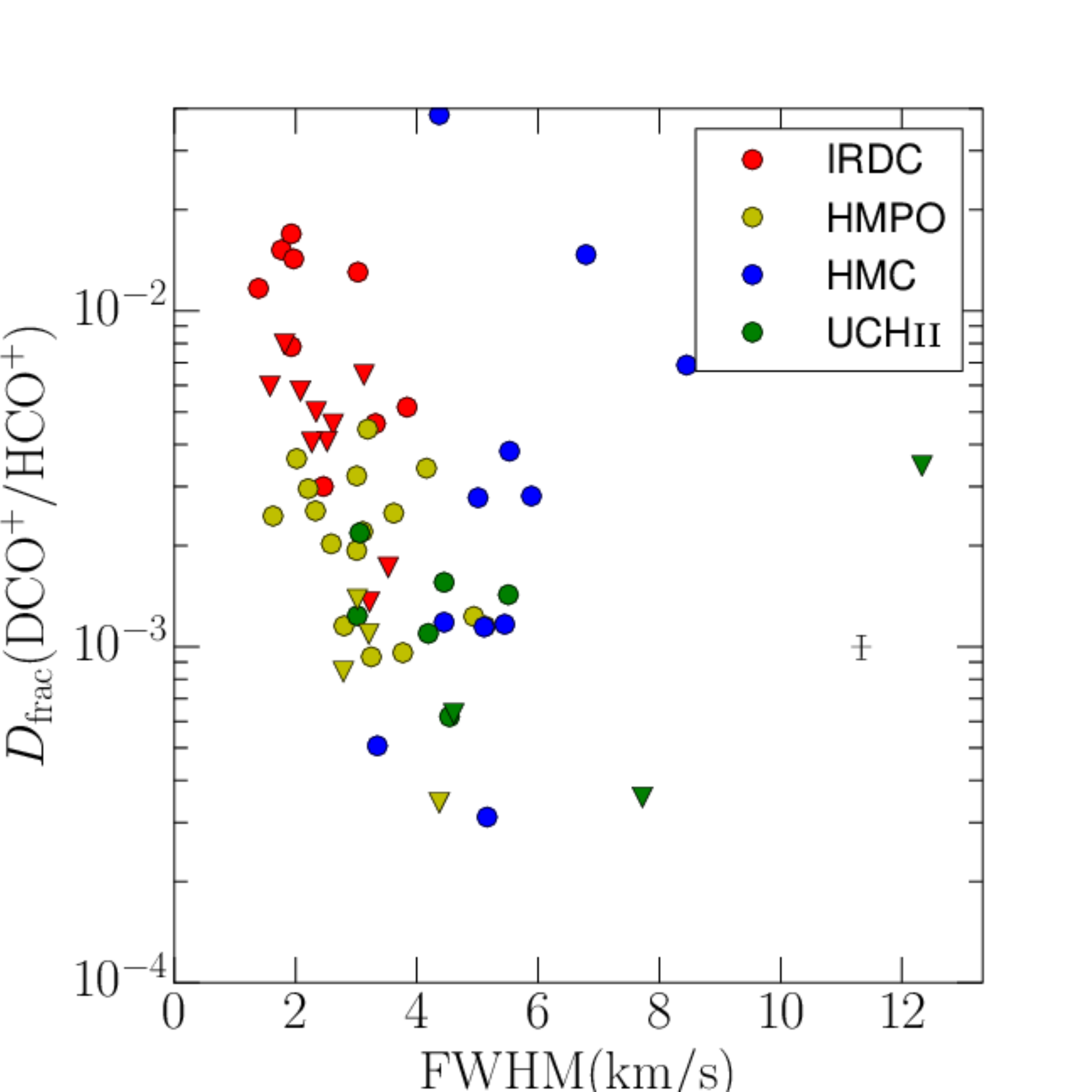}
\includegraphics[width=0.5\textwidth]{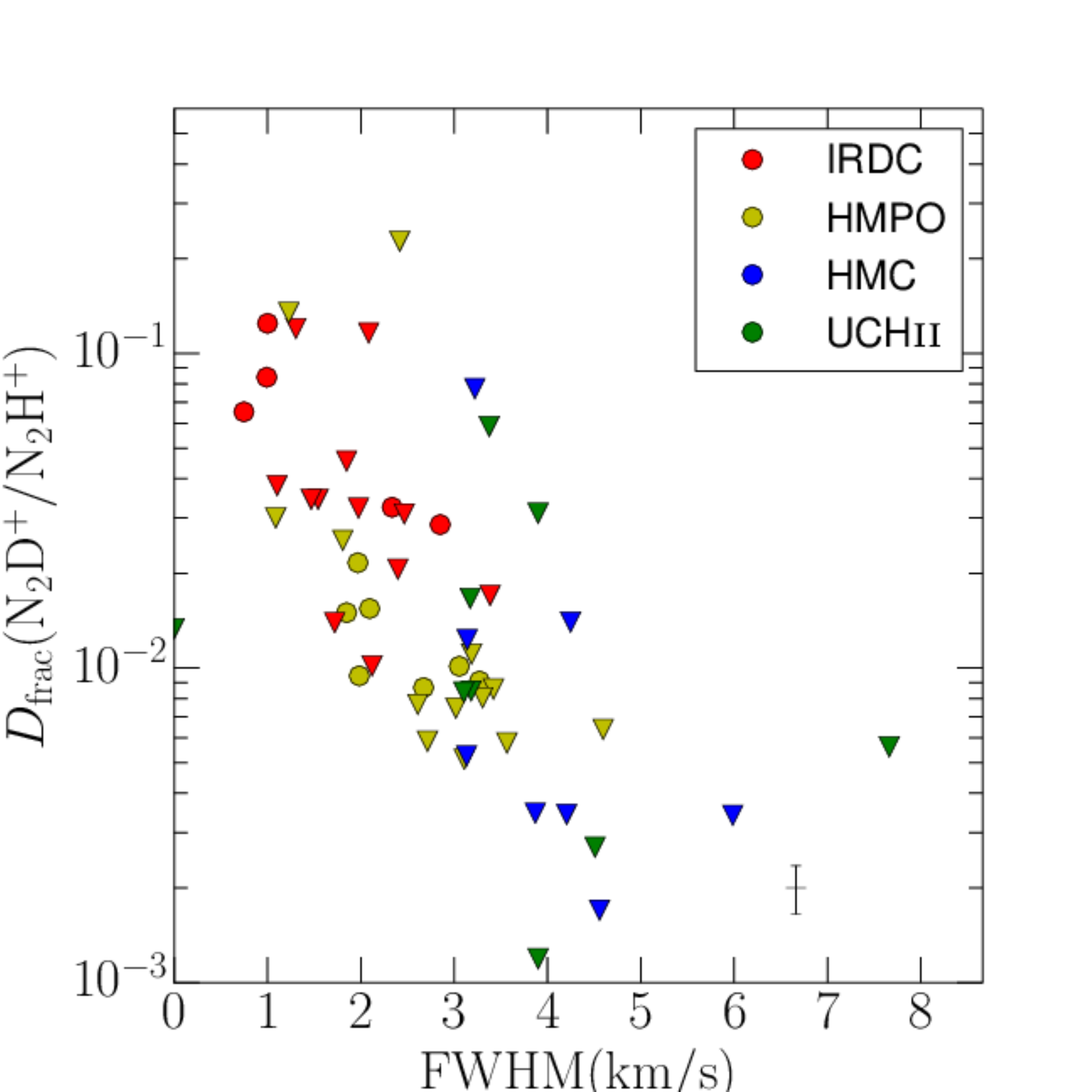}
\caption{Deuteration fractions of HCN, HNC, HCO$^+$ and N$_2$H$^+$
  vs. FWHM of non-deuterated species from HCN (or H$^{13}$CN in case
  HCN is not available), HN$^{13}$C, H$^{13}$CO$^+$ and N$_2$H$^+$,
  respectively. The dots mark detections, the triangles upper
  limits. The typical size of an error bar from the uncertainty in the
  integrated flux is given in the lower right.
\label{fig:deuteration_linewidths}}
\end{figure*}

We measured the full width at half maximum (FWHM) of HCN, HN$^{13}$C,
H$^{13}$CO$^+$ and N$_2$H$^+$. The isotopologues are assumed to be
optically thin. For N$_2$H$^+$ and HCN we corrected the measured
optically thick FWHM $\Delta \nu_{\rm thick}$ according to
\citet{phillips1979} using the relation:

\begin{equation}
\frac{{\rm FWHM}_{\rm thick}}{{\rm FWHM}_{\rm thin}}=\frac{1}{\sqrt{\ln 2}}\sqrt{\ln \frac{\tau}{\ln \frac {2}{\exp(-\tau)+1} }}
\end{equation}

where $\tau$ is the optical depth of the line center and FWHM$_{\rm
  thin}$ the optically thin FWHM.  We test for linear relationships
using the Pearson correlation coefficient. By definition, the absolute
value of $\rho$ is $\leq 1$ with a stronger correlation for larger
$\rho$. $\rho=0$ means no correlation. The sign indicates a positive
or negative correlation between the two quantities. In order to
estimate $\rho$ we randomly draw several times a set of data points
from the measured values with their corresponding errors as well as
the upper limits, and find for each of these drawn data sets the
underlying correlation coefficient. We then find the most frequent
correlation coefficient among all data sets as well as its 16\% and
84\% confidence values (Bailer-Jones et al. in prep.). The correlation
plots are shown in Figure~\ref{fig:deuteration_linewidths} and the
corresponding Pearson correlation coefficients $\rho$ are given in
Table~\ref{tbl:corr_coeff}. The plots show that earlier phases have
smaller FWHM.

\begin{table*}
%\tiny
\caption{Correlation coefficients $\rho$ with 16\% and 84\% confidence values.}
\label{tbl:corr_coeff}
\centering
\begin{tabular}{lccc}
\hline \hline
         & Luminosity & N(H$_2$) & FWHM \\
\hline \\
DCO$^+$/HCO$^+$          & -0.21$^{\hphantom{-}0.02}_{-0.35}$  & \hphantom{-}0.03$^{\hphantom{-}0.49}_{-0.31}$ & -0.05$^{\hphantom{-}0.06}_{-0.21}$ \\[4pt]
DCN/HCN                  & $\hphantom{-}0.19^{\hphantom{-}0.23}_{\hphantom{-}0.01}$  & -0.07$^{\hphantom{-}0.23}_{-0.13}$ & \hphantom{-}0.09$^{\hphantom{-}0.26}_{-0.03}$ \\[4pt]
DNC/HNC                  & -0.27$^{-0.14}_{-0.36}$  & -0.01$^{\hphantom{-}0.13}_{-0.15}$ & -0.23$^{-0.13}_{-0.32}$ \\[4pt]
N$_2$D$^+$/N$_2$H$^+$    & -0.47$^{-0.14}_{-0.51}$  & -0.05$^{\hphantom{-}0.04}_{-0.14}$ & -0.43$^{-0.26}_{-0.47}$ \\[4pt]
\hline
\end{tabular}
%\tablefoottext{}{{}}
\end{table*}

The resulting coefficient is consistent with the data being
uncorrelated for all four molecules. The FWHM of N$_2$H$^+$ might be
an exception and correlated, but is still $<0.5$ and no conclusive
answer can be given. Although the correlation with the FWHM is weak,
the decrease of the deuterium ratio with increasing FWHM is consistent
with the picture of more quiescent, early stages with lower FWHM and
higher deuteration fractions followed by more turbulent stages with
higher FWHM and lower deuterium ratios.

\begin{figure*}
\includegraphics[width=0.5\textwidth]{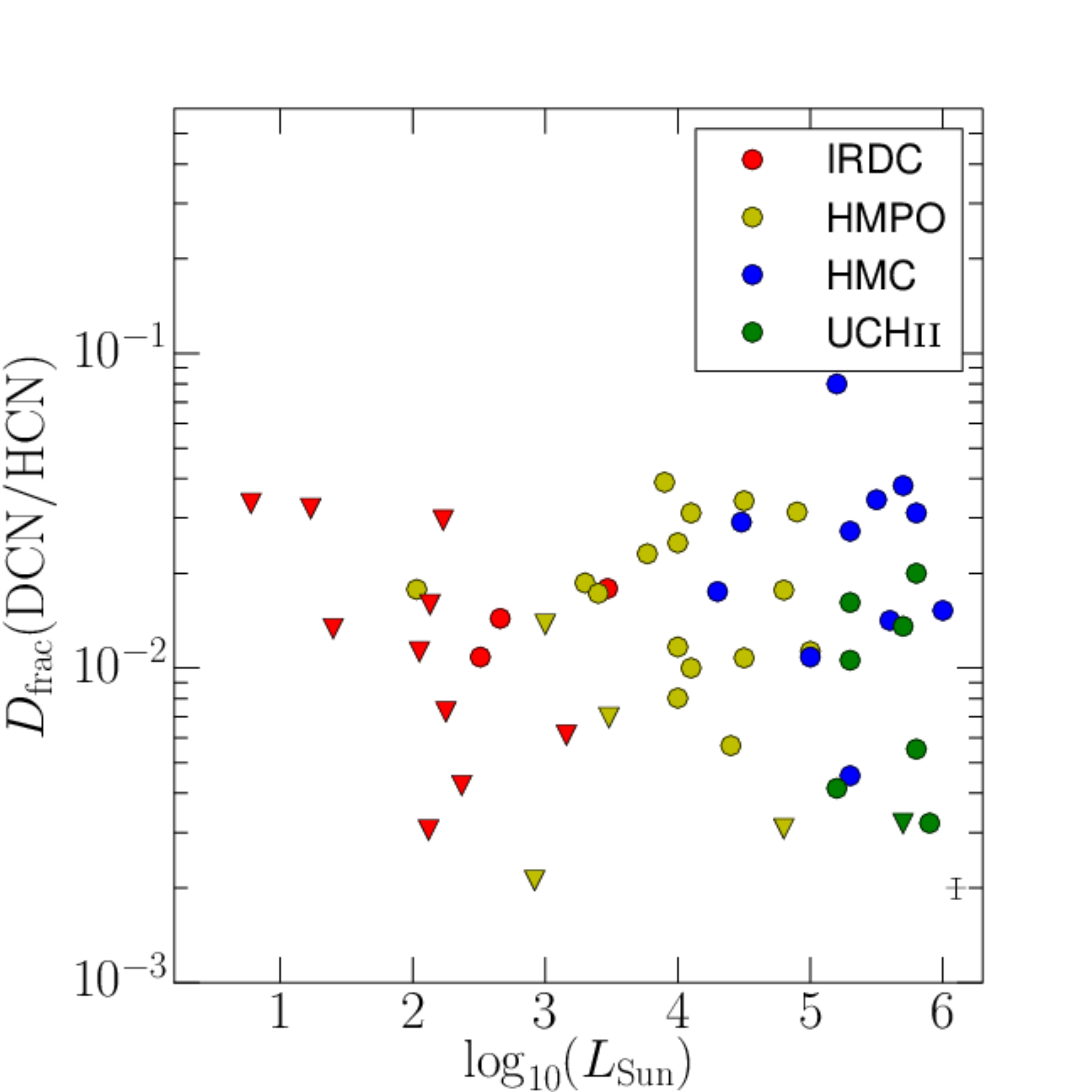}
\includegraphics[width=0.5\textwidth]{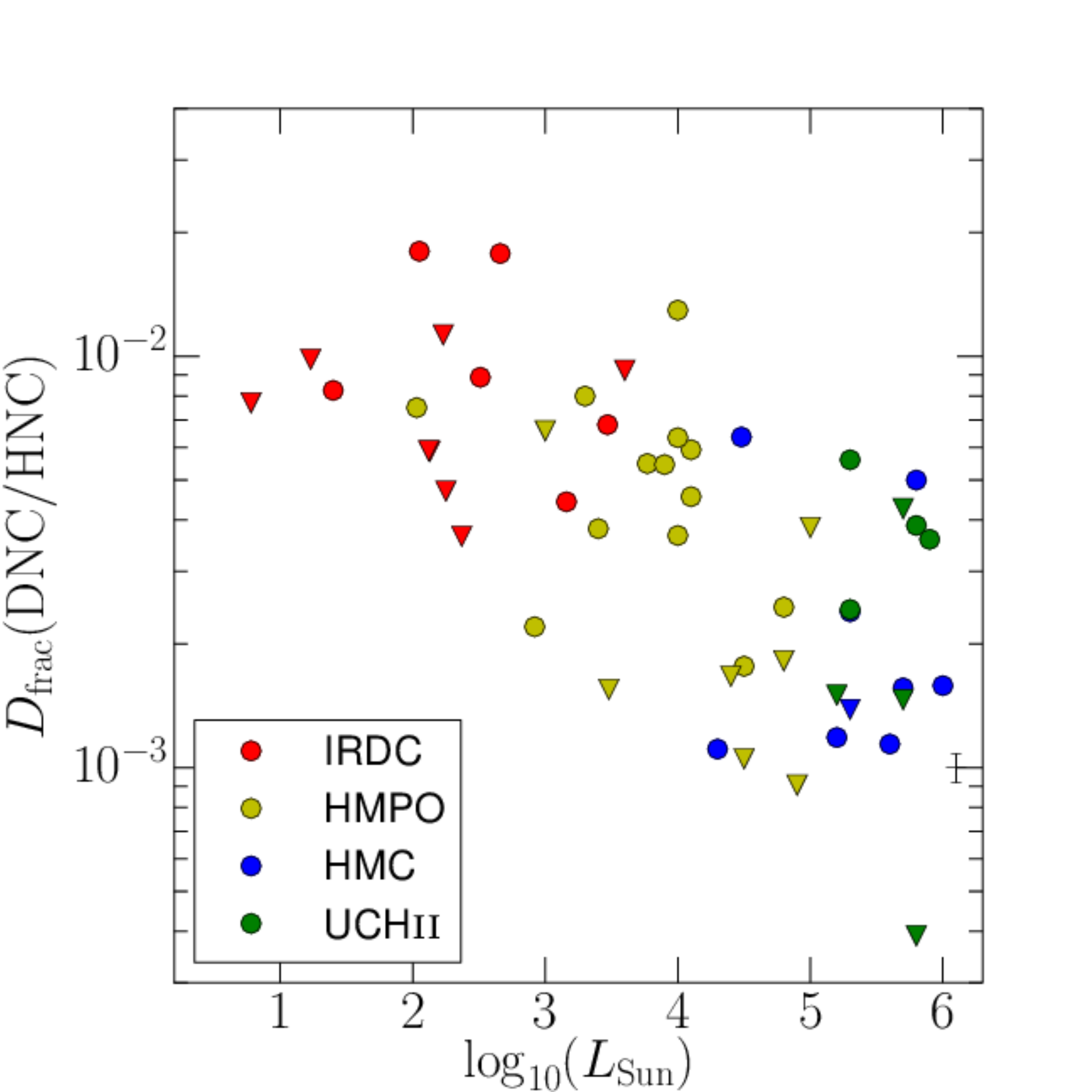}\\
\includegraphics[width=0.5\textwidth]{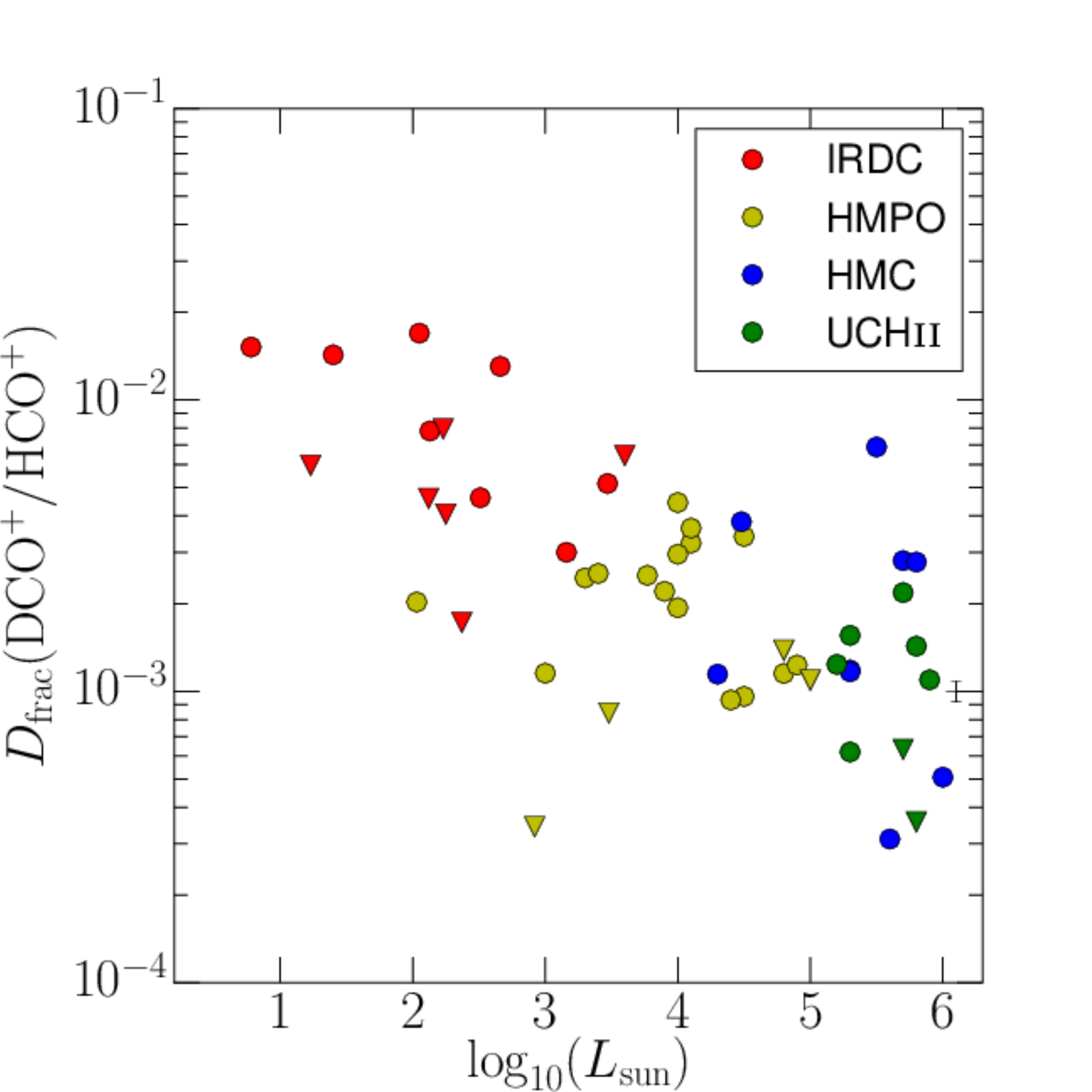}
\includegraphics[width=0.5\textwidth]{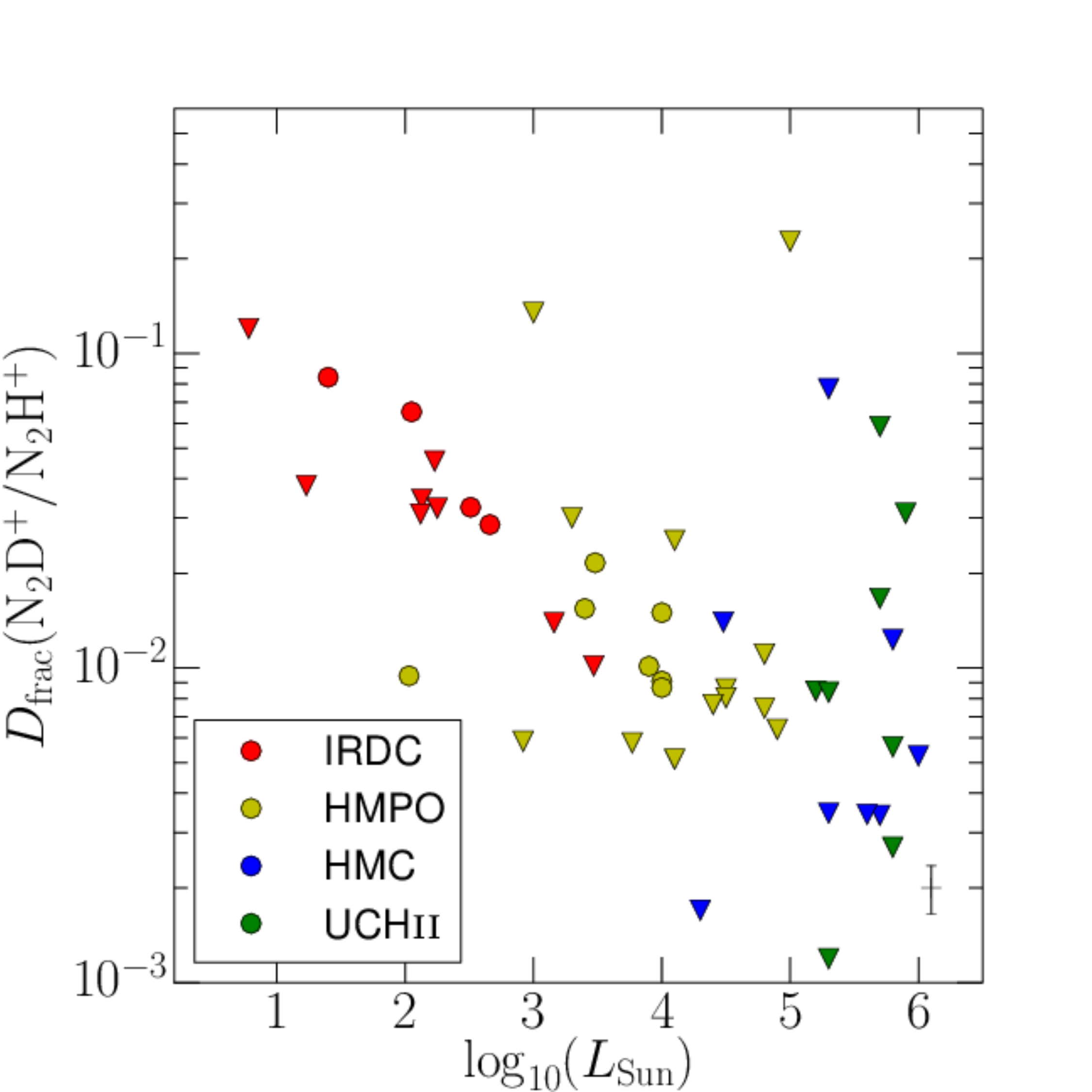}
\caption{Deuteration fractions of HCN, HNC, HCO$^+$ and N$_2$H$^+$
  vs. the luminosity of the source. The dots mark detections, the
  triangles upper limits. The typical size of an error bar from the
  uncertainty in the integrated flux is given in the lower right.
\label{fig:deuteration_lum}}
\end{figure*}

\begin{figure*}
\includegraphics[width=0.5\textwidth]{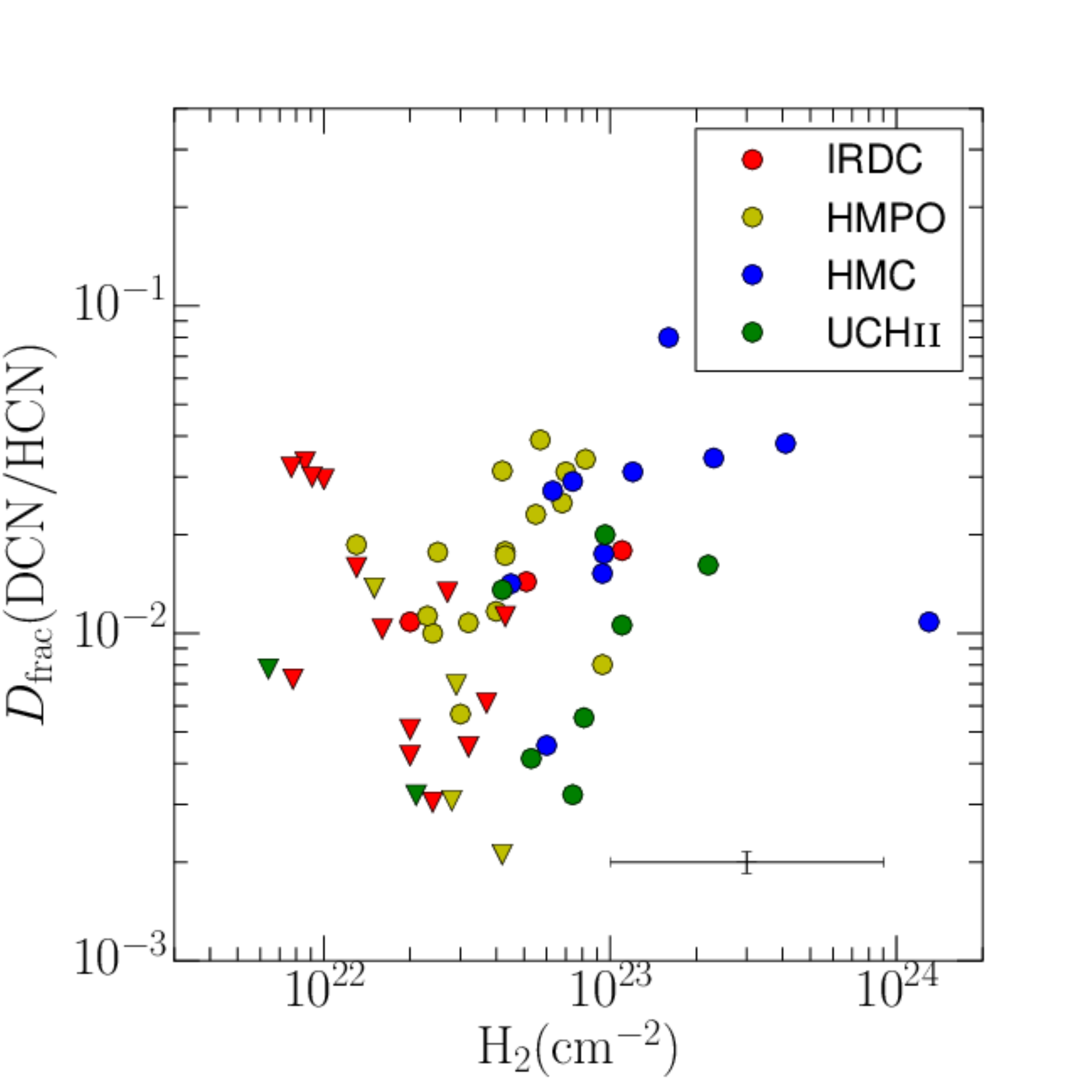}
\includegraphics[width=0.5\textwidth]{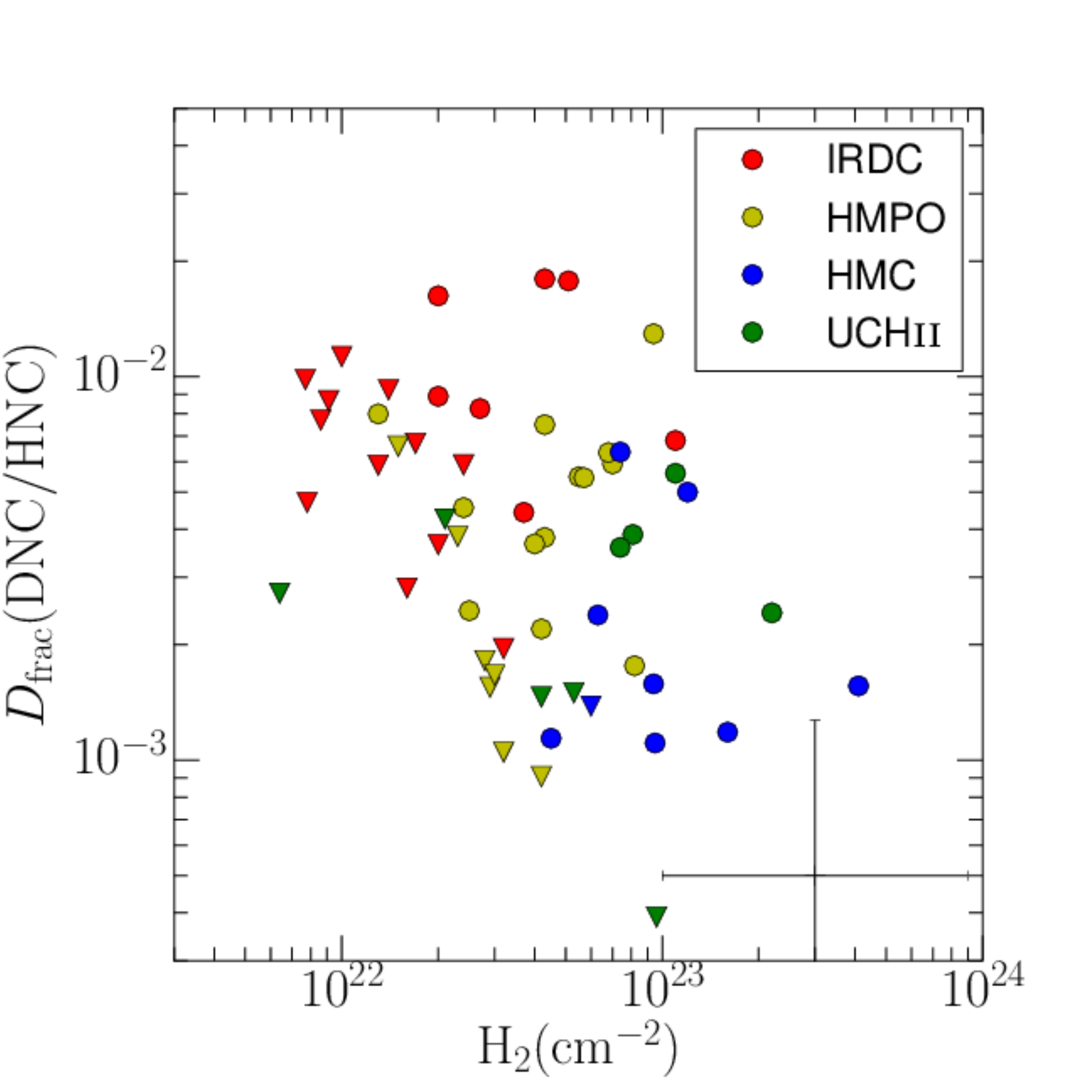}\\
\includegraphics[width=0.5\textwidth]{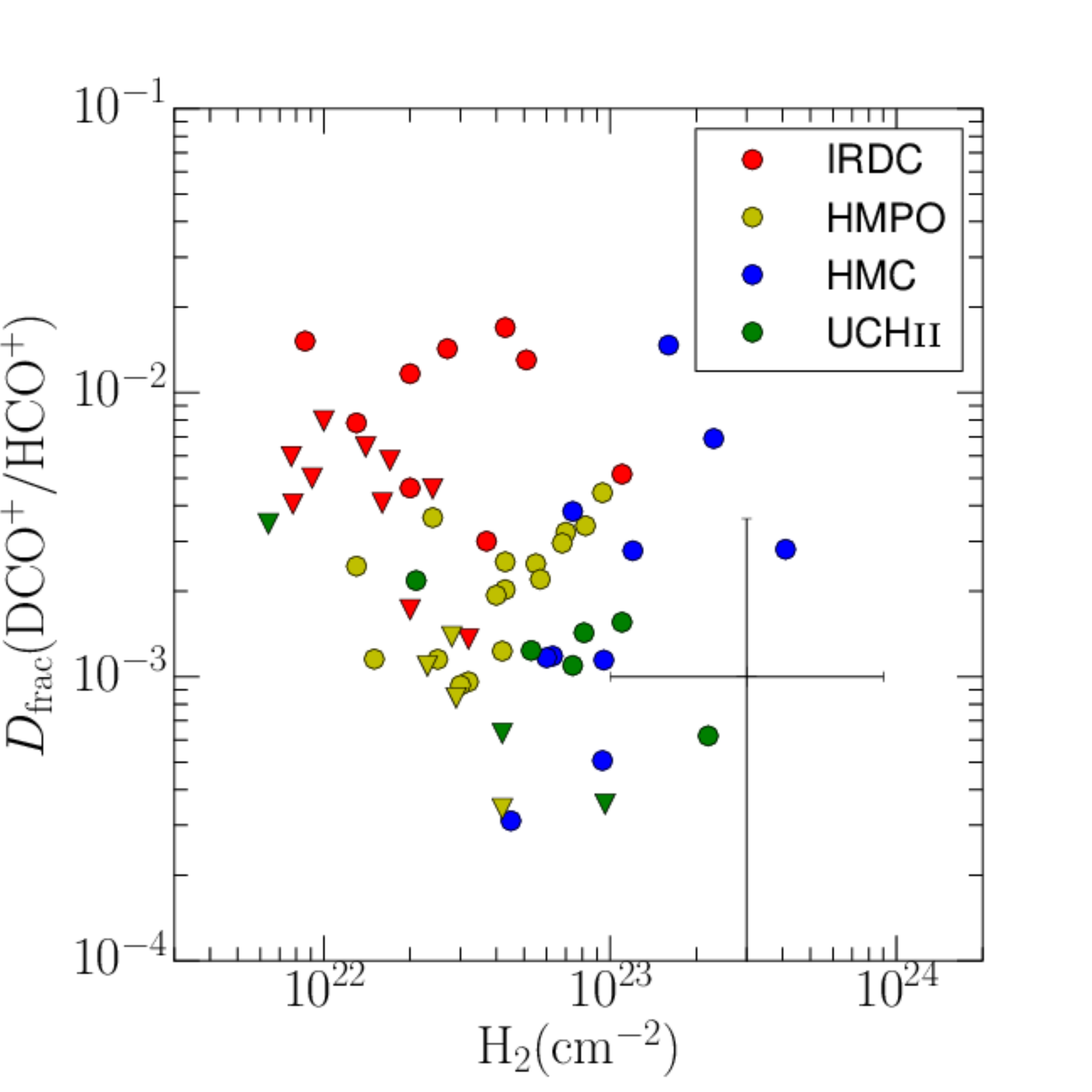}
\includegraphics[width=0.5\textwidth]{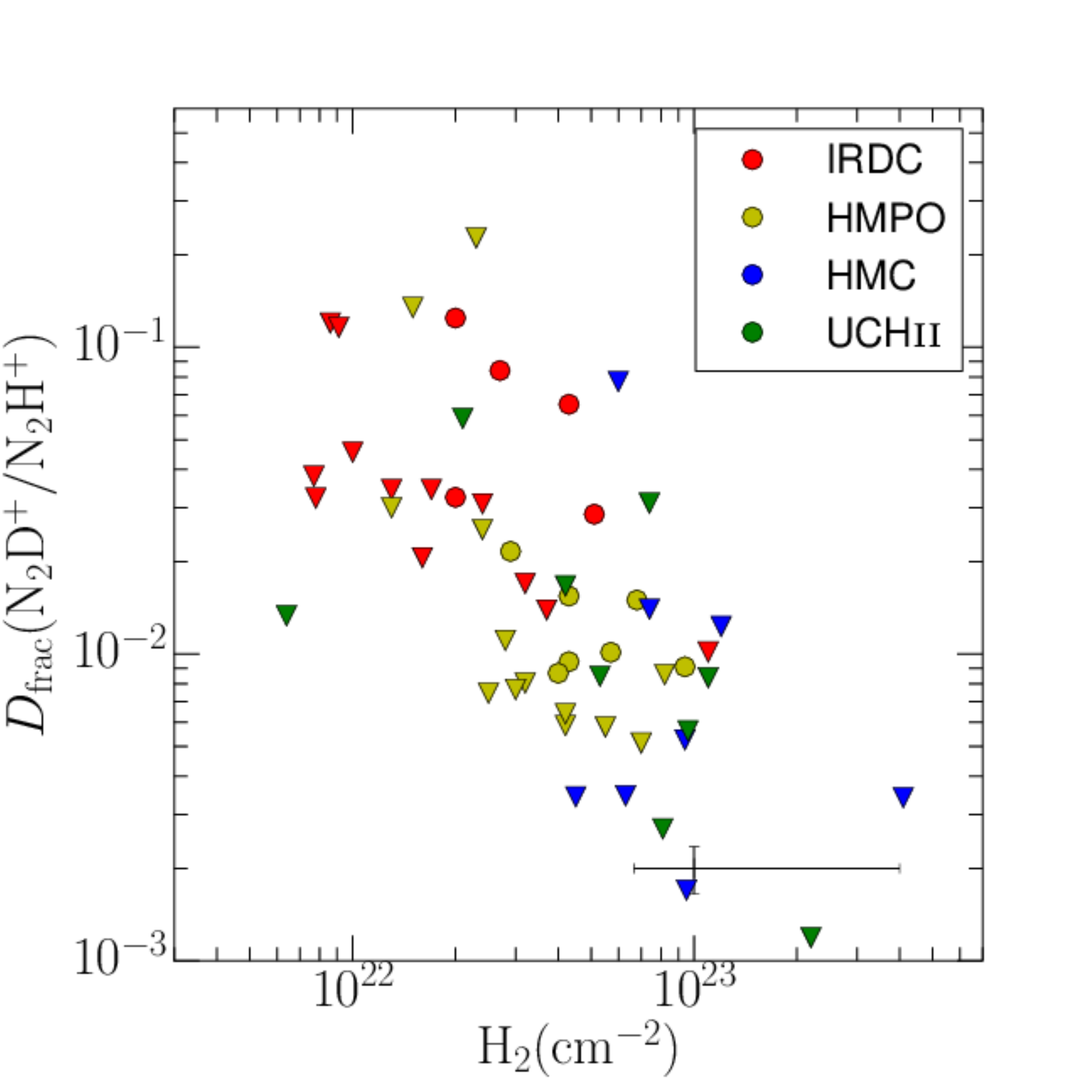}
\caption{Deuteration fractions of HCN, HNC, HCO$^+$ and N$_2$H$^+$
  vs. the H$_2$ column density of the source. The typical size of an
  error bar from the uncertainty in the integrated flux and from the
  uncertainty in the derived H$_2$ column densities (see
  Section~\ref{sec:column_density}) is given in the lower right.
\label{fig:deuteration_h2}}
\end{figure*}

Furthermore, we studied the correlation of the deuteration with the
H$_2$ column density and luminosity $L$. The luminosities and H$_2$
column densities are shown in
Tables~\ref{tbl:sourceparameters1}--\ref{tbl:sourceparameters2} and
the corresponding correlation plots in
Figure~\ref{fig:deuteration_lum} and
Figure~\ref{fig:deuteration_h2}. In general, the deuteration fraction
shows similarly weak correlations with the luminosities as with the
FWHM. Clearly, there is a lack of correlation with H$_2$ column
densities. Among the four deuteration fractions, the strongest
correlation is found for N$_2$D$^+$, but the correlation is still very
low. The luminosity of a source is not necessarily a tracer of its
evolutionary stage, but Figure~\ref{fig:deuteration_lum} shows that
the four different evolutionary stages are approximately separated
with respect to the luminosity. Taking into account that the
temperature of the objects increases with evolutionary stage, the
luminosity can be used as a proxy for the temperature in the case of
our source sample. Since higher luminosities indicate higher
temperatures and smaller regions with $T\lesssim20$~K, where CO is
frozen out, it reduces the overall abundance as well as the D/H ratio
of N$_2$H$^+$. Among all four molecular deuteration fractions, DCN and
DCO$^+$ show the weakest correlation with any of the shown
parameters. However, none of the studied correlations have correlation
coefficients $\rho>0.5$.

\subsection{Modeling the chemical evolution}
With the observed column densities of 18 species (14 from Paper {\sc
  I}) including 4 deuterated species in different stages of massive
star formation at hand, we applied the iterative physico-chemical
fitting model ``MUSCLE-D'' (``MUlti Stage CLoud codE with
D-chemistry'') to these data. A list of all fitted species is given in
Table~\ref{tbl:species_comp}. The model fits the evolution of the
observed chemical data and thereby constrains basic physical
properties in the assumed evolutionary path of high-mass star
formation, such as mean temperatures and mean chemical ages. The
chemical ages can be interpreted as typical lifetimes of the various
stages. This model is the extended version of the iterative fitting
model ``MUSCLE'' (``MUlti Stage CLoud codE'') already used and
described in \citet{gerner2014} and includes a physical and a chemical
model. Here we will give a short summary of the main characteristics
of the two parts of ``MUSCLE-D'', the physical model in
Section~\ref{sec:phys_model} and the chemical model ``ALCHEMIC'' in
Section~\ref{sec:chem_model}.

\begin{table}
%\tiny
\caption{List of species fitted with the model.}
\label{tbl:species_comp}
\centering
\begin{tabular}{c}
\hline \hline
Molecules \\
\\
\hline
CO\tablefootmark{a}, HNC\tablefootmark{a}, HCN, HCO$^+$\tablefootmark{a}, HNCO, H$_2$CO, N$_2$H$^+$, CS\tablefootmark{a}, SO\\
OCS, C$_2$H, SiO, CH$_3$CN, CH$_3$OH, DCO$^+$, DCN, DNC, N$_2$D$^+$\\
\hline
\end{tabular}
~
\tablefoottext{a}{Minor isotopologue observed}
\end{table}

\subsubsection{Physical model}\label{sec:phys_model}
The physical model treats the star-forming region as spherically
symmetric 1D-clouds with a fixed outer radius of $r_{\rm
  {out}}=0.5$~pc, based on the largest beam size of our observations
of $~30\arcsec$ and the typical size of high-mass star forming
regions. High angular resolution studies revealed in some of the
sources complex structures that deviate from the spherically symmetric
case \citep[e.g.,][]{beuther2002d}. However, in single-dish maps these
sources appear to be symmetric \citep[e.g.,][]{beuther2002a}. Since we
are analyzing single-dish data, a 1D-model physical model appears to
be sufficient. More sophisticated models, that take into account
complicated substructures, would increase the number of fitting
parameters and might lead to an over-interpretation of the data. The
radial density and temperature structure is modeled with

\begin{equation}
\label{eq:rho_r}
\begin{array}{ll}

 \rho(r) = \rho_{\rm {in}}(r/r_{\rm {in}})^{-p},  & r\ge r_{\rm {in}};\\
 \rho(r) = \rho_{\rm {in}},              &r<r_{\rm {in}}
 
\end{array}
\end{equation}

and

\begin{equation}
\label{eq:temp_r}
\begin{array}{ll}

 T(r) = T_{\rm {in}}(r/r_{\rm {in}})^{-q},  & r\ge r_{\rm {in}};\\
 T(r) = T_{\rm {in}},              &r<r_{\rm {in}},

\end{array}
\end{equation}
respectively. The temperature and density profiles do not change with
time. The fitted physical quantities are the inner radius $r_{\rm
  {in}}$, the temperature and density at the inner radius, $T_{\rm
  {in}}$ and $\rho_{\rm {in}}$ and the power law index of the density
$p$. The inner radius is limited between $5 \times 10^{-5}-5 \times
10^{-2}$ pc. While we assume the IRDC is an isothermal sphere, the
temperature structure of the more evolved stages is modeled by a inner
flat plateau with $T_{\rm {in}}$ and a power-law with slope $q=0.4$ as
a standard value for $r>r_{\rm {in}}$
\citep[see][]{vandertak2000c}. The radial density profile within
$r_{\rm {in}}$ is flat with $\rho_{\rm {in}}$ and decreases for
$r>r_{\rm {in}}$ as a power law with slope $p$. The value of $p$ is
limited to values between 1.5 and 2.0 in order to save computing
time. That range is supported by several observations,
e.g. \citet{guertler1991,beuther2002a,mueller2002,hatchell2003}. The
temperature and density profiles are simultaneously fitted. The model
does not take into account radiative transport. In the model, the
whole cloud is embedded in a larger diffuse low-density medium that
shields the high-mass star forming cloud from the interstellar FUV
radiation.

\subsubsection{Chemical model}\label{sec:chem_model}
The chemical model is an updated version of the time-dependent
gas-grain chemical model ``ALCHEMIC'' described in \citet{semenov2010}
and already used and described in \citet{gerner2014}. In addition, the
deuterium network from \citet{albertsson2013} was added and extended
with high-temperature reactions
\citep{harada2010,harada2012,albertsson2014} and ortho/para states of
H$_2$, H$_2^+$ and H$_3^+$ and their isotopologues
\citep{albertsson2014b}. In total, the chemical network comprises 15
elements that can form $~1260$ different species from $38\,500$
reactions.

The initial abundances prior to the IRDC stage are taken from the
``low metals'' set given in \citet{lee1998} with changed elemental
abundance of Si ($3 \times 10^{-9}$ with respect to H) and S ($8
\times 10^{-7}$ with respect to H). The changes were needed in order
to achieve proper fits to the IRDC phase. Initially, all metals
  (C, O, N, S, etc.) are in atomic form. Only H$_2$ is already in
  molecular form. The initial ortho-para ratio of H$_2$ is assumed to
be the statistical value of 3:1.
%In another realization of the model we assume all H$_2$ to be in the
%para state, due to the cold conditions. A comparison of these two
%extreme cases is given in Sect.~\ref{sec:opH2}.
The model includes nuclear spin state exchange reactions to account
for the evolution of the ortho-para ratio as well as freeze out of
CO. For the subsequent stages, the chemical outcome of the previous
best-fit model is used as input initial abundances.

\subsubsection{The fitting procedure}
The fitting for the different stages is done iteratively using the
physical and chemical model described above. In this sense we modeled
the observed column densities for the IRDC, HMPO and HMC stage by
varying the parameters $r_{\rm {in}}$, $T_{\rm {in}}$, $\rho_{\rm
  {in}}$ and $p$. While keeping all other parameters fixed, we vary
these four parameters and run the model over $10^5$\,yr for each of
the realizations. Then we compute the $\chi^2$-value for each time
step and model realization, given the observed mean column densities
and computed model column densities. We assumed the standard deviation
between modeled and observed values to be one order of magnitude as a
typical value. Molecules that were detected in less than $50\%$ of the
sources within one stage were considered as upper limits. Finally, the
model with the minimum $\chi^2$-value is found as the best-fit model
which matches best with the calculated mean column densities for each
stage.

Due to the iterative fitting along the evolutionary sequence, the
introduced uncertainties increase with evolutionary stage leading to a
limited confidence in the obtained results for the later stages.

\subsection{The modeling results}\label{sec:model_results}
\begin{figure*}
\includegraphics[width=0.5\textwidth]{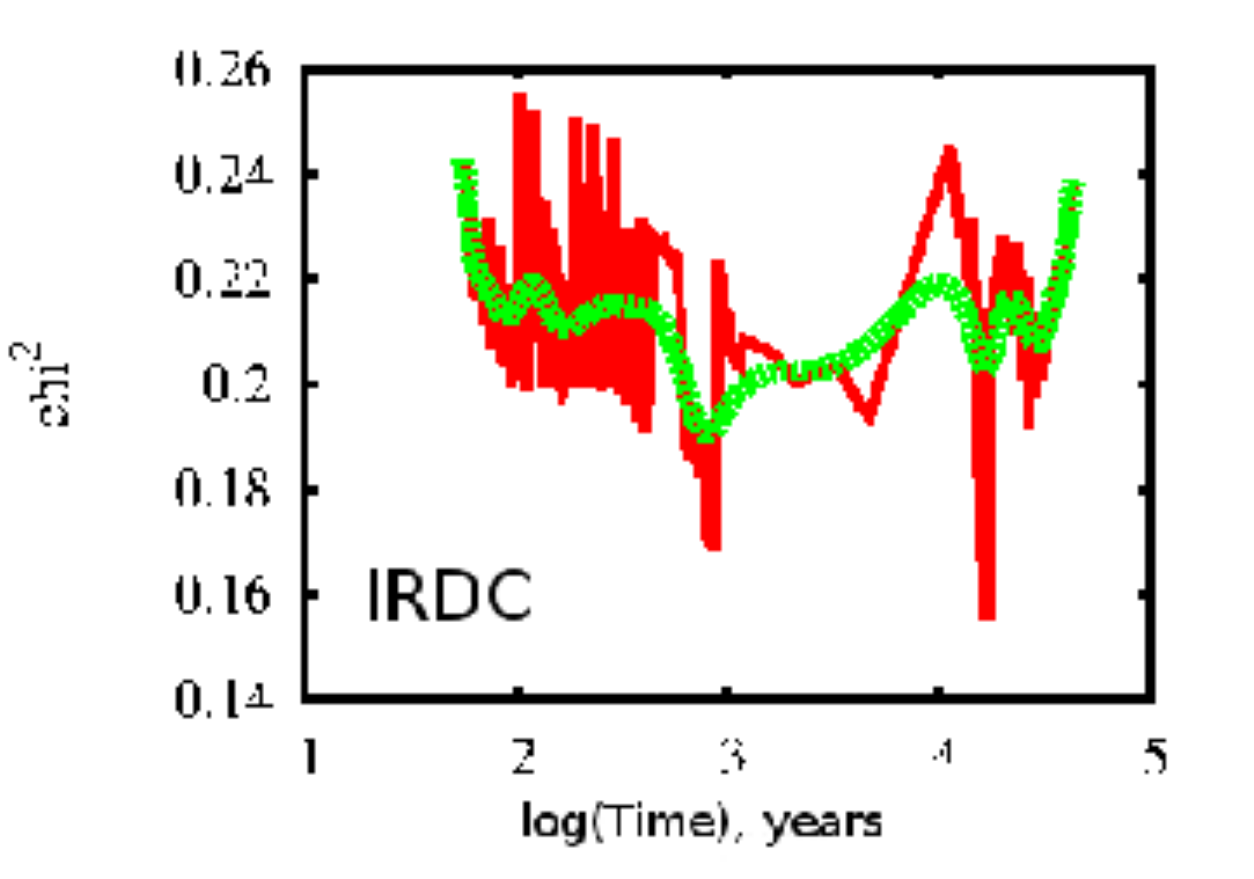}
\includegraphics[width=0.5\textwidth]{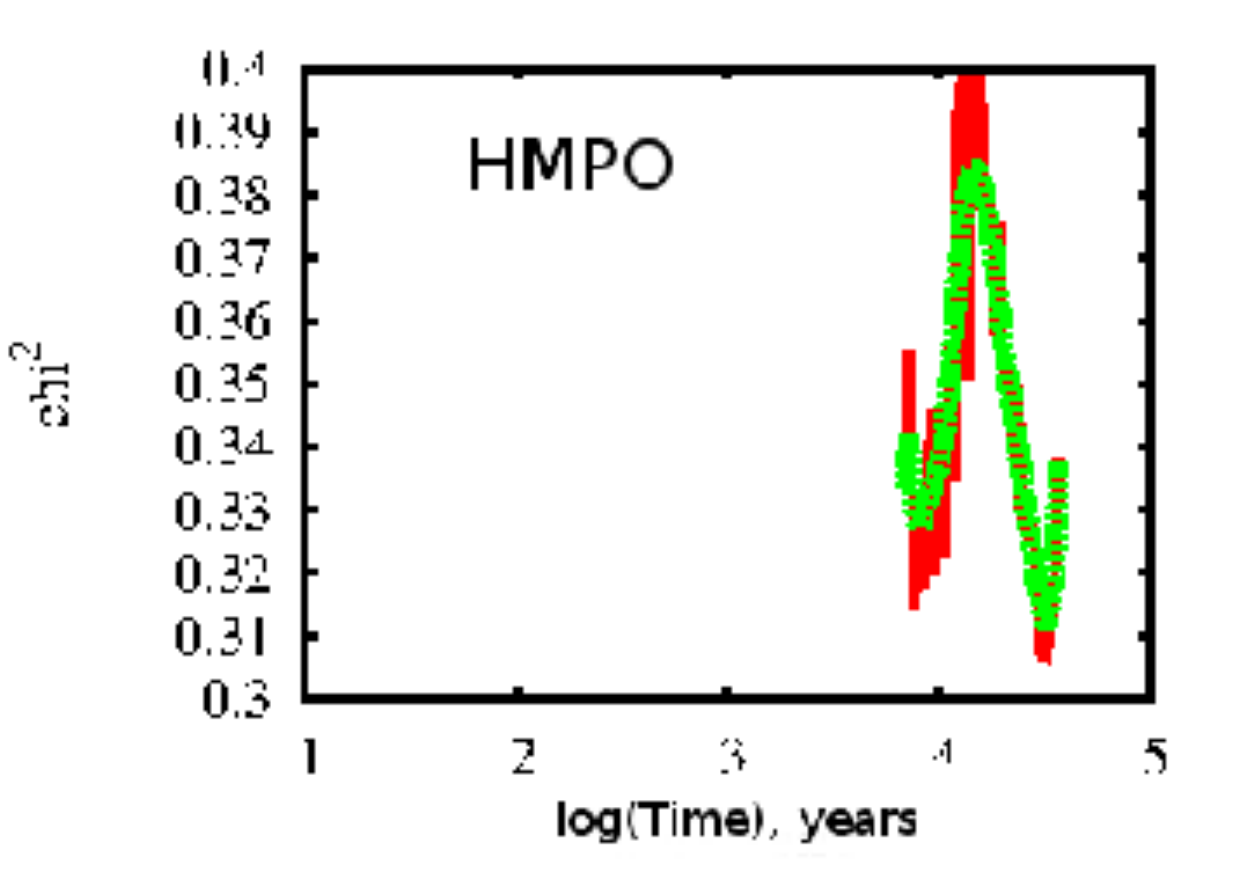}\\
\includegraphics[width=0.5\textwidth]{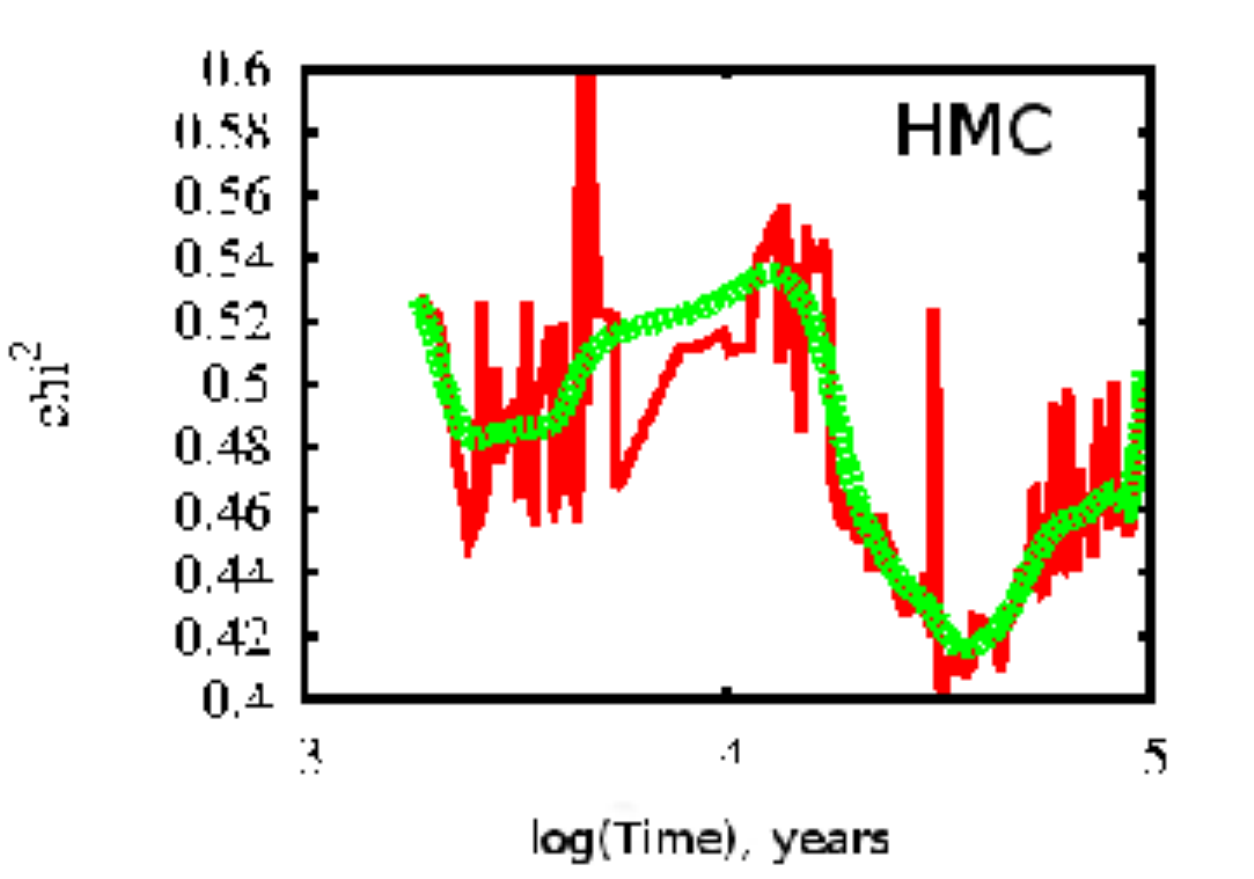}
\includegraphics[width=0.5\textwidth]{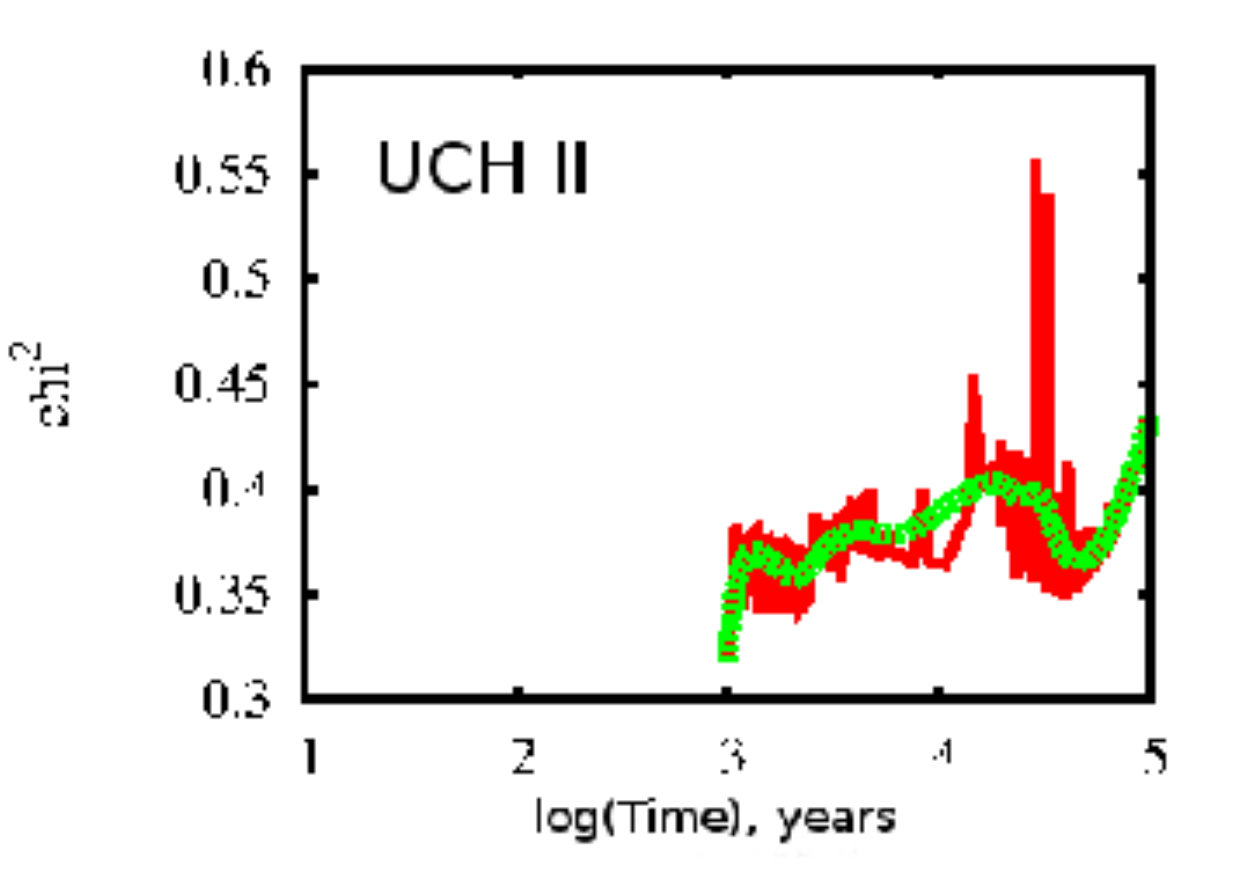}\\
\caption{Evolution of the minimum $\chi^2$ of the best-fit models with
  time. The four panels show the four different stages IRDC (upper
  left), HMPO (upper right), HMC (lower left) and UCH{\sc ii} (lower
  right). The red curve marks the calculated values at all 299 time
  moments, whereas the green curve shows their smoothed spline
  interpolation.}
\label{fig:evolution_chisquare_31}
\end{figure*}

We fitted the combined data from \citet{gerner2014} and this work with
the above described model assuming an ortho-para H$_2$ ratio
o/p=3:1. The resulting best-fit model parameters for the IRDC, HMPO,
HMC and UCH{\sc ii} stage are shown in Tables~\ref{tab:IRDC_fit_31} --
\ref{tab:UCH_fit_31}. The evolution of the best-fits with time is
shown in Figure~\ref{fig:evolution_chisquare_31}. These distributions
give a good sense of the uncertainty of the chemical age. The best-fit
lifetime is $16\,500$~years for the IRDC stage. A good fit was also
achieved with a chemical age of only $~ 1,000$ years. However, that
seems to be a rather short timescale and might be interpreted as a
lower limit. The HMPO best-fit age yielded $32\,000$~years with likely
values between $~10\,000-40\,000$~years. For the HMC stage we found a
best-fit age of $35\,000$~years with values below $~15\,000$~years
being unlikely. The best-fit age for the UCH{\sc ii} stage was found
rather short with $3\,000$~years, but quite unconstrained with likely
values up to several $10\,000$~years.

The observed and modeled column densities are shown in
Tables~\ref{tbl:bestfit_IRDC_31} -- \ref{tbl:bestfit_UCH_31}.  The
best-fit of the IRDC stage reproduces 18 of 18 molecules within the
assumed combined (observational + chemical) uncertainty of one order
of magnitude. In the HMPO stage the model is able to reproduce 16 of
18 molecules. SO is overproduced by a factor of $\sim20$ and CH$_3$OH
underproduced by a factor of $\sim40$. The underproduction of methanol
is possibly due to shock- or outflow-triggered enhanced desorption of
methanol ice from grains, which is not taken into account in the
model. The reason for the misfit of SO might be a poorly understood
chemistry of sulfur-bearing species in modern astrochemical models in
general. In the HMC stage the model could fit 14 of 18
species. Besides C$_2$H, which is $\sim20$~times underproduced, HNCO,
SO and CH$_3$OH are misfitted by more than a factor of 100. The
underproduction of CH$_3$OH and overproduction of SO in the HMPO stage
is continuing in the HMC phase. The overproduction of HNCO might be
connected to not well enough understood shock- and surface
chemistry. This difference in C$_2$H between the model and observations
might be influenced by UV-radiation of the central star(s) or a clumpy
structure of the environment, which is not considered in the
model. That is especially important for the UCH{\sc ii} regions, but
it is also already present in some HMCs.  For the last considered
stage of an UCH{\sc ii} region the model reproduces 13 of 18
species. As in the HMC stage, the molecules HNCO, SO and C$_2$H are
misfitted in the UCH{\sc ii} stage. In addition, SiO and DCO$^+$ are
slightly overproduced.

\begin{figure*}
\centering
\includegraphics[width=0.4\textwidth]{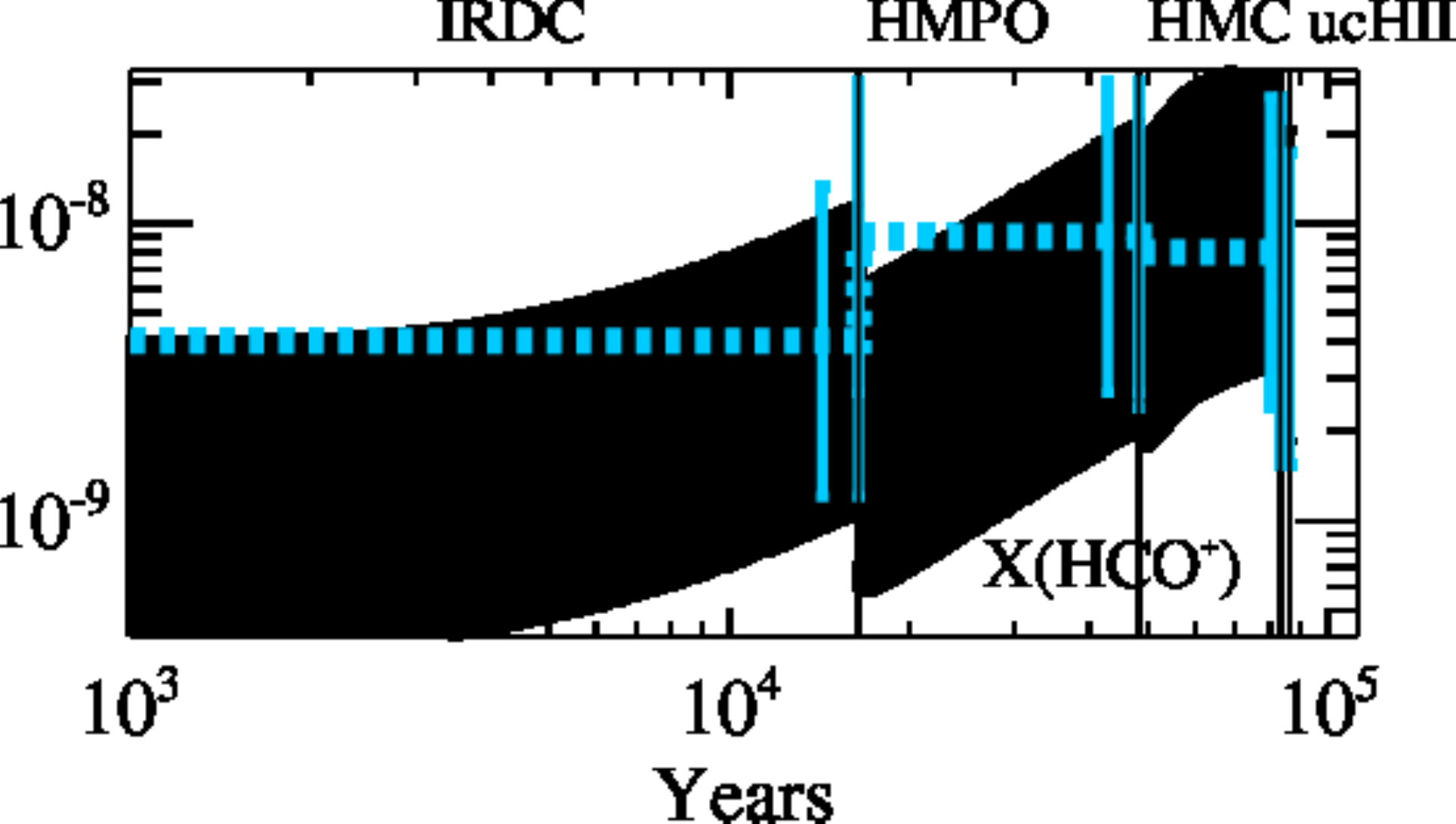}
\hspace{0.5cm}
\includegraphics[width=0.4\textwidth]{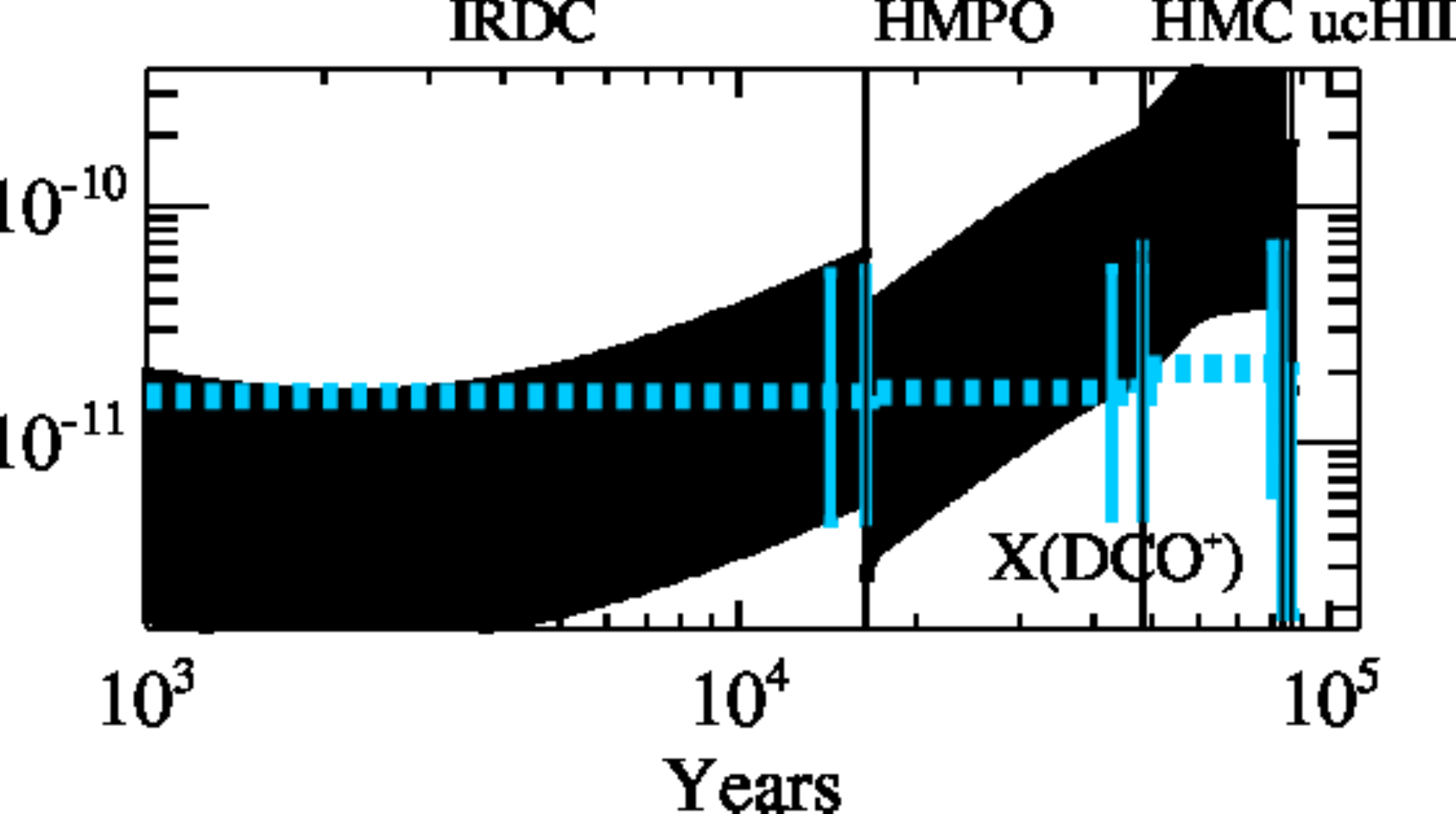}\\
\includegraphics[width=0.4\textwidth]{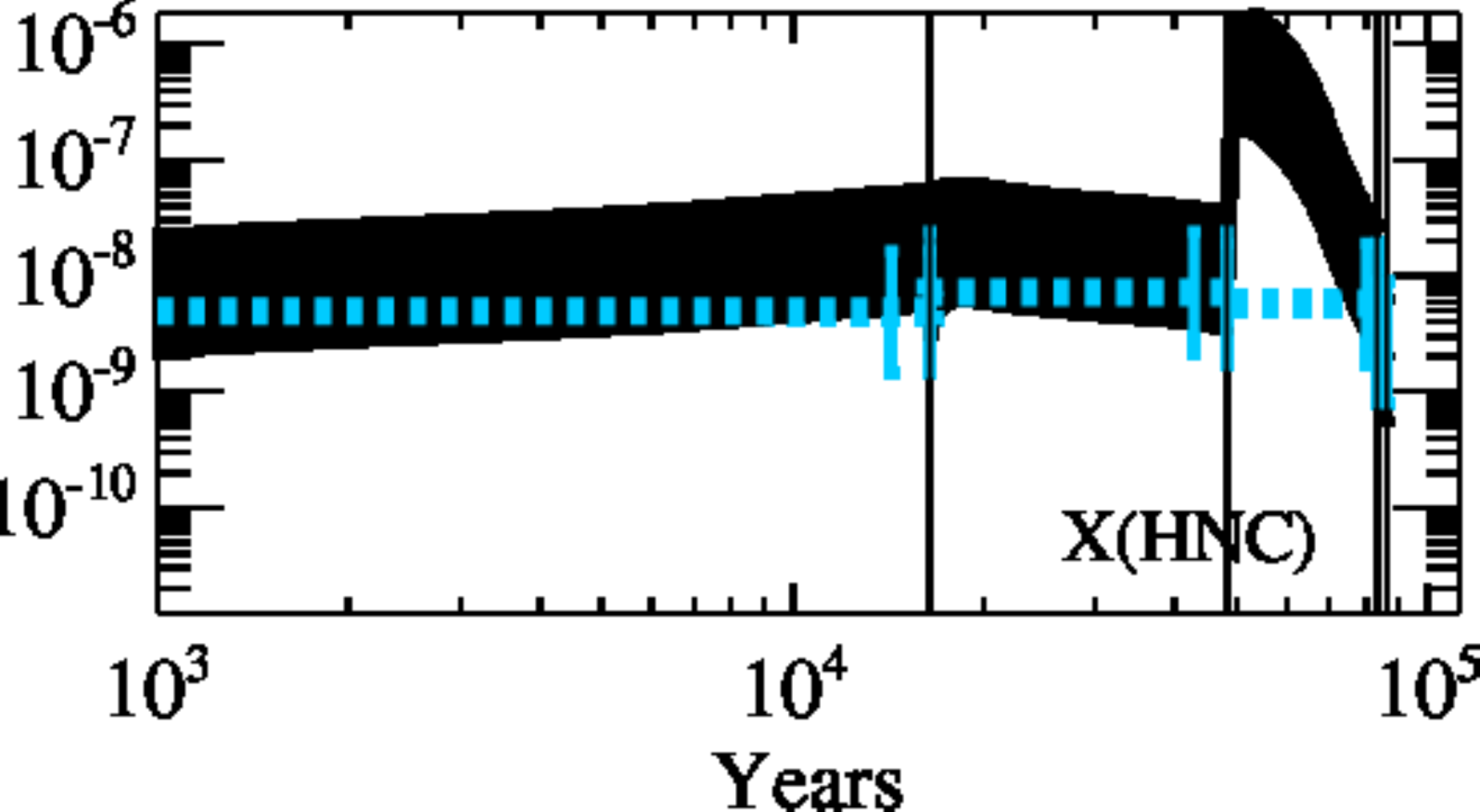}
\hspace{0.5cm}
\includegraphics[width=0.4\textwidth]{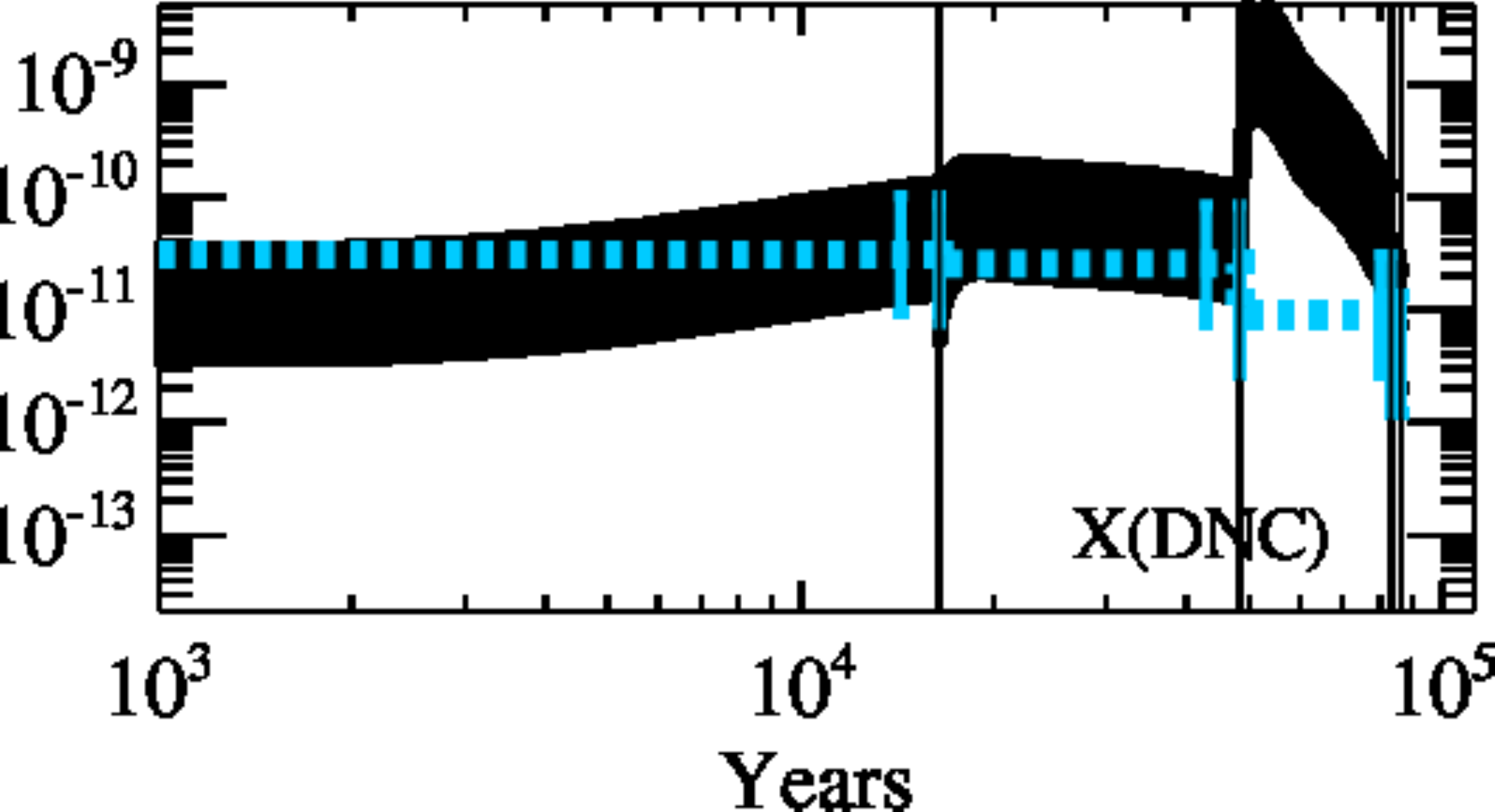}\\
\includegraphics[width=0.4\textwidth]{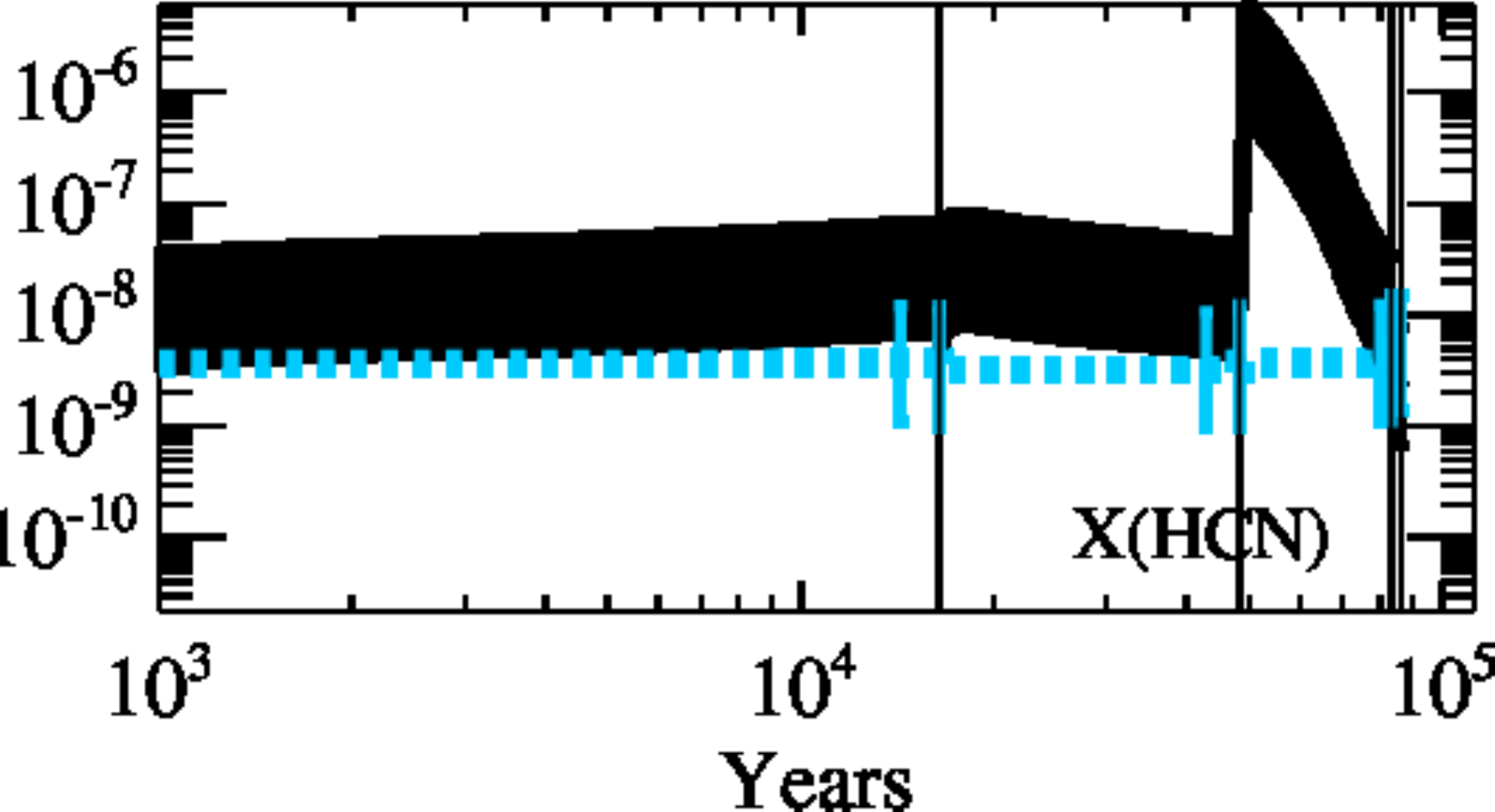}
\hspace{0.5cm}
\includegraphics[width=0.4\textwidth]{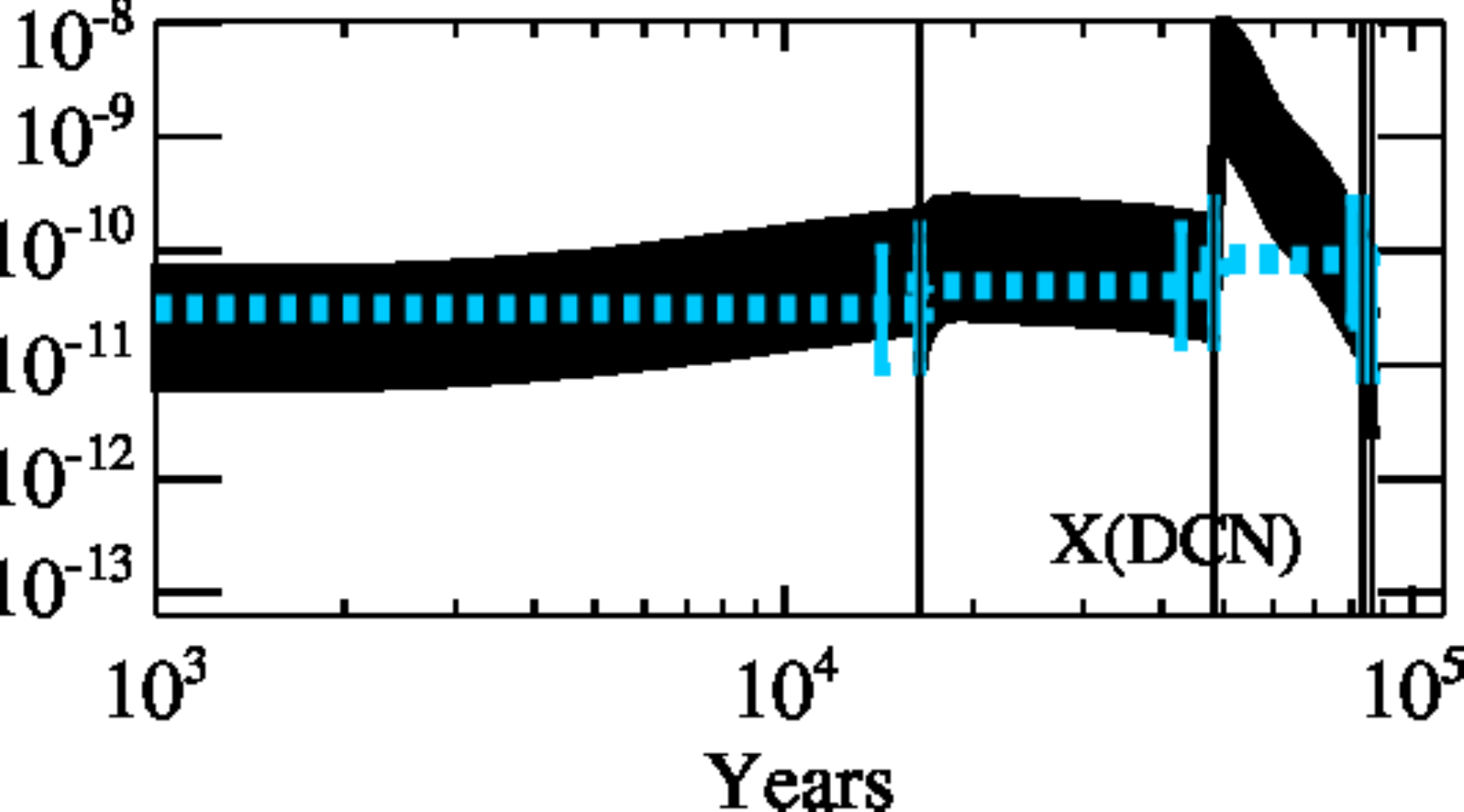}\\
\includegraphics[width=0.4\textwidth]{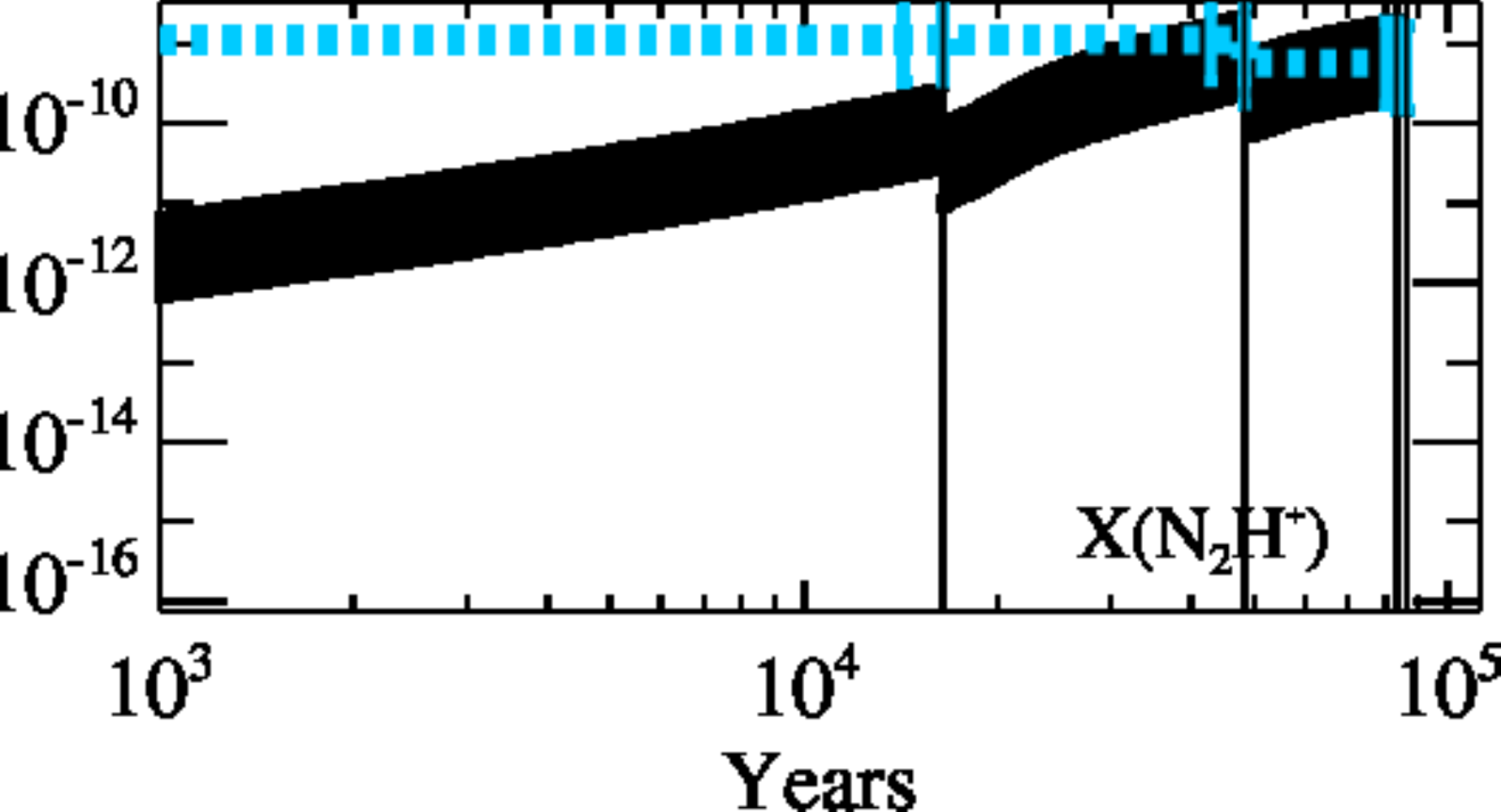}
\hspace{0.5cm}
\includegraphics[width=0.4\textwidth]{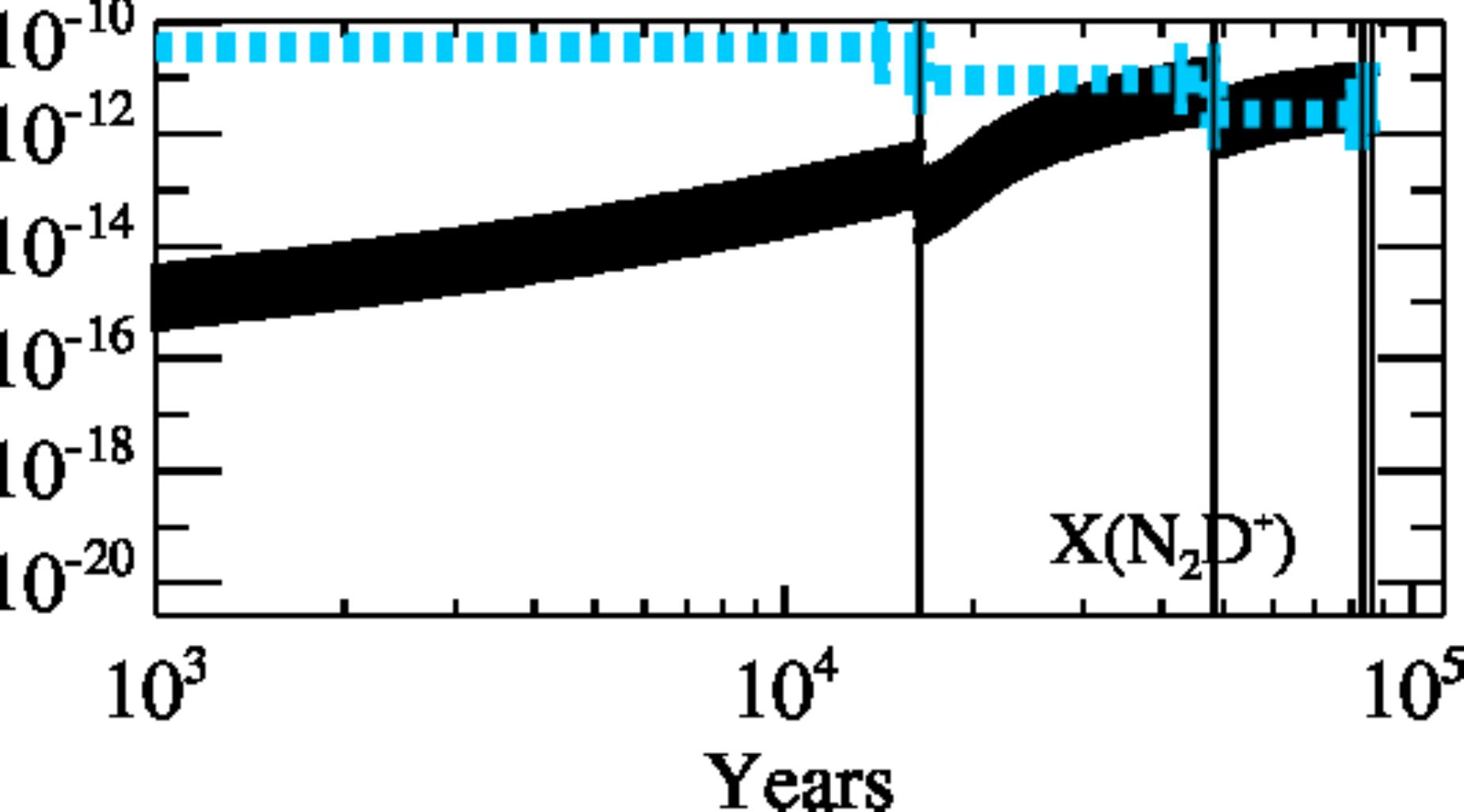}\\
%\vspace{0.8cm}\\
\caption{Modeled and observed relative abundances to H$_2$ are plotted
  for the IRDC-UCH{\sc ii} stages. The modeled values are shown by the
  black solid line, the observed values show the median of all
  detections and upper limits and are depicted by the blue dashed
  line. The error bars are indicated by the vertical marks.}
\label{fig:abundance_obs_model_31}
\end{figure*}

In general, the overall fit of all 18 molecules of the four fitted
phases is good. The specific time dependent evolution of the best-fit
abundances of molecules DCN, DNC, DCO$^+$, N$_2$D$^+$ and their
non-deuterated counterparts are shown in
Figure~\ref{fig:abundance_obs_model_31}. Between two consecutive
stages, the physical parameters are instantly changed and the
molecular species show a quick response to that change, followed by a
slower evolution under the new constant
conditions. Figures~\ref{fig:nx_stage1_31}--\ref{fig:nx_stage4_31}
show the modeled column densities in each stage separately. In total
the best-fit ages add up to $~85,000$~years, which is on the same
order as typical models of high-mass star formation of $~10^5$~years
\citep[e.g.,][]{mckee2003,tan2014}.

%\clearpage
\begin{figure*}
\centering
\includegraphics[width=0.3\textwidth]{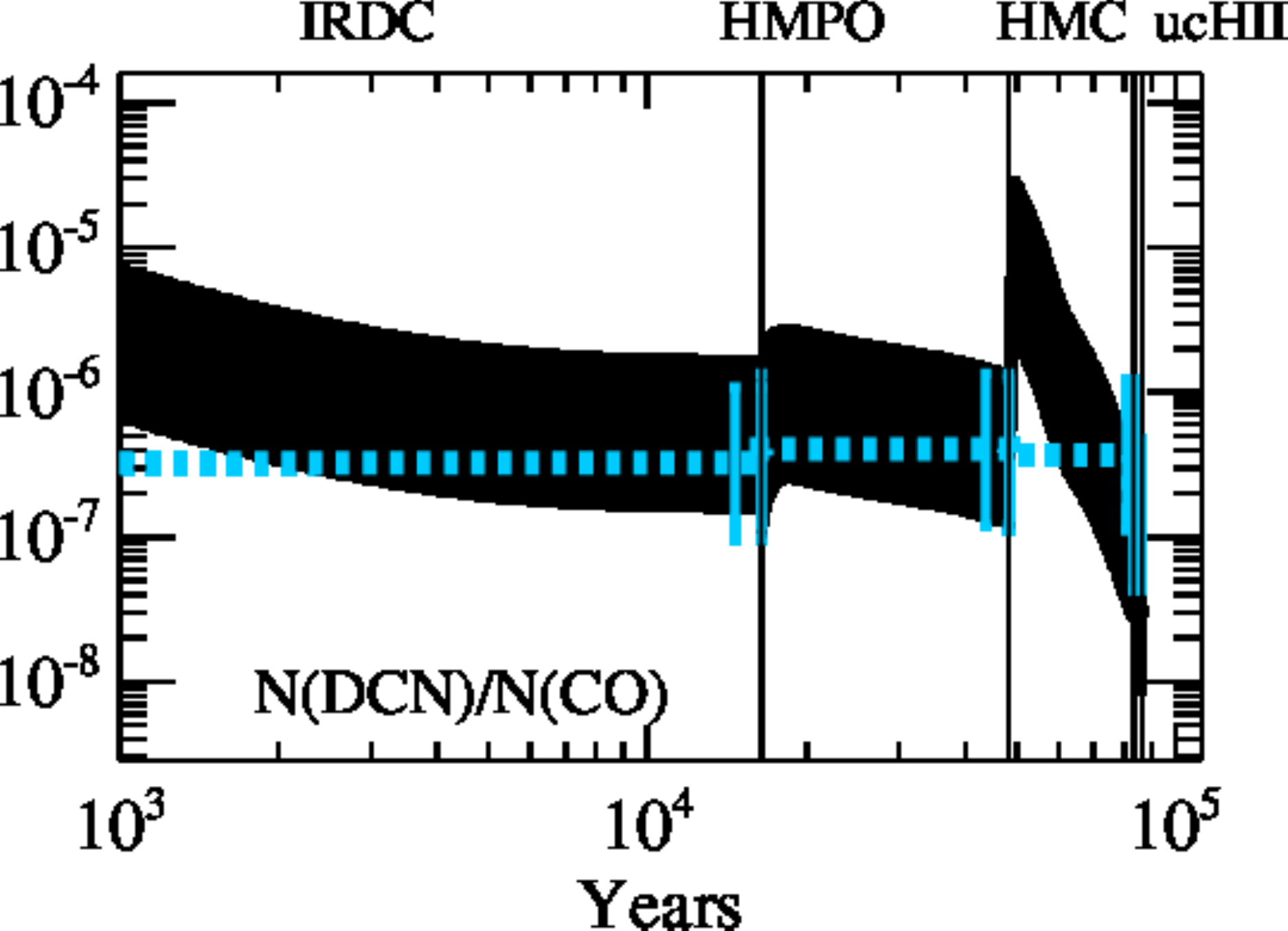}
\hspace{0.2cm}
\includegraphics[width=0.3\textwidth]{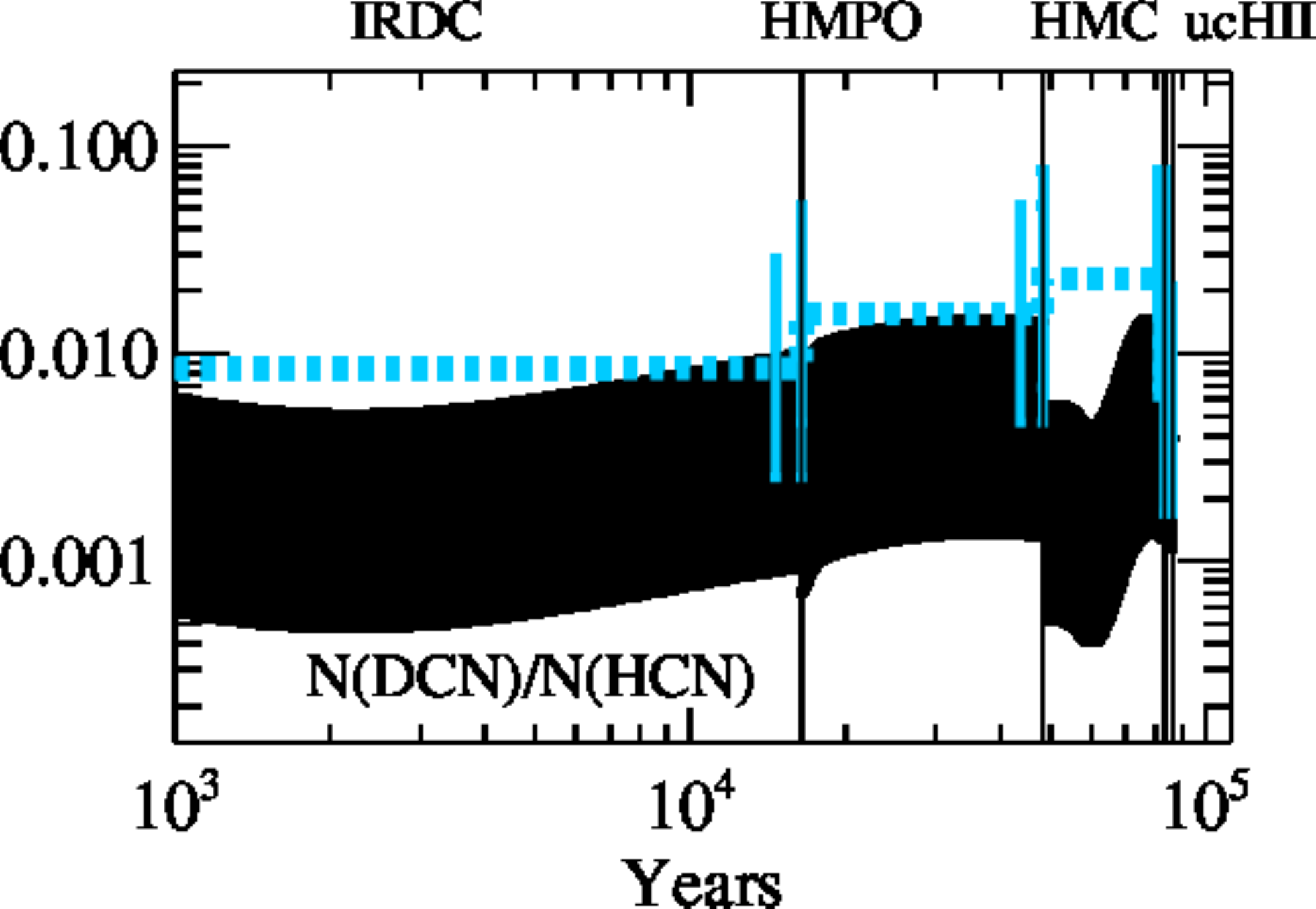}
\hspace{0.2cm}
\includegraphics[width=0.3\textwidth]{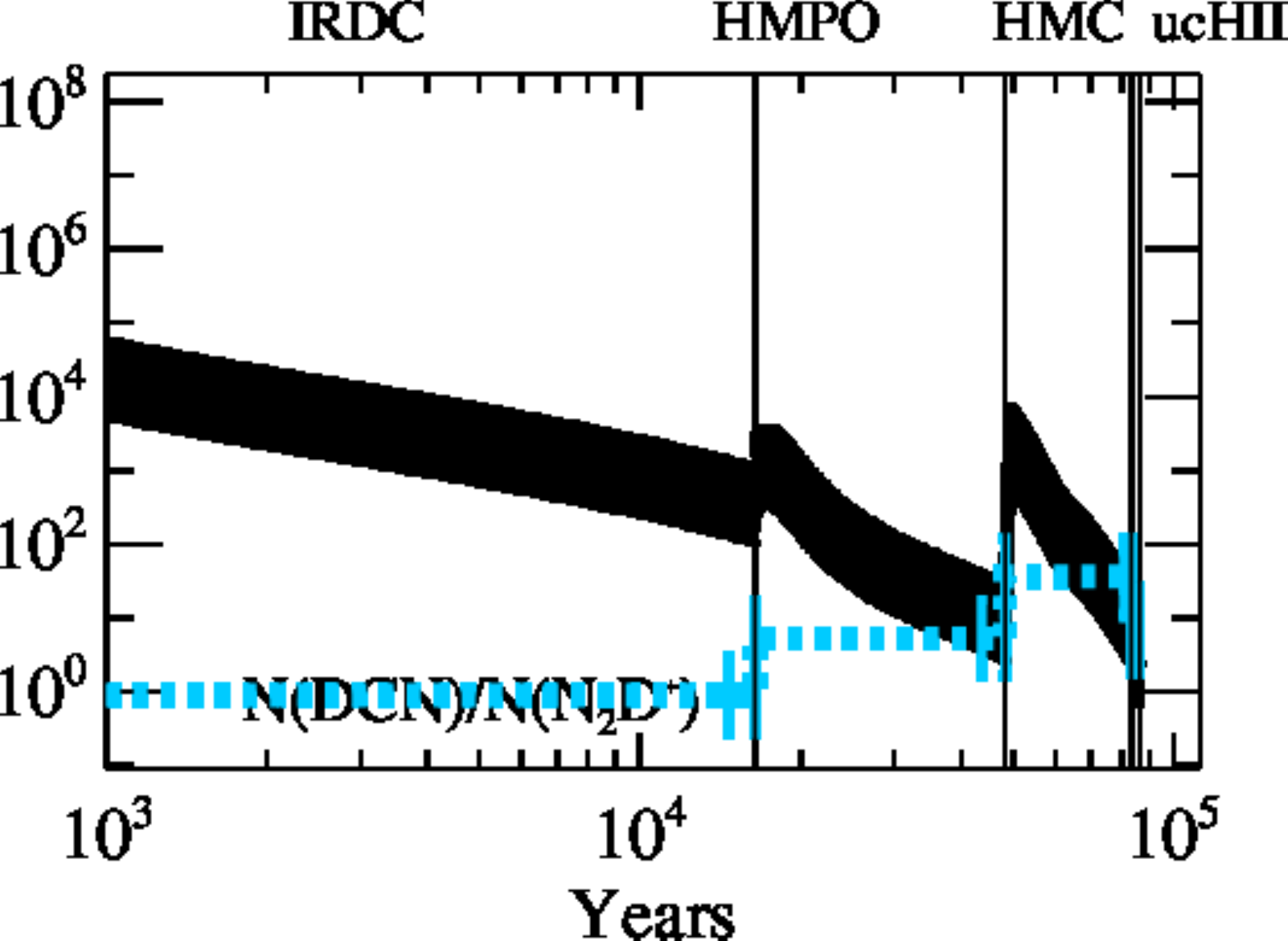}\\
\includegraphics[width=0.3\textwidth]{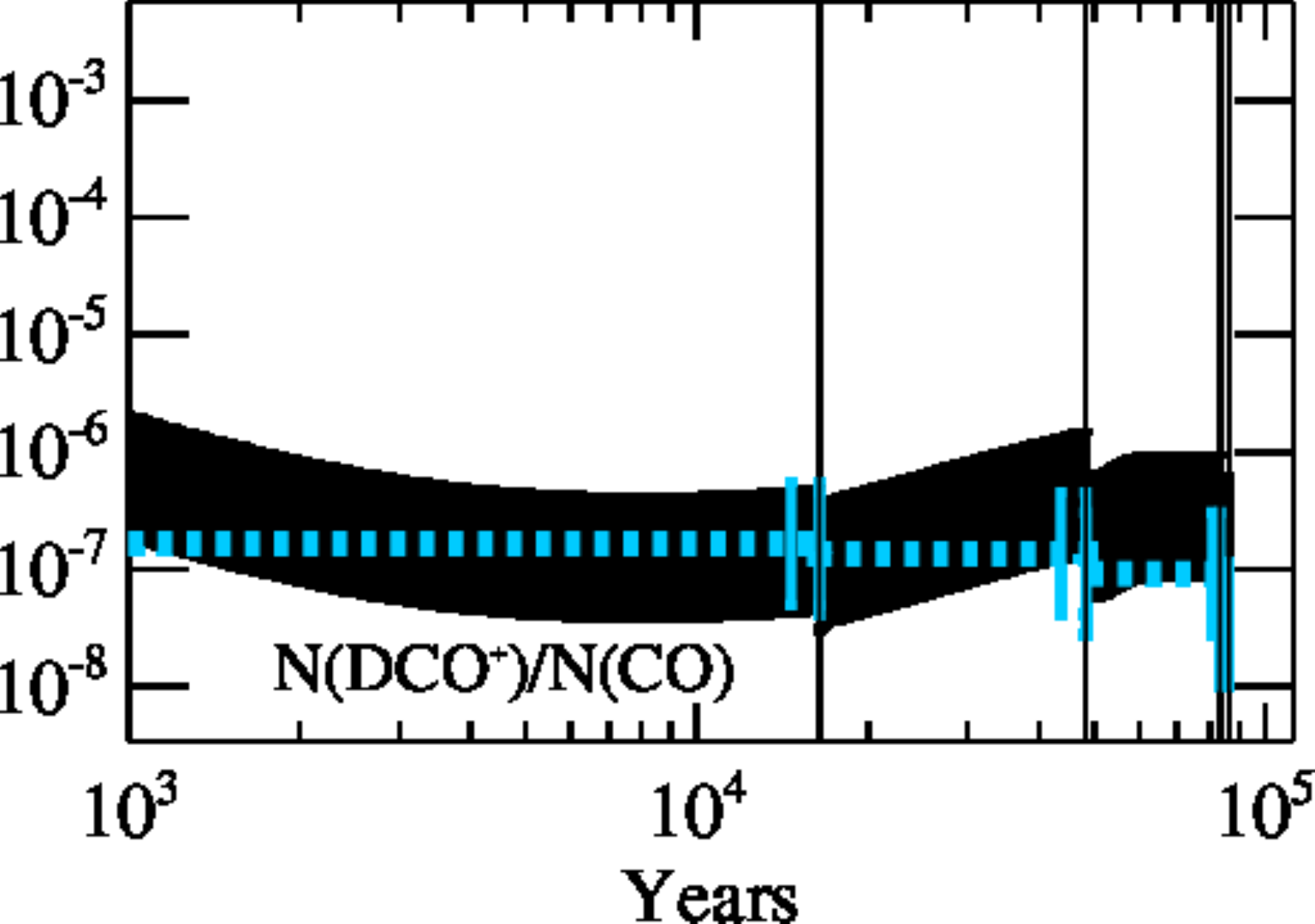}
\hspace{0.2cm}
\includegraphics[width=0.3\textwidth]{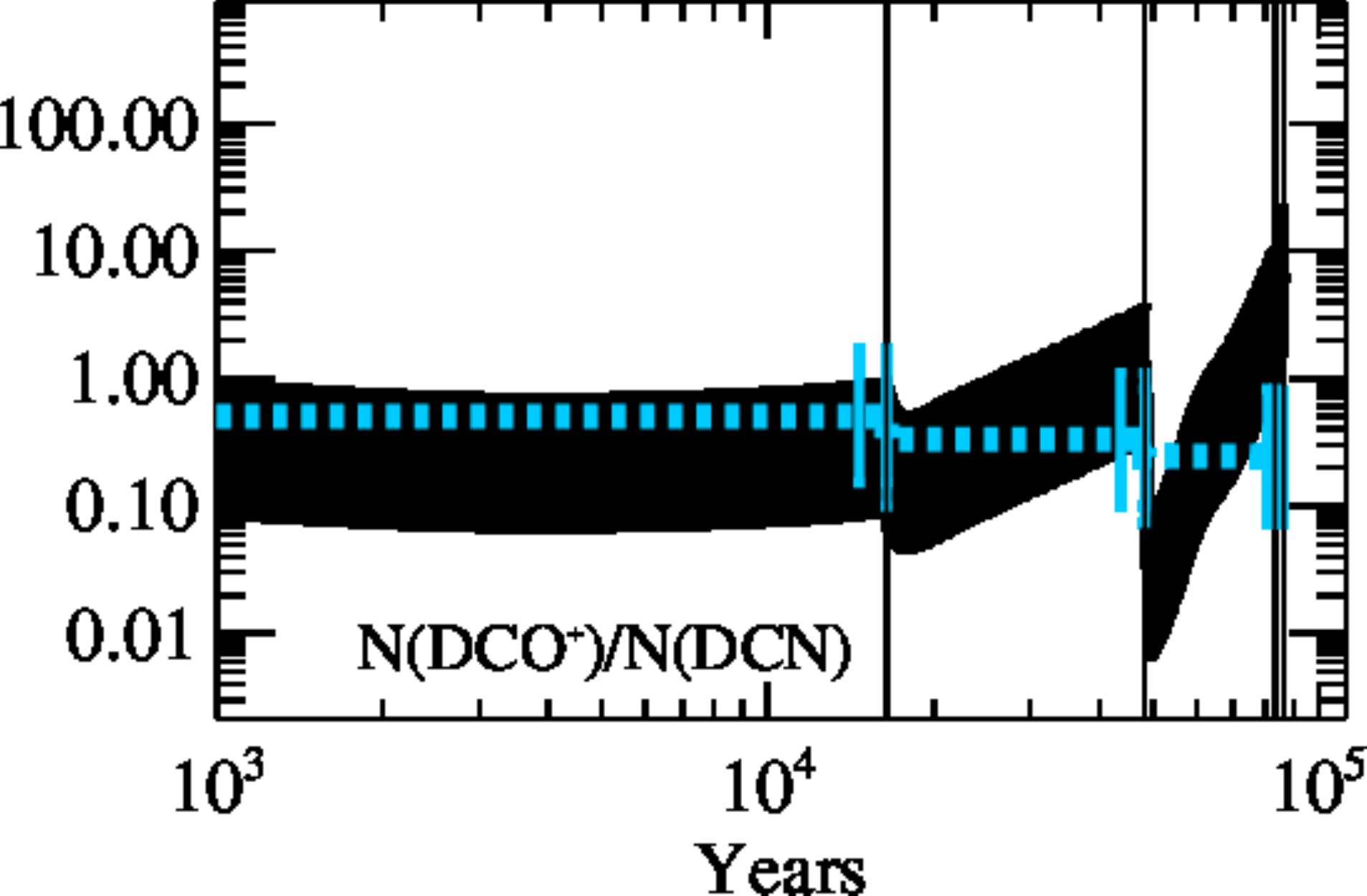}
\hspace{0.2cm}
\includegraphics[width=0.3\textwidth]{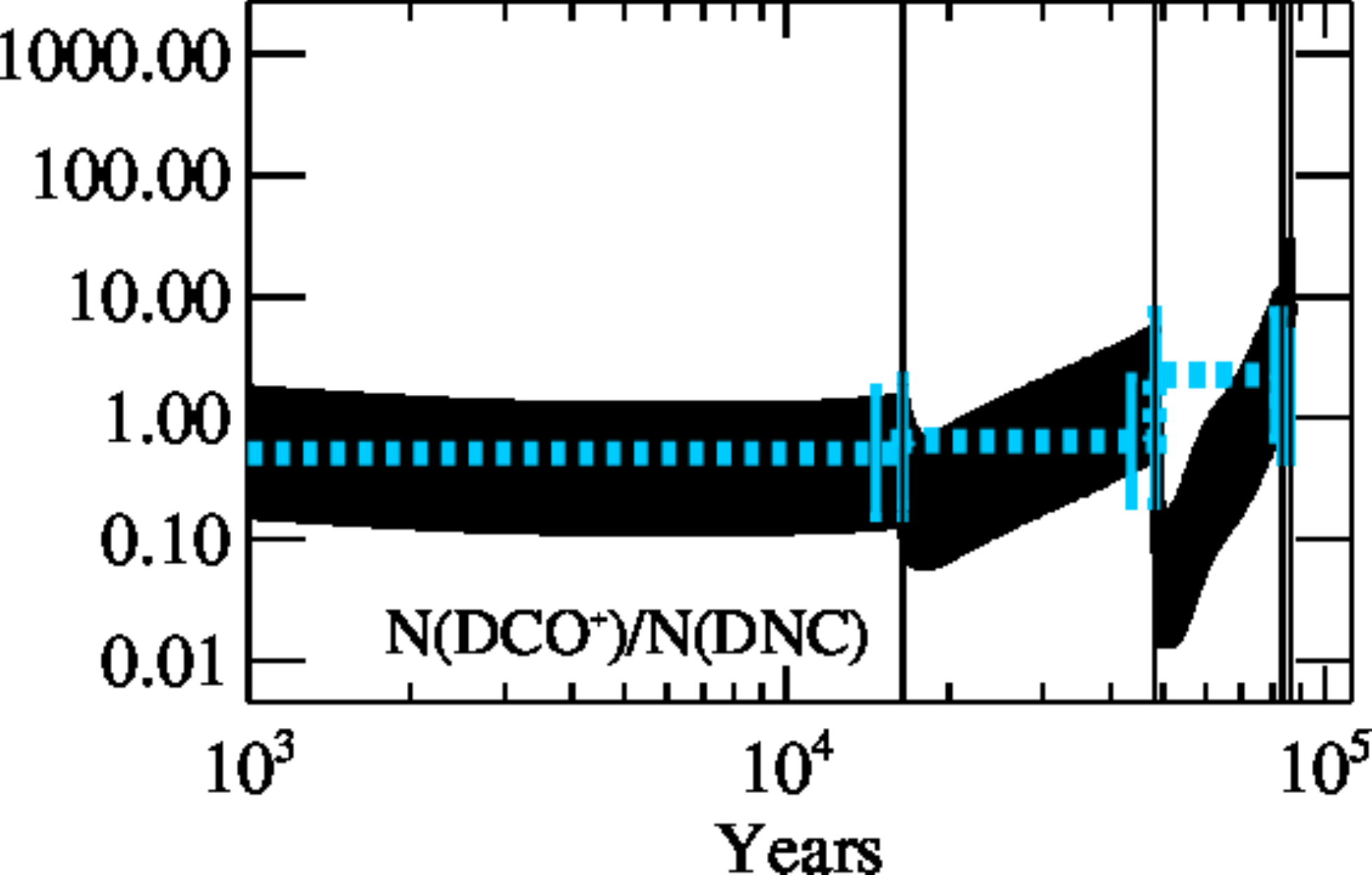}\\
\includegraphics[width=0.3\textwidth]{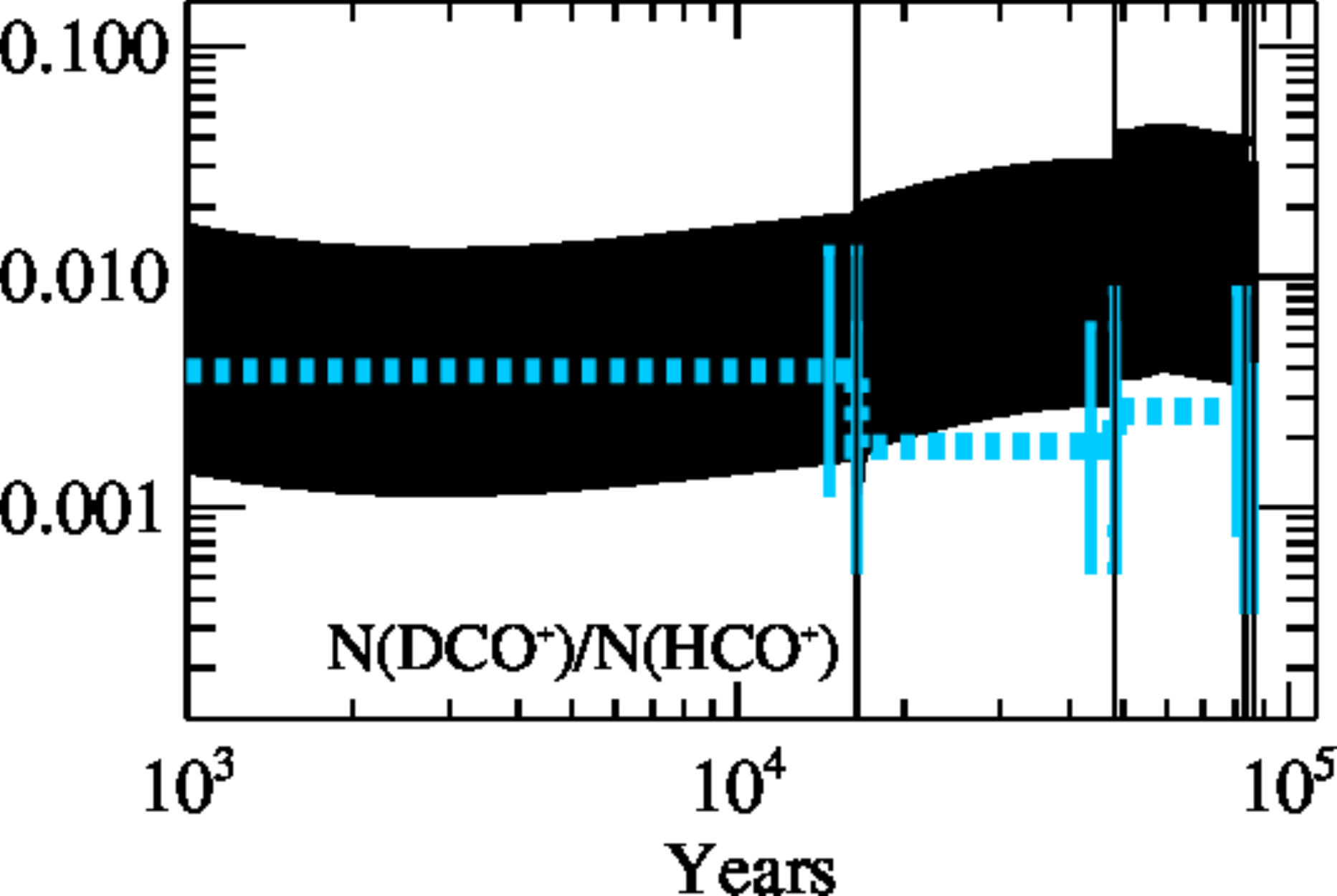}
\hspace{0.2cm}
\includegraphics[width=0.3\textwidth]{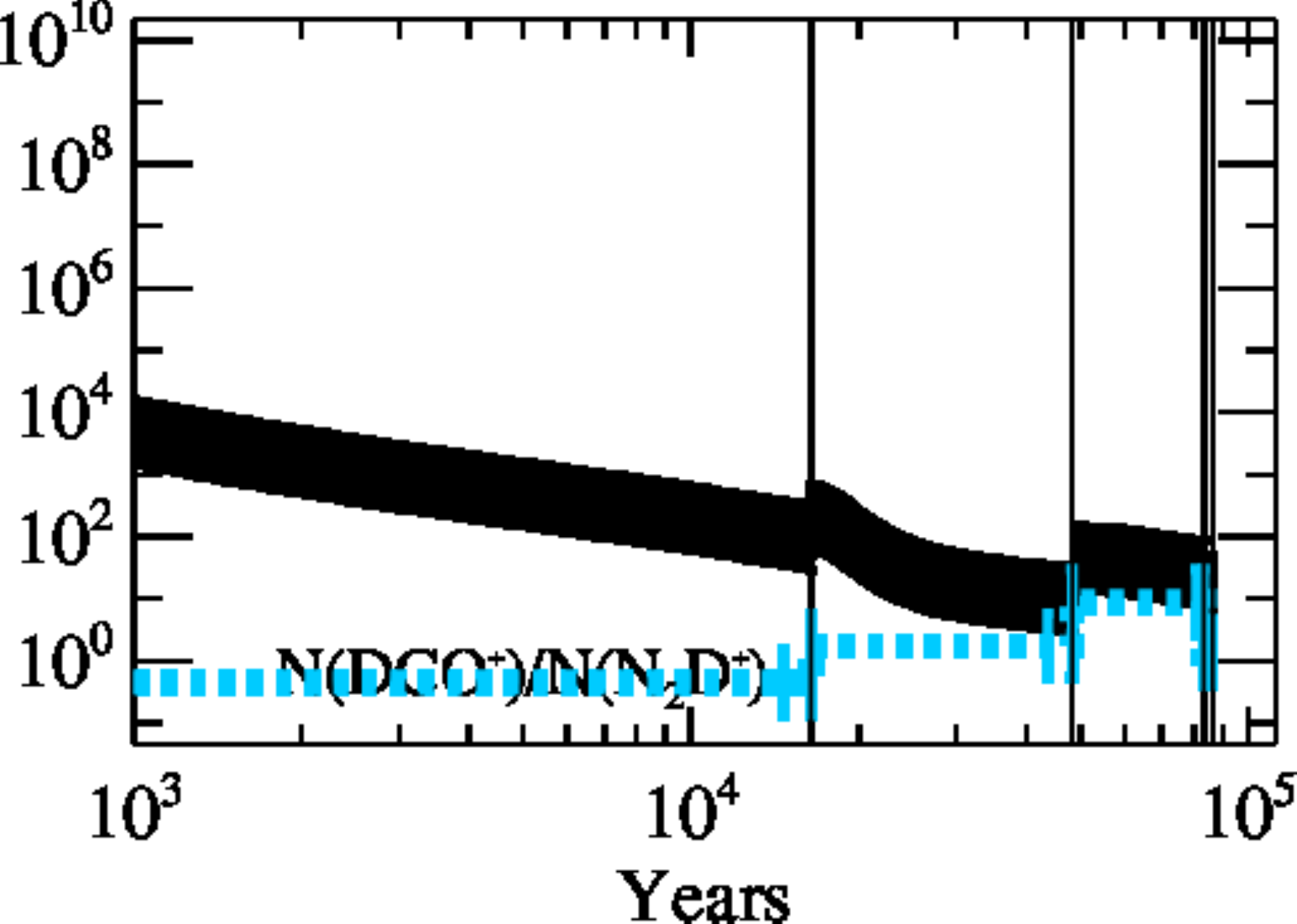}
\hspace{0.2cm}
\includegraphics[width=0.3\textwidth]{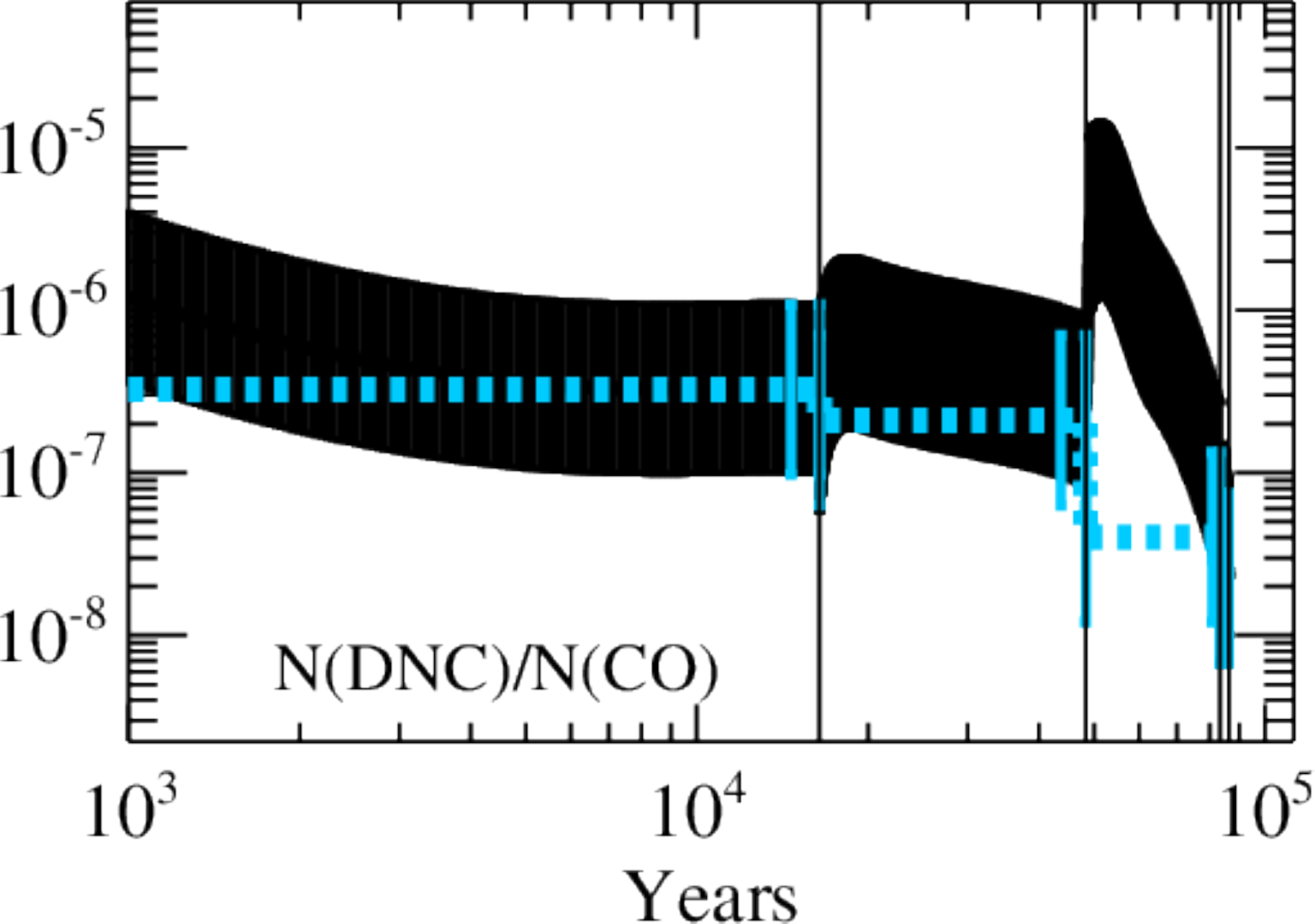}\\
\includegraphics[width=0.3\textwidth]{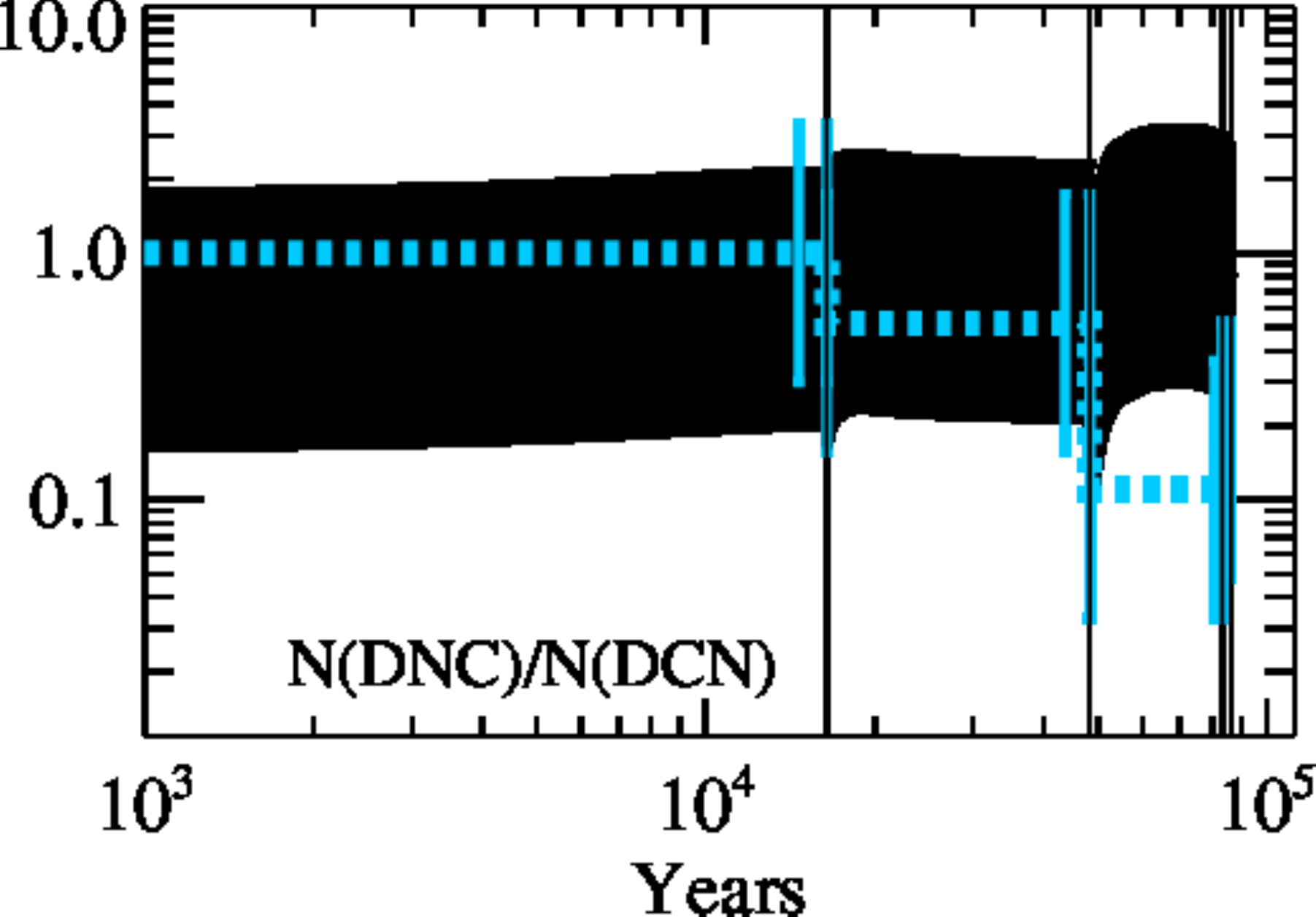}
\hspace{0.2cm}
\includegraphics[width=0.3\textwidth]{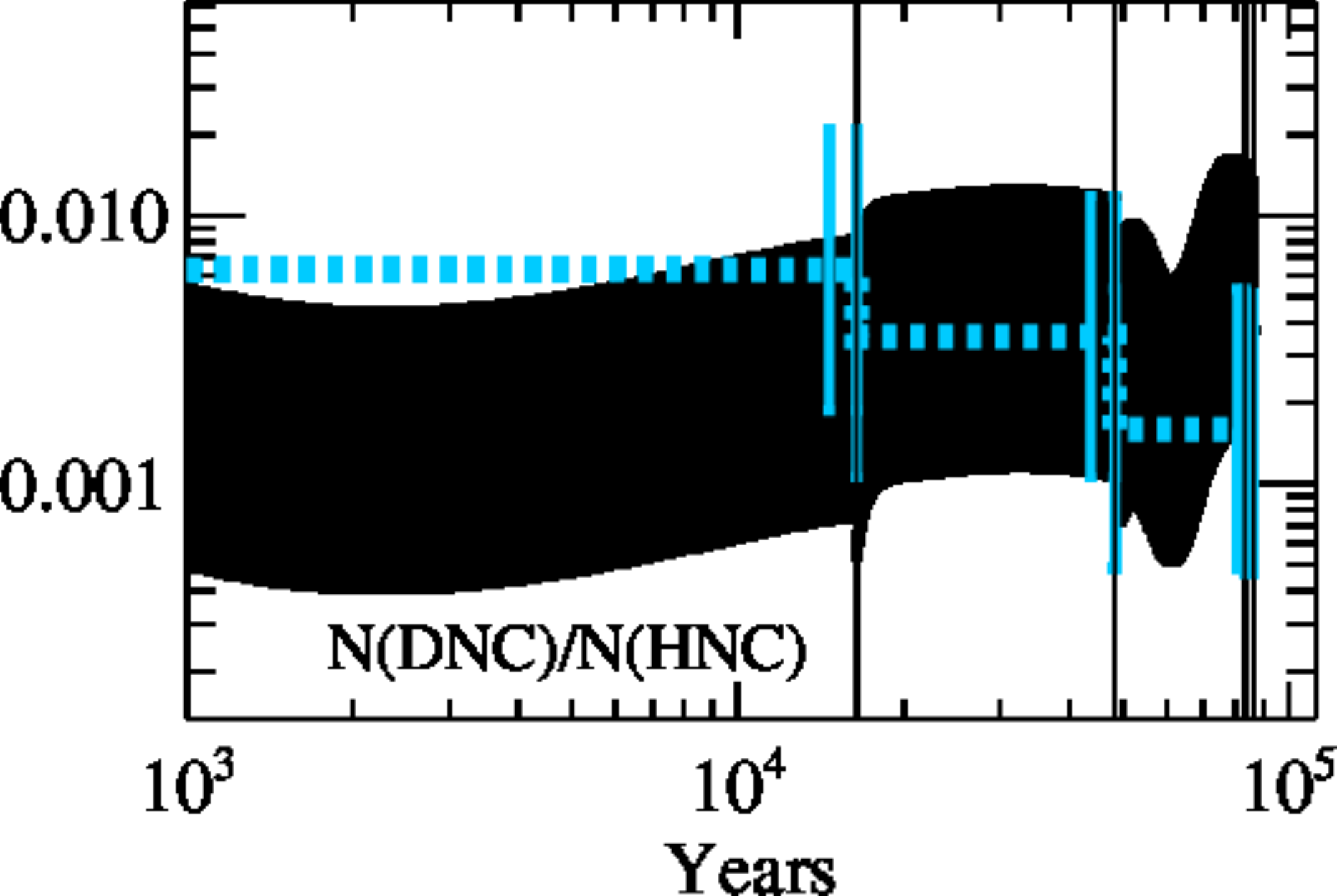}
\hspace{0.2cm}
\includegraphics[width=0.3\textwidth]{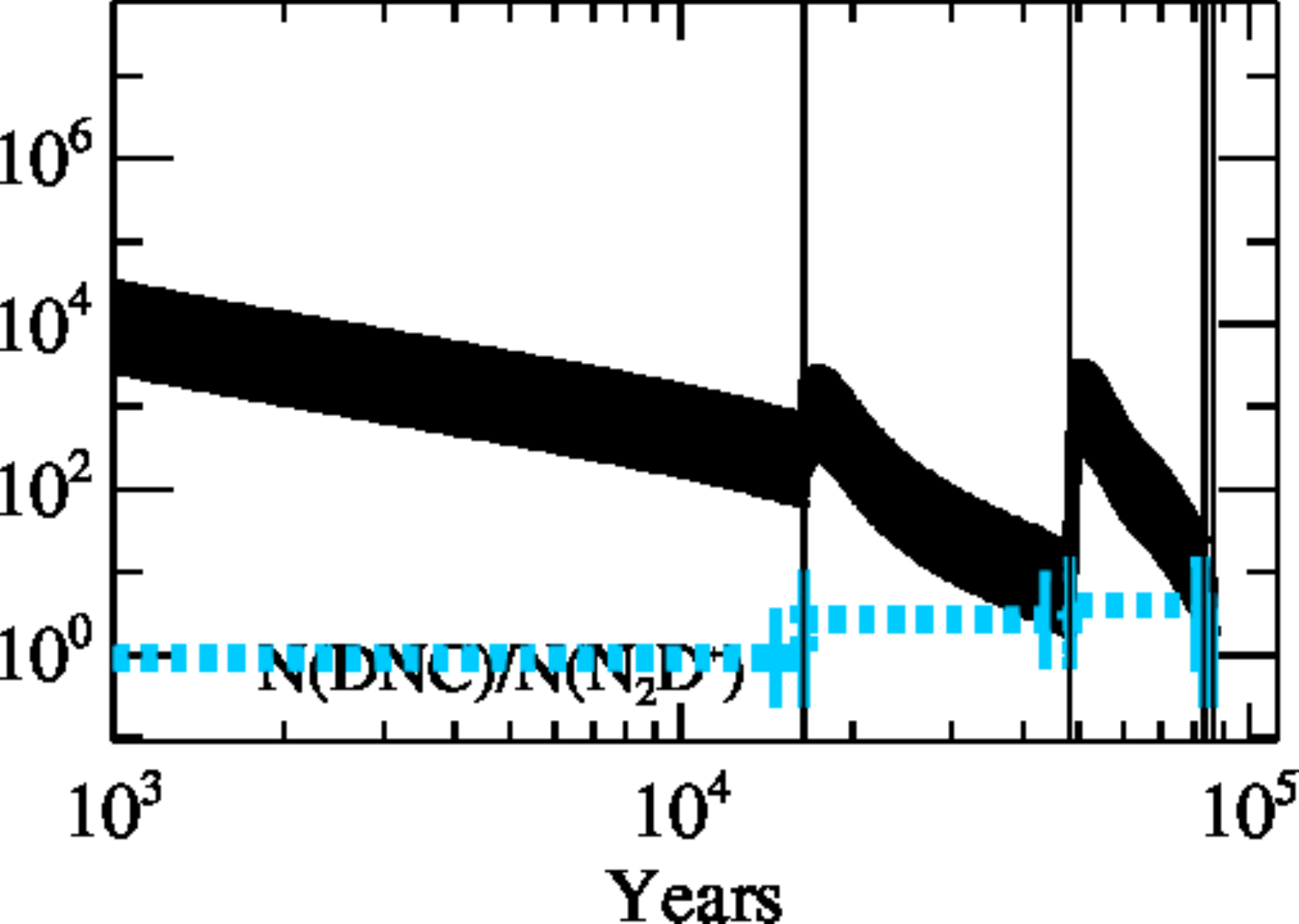}\\
\includegraphics[width=0.3\textwidth]{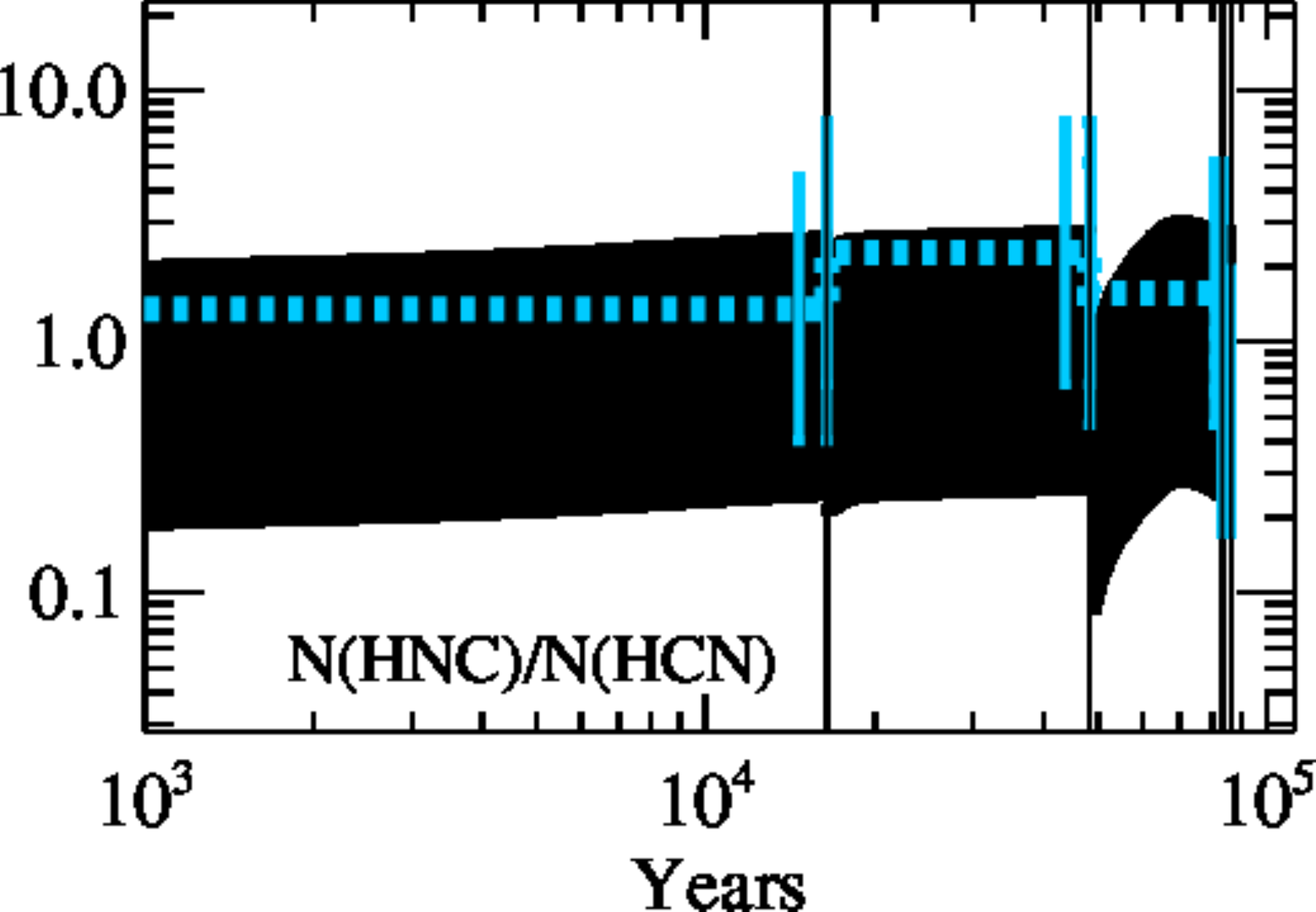}
\hspace{0.2cm}
\includegraphics[width=0.3\textwidth]{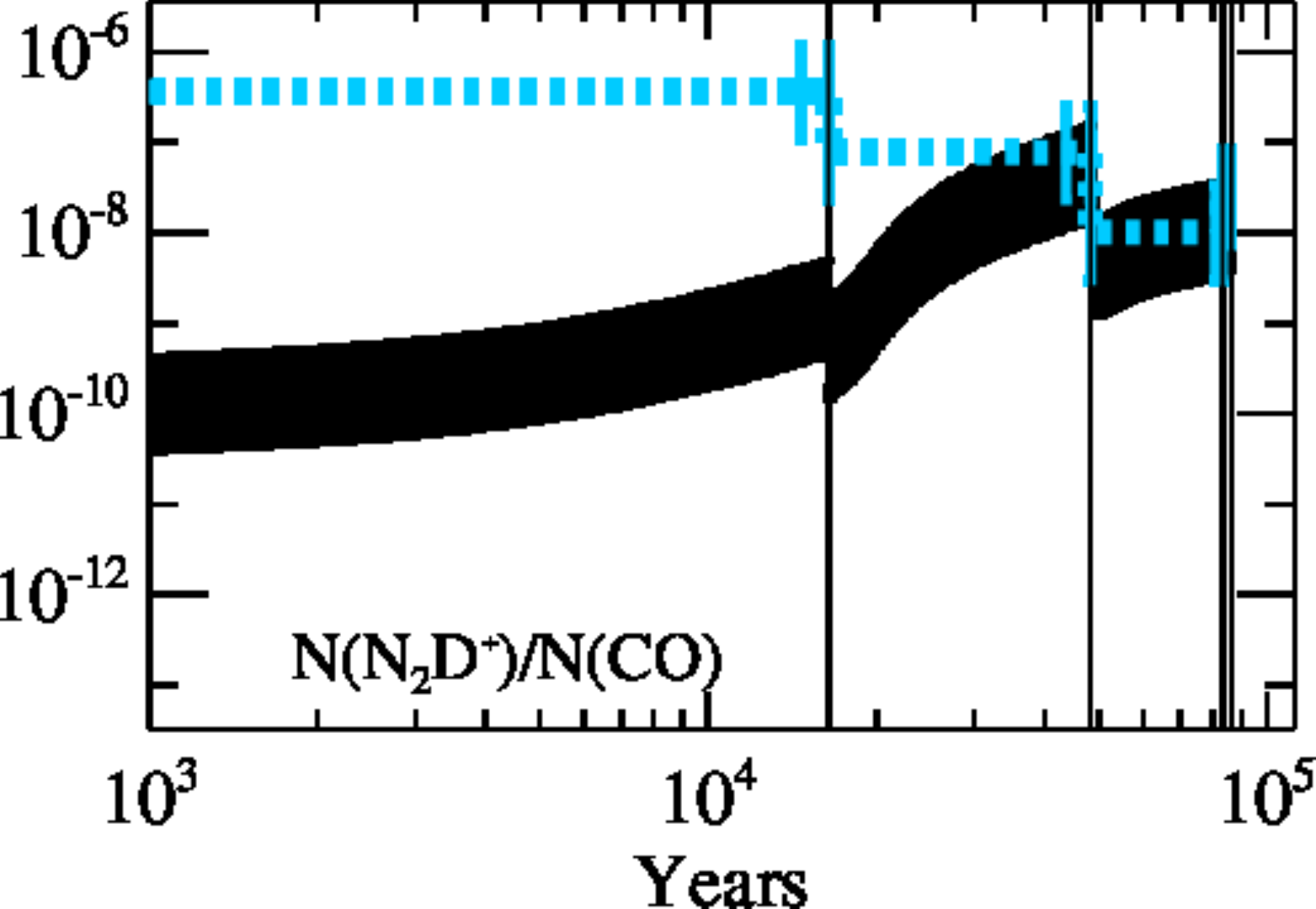}
\hspace{0.2cm}
\includegraphics[width=0.3\textwidth]{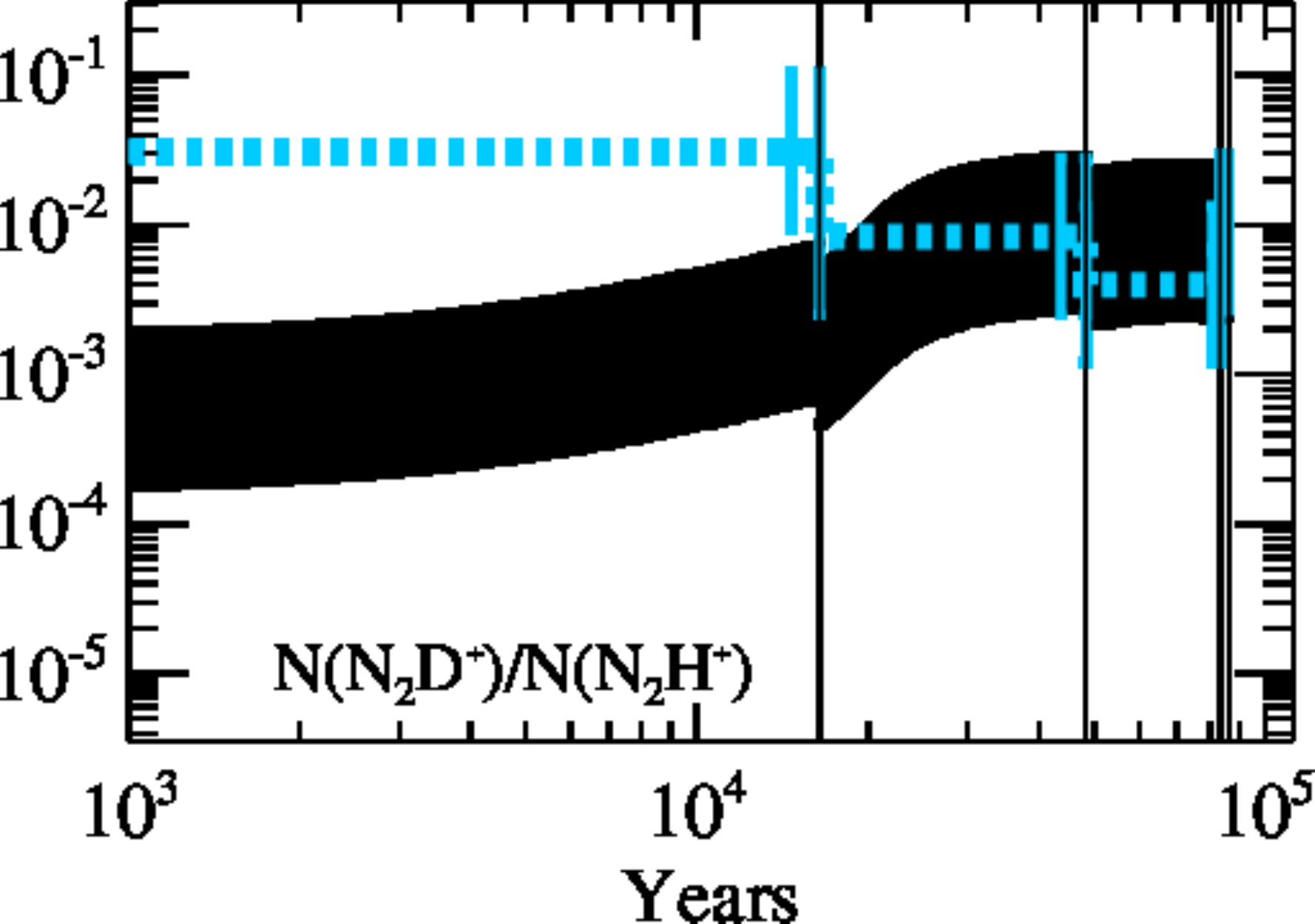}
\caption{Modeled and observed column density ratios are plotted for
  all stages. The modeled values are shown by the black solid line,
  the observed values show the median of all detections and upper
  limits and are depicted by the blue dashed line. The error bars are
  indicated by the vertical marks.}
\label{fig:ratio_obs_model_31}
\end{figure*}

The evolution for the different molecular ratios are shown in
Figure~\ref{fig:ratio_obs_model_31}. The modeled abundance ratios show
partly larger deviations from the observed values, since the
uncertainties from both molecules add up. In
Sect.~\ref{sec:ratios_observed} we discussed the ratio of DCN/DNC and
a possible chemical explanation based on a chemical network by
\citet{turner2001}. However, the chemical network used in our work
shows a different trend with a constant ratio rather than an increase
of DCN/DNC with temperature. Thus the observed trend disagrees with
our model predictions and the possibility of different chemical
formation pathways needs to be tested by future observations of more
deuterated molecules sharing the same formation pathways as DCN (e.g.,
light hydrocarbons with 2 or 3 C-atoms). The model agrees with the
constant ratio of HNC/HCN of unity with recent developments in the
collisional rates by \citet{sarrasin2010}.

\subsection{Comparison with best-fit models of Paper {\sc I}}
In Paper {\sc I} we iteratively fitted our model to an observational
dataset containing the median column densities of 14 different
non-deuterated species. The column densities were derived with typical
temperatures for each stage. The model followed the observed evolution
and successfully fit most of the species with few exceptions in
individual stages. The fits also constrained the physical structure of
the model and yielded mean best-fit temperatures for each stage. The
mean temperatures deviated from the temperatures used to derive the
observed column densities for the later stages. In a second step we
fitted the newly derived column densities. The temperature in the
best-fit models decreased again and we could not find a converging
solution.

In this work we used the best-fit temperatures of the first model from
Paper {\sc I} to recalculate the median column densities for the
previously analyzed 14 species and additional four deuterated
species. The temperature in the best-fit model of the IRDC stage is
lower by a factor of $~2$ compared to Paper {\sc I}, but the best-fit
temperatures of the IRDC stage for both, Paper {\sc I} and this work,
agree with typical IRDC temperatures. The mean density increased about
an order of magnitude. The best-fit lifetime in this work is a factor
of $~1.5$ longer, but agrees within the assumed uncertainties.  The
HMPO best-fit mean temperature in this work is lower by $~30$\%. The
mean density increased by a factor $3-4$. The best-fit lifetime is
shorter by a factor of $\sim2$.  The HMC best-fit mean temperature in
this work is higher by $~25$\%. The mean densities are similar. In
comparison to Paper {\sc I}, the best-fit lifetime is about $~20$\%
lower.  The UCH {\sc ii} best-fit mean temperatures are similar. The
best-fit ages deviate by a factor of $~4$, but are consistent within
their likely ranges. The mean density is smaller by one order of
magnitude.

The total lifetimes from the IRDC to the UCH{\sc ii} phase in Paper
{\sc I} adds up to $\sim 110,000$~years. This work yields a total
lifetime that is slightly shorter with $\sim 85,000$~years. The
molecules that could not be reproduced in Paper {\sc I} and this work
are not the same. Possible reasons are the additional four deuterated
molecules as free fit parameters that were treated with equal weight
as well as the revised excitation temperatures in order to derive
column densities compared to Paper {\sc I} (see
Section~\ref{sec:excitation_temp}). Although D-chemistry is seen as a
tracer of the thermal history and conditions of an object, and we see
weak correlations with luminosity as an evolutionary tracer, the model fits
did not improve substantially. Increasing the number of fitted
molecules from 14 to 18 species did not improve the achieved results
strongly, probably due to the less statistical importance of the
deuterated molecules within the total number of 18 species.

\subsection{Comparison with literature}
In low-mass star formation both observations and theory show that it
is possible to use the deuteration fraction of a molecule as an
evolutionary tracer
\citep[e.g.,][]{crapsi2005,aikawa2005,emprechtinger2009}. In a recent
work, \citet{fontani2011} observed 27 cores within high-mass
star-forming regions and derived the deuteration fraction of
N$_2$D$^+$/N$_2$H$^+$. They found a similar behavior as in the
low-mass regime. The deuteration fraction is the highest in the
starless cores with $D_{\rm {frac}}=0.26$ and decreases during the
formation of the protostellar objects to $D_{\rm {frac}}=0.04$. We
find in our work a similar trend, but the median deuteration fraction
we measure in our IRDC sample is lower by a factor of four compared to
their high-mass starless core sample. That might be due to the fact
that our IRDC sample contains starless as well as already more evolved
objects inhabiting 24$\mu$m sources. The same is true for the HMPO
samples of \citet{fontani2011} and this work. We find lower ratios by
about a factor of four. The largest D/H ratio we see is the
N$_2$D$^+$/N$_2$H$^+$ ratio in IRDC20081 of $D_{\rm
  {frac}}=0.12$. This source inhabits no point sources below $70\mu$m,
but it is close to a nearby source with extended emission. It is
presumably in a very early pre-stellar phase.  \citet{miettinen2011}
studied seven massive clumps within IRDCs and derived deuteration
fractions for N$_2$H$^+$ between $0.002-0.028$ and for HCO$^+$ between
$0.0002-0.014$. While the range of values for HCO$^+$ is comparable
with the IRDCs in our sample, the deuteration fractions for N$_2$H$^+$
are about one order of magnitude higher in this work. However, that
agrees with the high number of upper limits due to non-detections of
N$_2$D$^+$ in this work. \citet{chen2011} observed several cores in
various evolutionary stages and found deuteration values of N$_2$H$^+$
between $0.004-0.1$. The source lists have three targets in
common. For the HMPO18151 both works agree on a value of $D_{\rm
  {frac}}=0.01$. For the other two sources in common, IRDC18151 and
IRDC18223, we find a slightly higher value of $D_{\rm {frac}}=0.03$
instead of $D_{\rm {frac}}=0.02$. While the highest value they found
is similar to the maximum value we observe, they detected N$_2$D$^+$
in a larger number of sources and found also lower deuteration
fractions, especially in the more evolved sources for which we have no
detections of N$_2$D$^+$.  Work by \citet{crapsi2005} studying the
N$_2$D$^+$/N$_2$H$^+$ ratio in low-mass starless cores revealed
deuteration fractions on the percentage level up to almost $50\%$ in
the most extreme case of Oph D. The maximum ratio observed in our
survey is $\sim 0.1$.  \citet{sakai2012} measured deuteration
fractions of DNC/HNC for IRDCs and HMPOs. They find $D_{\rm
  {frac}}=0.003-0.03$ for MSX (Midcourse Space Experiment) dark
sources at $8\mu$m and $D_{\rm {frac}}=0.005-0.01$ for HMPOs. These
ratios agree well with $D_{\rm {frac}}=0.004-0.02$ for our IRDC sample
and $D_{\rm {frac}}=0.0015-0.015$ for our HMPO sample.  In a recent
study, \citet{kong2013} explored the effect of different parameters on
the deuteration fraction of N$_2$H$^+$ in dense and cold
environments. They found an increase of the deuteration fraction with
decreasing temperature. However, they also found a positive
correlation with the density, which is not clearly seen in our
data. They employed a 0D model to study deuteration in star-forming
clouds and found that the timescales to reach equilibrium in the
abundances is on the order of several free-fall times
($\sim10^6$~years) for typical densities of high-mass cores. In
contrast, the best-fit models in our work are not supposed to reach
chemical equilibrium and thus our model predicts a total timescale of
$\sim10^5$~years for the best-fit models. This is likely due to the
fact that in a $1$D model with a power law density profile with higher
densities in the inner region timescales in the chemical evolution are
shorter than in a $0$D model.

\section{Conclusion}\label{sec:conclusion}
In this work we extended the analysis from \citet{gerner2014} of the
chemical evolution in 59 high-mass star forming regions for deuterated
molecules. We measured beam averaged column densities and deuteration
fractions of the four deuterated species DCO$^+$, DCN, DNC, and
N$_2$H$^+$. We find an overall high detection fraction towards the
high-mass star-forming regions, except for N$_2$D$^+$, which is likely
due to the limited sensitivity of our survey. The detection fraction
of DCO$^+$, DCN and DNC increases from IRDCs to HMPOs and peaks at the
HMC sample, where we detect all three molecules in all sources but
one, and drops again towards the UCH{\sc ii} stage.

The (3-2) lines are sub-thermally populated. The median deuteration
fractions excluding upper limits are $~0.02$ for DCN, $~0.005$ for
DNC, $~0.0025$ for DCO$^+$ and $~0.02$ for N$_2$D$^+$. The deuteration
fractions of DNC, DCO$^+$ and N$_2$D$^+$ show a decreasing trend from
IRDCs over HMPOs and HMCs to UCH{\sc ii} regions, which supports the
hypothesis that deuterated molecules may be used as indicators of the
evolutionary stage in high-mass star-forming regions.

In general, we find no correlation between the deuteration fraction of
the various molecules and physical parameters for DCO$^+$/HCO$^+$ and
DCN/DNC. Only N$_2$D$^+$/N$_2$H$^+$ shows a slight anti-correlations
with the luminosity and the FWHM.  The total measured column density
of the gas does not correlate with the deuteration fraction. This
result hints towards the interpretation that, within the range of
probed densities, deuteration depends stronger on the temperature of
the environment than on the column density, and is enhanced in colder,
less luminous regions. However, for example \citet{albertsson2013}
found in their models differences in the deuteration with density at
lower densities.

The slight anti-correlation with luminosity and FWHM, which is related
to the evolutionary stage, indicates that evolutionary stage plays an
important role for the deuteration fraction. But its huge scatter
within single stages leads to the assumption that the evolution might
not be the only factor.

Furthermore, we fitted the observed data with a chemical model and
find reasonable physical model fits. Combining observations of
non-deuterated and deuterated species to obtain best-fits lead to
reasonable chemical and physical models. The best-fit models produce
reasonably good results for all stages. Due to large uncertainties in
the observations and the model and a large spread of observed values
within the 4 subsamples we could not substantially improve the
best-fit model results compared to Paper {\sc I}. Another reason is
that the combination of the 4 deuterated molecules with the 14
non-deuterated species probably reduced the statistical importance of
each molecule and thus the effect on the best-fit of the model.

\begin{acknowledgements}
The authors thank the anonymous referee for helping to improve the
paper. TG is supported by the Sonderforschungsbereich SFB 881 ‘The
Milky Way System’ (subproject B3) of the German Research Foundation
(DFG). T.G. is member of the IMPRS for Astronomy \& Cosmic Physics at
the University of Heidelberg.  This research made use of NASA's
Astrophysics Data System.  DS acknowledges support by the {\it
  Deutsche Forschungsgemeinschaft} through SPP~1385: ``The first ten
million years of the solar system - a planetary materials approach''
(SE 1962/1-2 and SE 1962/1-3).  TA acknowledges funding from the
European Community's Seventh Framework Programme [FP7/2007-2013] under
grant agreement no. 238258.  In the reduction and analysis of the data
we made use of the GILDAS software public available at
http://www.iram.fr/IRAMFR/GILDAS . The SMT and Kitt Peak 12m are
operated by the Arizona Radio Observatory (ARO), Steward Observatory,
University of Arizona, with support through the NSF University Radio
Observatories program (URO: AST-1140030).
\end{acknowledgements}
 
%\bibpunct{(}{)}{;}{a}{}{,}

%\bibliography{/home/gerner/Dropbox/MPIA/tex/bibliography} %im MPIA  

%\bibliographystyle{aa}    % this does the style, apj.bst necessary
%\bibliography{../bibliography}  %vom Laptop

%\bibliography{~/Dropbox/MPIA/paperdrafts}  %vom Laptop

\begin{thebibliography}{73}
\expandafter\ifx\csname natexlab\endcsname\relax\def\natexlab#1{#1}\fi

\bibitem[{{Aikawa} {et~al.}(2005){Aikawa}, {Herbst}, {Roberts}, \&
  {Caselli}}]{aikawa2005}
{Aikawa}, Y., {Herbst}, E., {Roberts}, H., \& {Caselli}, P. 2005, \apj, 620,
  330

\bibitem[{{Albertsson} {et~al.}(2014{\natexlab{a}}){Albertsson}, {Indriolo},
  {Kreckel}, {Semenov}, {Crabtree}, \& {Henning}}]{albertsson2014b}
{Albertsson}, T., {Indriolo}, N., {Kreckel}, H., {et~al.} 2014{\natexlab{a}},
  ArXiv e-prints

\bibitem[{{Albertsson} {et~al.}(2014{\natexlab{b}}){Albertsson}, {Semenov}, \&
  {Henning}}]{albertsson2014}
{Albertsson}, T., {Semenov}, D., \& {Henning}, T. 2014{\natexlab{b}}, \apj,
  784, 39

\bibitem[{{Albertsson} {et~al.}(2013){Albertsson}, {Semenov}, {Vasyunin},
  {Henning}, \& {Herbst}}]{albertsson2013}
{Albertsson}, T., {Semenov}, D.~A., {Vasyunin}, A.~I., {Henning}, T., \&
  {Herbst}, E. 2013, \apjs, 207, 27

\bibitem[{{Barnes} {et~al.}(2011){Barnes}, {Yonekura}, {Fukui}, {Miller},
  {M{\"u}hlegger}, {Agars}, {Miyamoto}, {Furukawa}, {Papadopoulos}, {Jones},
  {Hernandez}, {O'Dougherty}, \& {Tan}}]{barnes2011}
{Barnes}, P.~J., {Yonekura}, Y., {Fukui}, Y., {et~al.} 2011, \apjs, 196, 12

\bibitem[{{Beuther} {et~al.}(2007{\natexlab{a}}){Beuther}, {Churchwell},
  {McKee}, \& {Tan}}]{beuther2006b}
{Beuther}, H., {Churchwell}, E.~B., {McKee}, C.~F., \& {Tan}, J.~C.
  2007{\natexlab{a}}, in Protostars and Planets V, ed. B.~{Reipurth},
  D.~{Jewitt}, \& K.~{Keil}, 165--180

\bibitem[{{Beuther} {et~al.}(2002{\natexlab{a}}){Beuther}, {Schilke}, {Gueth},
  {McCaughrean}, {Andersen}, {Sridharan}, \& {Menten}}]{beuther2002d}
{Beuther}, H., {Schilke}, P., {Gueth}, F., {et~al.} 2002{\natexlab{a}}, \aap,
  387, 931

\bibitem[{{Beuther} {et~al.}(2002{\natexlab{b}}){Beuther}, {Schilke}, {Menten},
  {Motte}, {Sridharan}, \& {Wyrowski}}]{beuther2002a}
{Beuther}, H., {Schilke}, P., {Menten}, K.~M., {et~al.} 2002{\natexlab{b}},
  \apj, 566, 945

\bibitem[{{Beuther} {et~al.}(2002{\natexlab{c}}){Beuther}, {Schilke},
  {Sridharan}, {Menten}, {Walmsley}, \& {Wyrowski}}]{beuther2002b}
{Beuther}, H., {Schilke}, P., {Sridharan}, T.~K., {et~al.} 2002{\natexlab{c}},
  \aap, 383, 892

\bibitem[{{Beuther} {et~al.}(2012){Beuther}, {Tackenberg}, {Linz}, {Henning},
  {Krause}, {Ragan}, {Nielbock}, {Launhardt}, {Schmiedeke}, {Schuller},
  {Carlhoff}, {Nguyen-Luong}, \& {Sakai}}]{beuther2012}
{Beuther}, H., {Tackenberg}, J., {Linz}, H., {et~al.} 2012, \aap, 538, A11

\bibitem[{{Beuther} {et~al.}(2007{\natexlab{b}}){Beuther}, {Zhang}, {Bergin},
  {Sridharan}, {Hunter}, \& {Leurini}}]{beuther2007d}
{Beuther}, H., {Zhang}, Q., {Bergin}, E.~A., {et~al.} 2007{\natexlab{b}}, \aap,
  468, 1045

\bibitem[{{Beuther} {et~al.}(2004){Beuther}, {Zhang}, {Greenhill}, {Reid},
  {Wilner}, {Keto}, {Marrone}, {Ho}, {Moran}, {Rao}, {Shinnaga}, \&
  {Liu}}]{beuther2004g}
{Beuther}, H., {Zhang}, Q., {Greenhill}, L.~J., {et~al.} 2004, \apjl, 616, L31

\bibitem[{{Bourke} {et~al.}(2012){Bourke}, {Myers}, {Caselli}, {Di Francesco},
  {Belloche}, {Plume}, \& {Wilner}}]{bourke2012}
{Bourke}, T.~L., {Myers}, P.~C., {Caselli}, P., {et~al.} 2012, \apj, 745, 117

\bibitem[{{Campbell} {et~al.}(1995){Campbell}, {Butner}, {Harvey}, {Evans},
  {Campbell}, \& {Sabbey}}]{campbell1995}
{Campbell}, M.~F., {Butner}, H.~M., {Harvey}, P.~M., {et~al.} 1995, \apj, 454,
  831

\bibitem[{{Carey} {et~al.}(2000){Carey}, {Feldman}, {Redman}, {Egan},
  {MacLeod}, \& {Price}}]{carey2000}
{Carey}, S.~J., {Feldman}, P.~A., {Redman}, R.~O., {et~al.} 2000, \apjl, 543,
  L157

\bibitem[{{Caselli} \& {Ceccarelli}(2012)}]{caselli2012}
{Caselli}, P. \& {Ceccarelli}, C. 2012, \aapr, 20, 56

\bibitem[{{Caselli} {et~al.}(2002){Caselli}, {Stantcheva}, {Shalabiea},
  {Shematovich}, \& {Herbst}}]{caselli2002c}
{Caselli}, P., {Stantcheva}, T., {Shalabiea}, O., {Shematovich}, V.~I., \&
  {Herbst}, E. 2002, \planss, 50, 1257

\bibitem[{{Cesaroni} {et~al.}(1994){Cesaroni}, {Churchwell}, {Hofner},
  {Walmsley}, \& {Kurtz}}]{cesaroni1994}
{Cesaroni}, R., {Churchwell}, E., {Hofner}, P., {Walmsley}, C.~M., \& {Kurtz},
  S. 1994, \aap, 288, 903

\bibitem[{{Chen} {et~al.}(2011){Chen}, {Liu}, {Su}, \& {Wang}}]{chen2011}
{Chen}, H.-R., {Liu}, S.-Y., {Su}, Y.-N., \& {Wang}, M.-Y. 2011, \apj, 743, 196

\bibitem[{{Chen} {et~al.}(2006){Chen}, {Welch}, {Wilner}, \&
  {Sutton}}]{chen2006}
{Chen}, H.-R., {Welch}, W.~J., {Wilner}, D.~J., \& {Sutton}, E.~C. 2006, \apj,
  639, 975

\bibitem[{{Churchwell} {et~al.}(1990){Churchwell}, {Walmsley}, \&
  {Cesaroni}}]{churchwell1990}
{Churchwell}, E., {Walmsley}, C.~M., \& {Cesaroni}, R. 1990, \aaps, 83, 119

\bibitem[{{Crapsi} {et~al.}(2005){Crapsi}, {Caselli}, {Walmsley}, {Myers},
  {Tafalla}, {Lee}, \& {Bourke}}]{crapsi2005}
{Crapsi}, A., {Caselli}, P., {Walmsley}, C.~M., {et~al.} 2005, \apj, 619, 379

\bibitem[{{Dalgarno} \& {Lepp}(1984)}]{dalgarno1984}
{Dalgarno}, A. \& {Lepp}, S. 1984, \apjl, 287, L47

\bibitem[{{Daniel} {et~al.}(2007){Daniel}, {Cernicharo}, {Roueff}, {Gerin}, \&
  {Dubernet}}]{daniel2007}
{Daniel}, F., {Cernicharo}, J., {Roueff}, E., {Gerin}, M., \& {Dubernet}, M.~L.
  2007, \apj, 667, 980

\bibitem[{{Di Francesco} {et~al.}(2008){Di Francesco}, {Johnstone}, {Kirk},
  {MacKenzie}, \& {Ledwosinska}}]{difrancesco2008}
{Di Francesco}, J., {Johnstone}, D., {Kirk}, H., {MacKenzie}, T., \&
  {Ledwosinska}, E. 2008, \apjs, 175, 277

\bibitem[{{Emprechtinger} {et~al.}(2009){Emprechtinger}, {Caselli}, {Volgenau},
  {Stutzki}, \& {Wiedner}}]{emprechtinger2009}
{Emprechtinger}, M., {Caselli}, P., {Volgenau}, N.~H., {Stutzki}, J., \&
  {Wiedner}, M.~C. 2009, \aap, 493, 89

\bibitem[{{Fontani} {et~al.}(2006){Fontani}, {Caselli}, {Crapsi}, {Cesaroni},
  {Molinari}, {Testi}, \& {Brand}}]{fontani2006}
{Fontani}, F., {Caselli}, P., {Crapsi}, A., {et~al.} 2006, \aap, 460, 709

\bibitem[{{Fontani} {et~al.}(2011){Fontani}, {Palau}, {Caselli},
  {S{\'a}nchez-Monge}, {Butler}, {Tan}, {Jim{\'e}nez-Serra}, {Busquet},
  {Leurini}, \& {Audard}}]{fontani2011}
{Fontani}, F., {Palau}, A., {Caselli}, P., {et~al.} 2011, \aap, 529, L7

\bibitem[{{Friesen} {et~al.}(2013){Friesen}, {Kirk}, \&
  {Shirley}}]{friesen2013}
{Friesen}, R.~K., {Kirk}, H.~M., \& {Shirley}, Y.~L. 2013, \apj, 765, 59

\bibitem[{{Gerner} {et~al.}(2014){Gerner}, {Beuther}, {Semenov}, {Linz},
  {Vasyunina}, {Bihr}, {Shirley}, \& {Henning}}]{gerner2014}
{Gerner}, T., {Beuther}, H., {Semenov}, D., {et~al.} 2014, \aap, 563, A97

\bibitem[{{Guertler} {et~al.}(1991){Guertler}, {Henning}, {Kruegel}, \&
  {Chini}}]{guertler1991}
{Guertler}, J., {Henning}, T., {Kruegel}, E., \& {Chini}, R. 1991, \aap, 252,
  801

\bibitem[{{Harada} {et~al.}(2010){Harada}, {Herbst}, \& {Wakelam}}]{harada2010}
{Harada}, N., {Herbst}, E., \& {Wakelam}, V. 2010, \apj, 721, 1570

\bibitem[{{Harada} {et~al.}(2012){Harada}, {Herbst}, \& {Wakelam}}]{harada2012}
{Harada}, N., {Herbst}, E., \& {Wakelam}, V. 2012, \apj, 756, 104

\bibitem[{{Hatchell} {et~al.}(1998){Hatchell}, {Thompson}, {Millar}, \&
  {MacDonald}}]{hatchell1998b}
{Hatchell}, J., {Thompson}, M.~A., {Millar}, T.~J., \& {MacDonald}, G.~H. 1998,
  \aaps, 133, 29

\bibitem[{{Hatchell} \& {van der Tak}(2003)}]{hatchell2003}
{Hatchell}, J. \& {van der Tak}, F.~F.~S. 2003, \aap, 409, 589

\bibitem[{{Heitsch} {et~al.}(2008){Heitsch}, {Hartmann}, {Slyz}, {Devriendt},
  \& {Burkert}}]{heitsch2008}
{Heitsch}, F., {Hartmann}, L.~W., {Slyz}, A.~D., {Devriendt}, J.~E.~G., \&
  {Burkert}, A. 2008, \apj, 674, 316

\bibitem[{{Hiraoka} {et~al.}(2006){Hiraoka}, {Ushiama}, {Enoura}, {Unagiike},
  {Mochizuki}, \& {Wada}}]{hiraoka2006}
{Hiraoka}, K., {Ushiama}, S., {Enoura}, T., {et~al.} 2006, \apj, 643, 917

\bibitem[{{Kong} {et~al.}(2013){Kong}, {Caselli}, {Tan}, \&
  {Wakelam}}]{kong2013}
{Kong}, S., {Caselli}, P., {Tan}, J.~C., \& {Wakelam}, V. 2013, ArXiv e-prints

\bibitem[{{Lee} {et~al.}(1998){Lee}, {Roueff}, {Pineau des Forets},
  {Shalabiea}, {Terzieva}, \& {Herbst}}]{lee1998}
{Lee}, H.-H., {Roueff}, E., {Pineau des Forets}, G., {et~al.} 1998, \aap, 334,
  1047

\bibitem[{{Linsky}(2003)}]{linsky2003}
{Linsky}, J.~L. 2003, \ssr, 106, 49

\bibitem[{{Linz} {et~al.}(2005){Linz}, {Stecklum}, {Henning}, {Hofner}, \&
  {Brandl}}]{linz2005}
{Linz}, H., {Stecklum}, B., {Henning}, T., {Hofner}, P., \& {Brandl}, B. 2005,
  \aap, 429, 903

\bibitem[{{Lodders}(2003)}]{lodders2003}
{Lodders}, K. 2003, \apj, 591, 1220

\bibitem[{{Mangum} \& {Shirley}(2015)}]{mangum2015}
{Mangum}, J.~G. \& {Shirley}, Y.~L. 2015, ArXiv e-prints

\bibitem[{{McKee} \& {Tan}(2003)}]{mckee2003}
{McKee}, C.~F. \& {Tan}, J.~C. 2003, \apj, 585, 850

\bibitem[{{Miettinen} {et~al.}(2011){Miettinen}, {Hennemann}, \&
  {Linz}}]{miettinen2011}
{Miettinen}, O., {Hennemann}, M., \& {Linz}, H. 2011, \aap, 534, A134

\bibitem[{{Millar} {et~al.}(1989){Millar}, {Bennett}, \& {Herbst}}]{millar1989}
{Millar}, T.~J., {Bennett}, A., \& {Herbst}, E. 1989, \apj, 340, 906

\bibitem[{{Mueller} {et~al.}(2002){Mueller}, {Shirley}, {Evans}, \&
  {Jacobson}}]{mueller2002}
{Mueller}, K.~E., {Shirley}, Y.~L., {Evans}, N.~J., \& {Jacobson}, H.~R. 2002,
  \apjs, 143, 469

\bibitem[{{Narayanan} {et~al.}(2008){Narayanan}, {Cox}, {Shirley}, {Dav{\'e}},
  {Hernquist}, \& {Walker}}]{narayanan2008}
{Narayanan}, D., {Cox}, T.~J., {Shirley}, Y., {et~al.} 2008, \apj, 684, 996

\bibitem[{{Oliveira} {et~al.}(2003){Oliveira}, {H{\'e}brard}, {Howk}, {Kruk},
  {Chayer}, \& {Moos}}]{oliveira2003}
{Oliveira}, C.~M., {H{\'e}brard}, G., {Howk}, J.~C., {et~al.} 2003, \apj, 587,
  235

\bibitem[{{Ossenkopf} \& {Henning}(1994)}]{ossenkopf1994}
{Ossenkopf}, V. \& {Henning}, T. 1994, \aap, 291, 943

\bibitem[{{Parise} {et~al.}(2009){Parise}, {Leurini}, {Schilke}, {Roueff},
  {Thorwirth}, \& {Lis}}]{parise2009}
{Parise}, B., {Leurini}, S., {Schilke}, P., {et~al.} 2009, \aap, 508, 737

\bibitem[{{Phillips} {et~al.}(1979){Phillips}, {Huggins}, {Wannier}, \&
  {Scoville}}]{phillips1979}
{Phillips}, T.~G., {Huggins}, P.~J., {Wannier}, P.~G., \& {Scoville}, N.~Z.
  1979, \apj, 231, 720

\bibitem[{{Pillai} {et~al.}(2007){Pillai}, {Wyrowski}, {Hatchell}, {Gibb}, \&
  {Thompson}}]{pillai2007}
{Pillai}, T., {Wyrowski}, F., {Hatchell}, J., {Gibb}, A.~G., \& {Thompson},
  M.~A. 2007, \aap, 467, 207

\bibitem[{{Ragan} {et~al.}(2012){Ragan}, {Henning}, {Krause}, {Pitann},
  {Beuther}, {Linz}, {Tackenberg}, {Balog}, {Hennemann}, {Launhardt}, {Lippok},
  {Nielbock}, {Schmiedeke}, {Schuller}, {Steinacker}, {Stutz}, \&
  {Vasyunina}}]{ragan2012}
{Ragan}, S., {Henning}, T., {Krause}, O., {et~al.} 2012, \aap, 547, A49

\bibitem[{{Roberts} \& {Millar}(2000)}]{roberts2000}
{Roberts}, H. \& {Millar}, T.~J. 2000, \aap, 361, 388

\bibitem[{{Roueff} {et~al.}(2007){Roueff}, {Parise}, \& {Herbst}}]{roueff2007}
{Roueff}, E., {Parise}, B., \& {Herbst}, E. 2007, \aap, 464, 245

\bibitem[{{Sakai} {et~al.}(2012){Sakai}, {Sakai}, {Furuya}, {Aikawa}, {Hirota},
  \& {Yamamoto}}]{sakai2012}
{Sakai}, T., {Sakai}, N., {Furuya}, K., {et~al.} 2012, \apj, 747, 140

\bibitem[{{Sarrasin} {et~al.}(2010){Sarrasin}, {Abdallah}, {Wernli}, {Faure},
  {Cernicharo}, \& {Lique}}]{sarrasin2010}
{Sarrasin}, E., {Abdallah}, D.~B., {Wernli}, M., {et~al.} 2010, \mnras, 404,
  518

\bibitem[{{Sch{\"o}ier} {et~al.}(2005){Sch{\"o}ier}, {van der Tak}, {van
  Dishoeck}, \& {Black}}]{schoeier2005}
{Sch{\"o}ier}, F.~L., {van der Tak}, F.~F.~S., {van Dishoeck}, E.~F., \&
  {Black}, J.~H. 2005, \aap, 432, 369

\bibitem[{{Schuller} {et~al.}(2009){Schuller}, {Menten}, {Contreras},
  {Wyrowski}, {Schilke}, {Bronfman}, {Henning}, {Walmsley}, {Beuther},
  {Bontemps}, {Cesaroni}, {Deharveng}, {Garay}, {Herpin}, {Lefloch}, {Linz},
  {Mardones}, {Minier}, {Molinari}, {Motte}, {Nyman}, {Reveret}, {Risacher},
  {Russeil}, {Schneider}, {Testi}, {Troost}, {Vasyunina}, {Wienen}, {Zavagno},
  {Kovacs}, {Kreysa}, {Siringo}, \& {Wei{\ss}}}]{schuller2009}
{Schuller}, F., {Menten}, K.~M., {Contreras}, Y., {et~al.} 2009, \aap, 504, 415

\bibitem[{{Semenov} {et~al.}(2010){Semenov}, {Hersant}, {Wakelam}, {Dutrey},
  {Chapillon}, {Guilloteau}, {Henning}, {Launhardt}, {Pi{\'e}tu}, \&
  {Schreyer}}]{semenov2010}
{Semenov}, D., {Hersant}, F., {Wakelam}, V., {et~al.} 2010, \aap, 522, A42

\bibitem[{{Sridharan} {et~al.}(2005){Sridharan}, {Beuther}, {Saito},
  {Wyrowski}, \& {Schilke}}]{sridharan2005}
{Sridharan}, T.~K., {Beuther}, H., {Saito}, M., {Wyrowski}, F., \& {Schilke},
  P. 2005, \apjl, 634, L57

\bibitem[{{Sridharan} {et~al.}(2002){Sridharan}, {Beuther}, {Schilke},
  {Menten}, \& {Wyrowski}}]{sridharan2002}
{Sridharan}, T.~K., {Beuther}, H., {Schilke}, P., {Menten}, K.~M., \&
  {Wyrowski}, F. 2002, \apj, 566, 931

\bibitem[{{Tan} {et~al.}(2014){Tan}, {Beltran}, {Caselli}, {Fontani}, {Fuente},
  {Krumholz}, {McKee}, \& {Stolte}}]{tan2014}
{Tan}, J.~C., {Beltran}, M.~T., {Caselli}, P., {et~al.} 2014, ArXiv e-prints

\bibitem[{{Testi} {et~al.}(2000){Testi}, {Hofner}, {Kurtz}, \&
  {Rupen}}]{testi2000}
{Testi}, L., {Hofner}, P., {Kurtz}, S., \& {Rupen}, M. 2000, \aap, 359, L5

\bibitem[{{Turner}(2001)}]{turner2001}
{Turner}, B.~E. 2001, \apjs, 136, 579

\bibitem[{{van der Tak} {et~al.}(2007){van der Tak}, {Black}, {Sch{\"o}ier},
  {Jansen}, \& {van Dishoeck}}]{vandertak2007}
{van der Tak}, F.~F.~S., {Black}, J.~H., {Sch{\"o}ier}, F.~L., {Jansen}, D.~J.,
  \& {van Dishoeck}, E.~F. 2007, \aap, 468, 627

\bibitem[{{van der Tak} {et~al.}(2000){van der Tak}, {van Dishoeck}, {Evans},
  \& {Blake}}]{vandertak2000c}
{van der Tak}, F.~F.~S., {van Dishoeck}, E.~F., {Evans}, II, N.~J., \& {Blake},
  G.~A. 2000, \apj, 537, 283

\bibitem[{{Watson}(1974)}]{watson1974}
{Watson}, W.~D. 1974, \apj, 188, 35

\bibitem[{{Wilson} \& {Rood}(1994)}]{wilson1994}
{Wilson}, T.~L. \& {Rood}, R. 1994, \araa, 32, 191

\bibitem[{{Wood} \& {Churchwell}(1989{\natexlab{a}})}]{wc89}
{Wood}, D.~O.~S. \& {Churchwell}, E. 1989{\natexlab{a}}, \apj, 340, 265

\bibitem[{{Wood} \& {Churchwell}(1989{\natexlab{b}})}]{wc1989b}
{Wood}, D.~O.~S. \& {Churchwell}, E. 1989{\natexlab{b}}, \apjs, 69, 831

\bibitem[{{Zinnecker} \& {Yorke}(2007)}]{zinnecker2007}
{Zinnecker}, H. \& {Yorke}, H.~W. 2007, \araa, 45, 481

\end{thebibliography}

%\clearpage

\appendix

\section{Appendix material}

\subsection{Molecular column densities} \label{sec:column_density}
We calculated the molecular column densities of the upper level of a particular transition following the equation for optically thin emission:

\begin{equation}
N_{u}=\frac{8 \pi k \nu^2}{h c^3 A_{ul}} \frac{J_{\nu}(T_{\rm ex})}{[J_{\nu}(T_{\rm ex})-J_{\nu}(T_{\rm cmb})]} \cdot \int T_{mb} \delta \nu
\end{equation}

where the integrated intensity is in K\,km\,s$^{-1}$, and the Einstein coefficient $A_{ul}$ is in s$^{-1}$ and with:

\begin{equation}
J_{\nu}(T)=\frac{h \nu /k }{\exp[h \nu / k T] - 1}
\end{equation}

where $\nu$ is the frequency of the observed transition (e.g.,
\citealt{mangum2015}). Then the total column density (which we will
refer from now on as column density) can be calculated:

\begin{equation}
N_{tot} = N_{u} \cdot \frac{Q}{g_{u} \exp[-E_{u}/kT_{ex}]}
\end{equation}

In the cases where we could derive the optical depth $\tau$ due
  to hyperfine splitting we corrected the column density using:

\begin{equation}
N_{\rm corr} = N_{\rm tot} \cdot \frac{\tau}{1 - \exp[-\tau]}
\end{equation}

In order to obtain abundances we derived H$_2$ column densities either
from dust maps obtained with Mambo with the IRAM 30m telescope at
$1.2$~mm \citep{beuther2002a} and a resolution of $11\arcsec$, or the
galactic plane survey ATLASGAL \citep{schuller2009} at $870\mu$m and a
resolution of $19.2\arcsec$, or the SCUBA Legacy Catalog
\citep{difrancesco2008} at $850\mu$m and a resolution of $22.9\arcsec$
(see Table~\ref{tbl:sourcelist}). The continuum data were smoothed to
$29\arcsec$ resolution in order to be beam matched with the IRAM 30m
observations at 3\,mm and the SMT molecular line data. H$_2$ column
densities were calculated from the observed peak intensities assuming
optically thin emission and LTE following Equation~\ref{equ:dusttoh2}
\citep{schuller2009}. The dust opacities used were $\kappa_{850\mu \rm
  m}=1.48$, $\kappa_{870\mu \rm m}=1.42$, $\kappa_{1.2\rm mm}=0.97$,
interpolated values from \citet{ossenkopf1994}, assuming grains with
thin ice mantles, gas densities of $n=10^5$~cm$^{-3}$, and a
gas-to-dust mass ratio $R=100$. With these assumptions the H$_2$
column density is calculated as:

\begin{equation}
N_{H_2}=\frac{F_{\nu} \cdot R}{B_{\nu} \cdot \Omega  \cdot \kappa_{\nu} \cdot \mu \cdot m_{H}}  %*1e-23
\label{equ:dusttoh2}
\end{equation}

The uncertainties in the derived H$_2$ column densities are based
largely on the dust and temperature properties and are approximately
about a factor of 3. A more detailed description of the derivation is
given in \citet{gerner2014}.

The molecular column densities are then divided by the H$_2$ column
densities and the averaged abundances are derived.

\clearpage

\begin{table*}
\tiny
\caption{Source list showing the position, the distance, and the evolutionary stage of all high-mass star-forming regions.}             % title of Table
\label{tbl:sourcelist}      % is used to refer this table in the text
\centering                          % used for centering table
\begin{tabular}{lccccccccr}        % centered columns (4 columns)
\hline\hline                 % inserts double horizontal lines
source & $\alpha$ & $\delta$ & galactic $l$ & galactic $b$ & distance\tablefootmark{a} & type & 24~$\mu$m & 70~$\mu$m & continuum data\tablefootmark{b} \\
 & (J2000.0) & (J2000.0) & $[^{\circ}]$ & $[^{\circ}]$ & $[\rm {kpc}]$ &  &  &  &  \\
% table heading 
\hline                        % inserts single horizontal line
IRDC011.1    &   18:10:28.4 &    -19:22:34 & 11.108 & -0.115 & 3.6 &  IRDC & y & y  & ATLASGAL    \\
IRDC028.1    &   18:42:50.3 &    -04:03:20 & 28.343 & 0.060 &  4.8 &  IRDC & y & y  & ATLASGAL    \\
IRDC028.2    &   18:42:52.1 &    -03:59:54 & 28.397 & 0.080 &  4.8 &  IRDC & y & y  & ATLASGAL    \\
IRDC048.6    &   19:21:44.4 &    +13:49:24 & 48.657 & -0.285 &  2.5 &  IRDC & n & n  & ATLASGAL    \\
IRDC079.1    &   20:32:22.0 &    +40:20:10 & 79.338 & 0.341 &  1.0 &  IRDC & - & y & SCUBA   \\
\vspace{0.1cm}
IRDC079.3    &   20:31:57.7 &    +40:18:26 & 79.269 & 0.386 &  1.0 &  IRDC & - & y & SCUBA   \\
IRDC18151    &   18:17:50.3 &    -12:07:54 & 18.319 & 1.792 &  3.0 &  IRDC & - & y & Mambo       \\
IRDC18182    &   18:21:15.0 &    -14:33:03 & 16.578 & -0.081 &  3.6 &  IRDC & y & y & Mambo       \\
IRDC18223    &   18:25:08.3 &    -12:45:27 & 18.605 & -0.075 &  3.7 &  IRDC & y & y & Mambo       \\
IRDC18306    &   18:33:32.1 &    -08:32:28 & 23.297 & 0.0550 &  3.8 &  IRDC & n & n & Mambo       \\
\vspace{0.1cm}
IRDC18308    &   18:33:34.3 &    -08:38:42 & 23.209 & -0.001 &  4.9 &  IRDC & y & y & Mambo       \\
IRDC18310    &   18:33:39.5 &    -08:21:10 & 23.478 & 0.115 &  5.2 &  IRDC & n & n & Mambo       \\
IRDC18337    &   18:36:18.2 &    -07:41:00 & 24.374 & -0.158 &  4.0 &  IRDC & y & y & Mambo       \\
IRDC18385    &   18:41:17.4 &    -05:09:56 & 27.179 & -0.104 &  3.3 &  IRDC & y & y\tablefootmark{c} & Mambo       \\
IRDC18437    &   18:46:21.8 &    -02:12:21 & 30.390 & 0.123 &  (6.2) 7.3\tablefootmark{d} & IRDC & y & y & Mambo \\
\vspace{0.1cm}
IRDC18454.1    &   18:48:02.1 &    -01:53:56 & 30.854 & -0.109 &  (3.5) 6.4\tablefootmark{e} & IRDC & n & n & Mambo \\
IRDC18454.3    &   18:47:55.8 &    -01:53:34 & 30.848 & -0.083 &  6.0 (6.4)\tablefootmark{f} & IRDC & n & n & Mambo \\
IRDC19175    &   19:19:50.7 &    +14:01:23 & 48.617 & 0.214 &  1.1 &  IRDC & n & n & Mambo       \\
IRDC20081    &   20:10:13.0 &    +27:28:18 & 66.145 & -3.197 &  0.7 &  IRDC & - & n\tablefootmark{g} & Mambo       \\
HMPO18089    &   18:11:51.6 &    -17:31:29 & 12.889 & 0.489 &  3.6 &  HMPO & & & Mambo       \\
\vspace{0.1cm}
HMPO18102    &   18:13:11.3 &    -18:00:03 & 12.623 & -0.017 &  2.7 &  HMPO & & & Mambo       \\
HMPO18151    &   18:17:58.1 &    -12:07:26 & 18.341 & 1.768 &  3.0 &  HMPO & & & Mambo       \\
HMPO18182    &   18:21:09.2 &    -14:31:50 & 16.585 & -0.051 &  3.9\tablefootmark{h} &  HMPO & &  & Mambo       \\
HMPO18247    &   18:27:31.7 &    -11:45:56 & 19.755 & -0.129 &  6.7 &  HMPO & &  & Mambo       \\
HMPO18264    &   18:29:14.6 &    -11:50:22 & 19.884 & -0.535 &  3.28\tablefootmark{h}  &  HMPO & &  & Mambo       \\
\vspace{0.1cm}
HMPO18310    &   18:33:48.1 &    -08:23:50 & 23.455 & 0.063 &  5.2 (10.4)\tablefootmark{i} &  HMPO & &  & Mambo       \\
HMPO18488    &   18:51:25.6 &    +00:04:07 & 32.991 & 0.034 &  5.4 (8.9)\tablefootmark{i} &  HMPO & &  & Mambo       \\
HMPO18517    &   18:54:14.4 &    +04:41:40 & 37.430 & 1.517 &  2.9 &  HMPO & &  & Mambo       \\
HMPO18566    &   18:59:10.1 &    +04:12:14 & 37.554 & 0.200 &  6.7 &  HMPO & &  & Mambo       \\
HMPO19217    &   19:23:58.8 &    +16:57:44 & 51.679 & 0.720 & 10.5 &  HMPO & &  & Mambo       \\
\vspace{0.1cm}
HMPO19410    &   19:43:11.0 &    +23:44:10 & 59.784 & 0.066 &  2.1 &  HMPO & &  & Mambo       \\
HMPO20126    &   20:14:26.0 &    +41:13:32 & 78.122 & 3.633 &  1.7 &  HMPO & &  & Mambo       \\
HMPO20216    &   20:23:23.8 &    +41:17:40 & 79.127 & 2.279 &  1.7 &  HMPO & &  & Mambo       \\
HMPO20293    &   20:31:12.9 &    +40:03:20 & 78.982 & 0.352 &  1.3 (2.0)\tablefootmark{i} &  HMPO & &  & Mambo       \\
HMPO22134    &   22:15:09.1 &    +58:49:09 & 103.876 & 1.856 &  2.6 &  HMPO & &  & Mambo       \\
\vspace{0.1cm}
HMPO23033    &   23:05:25.7 &    +60:08:08 & 110.093 & -0.067 &  3.5 &  HMPO & &  & Mambo       \\
HMPO23139    &   23:16:10.5 &    +59:55:28 & 111.256 & -0.770 &  4.8 &  HMPO & &  & Mambo       \\
HMPO23151    &   23:17:21.0 &    +59:28:49 & 111.236 & -1.238 &  5.7 &  HMPO & &  & Mambo       \\
HMPO23545    &   23:57:06.1 &    +65:24:48 & 117.315 & 3.136 &  0.8 &  HMPO & &  & Mambo       \\
HMC009.62    &   18:06:15.2 &    -20:31:37 & 9.621 & 0.193 &  5.7 &   HMC & &  & ATLASGAL    \\
\vspace{0.1cm}
HMC010.47    &   18:08:38.2 &    -19:51:50 & 10.472 & 0.027 &  5.8 &   HMC & &  & ATLASGAL    \\
HMC029.96    &   18:46:04.0 &    -02:39:21 & 29.956 & -0.017 &  7.4 &   HMC & &  & ATLASGAL    \\
HMC031.41    &   18:47:34.2 &    -01:12:45 & 31.412 & 0.308 &  7.9 &   HMC & &  & ATLASGAL    \\
HMC034.26    &   18:53:18.5 &    +01:14:58 & 34.257 & 0.154 &  4.0 &   HMC & &  & ATLASGAL    \\
HMC045.47    &   19:14:25.7 &    +11:09:26 & 45.466 & 0.045 &  6.0 &   HMC & &  & ATLASGAL    \\
\vspace{0.1cm}
HMC075.78    &   20:21:44.1 &    +37:26:40 & 75.783 & 0.343 &  4.1 &   HMC & &  & SCUBA   \\
NGC7538B     &   23:13:45.4 &    +61:28:11 & 111.542 & 0.777 &  2.65 (5.61)\tablefootmark{j} &   HMC & & & SCUBA \\
Orion-KL     &   05:35:14.4 &    -05:22:31 & 208.993 & -19.385 &  0.44 &   HMC & &  & SCUBA   \\
W3IRS5       &   02:25:40.7 &    +62:05:52 & 133.715 & 1.215 &  1.8 &   HMC & &  & SCUBA   \\
W3H$_2$O     &   02:27:04.6 &    +61:52:25 & 133.949 & 1.065 &  2.0 &   HMC & &  & SCUBA   \\
\vspace{0.1cm}
UCH005.89    &   18:00:30.4 &    -24:04:00 & 5.886 & -0.392 &  2.5 &   UCH{\sc ii} & & & ATLASGAL    \\
UCH010.10    &   18:05:13.1 &    -19:50:35 & 10.099 & 0.739 &  4.4 &   UCH{\sc ii} & & & ATLASGAL    \\
UCH010.30    &   18:08:55.8 &    -20:05:55 & 10.300 & -0.147 &  6.0 &   UCH{\sc ii} & & & ATLASGAL    \\
UCH012.21    &   18:12:39.7 &    -18:24:20 & 12.208 & -0.102 & 13.5 &   UCH{\sc ii} & & & ATLASGAL    \\
UCH013.87    &   18:14:35.8 &    -16:45:43 & 13.872 & 0.280 &  4.4 &   UCH{\sc ii} & & & ATLASGAL    \\
\vspace{0.1cm}
UCH030.54    &   18:46:59.3 &    -02:07:24 & 30.535 & 0.021 &  6.1 &   UCH{\sc ii} & & & ATLASGAL    \\
UCH035.20    &   19:01:46.4 &    +01:13:25 & 35.200 & -1.741 &  3.2 &   UCH{\sc ii} & & & SCUBA   \\
UCH045.12    &   19:13:27.8 &    +10:53:37 & 45.122 & 0.132 &  6.9 &   UCH{\sc ii} & & & ATLASGAL    \\
UCH045.45    &   19:14:21.3 &    +11:09:14 & 45.454 & 0.060 &  6.0 &   UCH{\sc ii} & & & ATLASGAL    \\
\hline                                   %inserts single line
\end{tabular}
\tablefoot{For the IRDCs we indicate whether or not they show embedded 24 or 70$\mu$ m point sources with y(es) or n(o) (or ``-'' if there are no data available). In the last column the sources of the dust continuum data are presented.}
\tablefoottext{a}{Unbracketed values are preferred, bracketed values are alternative values}
\tablefoottext{b}{Dust continuum data. Either from IRAM 30m observations with Mambo (1.2mm), the galactic plane survey ATLASGAL (870 $\mu$m) \citep{schuller2009} or the SCUBA Legacy Catalog (850 $\mu$m) \citep{difrancesco2008}.}
\tablefoottext{c}{Very weak emission compared with the background located at the same position as 24$\mu$m emission source}
\tablefoottext{d}{For $v_{\rm lsr}=(97.6) 111.3$~km\,s$^{-1}$}
\tablefoottext{e}{For $v_{\rm lsr}=(52.8) 100.2$~km\,s$^{-1}$}
\tablefoottext{f}{For $v_{\rm lsr}=94.3 (98.4)$~km\,s$^{-1}$}
\tablefoottext{g}{No embedded central point source found, a nearby extended source with emission inside the beam is detected}
\tablefoottext{h}{Ellsworth-Bowers et al. subm.}
\tablefoottext{i}{For the near (far) kinematic solution}
\tablefoottext{j}{Parallactic (kinematic) distance}
\end{table*}

\clearpage

\begin{table*}
\tiny
\caption{Luminosity, H$_2$, DCO$^+$, DCN, DNC and N$_2$D+ column density and the corresponding error ($\Delta$) for each source. The errors show the uncertainties in the measured integrated flux.}
\label{tbl:sourceparameters1}
\centering
\tiny
\begin{tabular}{lcccccccccc}
\hline\hline                 % inserts double horizontal lines
Source & Luminosity & H$_2$ & DCO$^+$& $\Delta$(DCO$^+$) & DCN & $\Delta$(DCN) & DNC & $\Delta$(DNC) & N$_2$D+ & $\Delta$(N$_2$D+) \\
 & log L$_{\rm Sun}$ & cm$^{-2}$  & cm$^{-2}$ & cm$^{-2}$ & cm$^{-2}$ & cm$^{-2}$ \\  % table heading 
\hline                        % inserts single horizontal line
IRDC011.1 &  3.16\tablefootmark{a}   &  3.7(22) &  4.5(11) & 1.2(11) & $\leq$4.1(11) & 3.7(11) &  9.3(11) & 1.4(11) & $\leq$3.9(11)&      9.6(10) \\
IRDC028.1 &                          &  3.2(22) & $\leq$1.9(11) & 4.8(10) & $\leq$5.4(11) & 1.1(11) & $\leq$4.3(11) & 9.7(10) & $\leq$6.8(11)&      9.6(10) \\
IRDC028.2 &  3.47\tablefootmark{a}   &  1.1(23) &  9.3(11) & 1.1(11) &  3.4(12) & 3.2(11) &  1.5(12) &   2(11) & $\leq$6.6(11)&      1.9(11) \\
IRDC048.6 &  0.78\tablefootmark{a}   &  8.6(21) &  1.2(12) & 7.1(10) &   $\leq$9(11) & 1.9(11) & $\leq$4.7(11) &   2(11) & $\leq$9.6(11)&      2.5(11) \\
IRDC079.1 &  2.05\tablefootmark{a}   &  4.3(22) &  3.9(12) & 8.9(10) & $\leq$8.1(11) & 5.1(11) &  2.7(12) & 1.5(11) &  1.5(12)&      2.4(11) \\
\vspace{0.1cm}                                                                                                                         
IRDC079.3 &  1.4\tablefootmark{a}    &  2.7(22) &    3(12) &   1(11) & $\leq$9.6(11) & 6.9(11) &  1.9(12) & 1.5(11) &  2.1(12)&      3.3(11) \\
IRDC18151 &  2.66\tablefootmark{a}   &  5.1(22) &    3(12) & 1.3(11) &  2.3(12) & 2.8(11) &  3.2(12) & 2.3(11) &  1.4(12)&      2.1(11) \\
IRDC18182 &  2.13\tablefootmark{a}   &  1.3(22) &  5.4(11) &   8(10) & $\leq$5.4(11) & 4.3(11) & $\leq$5.7(11) & 1.2(11) & $\leq$5.5(11)&      8.2(10) \\
IRDC18223 &  2.51\tablefootmark{a}   &    2(22) &  9.7(11) & 1.4(11) &  1.3(12) & 3.4(11) &  3.2(12) & 2.5(11) &  1.2(12)&      1.8(11) \\
IRDC18306 &                          &  1.7(22) &   $\leq$3(11) & 9.8(10) & $\leq$5.7(11) &   2(11) & $\leq$6.3(11) &   1(11) & $\leq$5.5(11)&      2.1(11) \\
\vspace{0.1cm}                                                                                                                         
IRDC18308 &  2.12\tablefootmark{a}   &  2.4(22) & $\leq$2.8(11) & 9.3(10) & $\leq$6.1(11) & 1.5(11) & $\leq$5.3(11) &   2(11) & $\leq$6.8(11)&      3.6(11) \\
IRDC18310 &                          &  1.6(22) & $\leq$2.9(11) & 1.9(11) & $\leq$6.8(11) & 2.9(11) & $\leq$4.5(11) & 3.1(11) & $\leq$6.8(11)&      3.6(11) \\
IRDC18337 &  2.25\tablefootmark{a}   &  7.8(21) & $\leq$2.8(11) &   1(11) &   $\leq$5(11) & 2.9(11) & $\leq$3.9(11) & 2.2(11) & $\leq$5.8(11)&      1.5(11) \\
IRDC18385 &  1.23\tablefootmark{a}   &  7.7(21) & $\leq$3.1(11) & 9.8(10) & $\leq$6.1(11) & 1.2(11) & $\leq$5.9(11) & 2.2(11) & $\leq$5.7(11)&        1(11) \\
IRDC18437 &  2.23\tablefootmark{a}   &    1(22) & $\leq$3.1(11) & 1.4(11) & $\leq$5.9(11) & 1.2(11) & $\leq$5.3(11) & 1.6(11) &   $\leq$5(11)&      3.7(11) \\
\vspace{0.1cm}                                                                                                                         
IRDC18454.1  & 3.6\tablefootmark{b}  &  1.4(22) & $\leq$2.9(11) & 1.1(11) & $\leq$5.2(11) & 1.5(11) & $\leq$6.1(11) & 1.3(11) &         &      6.9(11) \\
IRDC18454.3  & 2.37\tablefootmark{a} &    2(22) & $\leq$3.1(11) & 7.3(10) & $\leq$5.5(11) & 2.5(11) &$\leq$7.3(11)  & 1.7(11) &         &      5.9(11) \\
IRDC19175 &  -1\tablefootmark{a}     &  9.1(21) & $\leq$2.7(11) & 9.2(10) & $\leq$5.4(11) & 2.8(11) & $\leq$4.5(11) & 1.3(11) & $\leq$6.5(11)&      1.6(11) \\
IRDC20081 &                          &    2(22) & 1.4(12)  & 7.4(10) & $\leq$6.1(11) &   4(11) &  9.1(11) & 1.3(11) &  8.1(11)&      1.3(11) \\
HMPO18089 & 4.5\tablefootmark{c}     &  8.2(22) &  1.7(12) &   9(10) &  6.8(12) & 2.5(11) &  9.7(11) & 1.3(11) & $\leq$4.2(11)&      9.1(10) \\
\vspace{0.1cm}                                                                                                                         
HMPO18102 & 2.92\tablefootmark{c}    &  4.2(22) & $\leq$1.2(11) & 9.7(10) & $\leq$3.8(11) & 2.6(11) &  1.1(12) & 1.2(11) & $\leq$3.8(11)&      2.1(11) \\
HMPO18151 & 2.03\tablefootmark{c}    &  4.3(22) &  8.1(11) & 4.4(10) &  3.2(12) & 1.3(11) &  2.1(12) & 9.5(10) &  3.4(11)&      7.6(10) \\
HMPO18182 & 3.77\tablefootmark{c}    &  5.5(22) &  9.5(11) & 7.8(10) &    3(12) & 2.1(11) &  1.7(12) & 1.6(11) & $\leq$3.7(11)&      6.4(10) \\
HMPO18247 & 4.8\tablefootmark{c}     &  2.8(22) & $\leq$1.3(11) & 2.2(10) & $\leq$3.7(11) & 1.8(11) &   $\leq$2(11) & 1.2(11) & $\leq$4.1(11)&      6.6(10) \\
HMPO18264 & 4\tablefootmark{c}       &  9.4(22) &  2.4(12) & 6.4(10) &  6.9(12) & 1.6(11) &  4.4(12) & 1.1(11) &    1(12)&      1.5(11) \\
\vspace{0.1cm}                                                                                                                         
HMPO18310 & 3.48\tablefootmark{c}    &  2.9(22) & $\leq$1.6(11) &   1(11) & $\leq$4.1(11) & 3.3(11) & $\leq$3.1(11) & 1.5(11) &  6.7(11)&      1.3(11) \\
HMPO18488 & 4.5\tablefootmark{c}     &  3.2(22) &  2.4(11) & 5.2(10) &  1.4(12) & 1.8(11) & $\leq$2.1(11) & 9.4(10) & $\leq$3.3(11)&      1.2(11) \\
HMPO18517 & 4.1\tablefootmark{c}     &    7(22) &    2(12) & 5.6(10) &  5.6(12) & 2.3(11) &  1.6(12) & 1.3(11) & $\leq$3.9(11)&      1.2(11) \\
HMPO18566 & 4.8\tablefootmark{c}     &  2.5(22) &  3.8(11) & 1.4(11) &  2.3(12) & 2.4(11) &  8.1(11) & 1.3(11) & $\leq$4.1(11)&      1.6(07) \\
HMPO19217 & 4.9\tablefootmark{c}     &  4.2(22) &  6.4(11) & 9.6(10) &  2.6(12) & 2.3(11) & $\leq$2.9(11) & 2.4(11) &   $\leq$3(11)&      7.6(10) \\
\vspace{0.1cm}                                                                                                                         
HMPO19410 & 4\tablefootmark{c}       &  6.8(22) &  1.3(12) & 6.2(10) &  3.5(12) & 1.5(11) &  2.6(12) & 1.1(11) &  1.5(12)&      1.1(11) \\
HMPO20126 & 3.9\tablefootmark{c}     &  5.7(22) &  1.3(12) &   8(10) &  7.4(12) & 1.8(11) &  3.6(12) & 1.3(11) &  7.7(11)&      1.2(11) \\
HMPO20216 & 3.3\tablefootmark{c}     &  1.3(22) &  4.9(11) & 6.4(10) &  1.1(12) & 1.4(11) &  1.2(12) & 9.8(10) &   $\leq$3(11)&      1.4(11) \\
HMPO20293 & 3.4\tablefootmark{c}     &  4.3(22) &  9.9(11) &   7(10) &  1.9(12) & 2.4(11) &  1.6(12) &   1(11) &  1.7(12)&      1.6(11) \\
HMPO22134 & 4.1\tablefootmark{c}     &  2.4(22) &  6.9(11) & 5.3(10) &  1.2(12) & 1.4(11) &  3.6(11) & 6.3(10) & $\leq$2.8(11)&      1.6(11) \\
\vspace{0.1cm}                                                                                                                         
HMPO23033 & 4\tablefootmark{c}       &   4(22)  &  1.2(12) & 7.9(10) &  2.1(12) & 2.6(11) &  1.1(12) & 1.6(11) &  3.9(11)&        1(11) \\
HMPO23139 & 4.4\tablefootmark{c}     &    3(22) &  2.8(11) & 7.9(10) &  1.7(12) & 1.5(11) & $\leq$2.5(11) & 5.6(10) & $\leq$2.3(11)&      1.6(11) \\
HMPO23151 & 5\tablefootmark{c}       &  2.3(22) & $\leq$1.2(11) & 5.4(10) &  6.1(11) & 1.1(11) & $\leq$2.3(11) &   9(10) & $\leq$2.5(11)&      1.6(11) \\
HMPO23545 & 3\tablefootmark{c}       &  1.5(22) &  1.5(11) & 5.2(10) & $\leq$3.3(11) & 1.6(11) & $\leq$2.7(11) &   9(10) & $\leq$3.1(11)&      1.7(11) \\
HMC009.62 & 4.3\tablefootmark{d}     &  9.5(22) &  8.7(11) & 9.6(10) &  5.6(12) & 1.1(11) &  6.1(11) & 7.6(10) & $\leq$1.7(11)&      1.4(11) \\
\vspace{0.1cm}                                                                                                                         
HMC010.47 & 5.5\tablefootmark{e}     &  2.3(23) &    6(12) &   7(10) &  1.2(13) & 2.9(11) &          &   3(11) &         &      5.4(11) \\
HMC029.96 & 6\tablefootmark{e}       &  9.4(22) &  3.6(11) & 1.2(11) &  6.4(12) & 1.7(11) &  8.7(11) & 8.6(10) & $\leq$3.1(11)&      1.4(11) \\
HMC031.41 & 5.2\tablefootmark{e}     &  1.6(23) &    5(12) & 7.8(10) &    8(12) & 1.4(11) &5.8(11)   & 1.3(11) &         &      1.8(11) \\
HMC034.26 & 5.7\tablefootmark{e}     &  4.1(23) &  5.9(12) & 8.7(10) &  1.9(13) & 1.7(11) &  2.5(12) & 1.5(11) & $\leq$3.3(11)&      1.7(11) \\
HMC045.47 & 5.6\tablefootmark{e}     &  4.5(22) &  2.8(11) & 5.5(10) &  1.7(12) & 1.3(11) &  7.3(11) & 1.3(11) & $\leq$2.5(11)&        2(11) \\
\vspace{0.1cm}                                                                                                                         
HMC075.78 & 5.3\tablefootmark{e}     &  6.3(22) &  1.1(12) & 5.5(10) &  4.9(12) & 1.3(11) &  1.1(12) & 7.7(10) & $\leq$1.9(11)&      1.2(11) \\
W3H2O     & 4.48\tablefootmark{f}    &  7.4(22) &  3.4(12) & 9.1(10) &  1.6(13) & 1.5(11) &  2.1(12) & 7.8(10) & $\leq$2.8(11)&      5.8(10) \\
W3IRS5    & 5.3\tablefootmark{g}     &    6(22) &  4.2(11) & 7.6(10) &  1.5(12) & 1.1(11) & $\leq$1.8(11) & 1.2(11) & $\leq$1.7(11)&      7.1(10) \\
NGC7538B  & 5.8\tablefootmark{e}     &  1.2(23) &    2(12) & 7.8(10) &  1.4(13) & 1.5(11) &  1.5(12) &   1(11) & $\leq$1.6(11)&      7.1(10) \\
Orion-KL  & 5\tablefootmark{h}       &  1.3(24) &  2.8(13) & 1.6(11) &  1.3(14) & 1.6(12) &          & 4.3(11) &         &      1.2(12) \\
\vspace{0.1cm}                                                                                                                         
UCH005.89 & 5.3\tablefootmark{e}     &  2.2(23) &  1.3(12) & 1.6(11) &  2.1(13) & 1.5(11) &  6.3(12) & 2.3(11) & $\leq$2.5(11)&      5.8(10) \\
UCH010.10 &                          &  6.4(21) & $\leq$1.9(11) & 2.2(10) & $\leq$2.8(11) & 1.5(11) &   $\leq$3(11) & 1.8(11) & $\leq$2.8(11)&      6.9(10) \\
UCH010.30 & 5.8\tablefootmark{e}     &  8.1(22) &    7(11) & 1.1(11) &  3.7(12) & 2.1(11) &  1.9(12) & 1.4(11) & $\leq$3.5(11)&      1.2(11) \\
UCH012.21 & 5.8\tablefootmark{e}     &  9.6(22) & $\leq$2.1(11) & 2.5(10) &  5.2(12) & 2.6(11) & $\leq$2.3(11) & 1.3(11) & $\leq$4.1(11)&      1.6(11) \\
UCH013.87 & 5.2\tablefootmark{e}     &  5.3(22) &  4.7(11) & 6.4(10) &  1.7(12) & 1.6(11) &   $\leq$3(11) & 5.5(10) & $\leq$3.3(11)&      3.2(11) \\
\vspace{0.1cm}                                                                                                                         
UCH030.54 & 5.7\tablefootmark{i}     &  2.1(22) &  2.4(11) & 8.3(10) & $\leq$3.2(11) & 2.2(11) & $\leq$2.3(11) & 1.1(11) & $\leq$3.4(11)&      1.8(11) \\
UCH035.20 & 5.3\tablefootmark{e}     &  1.1(23) &  5.6(11) &   1(11) &  3.6(12) & 1.9(11) &  1.4(12) &   2(11) & $\leq$3.1(11)&      1.3(11) \\
UCH045.12 & 5.9\tablefootmark{e}     &  7.4(22) &  4.6(11) & 8.5(10) &  1.8(12) & 2.1(11) &  6.1(11) & 1.3(11) & $\leq$2.7(11)&      1.9(11) \\
UCH045.45 & 5.7\tablefootmark{e}     &  4.2(22) & $\leq$1.2(11) &   1(11) &  1.9(12) & 2.2(11) & $\leq$1.9(11) & 1.8(11) & $\leq$3.5(11)&      1.9(11) \\
\hline
\end{tabular}
\tablefoot{Column densities written as $a(x)=a \times 10^x$. H$_2$ column density is averaged over a 29$\arcsec$-beam and for all other molecules over a 30$\arcsec$-beam.}
\tablefoottext{a}{\citet{ragan2012}}
\tablefoottext{b}{\citet{beuther2012}}
\tablefoottext{c}{\citet{sridharan2002}}
\tablefoottext{d}{\citet{linz2005}}
\tablefoottext{e}{\citet{churchwell1990}}
\tablefoottext{f}{\citet{chen2006}}
\tablefoottext{g}{\citet{campbell1995}}
\tablefoottext{h}{\citet{beuther2004g}}
\tablefoottext{i}{\citet{wc89}}
\end{table*}

\clearpage

\begin{table*}
\tiny
\caption{HCO$^+$, HCN, HNC and N$_2$H$^+$ column density and corresponding error ($\Delta$) for each source. The errors show the uncertainties in the measured integrated fluxes and determined optical depths.}
\label{tbl:sourceparameters2}
\centering
\tiny
\begin{tabular}{lcccccccc}
\hline\hline                 % inserts double horizontal lines
Source & HCO$^+$  & $\Delta$(HCO$^+$) & HCN & $\Delta$(HCN) & HNC & $\Delta$(HNC) & N$_2$H$^+$ & $\Delta$(N$_2$H$^+$) \\
 & cm$^{-2}$ & cm$^{-2}$ & cm$^{-2}$ & cm$^{-2}$ \\  % table heading 
\hline                        % inserts single horizontal line
IRDC011.1  &  1.5(14)&  2.0(12) &  6.7(13) &  4.2(12) &  2.1(14) &  4.2(12) &  2.8(13) &  9.5(11) \\
IRDC028.1  &  1.4(14)&  2.3(12) &  1.2(14) &  4.9(12) &  2.2(14) &  4.8(12) &  4.0(13) &  3.2(11)  \\
IRDC028.2  &  1.8(14)&  2.3(12) &  1.9(14) &  8.0(12) &  2.2(14) &  4.8(12) &  6.5(13) &  2.5(12)  \\
IRDC048.6  &  7.9(13)&  2.4(12) &  2.7(13) &  3.0(13) &  6.1(14) &  5.0(12) &  8.0(12) &  8.4(11)  \\
IRDC079.1  &  2.3(14)&  4.1(12) &  7.2(13) &  2.4(12) &  1.5(14) &  8.5(12) &  2.3(13) &  1.8(12)  \\
\vspace{0.1cm}	               					           
IRDC079.3  &  2.1(14)&  4.0(12) &  7.2(13) &  7.1(13) &  2.3(14) &  8.2(12) &  2.5(13) &  2.2(12)  \\
IRDC18151  &  2.3(14)&  2.2(12) &  1.6(14) &  2.8(14) &  1.8(14) &  4.5(12) &  4.9(13) &  1(12)  \\
IRDC18182  &  6.9(13)&  2.1(12) &  3.4(13) &  2.4(12) &  9.7(13) &  4.3(12) &  1.6(13) &  5.8(11)  \\
IRDC18223  &  2.1(14)&  2.0(12) &  1.2(14) &  1.4(14) &  3.6(14) &  4.2(12) &  3.7(13) &  2.2(11)  \\
IRDC18306  &  5.2(13)&  2.0(12) &          &          &  9.4(13) &  4.1(12) &  1.6(13) &  1.8(12)  \\
\vspace{0.1cm}	               					           
IRDC18308  &  6.1(13)&  1.6(12) &  2.0(14) &  1.1(14) &  9.0(13) &  3.3(12) &  2.2(13) &  3.9(10)  \\
IRDC18310  &  7.1(13)&  1.7(12) &  6.6(13) &  3.4(12) &  1.6(14) &  3.5(12) &  3.3(13) &  6.9(11)  \\
IRDC18337  &  6.9(13)&  1.2(12) &  6.9(13) &  3.0(12) &  8.3(13) &  2.4(12) &  1.8(13) &  3.6(11)  \\
IRDC18385  &  3.9(13)&  2.8(12) &  2.0(13) &  7.2(13) &  4.7(13) &  5.8(12) &  1.1(13) &  1.8(13)  \\
IRDC18437  &  5.2(13)&  2.1(12) &  2.0(13) &  3.0(12) &  6.0(13) &  4.4(12) &  1.5(13) &  1.5(12)  \\
\vspace{0.1cm}	               					           
IRDC18454.1 &  4.5(13)& 2.2(12) &          &          &  6.6(13) &  4.6(12) &  1.9(13) &  2.2(12)  \\
IRDC18454.3 &  1.8(14)& 2.2(12) &  1.3(14) &  5.0(12) &  2.0(14) &  4.6(12) &  3.2(13) &  7.9(11)  \\
IRDC19175  &  5.4(13)&  3.2(12) &  1.8(13) &  1.6(12) &  5.2(13) &  6.6(12) &  5.6(12) &  2.3(12)  \\
IRDC20081  &  1.2(14)&  3.9(12) &  1.2(14) &  1.1(14) &  5.6(13) &  8.1(12) &  6.5(12) &  1.5(12)  \\
HMPO18089  &  5.0(14)&  2.6(12) &  2.0(14) &  1.5(12) &  5.5(14) &  5.4(12) &  4.9(13) &  3.7(11)  \\
\vspace{0.1cm}	               					           
HMPO18102  &  3.5(14)&  3.0(12) &  1.8(14) &  1.7(12) &  5.0(14) &  6.2(12) &  6.5(13) &  1.2(12) \\
HMPO18151  &  4.0(14)&  2.7(12) &  1.8(14) &  1.6(12) &  2.8(14) &  5.6(12) &  3.6(13) &  4.1(12)  \\
HMPO18182  &  3.8(14)&  2.4(12) &  1.3(14) &  7.0(14) &  3.1(14) &  5.1(12) &  6.4(13) &  2.4(11)  \\
HMPO18247  &  9.4(13)&  1.7(12) &  1.2(14) &  8.0(13) &  1.1(14) &  3.6(12) &  3.7(13) &  4.4(12)  \\
HMPO18264  &  5.4(14)&  2.7(12) &  8.6(14) &  4.1(14) &  3.4(14) &  5.6(12) &  1.1(14) &  3.1(12)  \\
\vspace{0.1cm}	               					           
HMPO18310  &  1.9(14)&  2.1(12) &  5.9(13) &  1.2(12) &  2.0(14) &  4.3(12) &  3.1(13) &  1.5(11)  \\
HMPO18488  &  2.5(14)&  3.1(12) &  1.3(14) &  1.4(12) &  2.0(14) &  6.3(12) &  4.1(13) &  1.5(13)  \\
HMPO18517  &  6.2(14)&  3.7(12) &  1.8(14) &  8.2(11) &  2.7(14) &  7.7(12) &  7.6(13) &  9.6(12)  \\
HMPO18566  &  3.3(14)&  3.1(12) &  1.3(14) &  1.5(12) &  3.3(14) &  6.4(12) &  5.5(13) &  1.2(12)  \\
HMPO19217  &  5.2(14)&  4.4(12) &  8.3(13) &  1.8(12) &  3.2(14) &  9.1(12) &  4.7(13) &  5.2(11)  \\
\vspace{0.1cm}	               					           
HMPO19410  &  4.4(14)&  3.7(12) &  1.4(14) &  7.5(11) &  4.1(14) &  7.7(12) &  1.0(14) &  1.3(12)  \\
HMPO20126  &  5.9(14)&  7.0(12) &  1.9(14) &  1.2(15) &  6.6(14) &  1.4(13) &  7.6(13) &  6.4(11)  \\
HMPO20216  &  2.0(14)&  6.2(12) &  5.9(13) &  1.4(12) &  1.5(14) &  1.3(13) &  1.0(13) &  3.0(12)  \\
HMPO20293  &  3.9(14)&  5.0(12) &  1.1(14) &  3.7(14) &  4.2(14) &  1.0(13) &  1.1(14) &  3.7(12)  \\
HMPO22134  &  1.9(14)&  3.8(12) &  1.2(14) &  4.7(13) &  7.9(13) &  7.9(12) &  1.1(13) &  4.0(11)  \\
\vspace{0.1cm}	               					           
HMPO23033  &  6.2(14)&  4.1(12) &  1.8(14) &  1.9(13) &  3.0(14) &  8.5(12) &  4.5(13) &  1.2(12)  \\
HMPO23139  &  3.0(14)&  4.2(12) &  3.0(14) &  1.4(14) &  1.5(14) &  8.8(12) &  3.0(13) &  1.3(12)  \\
HMPO23151  &  1.1(14)&  4.7(12) &  5.4(13) &    4(13) &  6.0(13) &  9.7(12) &  1.1(12) &  8.0(12)  \\
HMPO23545  &  1.3(14)&  3.8(12) &  2.4(13) &    5(12) &  4.1(13) &  7.8(12) &  2.3(12) &  1.2(12)  \\
HMC009.62  &  7.6(14)&  2.2(12) &  3.2(14) &  1.1(13) &  5.5(14) &  4.5(12) &  1.0(14) &  3.8(11)  \\
\vspace{0.1cm}	               					           
HMC010.47  &  8.7(14)&  2.3(12) &  3.5(14) &  1.9(12) &  8.1(14) &  4.8(12) &  1.6(14) &  3.9(12)  \\
HMC029.96  &  7.1(14)&  2.7(12) &  4.2(14) &  1.8(12) &  5.5(14) &  5.6(12) &  5.9(13) &  5.6(11)  \\
HMC031.41  &  3.4(14)&  2.9(12) &  1.0(14) &  3.9(12) &  4.9(14) &  6.1(12) &  4.4(13) &  2.3(12)  \\
HMC034.26  &  2.1(15)&  3.8(12) &  5.0(14) &  1.4(13) &  1.6(15) &  8.0(12) &  9.7(13) &  1.5(12)  \\
HMC045.47  &  9.0(14)&  3.6(12) &  1.2(14) &    1(12) &  6.4(14) &  7.5(12) &  7.3(13) &  6.4(11)  \\
\vspace{0.1cm}	               					           
HMC075.78  &  9.3(14)&  6.7(12) &  1.8(14) &  1.3(12) &  4.6(14) &  1.4(13) &  5.5(13) &  9.5(11)  \\
W3H2O      &  8.9(14)&  5.5(12) &  5.5(14) &  5.6(14) &  3.3(14) &  1.1(13) &  2.0(13) &  2.0(12)  \\
W3IRS5     &  3.6(14)&  3.4(12) &  3.3(14) &  1.6(12) &  1.3(14) &    7(12) &  2.2(12) &  4.4(11)  \\
NGC7538B   &  7.2(14)&  5.1(12) &  4.5(14) &  3.1(12) &  3.0(14) &  1.1(13) &  1.3(13) &  1.0(12)  \\
Orion-KL   &  7.3(14)&    1(13) &  1.2(16) &  3.1(13) &  7.3(14) &  2.1(13) &  4.3(12) &  1.5(12)  \\
\vspace{0.1cm}	               					           
UCH005.89  &  2.1(15)&  4.1(12) &  1.3(15) &  5.3(12) &  2.6(15) &  8.5(12) &  2.1(14) &  7.2(11)  \\
UCH010.10  &  5.5(13)&  1.9(12) &  3.6(13) &  2.8(13) &  1.1(14) &    4(12) &  2.1(13) &  7.3(13)  \\
UCH010.30  &  4.9(14)&  1.9(12) &  6.7(14) &  2.9(12) &  4.9(14) &    4(12) &  1.3(14) &  6.0(11)  \\
UCH012.21  &  5.9(14)&  3.2(12) &  2.6(14) &  3.7(12) &  5.9(14) &  6.6(12) &  7.3(13) &  3.1(13)  \\
UCH013.87  &  3.8(14)&  2.6(12) &  4.1(14) &  2.5(12) &  2.0(14) &  5.4(12) &  3.9(13) &  3.3(11)  \\
\vspace{0.1cm}	               					           
UCH030.54  &  1.1(14)&  2.7(12) &  1.0(14) &  8.5(11) &  5.4(13) &  5.5(12) &  5.8(12) &  2.0(12)  \\
UCH035.20  &  3.6(14)&  3.4(12) &  3.4(14) &  1.3(12) &  2.5(14) &  6.9(12) &  3.7(13) &  3.4(11)  \\
UCH045.12  &  4.2(14)&  3.5(12) &  5.6(14) &  2.5(12) &  1.7(14) &  7.2(12) &  8.7(12) &  1.7(12)  \\
UCH045.45  &  1.9(14)&  3.3(12) &  1.4(14) &  1.1(12) &  1.3(14) &  6.8(12) &  2.1(13) &  4.3(11)  \\
\hline
\end{tabular}
\tablefoot{Column densities written as $a(x)=a \times 10^x$. H$_2$
  column density is averaged over a 29$\arcsec$-beam and for all other
  molecules over a 30$\arcsec$-beam. The given errors are
  uncertainties in the measured integrated fluxes and the optical
  depth $\tau$. The high uncertainties in some cases are due to high
  uncertainties in the determined optical depth.}
\end{table*}

\clearpage

%Merged Table
\begin{table*}
\tiny
\caption{Parameters of the best-fit IRDC model.}
% title of Table
\label{tab:IRDC_fit_31}      % is used to refer this table in the text
%\centering                          % used for centering table
\begin{tabular}{lll}        % centered columns (4 columns)
\hline\hline                 % inserts double horizontal lines
Parameter & Symbol & Model
\\    % table heading 
\hline                        % inserts single horizontal line
Inner radius & $r_0$ & $12\,700$~AU \\
Outer radius & $r_1$ & $0.5$~pc\tablefootmark{\rm \bf a}\\
Density at the inner radius & $\rho_0$ & $1.4\times10^{5}$~cm$^{-3}$ \\
Average density with a beam of $26\,000$~AU& $\bar{\rho}$ & $8.9\times10^{4}$~cm$^{-3}$ \\
Average density with a beam of $54\,000$~AU& $\bar{\rho}$ & $5.2\times10^{4}$~cm$^{-3}$ \\
Density profile & $p$ & $1.5$ \\
Temperature at the inner radius & $T_0$ & $11.3$~K \\
Average temperature & $\bar{T}$ & $11.3$~K  \\
Temperature profile & $q$ & $0$ \\
Lifetime & & $16\,500$~years\\
\hline                                   %inserts single line
\end{tabular}
\tablefoot{}
\tablefoottext{a}{This value is limited by the largest $29\arcsec$ IRAM beam size used in
our observations.}
\end{table*}

% Best-fit HMPOs parameters:
\begin{table*}
\tiny
\caption{Parameters of the best-fit HMPO model.}
% title of Table
\label{tab:HMPO_fit_31}      % is used to refer this table in the text
%\centering                          % used for centering table
\begin{tabular}{lll}        % centered columns (4 columns)
\hline\hline                 % inserts double horizontal lines
Parameter & Symbol & Model \\    % table heading 
\hline                        % inserts single horizontal line
Inner radius & $r_0$ & $103$~AU \\
Outer radius & $r_1$ & $0.5$~pc\tablefootmark{\rm \bf a} \\
Density at the inner radius & $\rho_0$ & $1.5\times10^{9}$~cm$^{-3}$ \\
Average density with a beam of $21\,700$~AU& $\bar{\rho}$ & $1.6\times10^{6}$~cm$^{-3}$ \\
Average density with a beam of $57\,300$~AU& $\bar{\rho}$ & $3.7\times10^{5}$~cm$^{-3}$ \\
Density profile & $p$ & $1.8$ \\
Temperature at the inner radius & $T_0$ & $75.8$~K \\
Average temperature & $\bar{T}$ & $21.5$~K \\
Temperature profile & $q$ & $0.4$ \\
Lifetime & & $32\,000$~years \\
\hline                                   %inserts single line
\end{tabular}
\tablefoot{}
\tablefoottext{a}{This value is limited by the largest $29\arcsec$ IRAM beam size used in
our observations.}
\end{table*}

% Best-fit HMC parameters:
\begin{table*}
\tiny
\caption{Parameters of the best-fit HMC model.}
% title of Table
\label{tab:HMC_fit_31}      % is used to refer this table in the text
%\centering                          % used for centering table
\begin{tabular}{lll}        % centered columns (4 columns)
\hline\hline                 % inserts double horizontal lines
Parameter & Symbol & Model \\    % table heading 
\hline                        % inserts single horizontal line
Inner radius & $r_0$ & $1\,140$~AU \\
Outer radius & $r_1$ & $0.5$~pc\tablefootmark{\rm \bf a} \\
Density at the inner radius & $\rho_0$ & $1.3\times10^{8}$~cm$^{-3}$ \\
Average density with a beam of $45\,400$~AU& $\bar{\rho}$ & $1.7\times10^{6}$~cm$^{-3}$ \\
Average density with a beam of $63\,100$~AU& $\bar{\rho}$ & $9.6\times10^{5}$~cm$^{-3}$ \\
Density profile & $p$ & $2.0$ \\
Temperature at the inner radius & $T_0$ & $162.9$~K \\
Average temperature & $\bar{T}$ & $50-55$~K \\
Temperature profile & $q$ & $0.4$ \\
Lifetime & & $35\,000$~years \\
\hline                                   %inserts single line
\end{tabular}
\tablefoot{}
\tablefoottext{a}{This value is limited by the largest $29\arcsec$ IRAM beam size used in our observations.}
\end{table*}

% Best-fit UCH parameters:
\begin{table*}
\tiny
\caption{Parameters of the best-fit UCH{\sc ii} model.}
% title of Table
\label{tab:UCH_fit_31}      % is used to refer this table in the text
%\centering                          % used for centering table
\begin{tabular}{lll}        % centered columns (4 columns)
\hline\hline                 % inserts double horizontal lines
Parameter & Symbol & Model \\    % table heading 
\hline                        % inserts single horizontal line
Inner radius & $r_0$ & $103$~AU \\
Outer radius & $r_1$ & $0.5$~pc\tablefootmark{\rm \bf a} \\
Density at the inner radius & $\rho_0$ & $1.0\times10^{10}$~cm$^{-3}$ \\
Average density with a beam of $57\,800$~AU& $\bar{\rho}$ & $1.3\times10^{6}$~cm$^{-3}$ \\
Average density with a beam of $85\,400$~AU& $\bar{\rho}$ & $7.1\times10^{5}$~cm$^{-3}$ \\
Density profile & $p$ & $2.0$ \\
Temperature at the inner radius & $T_0$ & $244.3$~K \\
Average temperature & $\bar{T}$ & $31-34$~K \\
Temperature profile & $q$ & $0.4$ \\
Lifetime & & $3\,000$~years \\
\hline                                   %inserts single line
\end{tabular}
\tablefoot{}
\tablefoottext{a}{This value is limited by the largest $29\arcsec$ IRAM beam size used in our observations.}
\end{table*}

\clearpage

\begin{table}
\tiny
\caption{Median column densities in $a(x)=a \times 10^x$ for
  observations (including detections and upper limits) and best-fit
  IRDC model. Modeled best-fit values in italics do not agree with the
  observed values within one order of magnitude. If the molecule is
  detected in less than $50\%$ of the sources we marked it as an upper
  limit.}
% title of Table
\label{tbl:bestfit_IRDC_31}      % is used to refer this table in the text
\centering                          % used for centering table
\begin{tabular}{lcc}        % centered columns (4 columns)
\hline\hline                 % inserts double horizontal lines
Molecule & Obs. col. den. & Mod. col. den. \\
 & $[{\rm cm}^{-2}]$ & $[{\rm cm}^{-2}]$ \\    % table heading 
\hline                        % inserts single horizontal line
CO & 1.9(18) & 2.8(18)  \\
HNC & 9.7(13)& 5.8(14)  \\
HCN & 7.2(13) & 7.1(14) \\
HCO$^+$ & 7.9(13) & 8.9(13) \\
HNCO & 2.4(12) & 2.7(12)  \\
H$_2$CO & 1.7(13) & 5.1(13) \\
N$_2$H$^+$ & 2.2(13) & 2.2(12) \\
CS & $\leq$4.7(14) & 6.8(14)  \\
SO & $\leq$6.6(12) & 5.4(13)  \\
OCS & $\leq$4.2(14) & 4.0(12) \\
C$_2$H & 4.8(14) & 2.5(14) \\
SiO & 2.3(12) & 3.5(12) \\
CH$_3$CN & $\leq$3.7(12) & 1.3(13) \\
CH$_3$OH & $\leq$4.1(13) & 2.8(13) \\
DCO$^+$ & $\leq$3.1(11) & 5.9(11)  \\
DCN & $\leq$6.1(11) & 2.4(12) \\
DNC & $\leq$6.1(11) & 1.6(11) \\
N$_2$D$^+$ & $\leq$6.8(11) & 4.0(8) \\
\hline
Agreement & & 18/18 = 100\%  \\ 
\hline                                   %inserts single line
\end{tabular}
\end{table}

\begin{table}
\tiny
\caption{Median column densities in $a(x)=a \times 10^x$ for
  observations (including detections and upper limits) and best-fit
  HMPO model. Modeled best-fit values in italics do not agree with the
  observed values within one order of magnitude. If the molecule is
  detected in less than $50\%$ of the sources we marked it as an upper
  limit.}
% title of Table
\label{tbl:bestfit_HMPO_31}      % is used to refer this table in the text
\centering                          % used for centering table
\begin{tabular}{lcc}        % centered columns (4 columns)
\hline\hline                 % inserts double horizontal lines
Molecule  & Obs. col. den. & Mod. col. den. \\
 & $[{\rm cm}^{-2}]$ & $[{\rm cm}^{-2}]$ \\    % table heading 
\hline                        % inserts single horizontal line
CO & 5.0(18) & 6.9(18) \\
HNC & 2.9(14) & 5.3(14) \\
HCN & 1.3(14) & 6.4(14) \\
HCO$^+$ & 3.7(14) & 3.6(14) \\
HNCO & 3.8(12) & 3.6(13)   \\
H$_2$CO & 4.1(13) & 3.8(13) \\
N$_2$H$^+$ & 4.6(13) & 6.6(13) \\
CS & 1.1(15) & 3.9(14) \\
SO & 8.2(13) & {\it 1.8(15)} \\
OCS & $\leq$2.0(14) & 3.7(13) \\
C$_2$H & 1.9(15) & 2.0(14) \\
SiO & 5.9(12) & 1.2(13) \\
CH$_3$CN & 3.4(12) & 7.7(12) \\
CH$_3$OH & 1.4(14) & {\it 3.3(12)} \\
DCO$^+$ & 6.7(11) & 3.9(12) \\
DCN & 2.0(12) & 3.0(12) \\
DNC & 1.0(12) & 2.0(12) \\
N$_2$D$^+$ & $\leq$3.8(11) & 6.3(11) \\
\hline
Agreement & & 16/18 = 89\% \\  
\hline                                   %inserts single line
\end{tabular}
\end{table}

\begin{table}
\tiny
\caption{Median column densities in $a(x)=a \times 10^x$ for
  observations (including detections and upper limits) and best-fit
  HMC model. Modeled best-fit values in italics do not agree with the
  observed values within one order of magnitude. If the molecule is
  detected in less than $50\%$ of the sources we marked it as an upper
  limit.}
 % title of Table
\label{tbl:bestfit_HMC_31}      % is used to refer this table in the text
\centering                          % used for centering table
\begin{tabular}{lcc}        % centered columns (4 columns)
\hline\hline                 % inserts double horizontal lines
Molecule & Obs. col. den. & Mod. col. den. \\
 & $[{\rm cm}^{-2}]$ & $[{\rm cm}^{-2}]$ \\    % table heading 
\hline                        % inserts single horizontal line
CO & 2.2(19) & 6.8(19) \\
HNC & 5.5(14) & 1.7(15) \\
HCN & 3.5(14) & 2.0(15) \\
HCO$^+$ & 7.6(14) & 1.5(15) \\
HNCO & 1.7(13) & {\it 5.0(16)} \\
H$_2$CO & 2.7(14) & 2.6(14) \\
N$_2$H$^+$ & 5.5(13) & 7.8(13) \\
CS & 1.3(16) & 7.5(15) \\
SO & 5.2(14) & {\it 6.6(16)} \\
OCS & 1.7(15) & 1.3(15) \\
C$_2$H & 4.3(15) & {\it 2.4(14)} \\
SiO & 2.2(13) & 5.6(13) \\
CH$_3$CN & 5.2(13) & 6.0(13) \\
CH$_3$OH & 8.7(14) & {\it 3.2(12)} \\
DCO$^+$ & 2.0(12) & 1.7(13) \\
DCN & 8.0(12) & 8.9(12) \\
DNC & 8.7(11) & 8.1(12) \\
N$_2$D$^+$ & $\leq$2.2(11) & 5.5(11) \\
\hline
Agreement & & 14/18 = 78\%  \\ 
\hline                                   %inserts single line
\end{tabular}
\end{table}

\begin{table}
\tiny
\caption{Median column densities in $a(x)=a \times 10^x$ for
  observations (including detections and upper limits) and best-fit
  UCH{\sc ii} model. Modeled best-fit values in italics do not agree
  with the observed values within one order of magnitude. If the
  molecule is detected in less than $50\%$ of the sources we marked it
  as an upper limit.}
 % title of Table
\label{tbl:bestfit_UCH_31}      % is used to refer this table in the text
\centering                          % used for centering table
\begin{tabular}{lcc}        % centered columns (4 columns)
\hline\hline                 % inserts double horizontal lines
Molecule & Obs. col. den. & Mod. col. den. \\
 & $[{\rm cm}^{-2}]$ & $[{\rm cm}^{-2}]$ \\    % table heading 
\hline                        % inserts single horizontal line
CO & 1.4(19) & 3.7(19) \\
HNC & 2.0(14) & 2.2(14) \\
HCN & 3.4(14) & 2.7(14) \\
HCO$^+$ & 3.8(14) & 6.6(14) \\
HNCO & 2.7(12) & {\it 3.9(15)} \\
H$_2$CO & 7.5(13) & 3.5(13) \\
N$_2$H$^+$ & 3.7(13) & 4.9(13) \\
CS & 2.3(15) & 1.9(15) \\
SO & 7.7(13) & {\it 3.5(16)} \\
OCS & $\leq$5.1(13) & 2.9(14) \\
C$_2$H & 3.0(15) & {\it 6.0(13)} \\
SiO & 3.4(12) & {\it 3.7(13)} \\
CH$_3$CN & 8.6(12) & 1.1(13) \\
CH$_3$OH & 6.6(13) & 5.6(13) \\
DCO$^+$ & 4.6(11) & {\it 7.0(12)} \\
DCN & 1.9(12) & 1.3(12) \\
DNC & $\leq$3.0(11) & 1.0(12) \\
N$_2$D$^+$ & $\leq$3.3(11) & 3.9(11) \\
\hline
Agreement & & 13/18 = 72\% \\ 
\hline                                   %inserts single line
\end{tabular}
\end{table}

\newpage

\begin{figure*}
\includegraphics[width=0.4\textwidth]{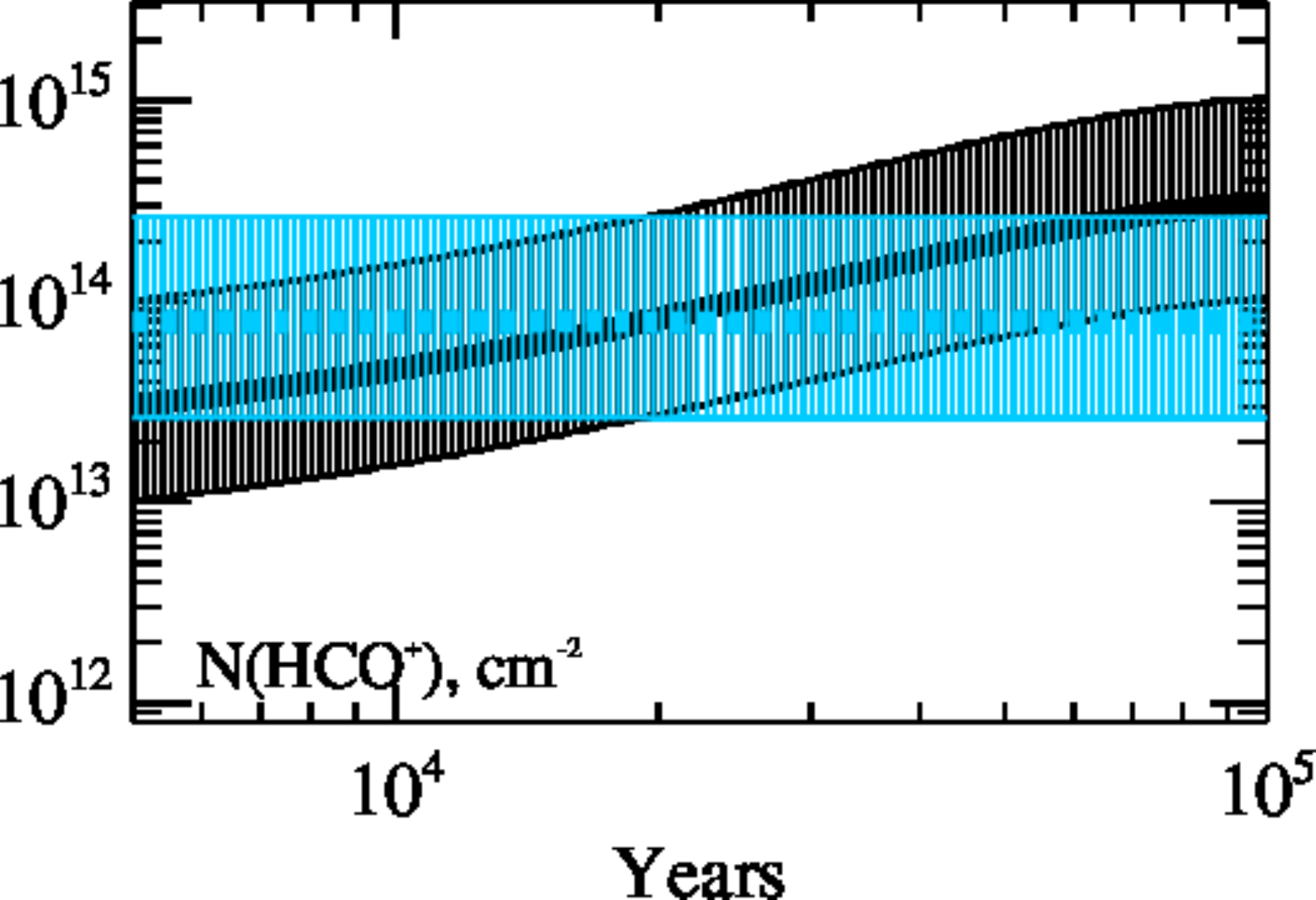}
\includegraphics[width=0.4\textwidth]{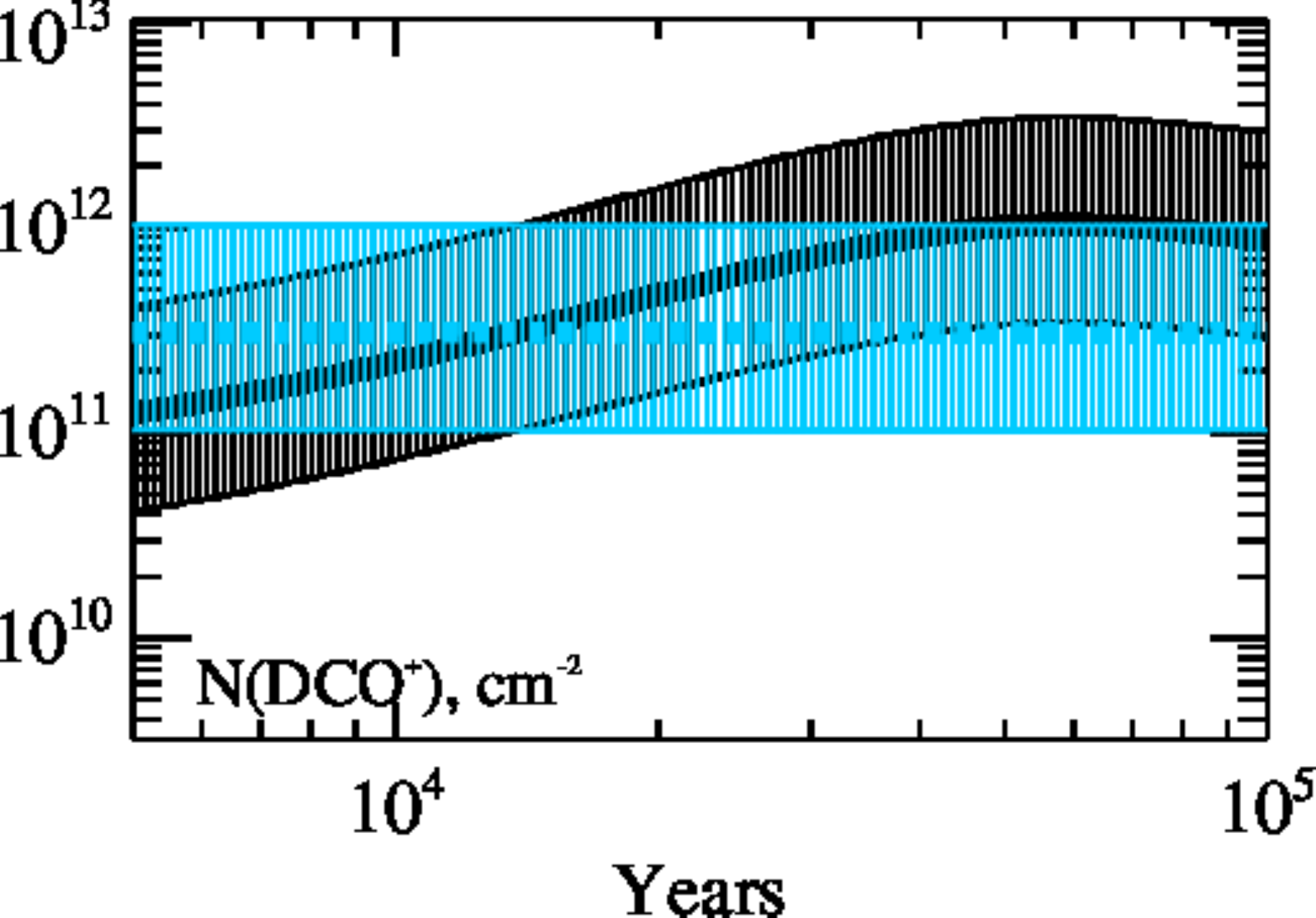}\\
\includegraphics[width=0.4\textwidth]{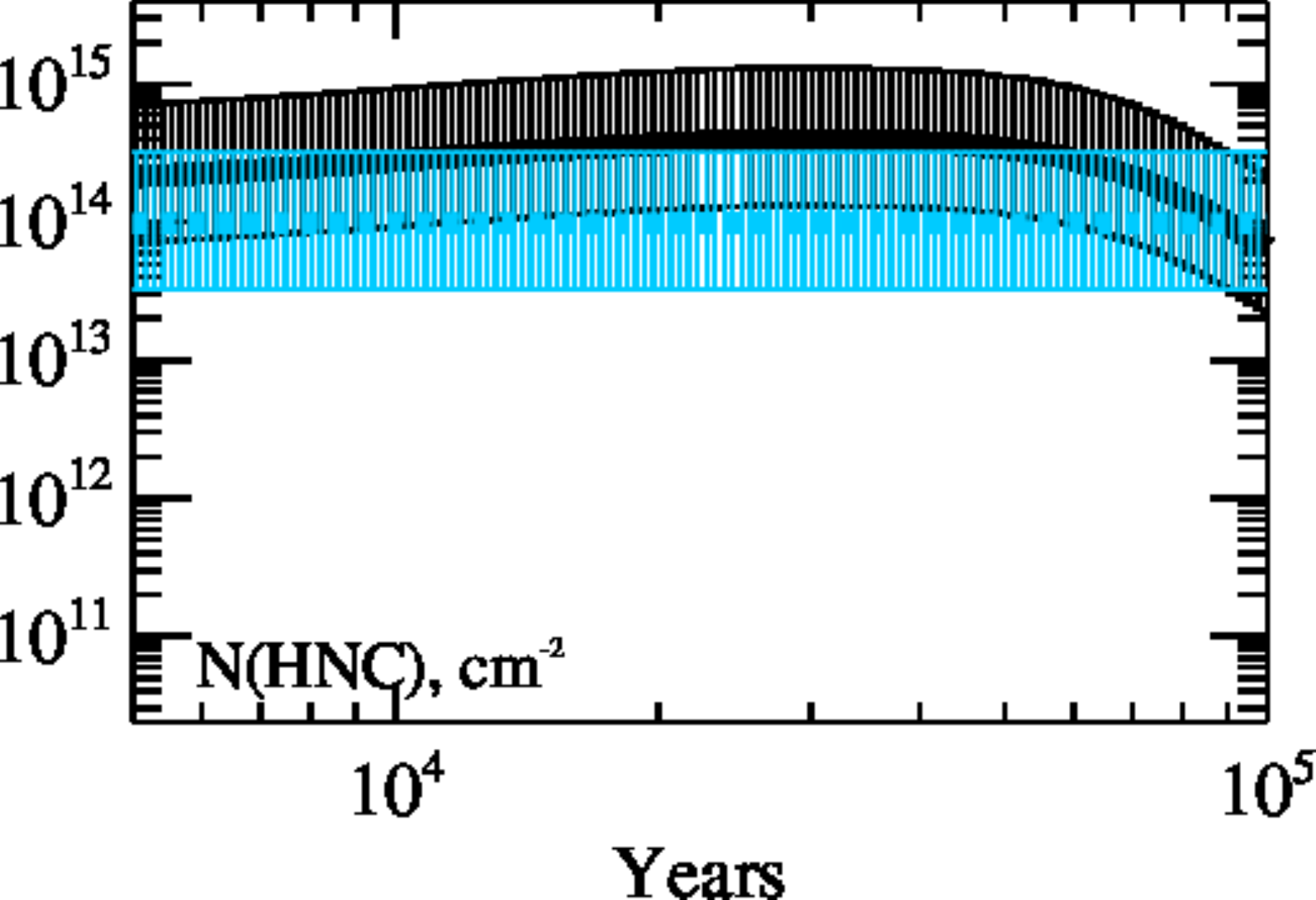}
\includegraphics[width=0.4\textwidth]{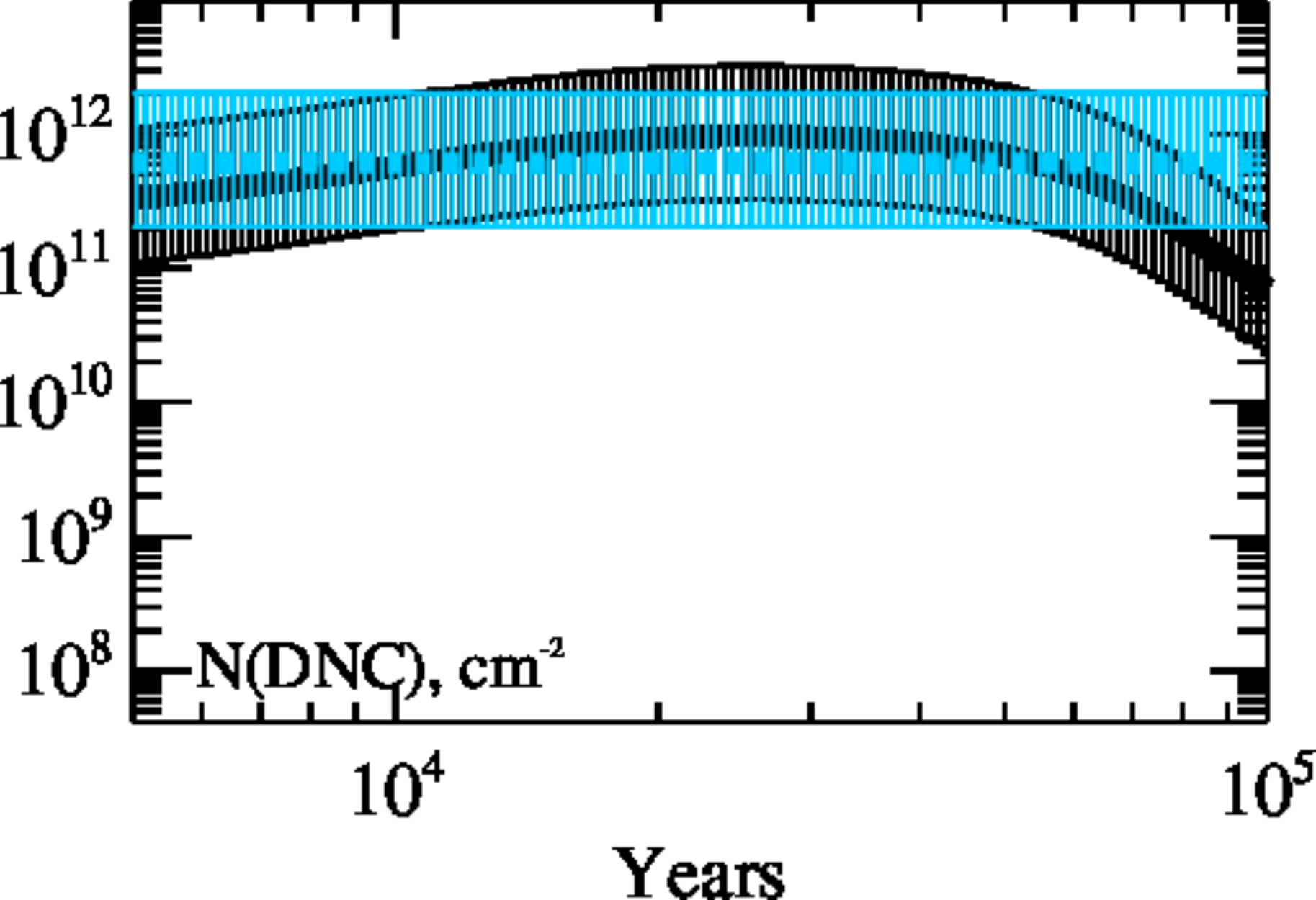}\\
\includegraphics[width=0.4\textwidth]{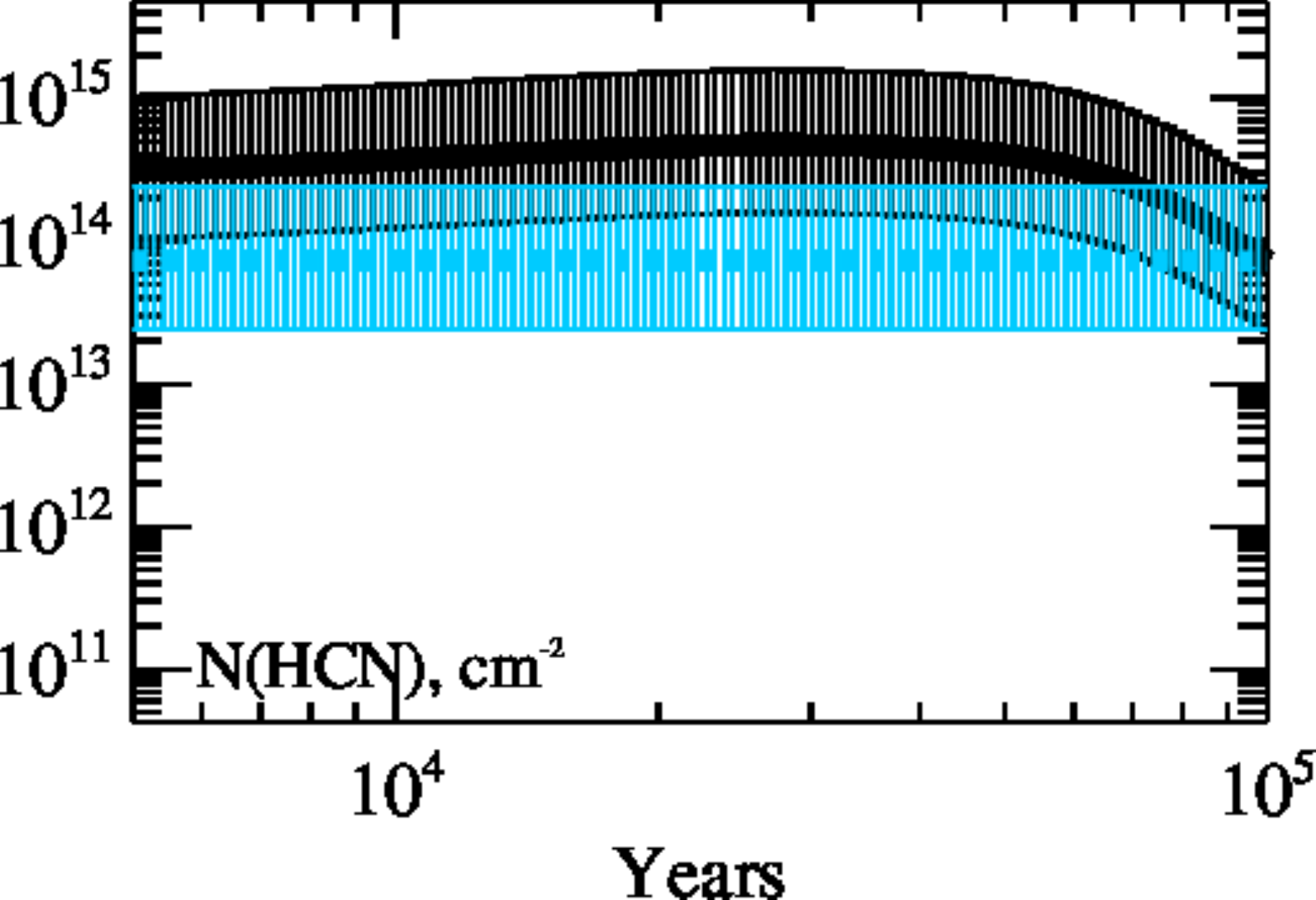}
\includegraphics[width=0.4\textwidth]{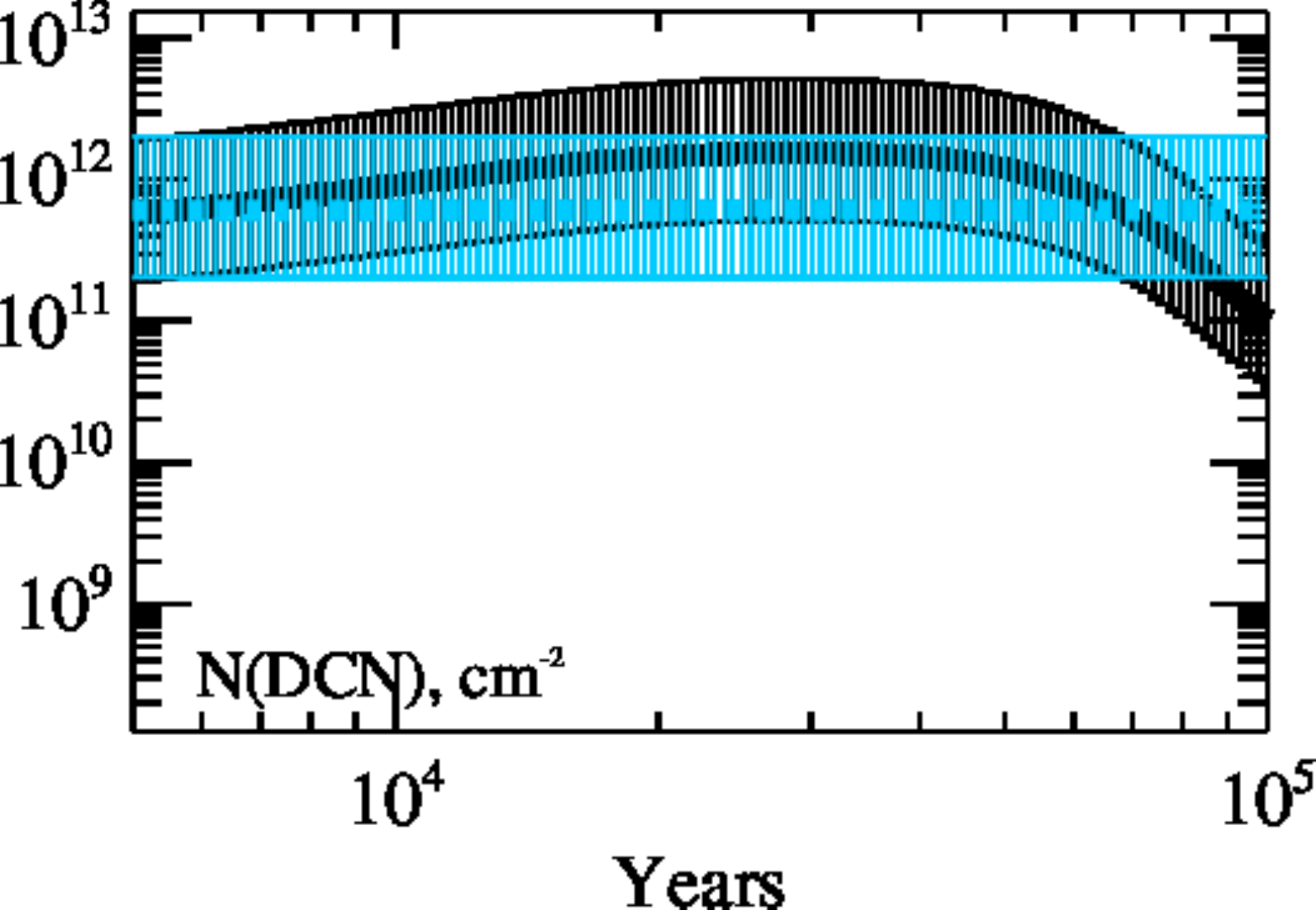}\\
\includegraphics[width=0.4\textwidth]{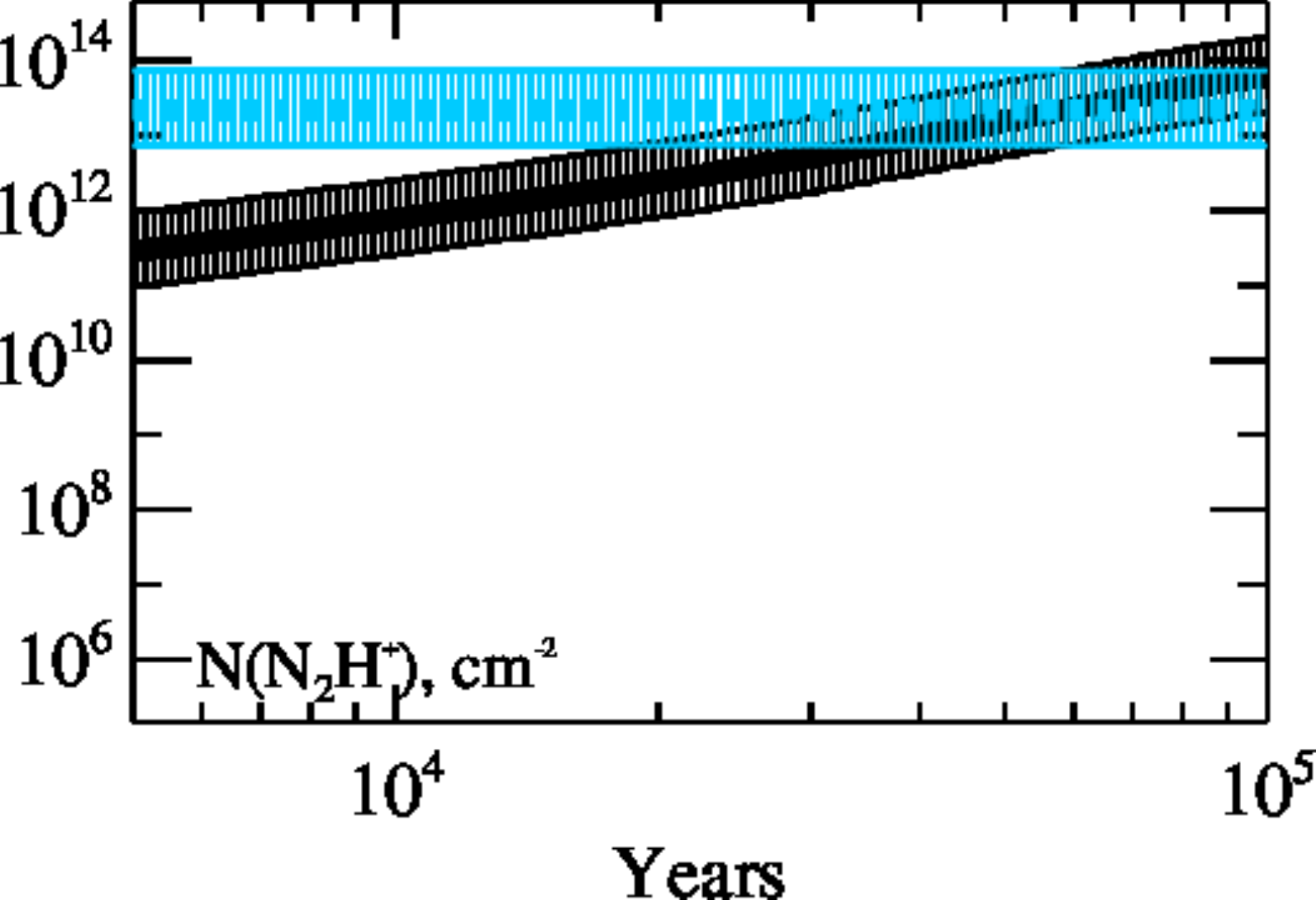}
\includegraphics[width=0.4\textwidth]{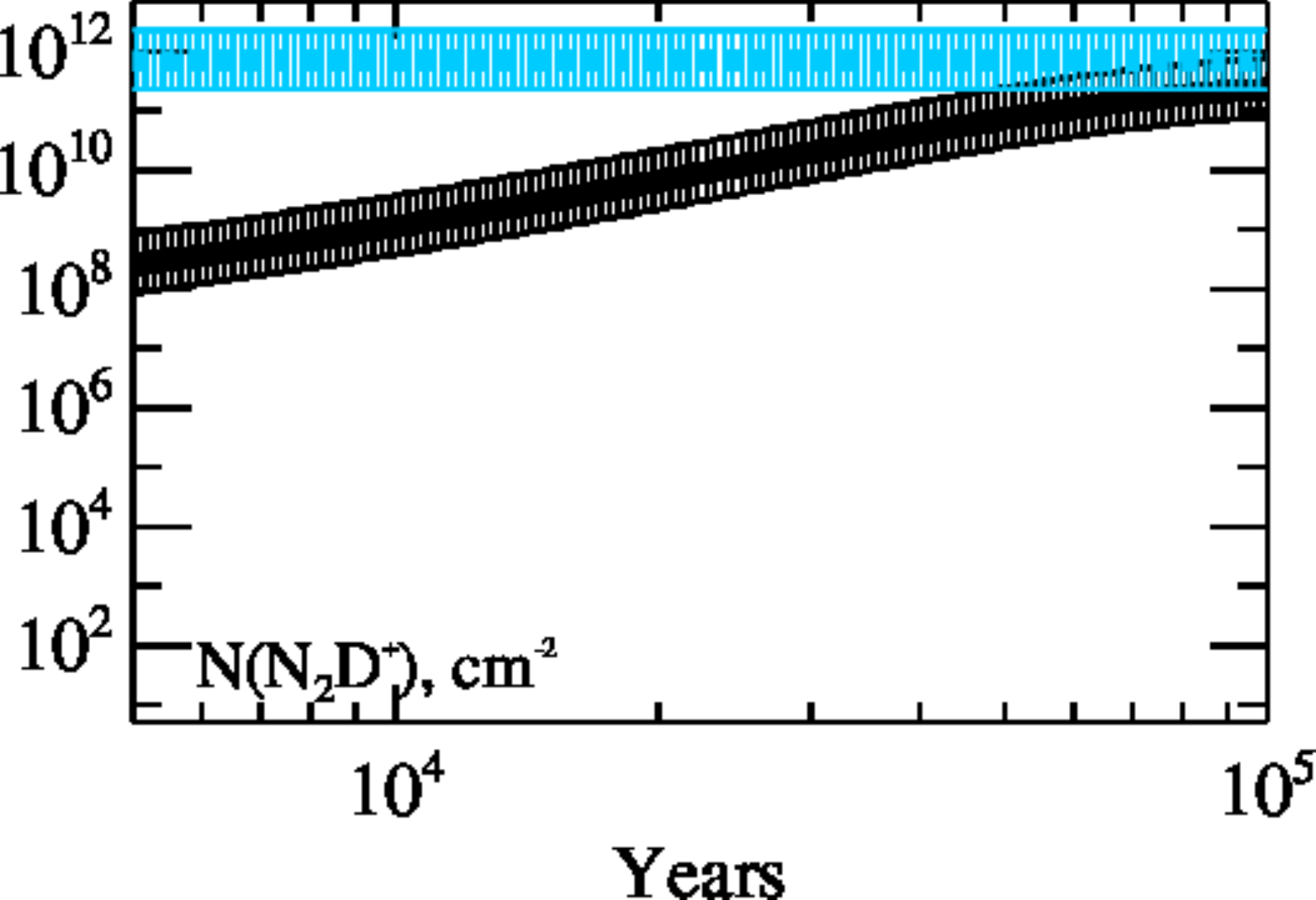}\\
\caption{Observed and modeled column densities in cm$^{-2}$ in the IRDC stage. The observed values are shown in blue, the modeled values in black. The error bars are indicated by the vertical marks. Molecules are labeled in the plots.}
\label{fig:nx_stage1_31}
\end{figure*}

% Stage2
\newpage
\begin{figure*}
\includegraphics[width=0.4\textwidth]{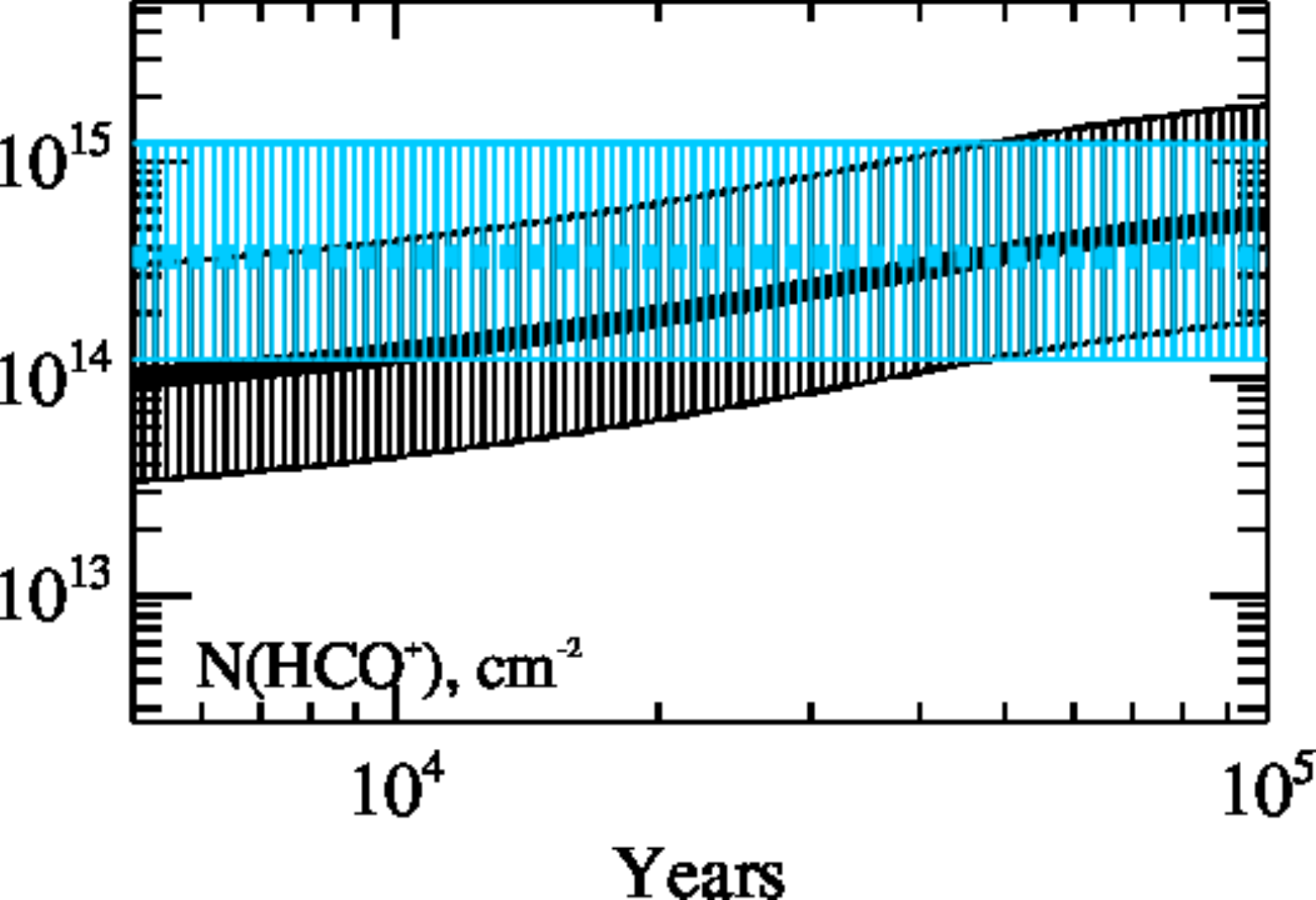}
\includegraphics[width=0.4\textwidth]{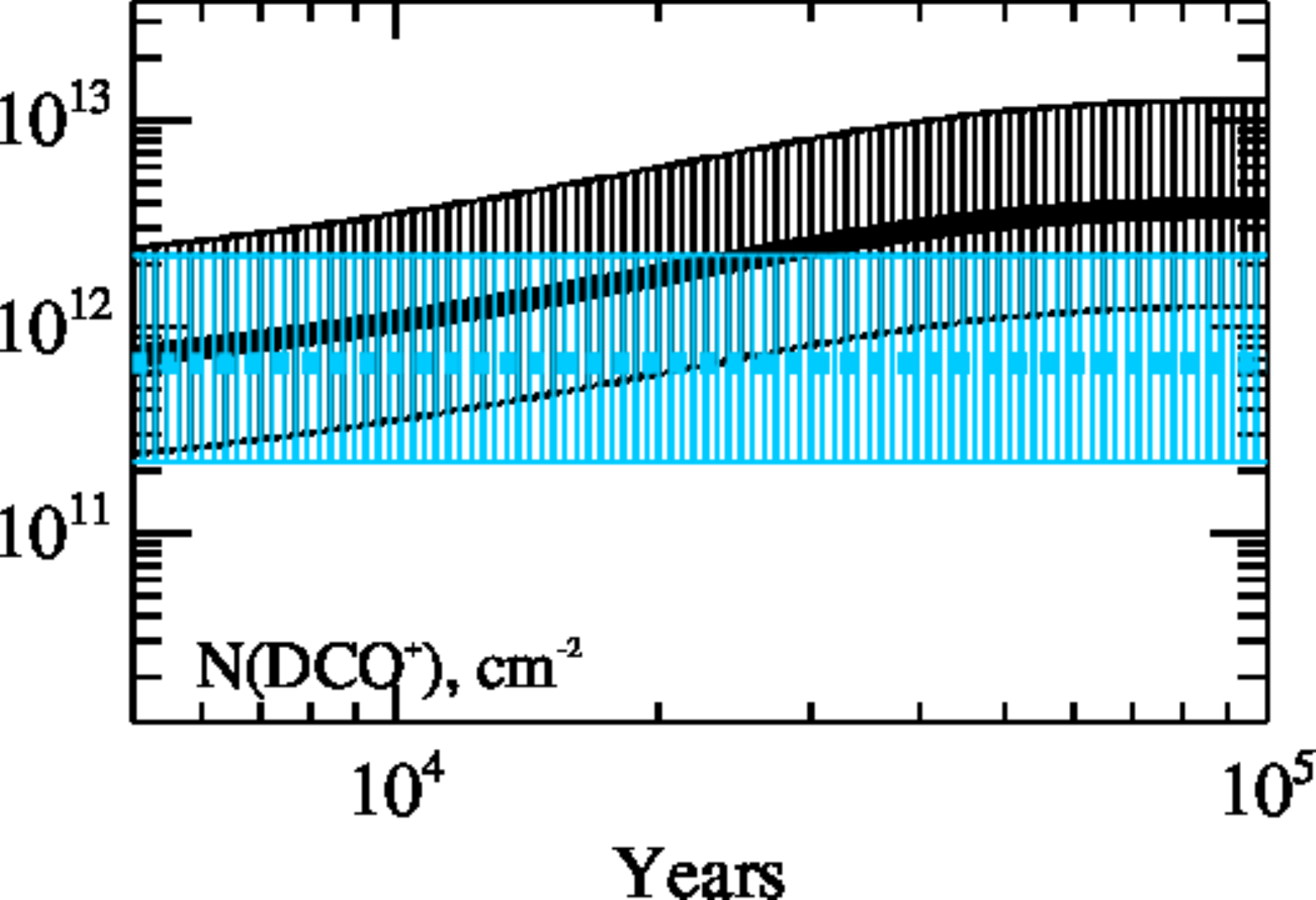}\\
\includegraphics[width=0.4\textwidth]{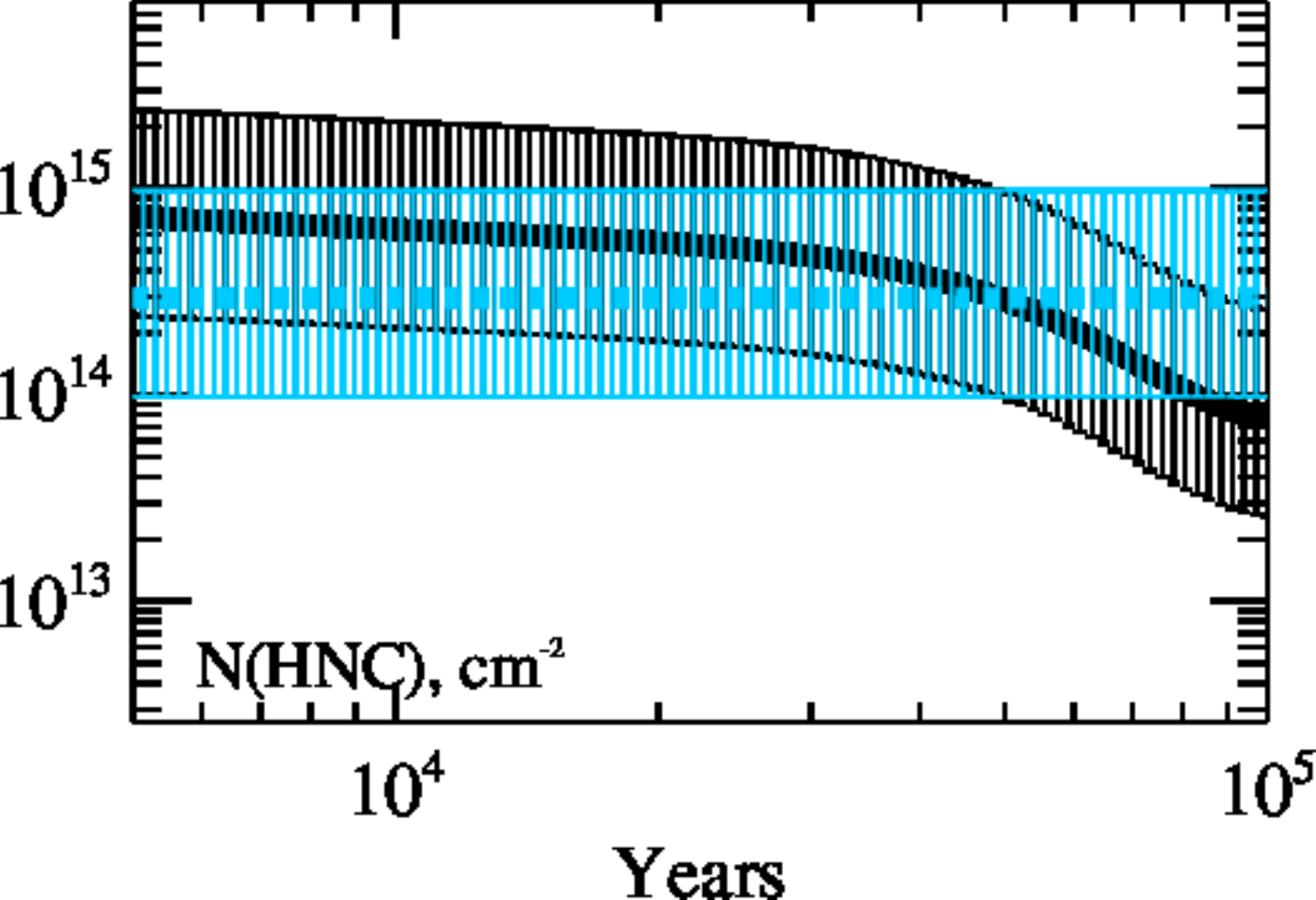}
\includegraphics[width=0.4\textwidth]{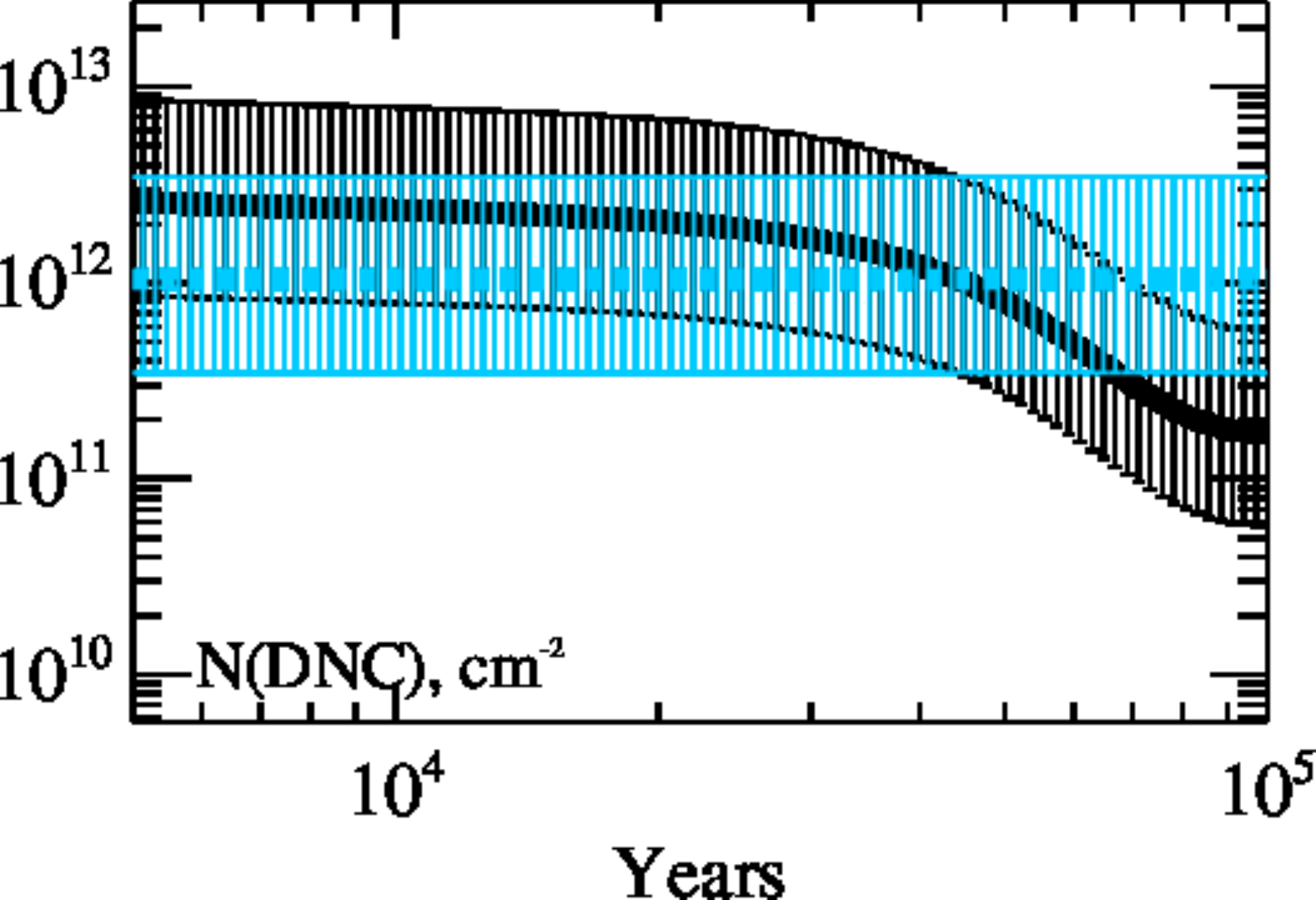}\\
\includegraphics[width=0.4\textwidth]{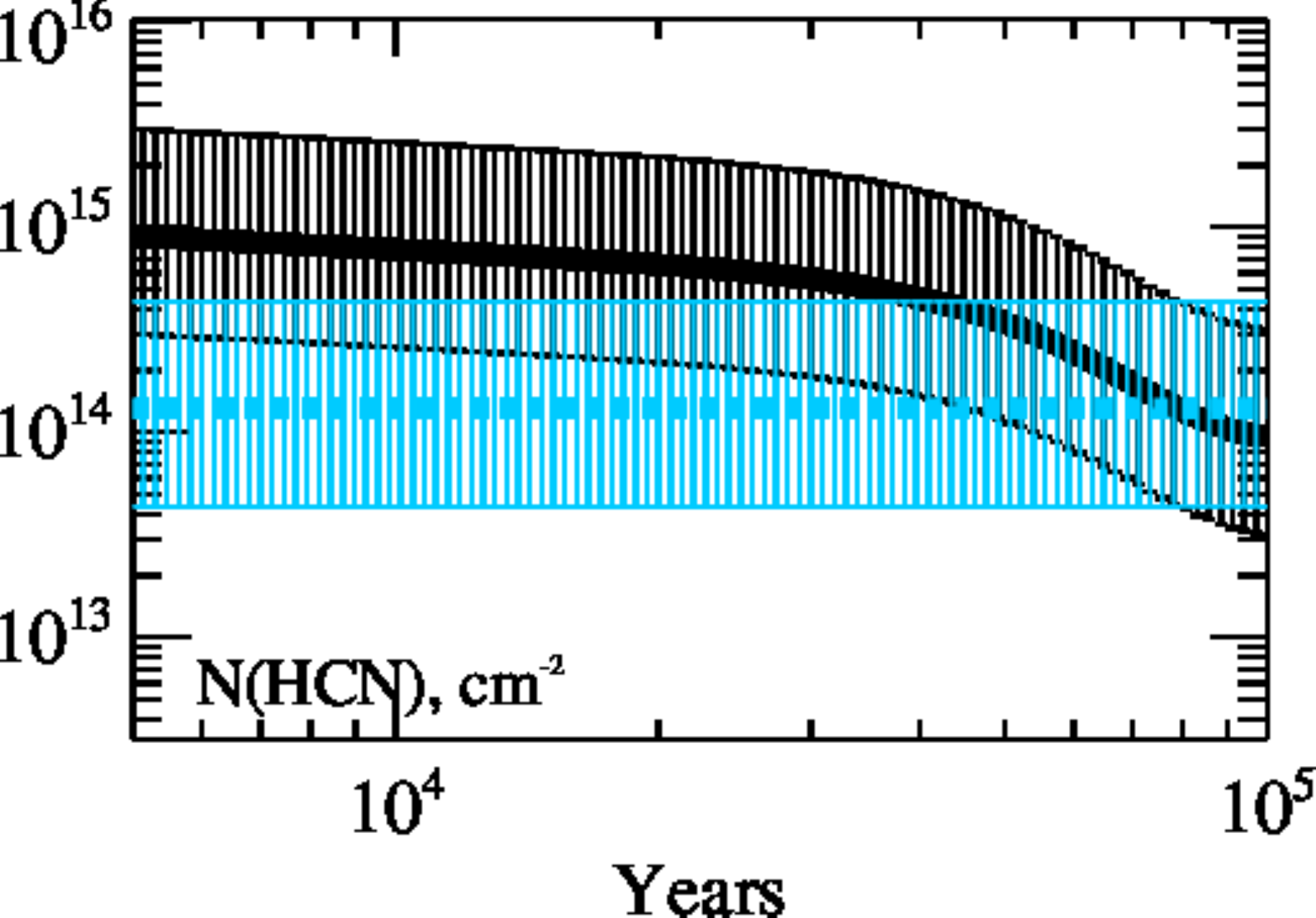}
\includegraphics[width=0.4\textwidth]{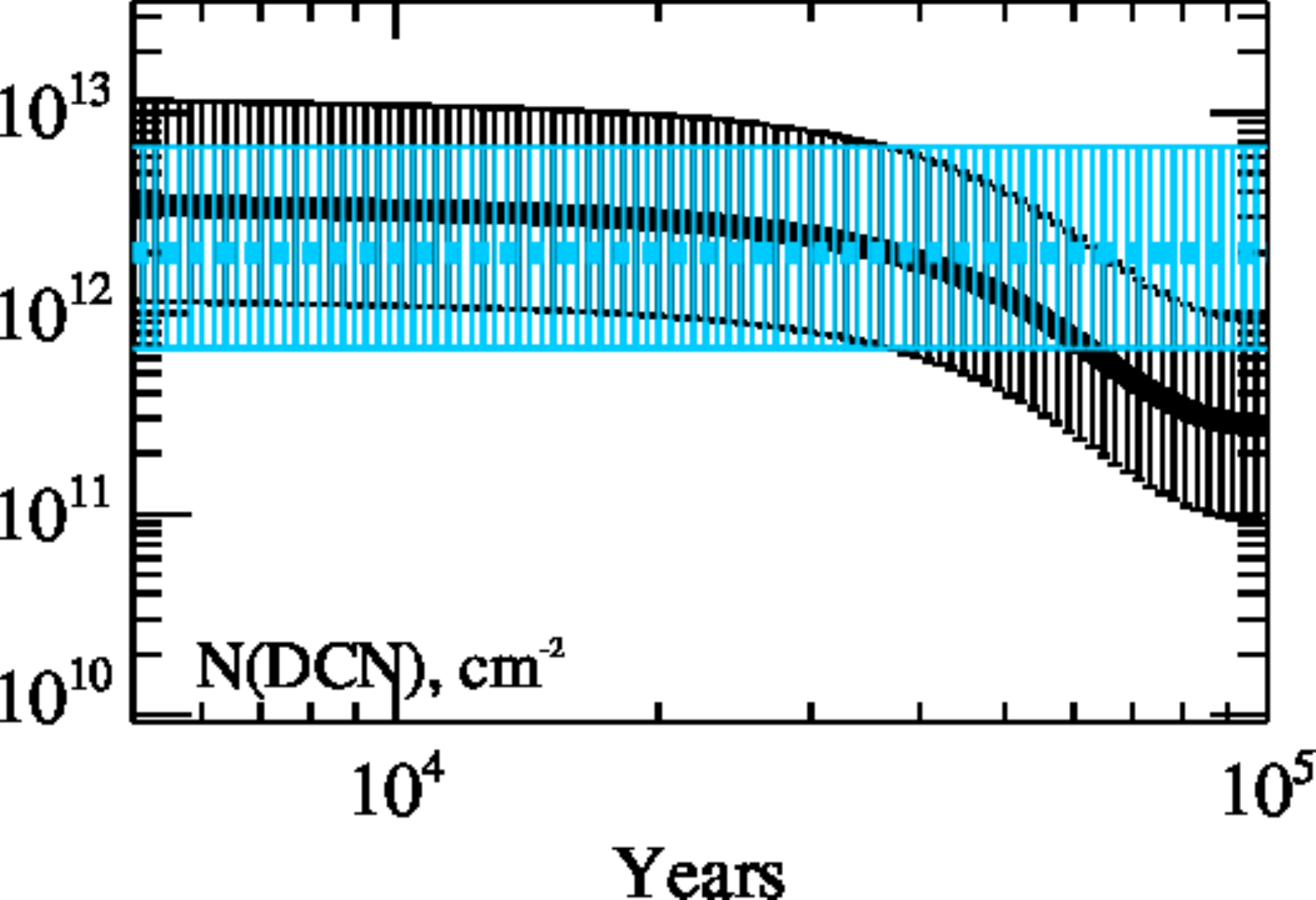}\\
\includegraphics[width=0.4\textwidth]{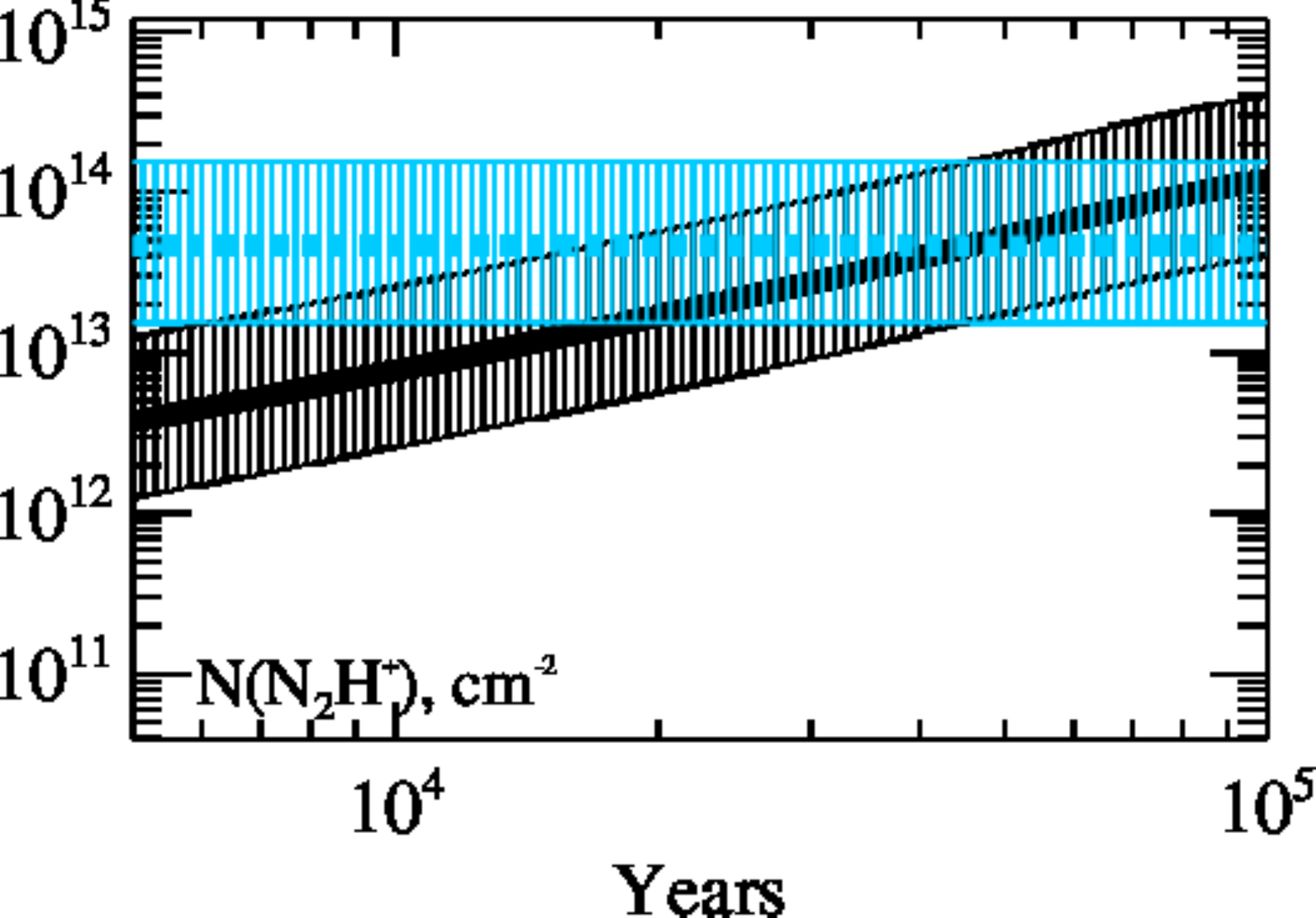}
\includegraphics[width=0.4\textwidth]{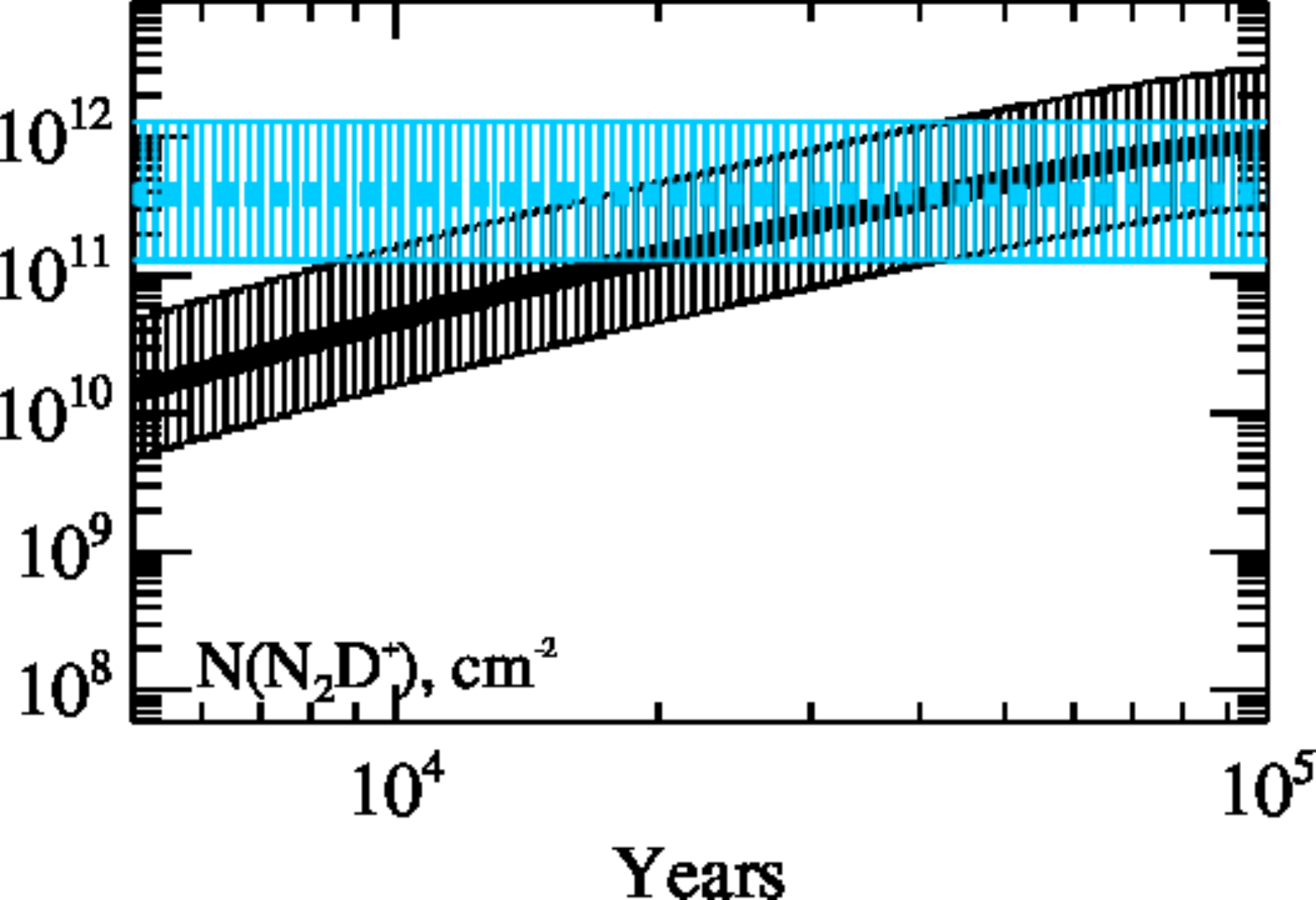}\\
\caption{Observed and modeled column densities in cm$^{-2}$ in the HMPO stage. The observed values are shown in blue, the modeled values in black. The error bars are indicated by the vertical marks. Molecules are labeled in the plots.}
\label{fig:nx_stage2_31}
\end{figure*}

% Stage3
%\clearpage
\newpage
\begin{figure*}
\includegraphics[width=0.4\textwidth]{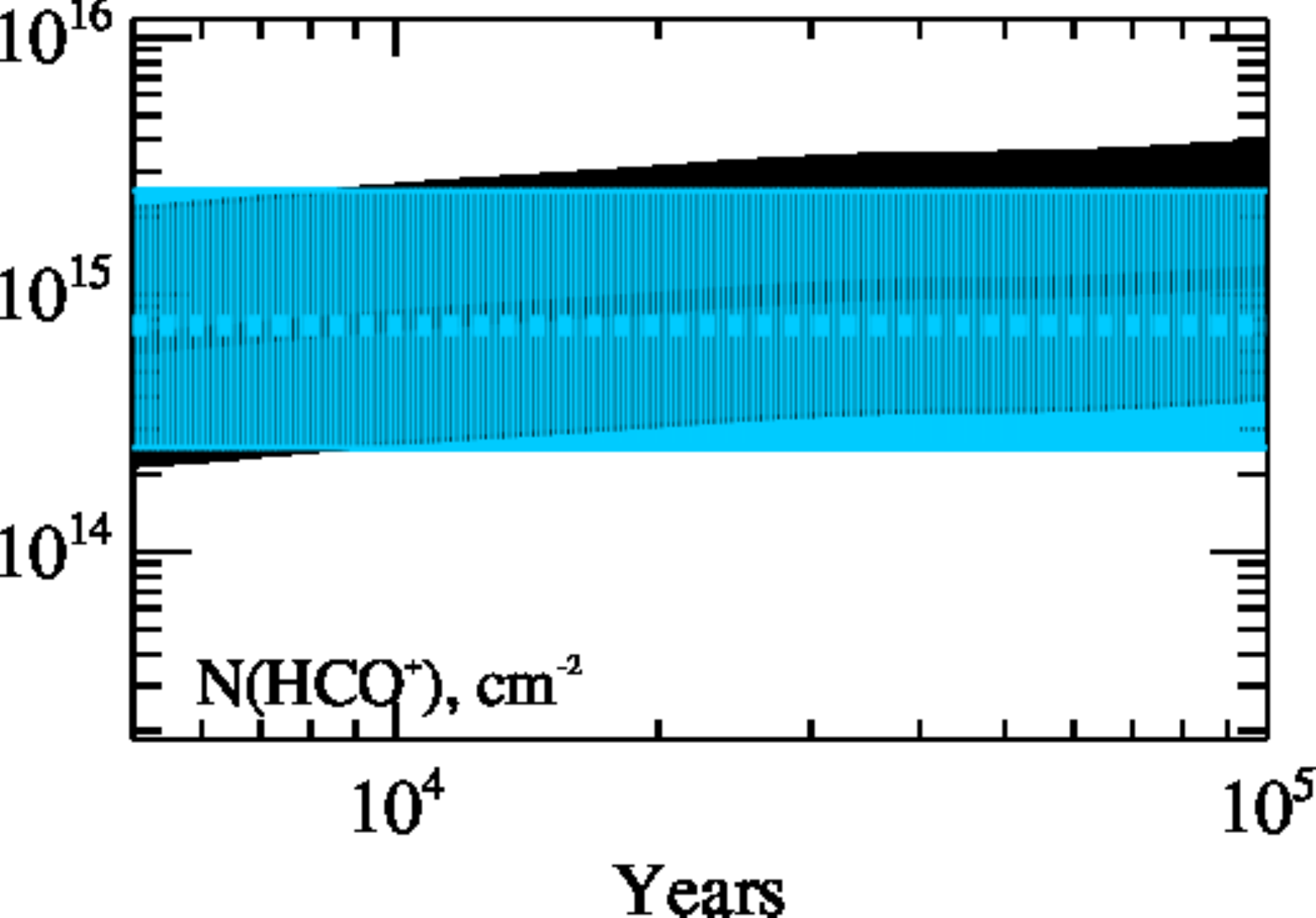}
\includegraphics[width=0.4\textwidth]{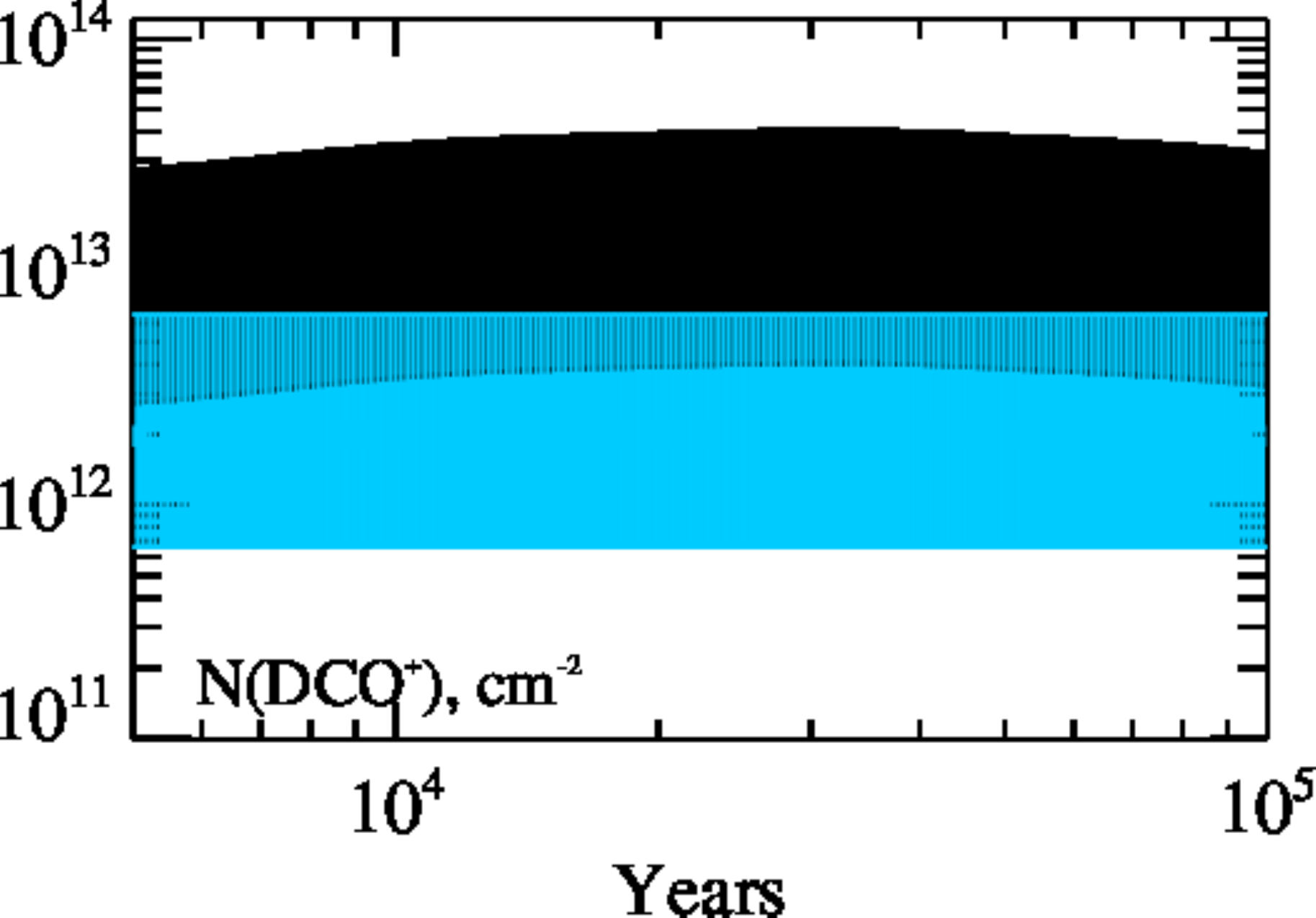}\\
\includegraphics[width=0.4\textwidth]{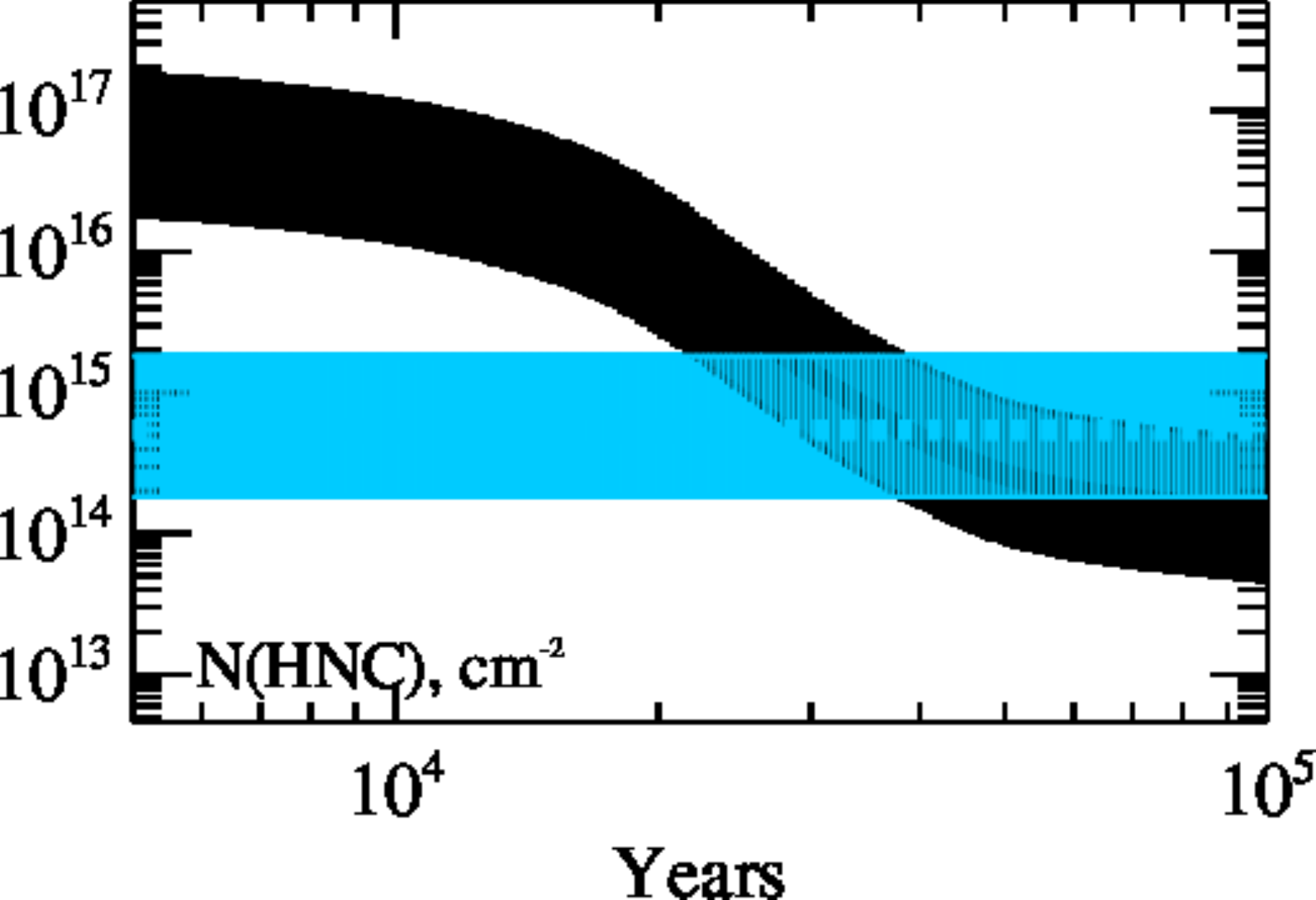}
\includegraphics[width=0.4\textwidth]{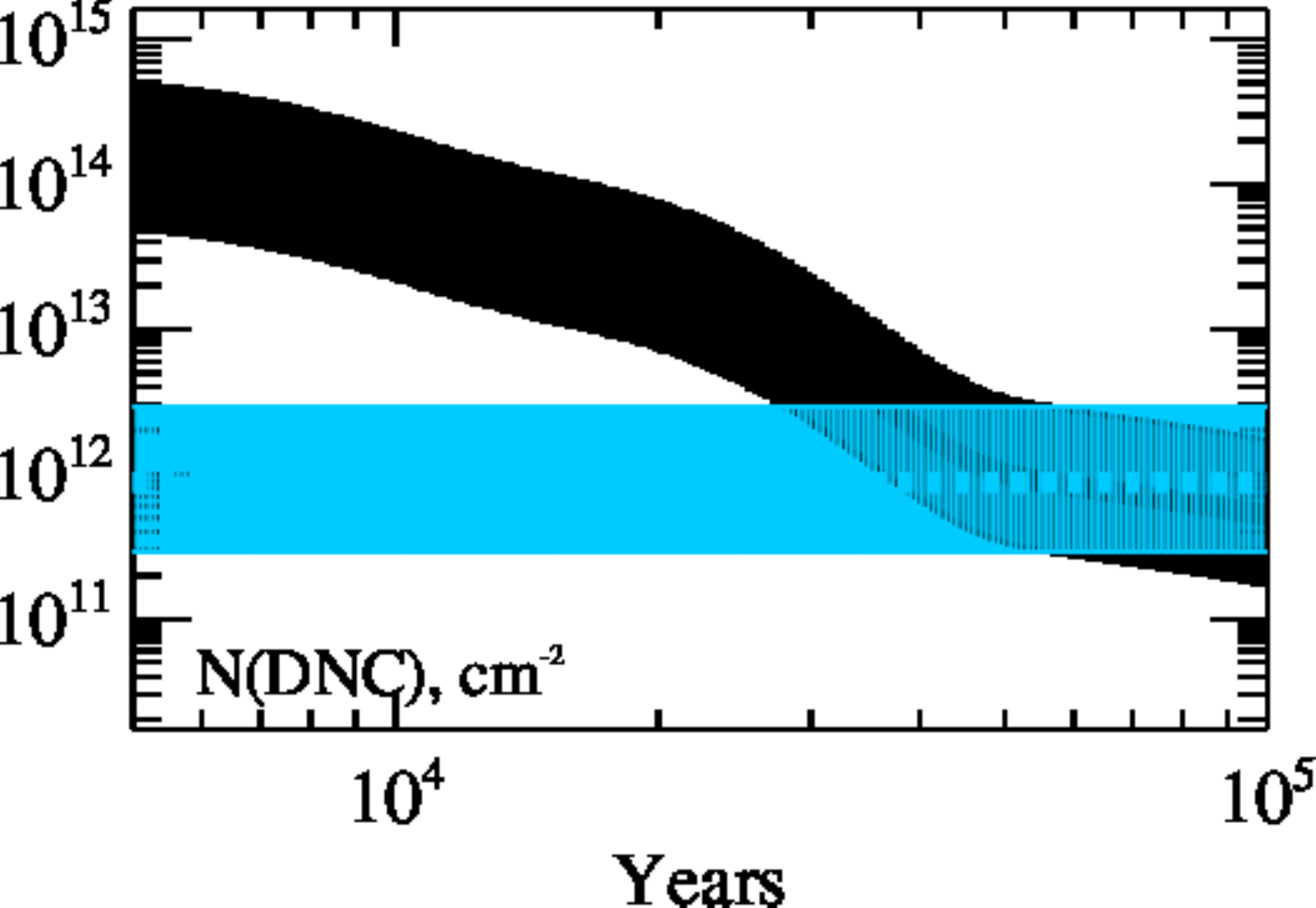}\\
\includegraphics[width=0.4\textwidth]{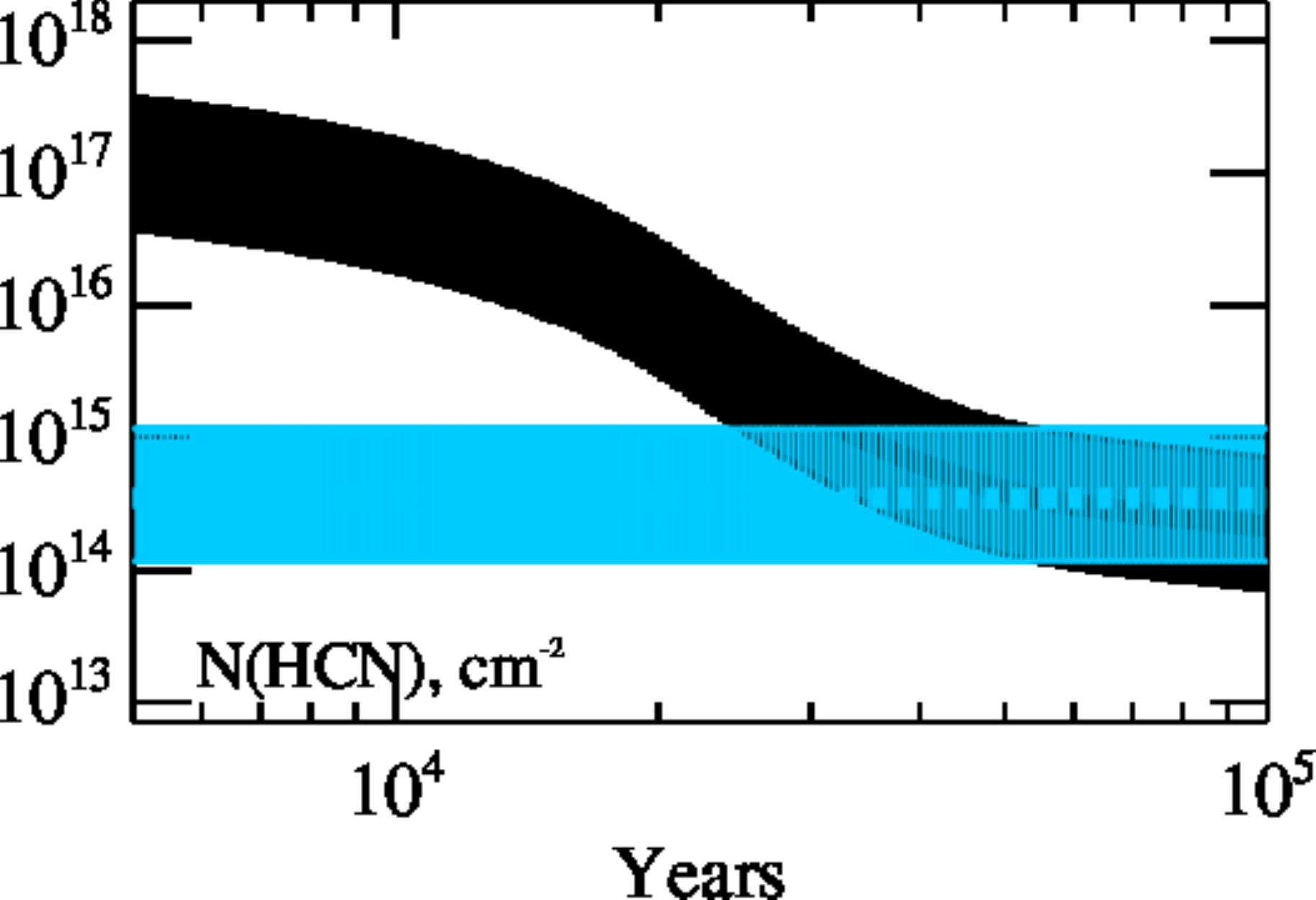}
\includegraphics[width=0.4\textwidth]{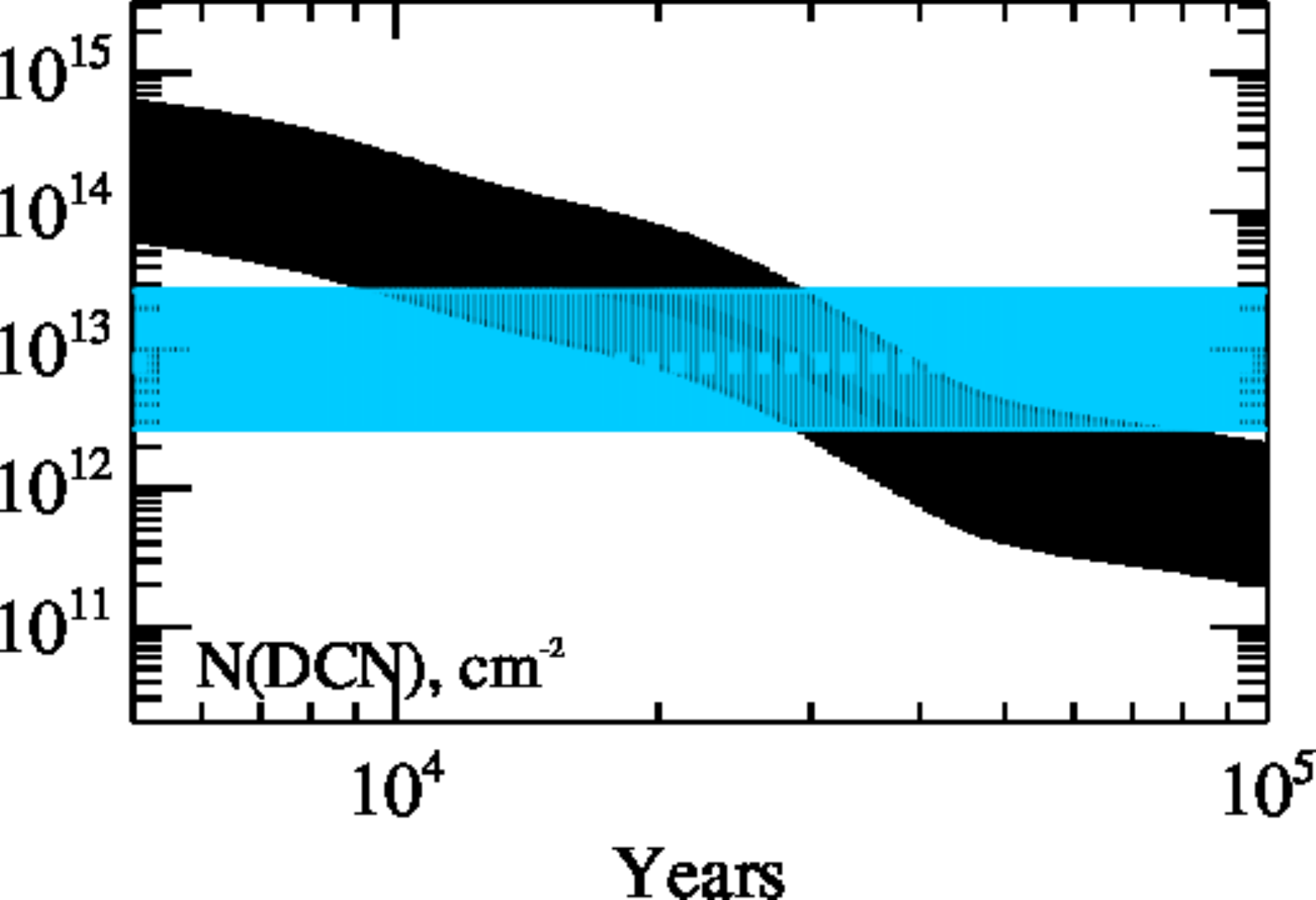}\\
\includegraphics[width=0.4\textwidth]{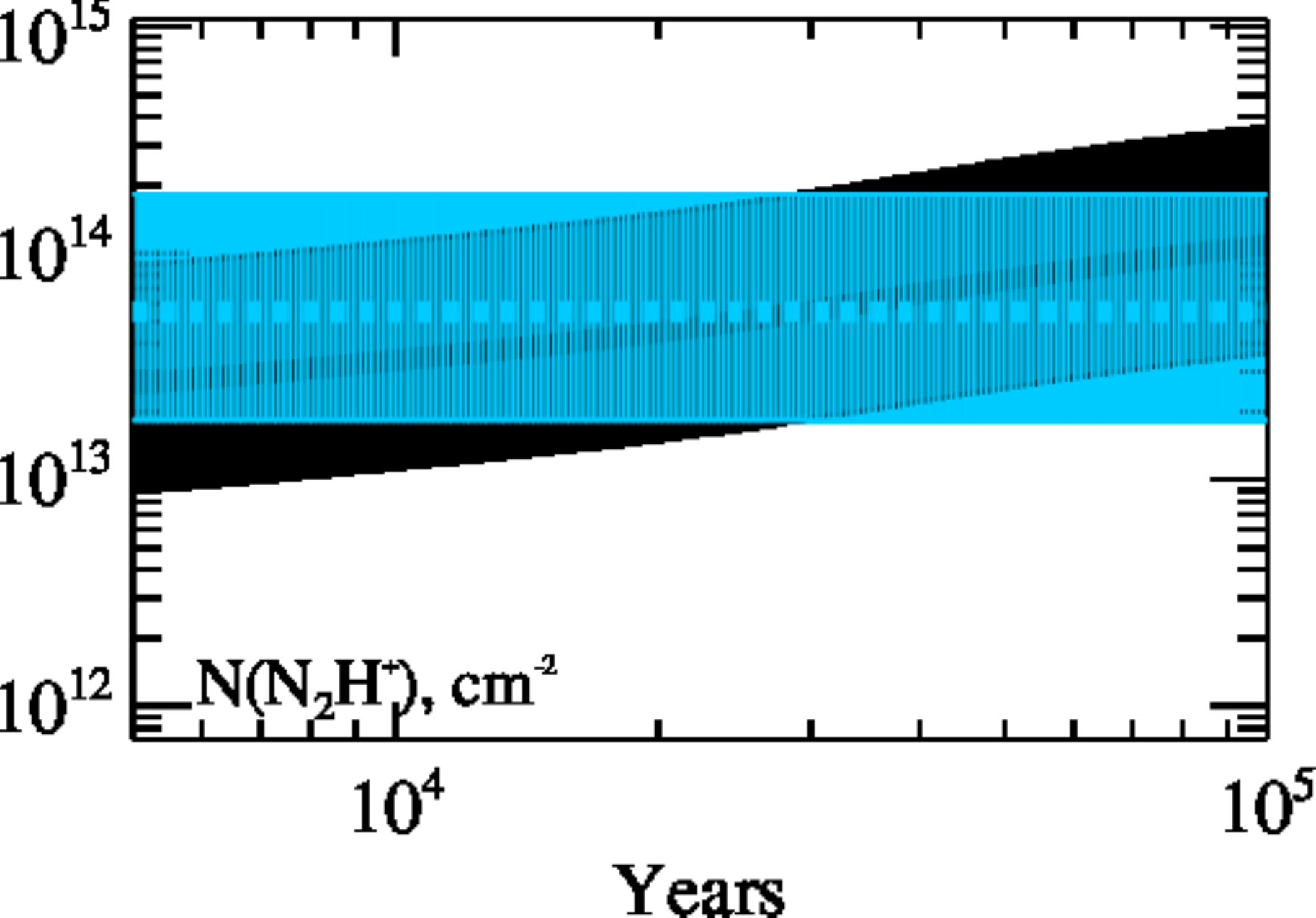}
\includegraphics[width=0.4\textwidth]{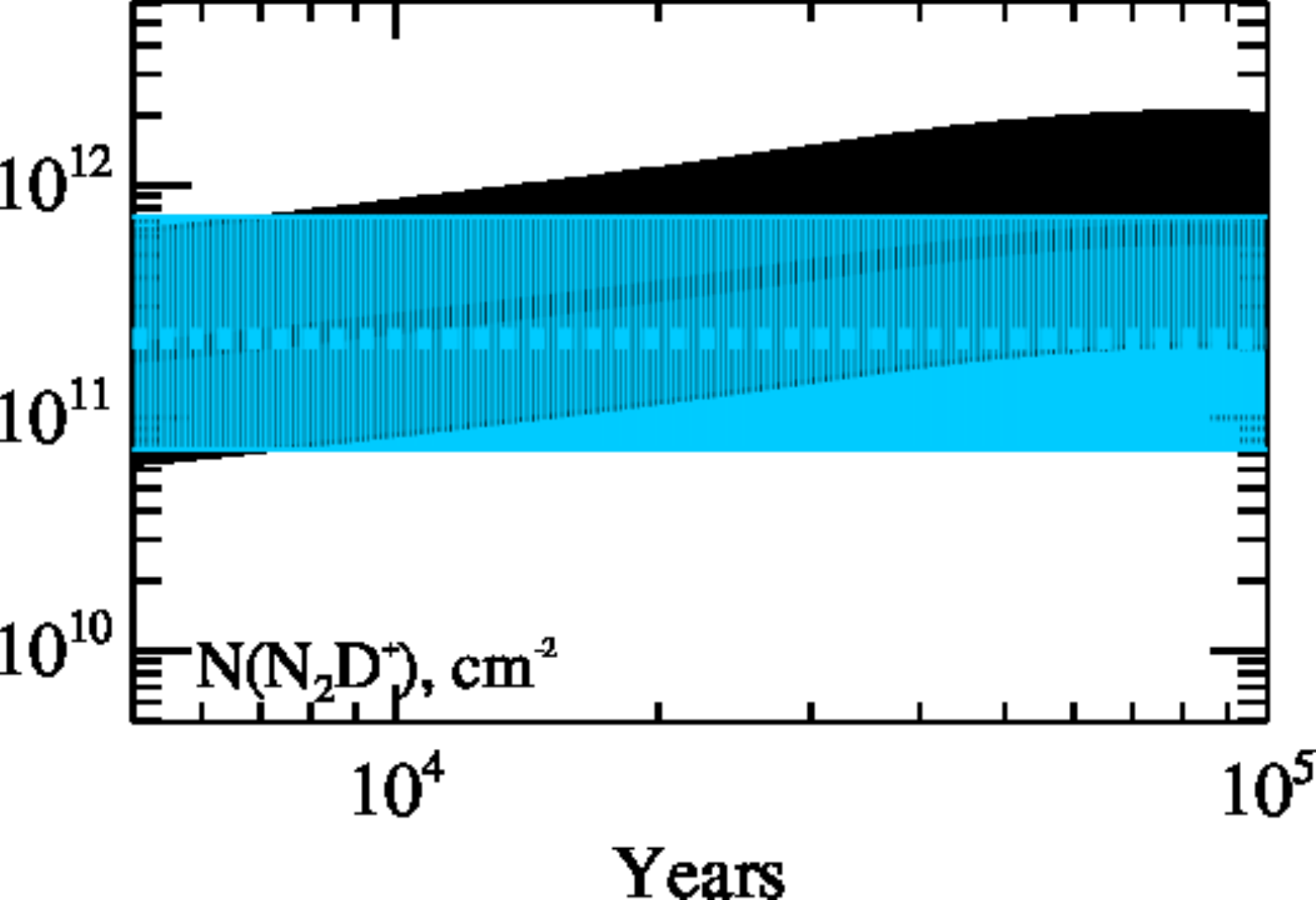}\\
\caption{Observed and modeled column densities in cm$^{-2}$ in the HMC stage. The observed values are shown in blue, the modeled values in black. The error bars are indicated by the vertical marks. Molecules are labeled in the plots.}
\label{fig:nx_stage3_31}
\end{figure*}

% Stage4
%\clearpage
\newpage
\begin{figure*}
\includegraphics[width=0.4\textwidth]{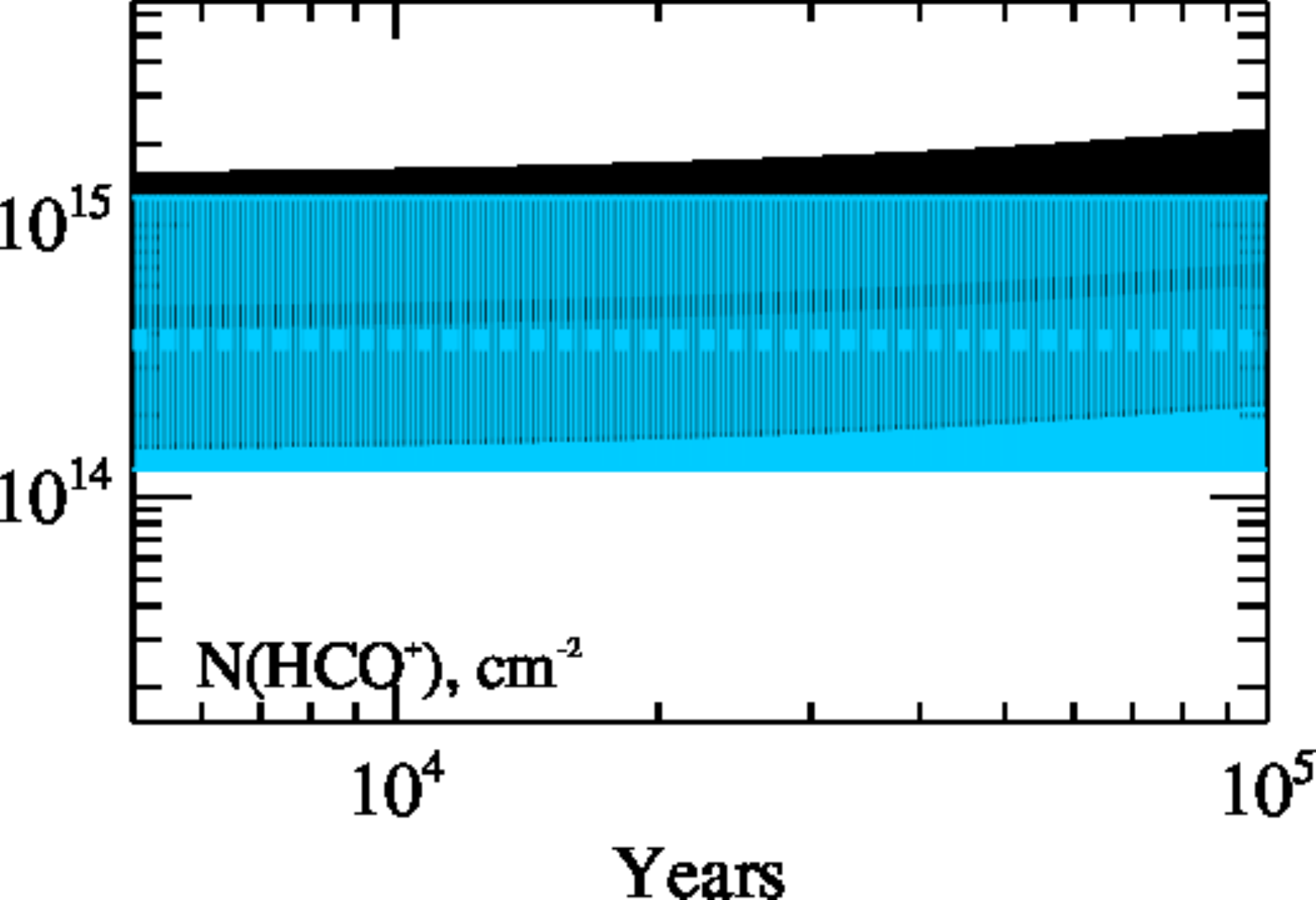}
\includegraphics[width=0.4\textwidth]{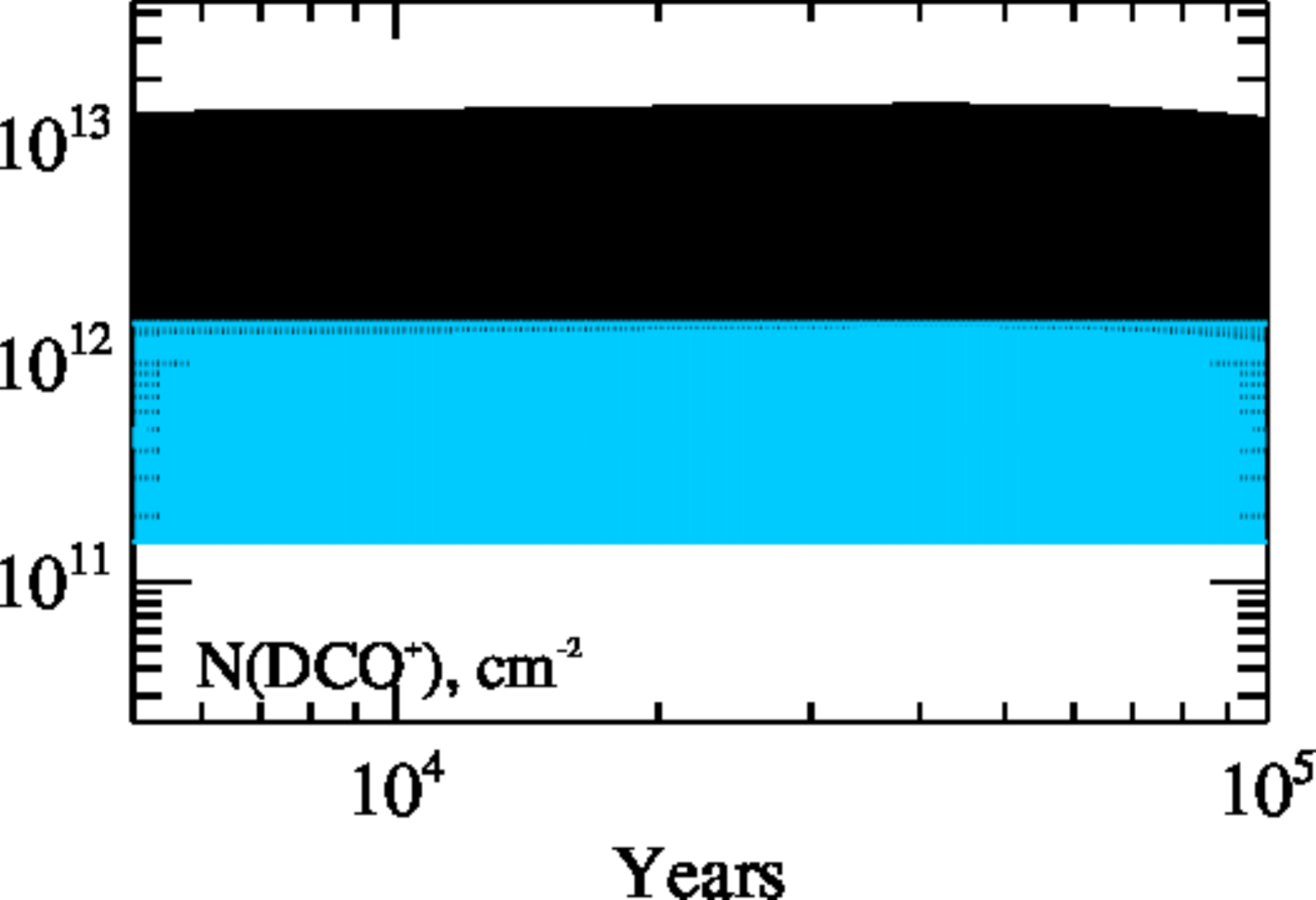}\\
\includegraphics[width=0.4\textwidth]{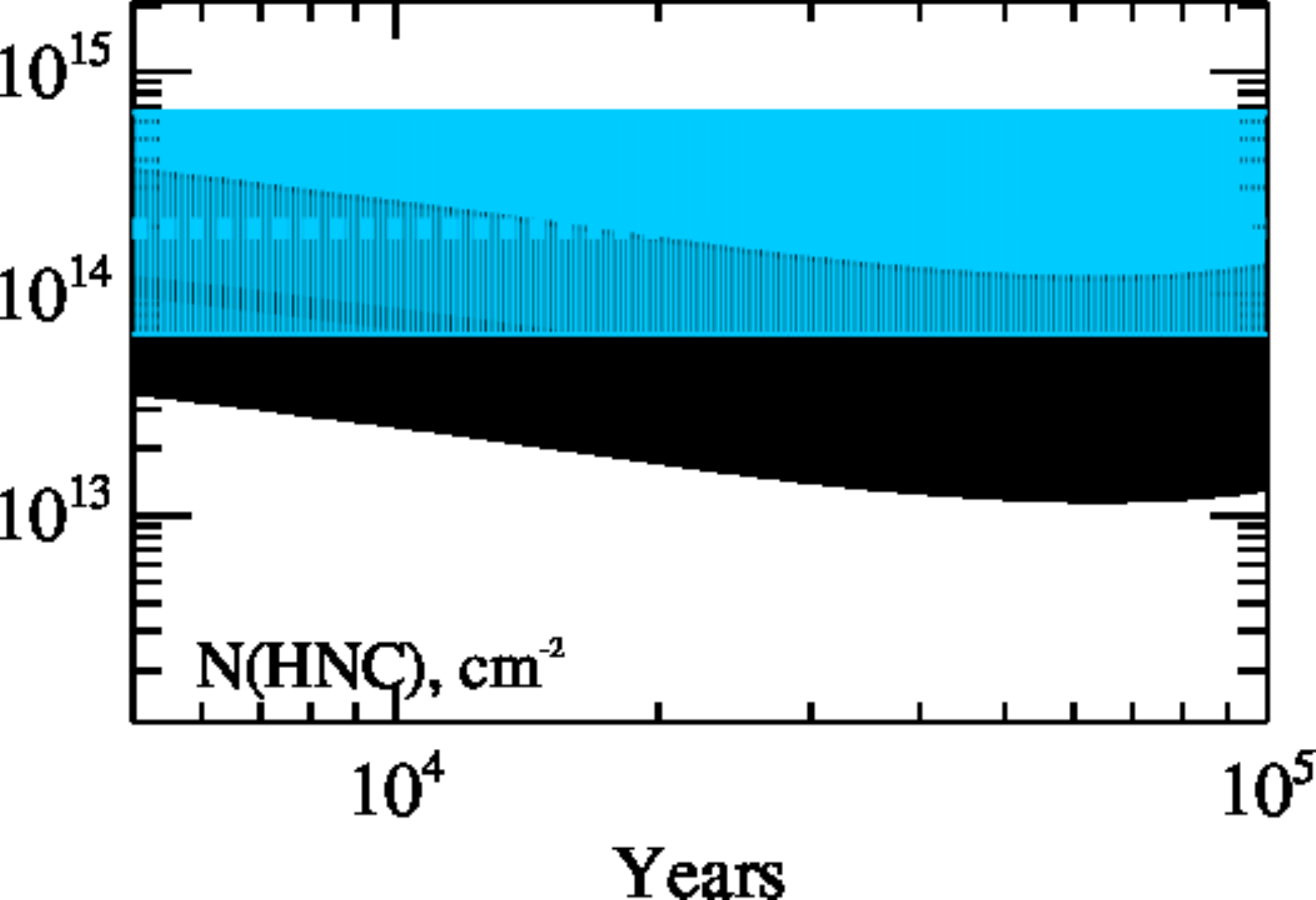}
\includegraphics[width=0.4\textwidth]{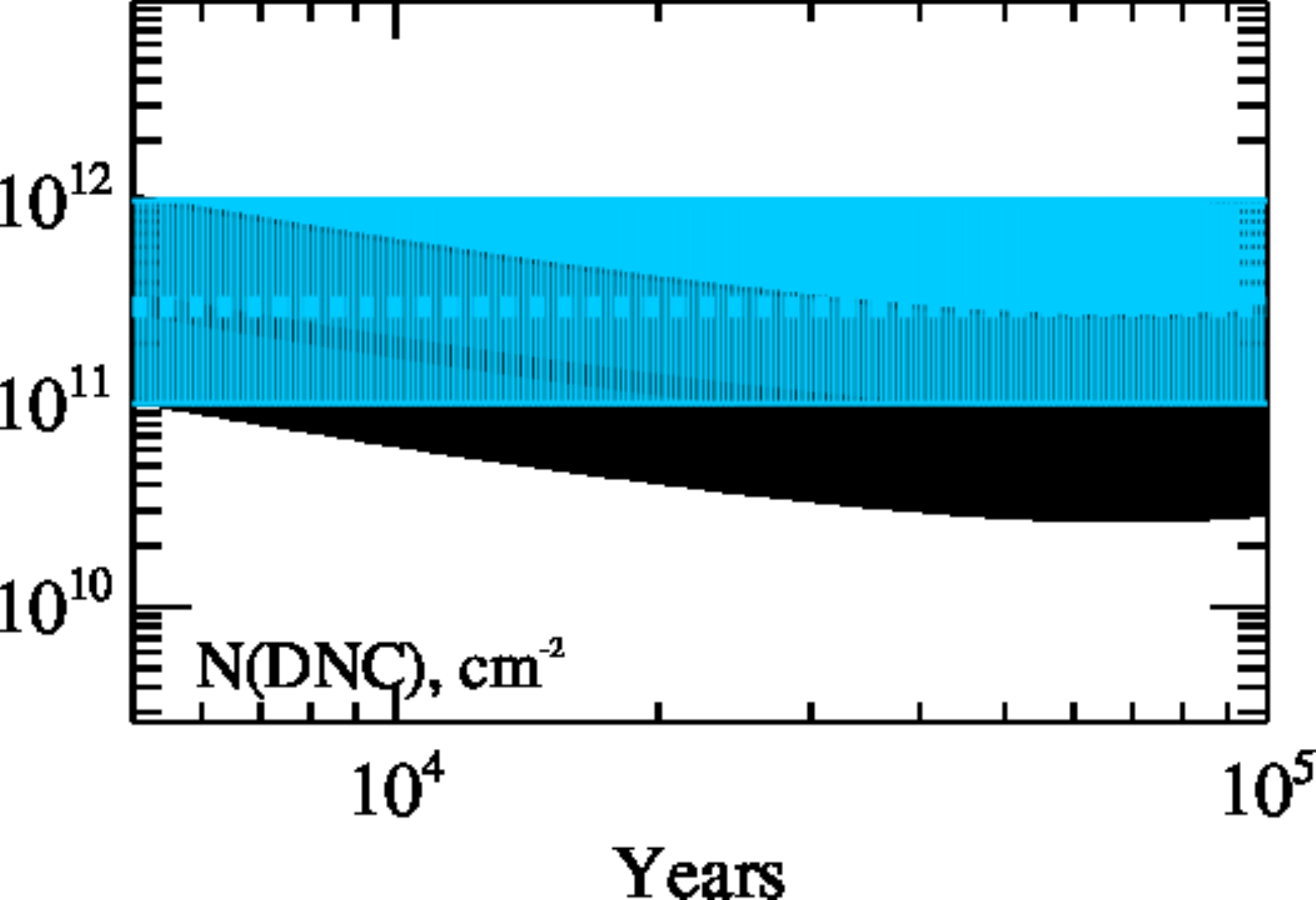}\\
\includegraphics[width=0.4\textwidth]{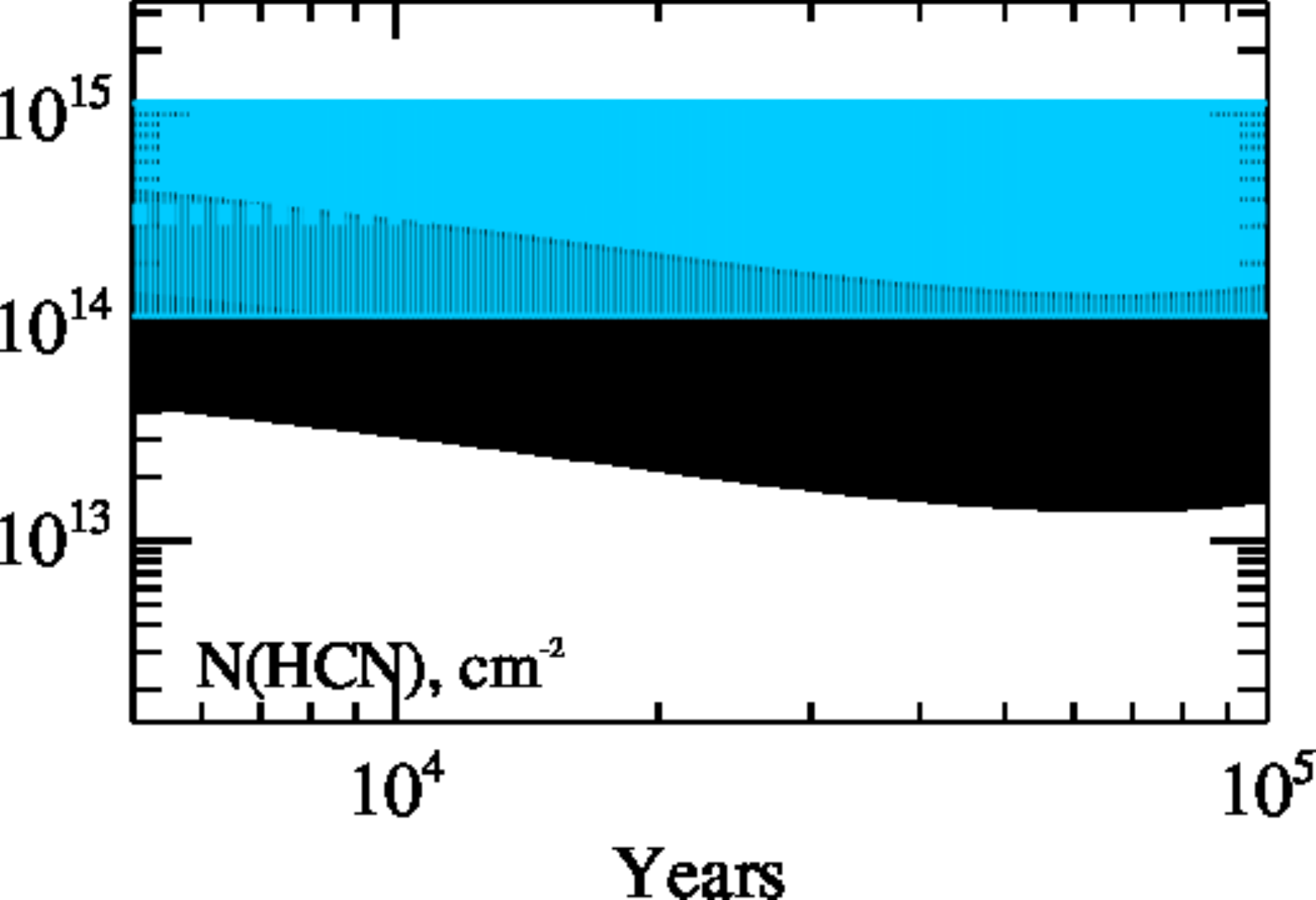}
\includegraphics[width=0.4\textwidth]{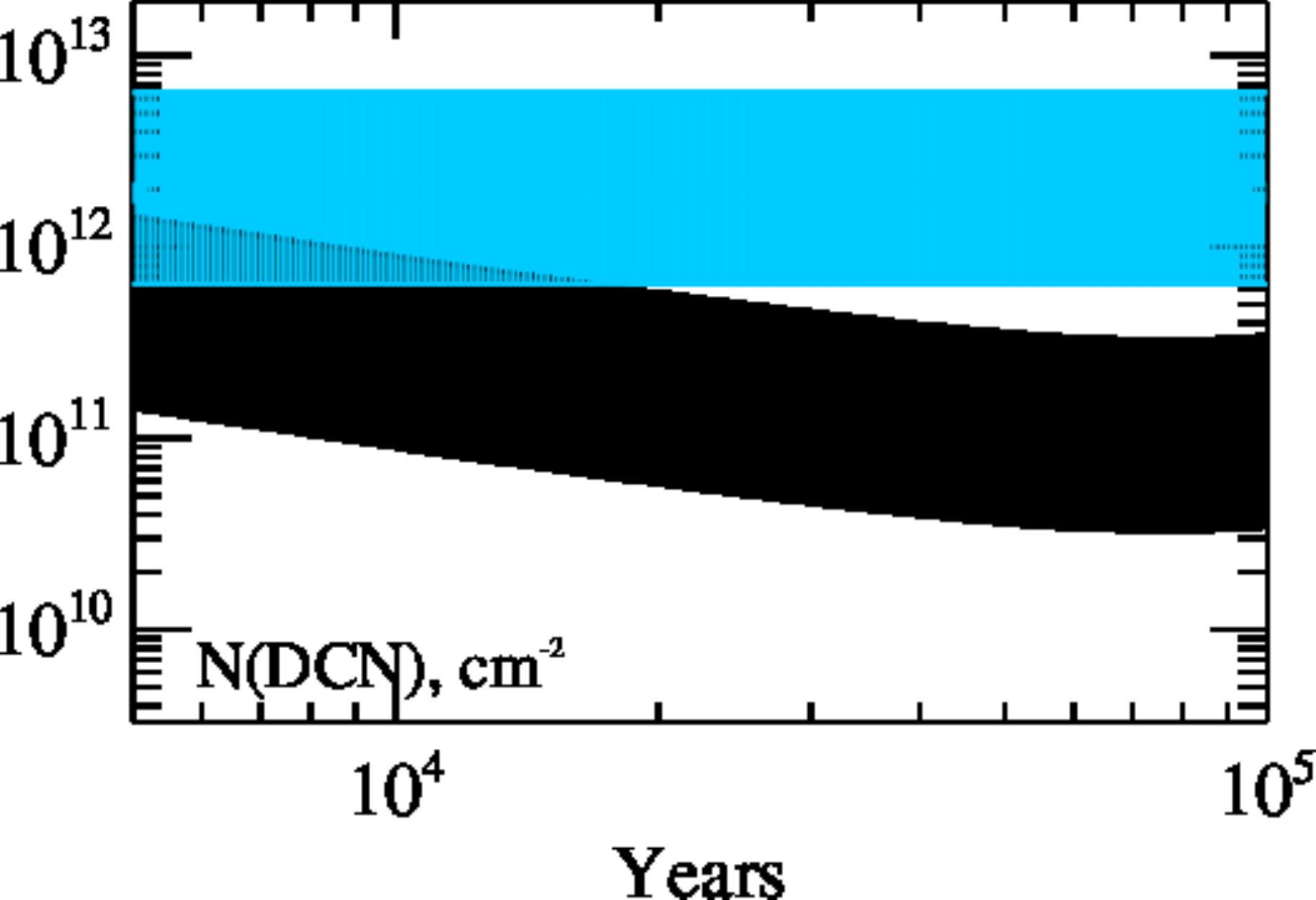}\\
\includegraphics[width=0.4\textwidth]{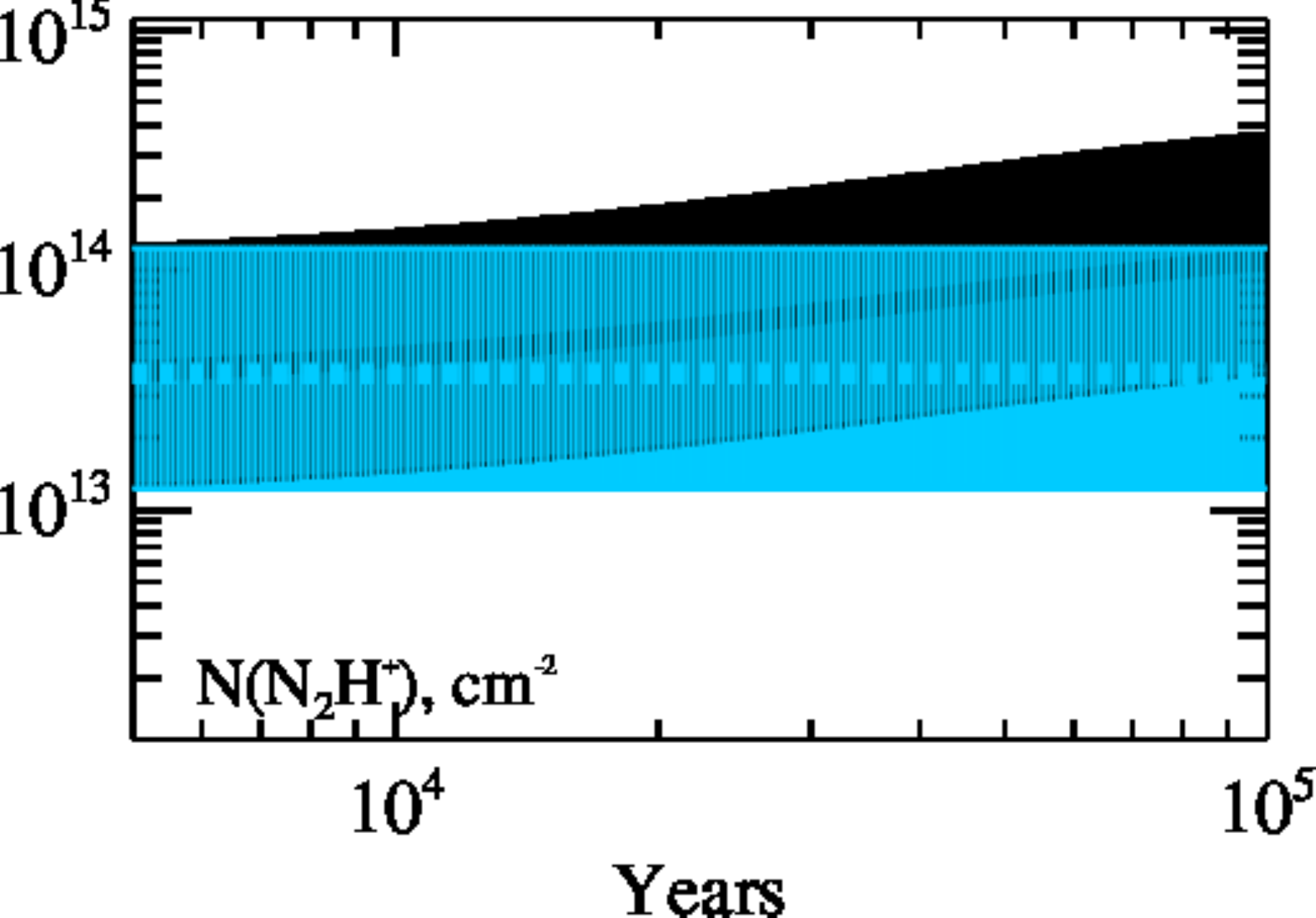}
\includegraphics[width=0.4\textwidth]{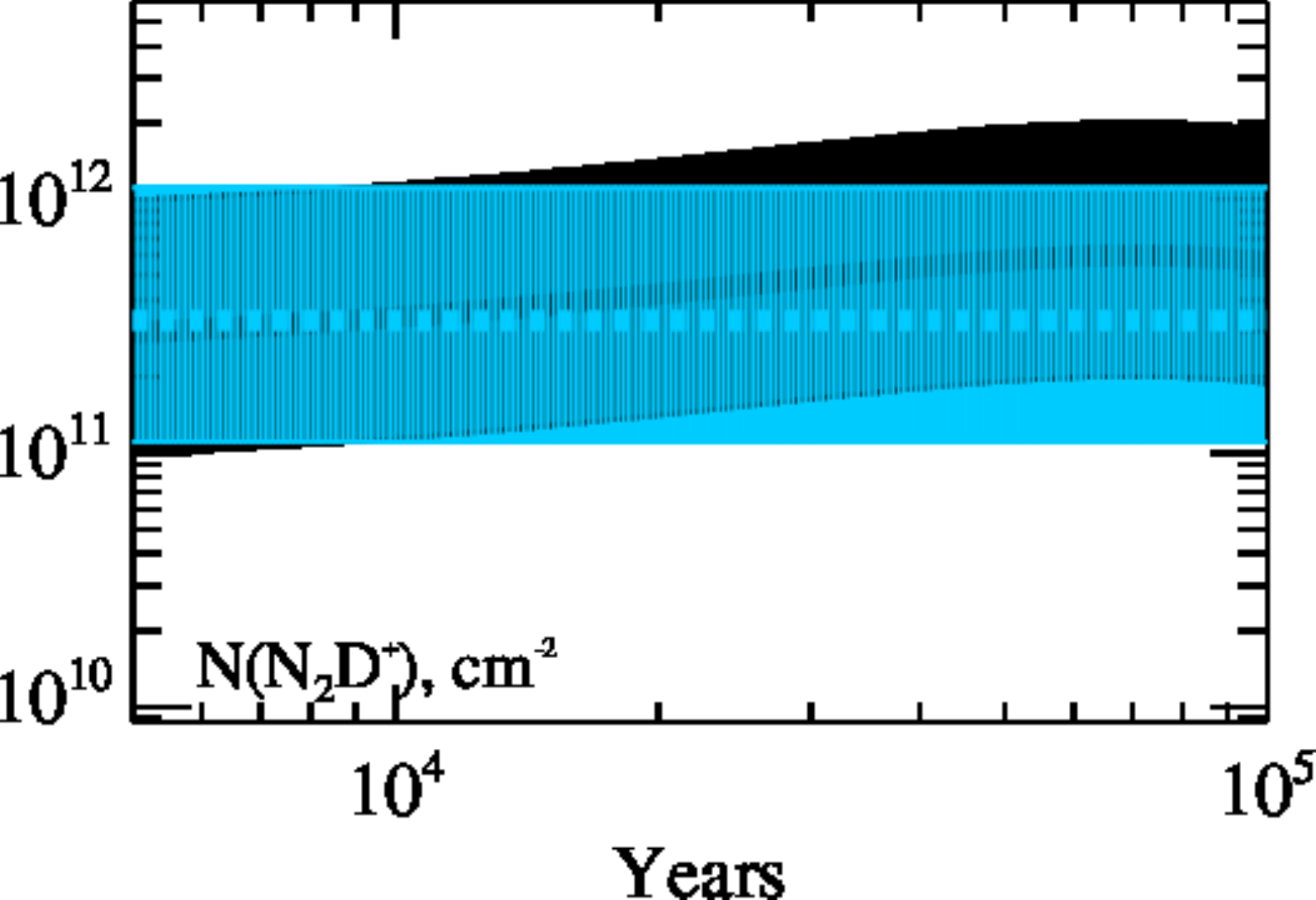}\\
\caption{Observed and modeled column densities in cm$^{-2}$ in the UCH{\sc ii} stage. The observed values are shown in blue, the modeled values in black. The error bars are indicated by the vertical marks. Molecules are labeled in the plots.}
\label{fig:nx_stage4_31}
\end{figure*}

\end{document}